\documentclass[toc]{PoS}
\usepackage{url}
\usepackage{breakurl}
\usepackage{textgreek}
\newcommand{\lumi}[2]{${#1} \times 10^{#2}~$cm$^{-2}$s$^{-1}$}
\def\ifb{fb$^{-1}$}
\def\iab{ab$^{-1}$}

\def \gsim{\mathrel{\vcenter
     {\hbox{$>$}\nointerlineskip\hbox{$\sim$}}}}
\def \lsim{\mathrel{\vcenter
     {\hbox{$<$}\nointerlineskip\hbox{$\sim$}}}}

\newcommand{\beq}{\begin{equation}}
\newcommand{\eeq}{\end{equation}}
\newcommand{\beqn}{\begin{eqnarray}}
\newcommand{\eeqn}{\end{eqnarray}}

\newcommand{\beqa}{\begin{eqnarray}}
\newcommand{\eeqa}{\end{eqnarray}}

\renewcommand\arraystretch{1.2}

\newcommand{\be}{\begin{equation}}
\newcommand{\ee}{\end{equation}}
\newcommand{\ba}{\begin{eqnarray}}
\newcommand{\ea}{\end{eqnarray}}
\newcommand{\bi}{\begin{itemize}}
\newcommand{\ei}{\end{itemize}}
\newcommand{\ben}{\begin{enumerate}}
\newcommand{\een}{\end{enumerate}}

\title{TASI Lectures on Future Colliders}

\ShortTitle{Future Colliders}

\author{\speaker{Michelangelo L. Mangano} 
        \\
        TH Department, CERN, 1211 Geneva, Switzerland\\
        E-mail: \email{michelangelo.mangano@cern.ch}}

\abstract{These lectures review the main motivations for future
  high-energy colliders, focusing on the understanding of electroweak
  symmetry breaking and on the search for physics beyond the Standard
  Model. The open questions and the challenges are common to all
  future projects; for concreteness, I will use studies of the
  potential of $\rm e^+e^-$ and pp circular colliders to provide
  examples of the anticipated physics reach.
  
  \vskip 12cm
\noindent  \hskip -0.8cm CERN-TH-2019-073}

\FullConference{Theoretical Advanced Study Institute Summer School
  2018 'Theory in an Era of Data'(TASI2018)\\ 
		4 - 29 June, 2018\\
		Boulder, Colorado}

\begin{document}

\section{Introduction}
Several projects for  future
high-energy particle colliders are under consideration in
various regions worldwide, to complement and extend the physics reach
of CERN's Large Hadron Collider (LHC). These include:
\begin{itemize}
  \item the $\rm e^+e^-$ International Linear Collider
    (ILC~\cite{Behnke:2013xla,Evans:2017rvt,Barklow:2017suo,Fujii:2017vwa}),
    exploiting an established acceleration technology based on
    superconducting (SC) radio frequency (RF) cavities, operating initially
    at $\sqrt{s}=250$~GeV, with an upgrade path in principle up to 1000~GeV.
    \item the $\rm e^+e^-$ Compact Linear Collider
      (CLIC~\cite{Aicheler:1500095,CLIC:2016zwp,Charles:2018vfv,deBlas:2018mhx}),
      relying on the RF field produced by a drive beam, to induce the
      accelerating gradients required to push the center of mass
      energy from an initial value of $\sqrt{s}=380$~GeV, up to 1.5
      and eventually 3~TeV.
    \item the Future Circular Collider
      (FCC~\cite{Mangano:2651294,Benedikt:2651299,Benedikt:2651300,Zimmermann:2651305})
      and the CEPC/SppC~\cite{CEPCStudyGroup:2018rmc,CEPCStudyGroup:2018ghi}, two
      similar projects promoted by CERN and China respectively,
      envisioning a staged facility, enabled by a 100~km circular ring,
      designed to deliver $\rm e^+e^-$ collisions at energies in the range
      $M_Z < \sqrt{s} < 365$~GeV, pp collisions up to
      $\sqrt{S}\sim 100$~TeV, ep collisions at $\sqrt{S}=3.5$~TeV, as
      well as heavy ion collisions.
\end{itemize}
This variety of layouts (circular or linear), beam types (electrons or
protons) and energies, reflects slightly different priorities for the
physics targets and observables, as well as a different judgement on
the overall balance between physics returns, technological challenges
and feasibility, time scales for completion and exploitation, and
financial/political realities.

If approved today, the $\rm e^+e^-$ projects in this list could in
principle begin delivering physics results at some point during the
decade 2030-40, and operate for the following 15-25 years, depending
on the technology and upgrade path. Beyond this, but with an
unspecified time scale, ideas are on the table for a possible
subsequent generation of even more ambitious lepton-collider projects,
which I only mention here: linear electron accelerators based on
multi-GeV/m gradient technologies like plasma wake fields or
lasers~\cite{ALEGRO:2019alc}, and muon circular
colliders~\cite{MuonCollider,Antonelli:2015nla}.

The FCC and SppC proton colliders would face a preparatory phase
longer than the $\rm e^+e^-$ colliders, mostly because of the R\&D period
required to produce reliable and affordable SC bending magnets with
the  16\,T magnetic field needed to keep 50~TeV protons in orbit
in the 100~km ring (A 12\,T option, based on high-T$_c$ SCs, is
considered for SppC, but is itself far from being established).

No matter what the energy or the technology, all these projects share
common goals, driven by the need to clarify several outstanding open
issues in particle physics. This need singles out the next generation
of colliders beyond the LHC as unique and indispensable exploratory
tools to continue driving the progress in our understanding of nature.

In these lectures, I will review the main motivations for future
high-energy colliders and discuss their physics potential. I will
mostly cover topics such as Higgs and physics beyond the Standard
Model (BSM), and, while I will recall some basic theoretical
background, I will give for granted that students know those topics
from their studies, or from other lectures in the TASI program. In
particular, in 2018 students have been exposed, among others, to
excellent lectures on the Higgs boson~\cite{tasi7}), supersymmetry and
dark matter~\cite{tasi4}, QCD at colliders~\cite{tasi3}, effective
field theory~\cite{Cohen:2019wxr} and
flavour~\cite{tasi8}. I refer to these, and to the great lectures on
LHC physics by our host Tilman Plehn~\cite{Plehn:2015dqa}, for the
necessary background material.

I will illustrate the value of the physics reach through concrete
examples of the FCC physics potential.  No attempt is made here to
compare the FCC against the other projects, as the point of these
lectures is not to promote one project over another: I choose here the
FCC since it is the project that I know better, and the one that, in
terms of breadth, variety and physics performance, best illustrates
how ambitious the targets of a future collider can be.

I will include also a few exercises here and there. They are meant to
stimulate your thinking, they are simple, do not necessarily require
calculations and are mostly for a qualitative discussion.
But if you take them seriously, some of them could be
the seed for interesting work!

\section{Where we stand}
In almost ten years of studies at the LHC, the picture of the particle
physics landscape has greatly evolved. The legacy of this first phase
of the LHC physics programme can be briefly summarised in three points: a)
the discovery of the Higgs boson, and the start of a new phase of
detailed studies of its properties, aimed at revealing the deep origin
of electroweak (EW) symmetry breaking; b) the indication that signals
of new physics around the TeV scale are, at best, elusive; c) the
rapid advance of theoretical calculations, whose constant progress and
reliability underline the key role of ever improving
precision measurements, from the Higgs to the flavour sectors. Last
but not least, the LHC success has been made possible by the
extraordinary achievements of the accelerator and of the detectors,
whose performance is exceeding all expectations, supporting the
confidence in the ability of the next generation of colliders to
achieve what they promise.

\subsection{The puzzling origin of the Higgs field}
  The years that preceded the discovery of the Higgs boson have been
characterized by a general strong belief in its existence, justified
by the success of the Standard Model (SM), and by the confidence that
EW symmetry breaking (EWSB) is indeed driven by the basic dynamics of the
Higgs mechanism, as described in the SM. Starting from this
assumption, the theoretical speculations focused on identifying
possible solutions to the hierarchy problem, namely the extreme fine
tuning required to achieve the decoupling of the Higgs and EW mass
scale from the phenomena expected to emerge at much higher energy
scales, up to the Planck scale. These speculations led to the
consideration of several possible scenarios of new physics, from
supersymmetry to large extra dimensions, which would provide {\it
  natural} solutions to the hierarchy problem by introducing new
degrees of freedom, new symmetries, or new dynamics at the TeV scale. The opportunity
to combine the solution of the hierarchy problem with the
understanding of experimental facts such as the existence of dark
matter, or of the features of flavour phenomena, gave further impetus
to these theoretical efforts, and to the many experimental studies
dedicated to the search for BSM manifestations.

The conceptual simplicity and appeal of several of these scenarios,
justified optimism that their concrete manifestations would appear
``behind the corner''. After all, the SM itself was born as the
simplest possible model in which to embed an elegant explanation (the
Higgs mechanism) to the problem of justifying the mass of the weak
force carrier and parity non-conservation, and this simple framework
DID work! Why shouldn't the next step beyond the SM be accomplished by
similarly elegant and ``simple'' proposals?
 
Lack of evidence of new physics at the TeV scale, made even more
compelling by the hundreds of inconclusive searches scrupolously
carried out by the LHC experiments, has not removed, however, the need
to continue addressing the original motivations for a BSM extension of
the SM. If anything, this has
made the open issues even more intriguing, and challenging. But, while
the efforts to review the underlying perspective on the hierarchy
problem and naturalness continue, we should focus on a perhaps even
more basic question: who ordered the Higgs? Where does the famous
``mexican hat'' Higgs potential come from? This appears like a
pointless trivial question. A sort of mexican-hat potential must be there, it's
a necessary ingredient in the realization of EWSB,
without it we would not be here discussing it. But what is its true
origin? 

To understand the value of this question, it is useful to compare the
dynamics of the Higgs field with that of electromagnetism (EM), or of any
of the other known fundamental forces in nature. All properties of EM
arise from a simple principle, the gauge principle. Coulomb's law has
no free parameter, except the overall scale of the electric charge,
absorbed in the definition of the charge unit. The quantization of the
charge may have a deep origin in quantum mechanical properties such as
anomaly cancellation, or in the algebraic structure of the
representations of larger gauge groups in which EM is embedded.
The sign of the
electric force, positive or negative depending on the relative sign of
the interacting charges, follows from the spin-1 nature of the
photon. The $1/r^2$ behaviour follows from the Gauss theorem, or
charge conservation, or gauge invariance, depending on how we want to
phrase it. We do not know {\it why} nature has chosen gauge symmetry as
a guiding principle, although this appears as an unavoidable
consequence of the existence of interactions mediated by massless
particles, which are the basis of the long-range forces needed to
sustain our existence. But gauge symmetry appears everywhere,
e.g. in the zero modes of a string theory, or as a result of
compactification in Kaluza-Klein gravitational theories. Gauge
symmetry is therefore intimately related to possible deeper properties
of nature.

On the contrary, nothing is fundamental
in the Higgs potential, there doesn't seem to be any fundamental
symmetry or underlying principle that controls its structure.
To fix the notation for further use, we shall
write this potential as
\begin{equation} \label{eq:Hpotential}
  V(\Phi)=-\frac{\mu^2}{2} \vert
  \Phi\vert^2 +\frac{\lambda}{4} \vert\Phi\vert^4 \; ,
\end{equation}
where $\Phi$ is the SU(2) doublet scalar field and $\phi=v+H$ is the real
part of the neutral component ($v$ being its vacuum expectation
value).
The condition of minimum of the
potential ($\partial V/\partial\phi\vert_{\phi=v}=0$) and the Higgs mass
definition ($m_H^2=\partial^2 V/\partial \phi^2$) lead to the relations
$\mu^2=m_H^2/2$ and $\lambda=m_H^2/2v^2$, expressed in terms of the
measured Higgs mass and of Fermi's coupling $v=(\sqrt{2}
G_F)^{-1/2}\sim 246$~GeV.
The sign and value of the parameters
$\mu^2$ and $\lambda$ are a priori arbitrary. A negative sign in front
of the quadratic term is required to achieve symmetry breaking, but is
not required by any symmetry. The positive sign of $\lambda$ is
necessary for the stability of the potential at large $\phi$ but,
again, is not dictated by anything: it could be negative, and the
potential could be stabilized at larger $\phi$ values by higher-order
terms. Even the functional form is not fundamental: the underlying
gauge symmetry only requires the potential to depend on $\vert \Phi
\vert^2$, and the quartic form could simply represent the leading
terms in the power expansion of a more complex functional dependence of the
potential.

The SM Higgs potential has therefore the features of an effective
potential, as in other natural phenomena. Spontaneous symmetry
breaking, in fact, is not a process unique to the EW theory. There are
many other examples in nature where the potential energy of a system
is described by a mexican-hat functional form, leading to some order
parameter acquiring a non-zero expectation value. A well know case is
that of superconductivity. The Landau-Ginzburg theory (L-G)
\cite{Ginzburg:1950sr} is a phenomenological model that describes the
macroscopic behaviour of type-1 superconductors.  This model contains
a scalar field $\phi$, with free energy given by
\begin{equation}
F = \frac{1}{2 m} |(-i \hbar \nabla - 2 e A) \phi|^2+\alpha |\phi|^2 + \frac{\beta}{2} |\phi|^4 +...
\end{equation}
where the ellipsis denote additional terms not relevant to this
discussion.  This equation describes a scalar field of charge $Q=2$
with a mass and a quartic interaction.  These parameters are
temperature dependent.  At high temperature the mass-squared is
positive and the scalar field has a vanishing expectation value
throughout the superconductor.  However, below the critical
temperature $T_c$ the mass-squared is negative, leading to a
non-vanishing expectation value of $\phi$ throughout the
superconductor.  This expectation value essentially generates a mass
for the photon within the superconductor, leading to the basic
phenomenology of superconductivity.

Similarly to the Higgs mechanism, the L-G theory is a phenomenological
model, which offers no explanation as to the fundamental origin of the
parameters of the model. It also does not explain the fundamental
origin of the scalar field itself. Ultimately, these questions were
answered by Bardeen, Cooper, and Schrieffer, in the celebrated BCS
theory of superconductivity~\cite{Bardeen:1957mv}.  The scalar field
is a composite of electrons, the Cooper pairs, and its mass relates to
the fundamental microscopic parameters describing the material. The
$\alpha$ and $\beta$ parameters can therefore be calculated from first
principles, starting from the underlying dynamics, namely the
electromagnetic interactions inside the metal, subject to the rules of
quantum mechanics and to the phonon interactions within the solid.

The analogy of the Higgs with the L-G model is striking, with the
exception that the model is relativistically invariant and the gauge
forces non-Abelian.  Unlike with superconductivity, currently neither
the fundamental origin of the SM scalar field nor the origin of the
mass and self-interaction parameters in the Higgs scalar potential are
known.  The SM itself does not provide a dynamical framework that
allows us to predict the shape of the Higgs potential. This {\em must}
follow from a theory beyond the SM. What are possible scenarios? An
obvious option is a mechanism analogous to BCS: the Higgs could be the
bound state of a pair of fermions, strongly coupled by a new
fundamental (possibly gauge-) interaction, whose dynamics determines
the properties of the Higgs field. Another well known framework is
supersymmetry: elementary scalar fields appear as a result of the
symmetry itself, and the Higgs potential is likewise determined by the
symmetry. For example, in the minimal supersymmetric SM (MSSM) the
Higgs self-coupling is not a free parameter, but is related to the
weak gauge coupling. The parameters that characterize supersymmetry
breaking modify the supersymmetric predictions for the Higgs
interactions, and ultimately the dynamics of EW symmetry breaking
would be calculable from the fundamental properties of supersymmetry breaking.

Now that the Higgs boson has been discovered,
and the basic phenomenology of EWSB established,
the next stage of exploration for any future high energy physics
programme is to determine their microscopic origins. And the obvious
place where one should look for hints is the Higgs boson itself,
exploring in detail all of its properties. As of today, we are not
aware of any other experimental context except particle colliders,
where the question of the origin of the Higgs potential can be studied.

Another aspect of the Higgs dynamics makes it appear very different
than EM or other gauge forces. This is the (lack of) decoupling
between short- and long-distance interactions, or between low- and
high-energy modes. The Gauss theorem, or gauge invariance, teaches us
that the charge of an electron can be obtained by measuring the
integral of the flux of its field through any closed surface
surrounding the electron. Using the surface of a sphere of small or
large radius will give the same value. Possible additional unknown
interactions of the electron at very short distances do not modify the
charge we measure at large distances. In the case of the Higgs
potential, its parameters $\mu^2$ and $\lambda$ receive instead
dominant contributions from any short-distance Higgs
interaction. Self-energy loop diagrams for the Higgs boson shift the
Higgs mass squared by amounts proportional to the mass scale of the
particles in the loop. Given that we anticipate the existence in
nature of other fundamental mass scales much larger than the weak
scale, notably the Planck scale, the mass of the Higgs boson is
intrinsically unpredictable, and its small value rather unnatural:
this is the so called hierarchy problem. This puzzle could be resolved
if there were an additional new microscopic scale near the weak scale,
involving new particles and interactions governed by symmetries that
decouple the Higgs mass from short-distance contributions.

\begin{figure}[t]
\centering
\includegraphics[width=0.49\textwidth]{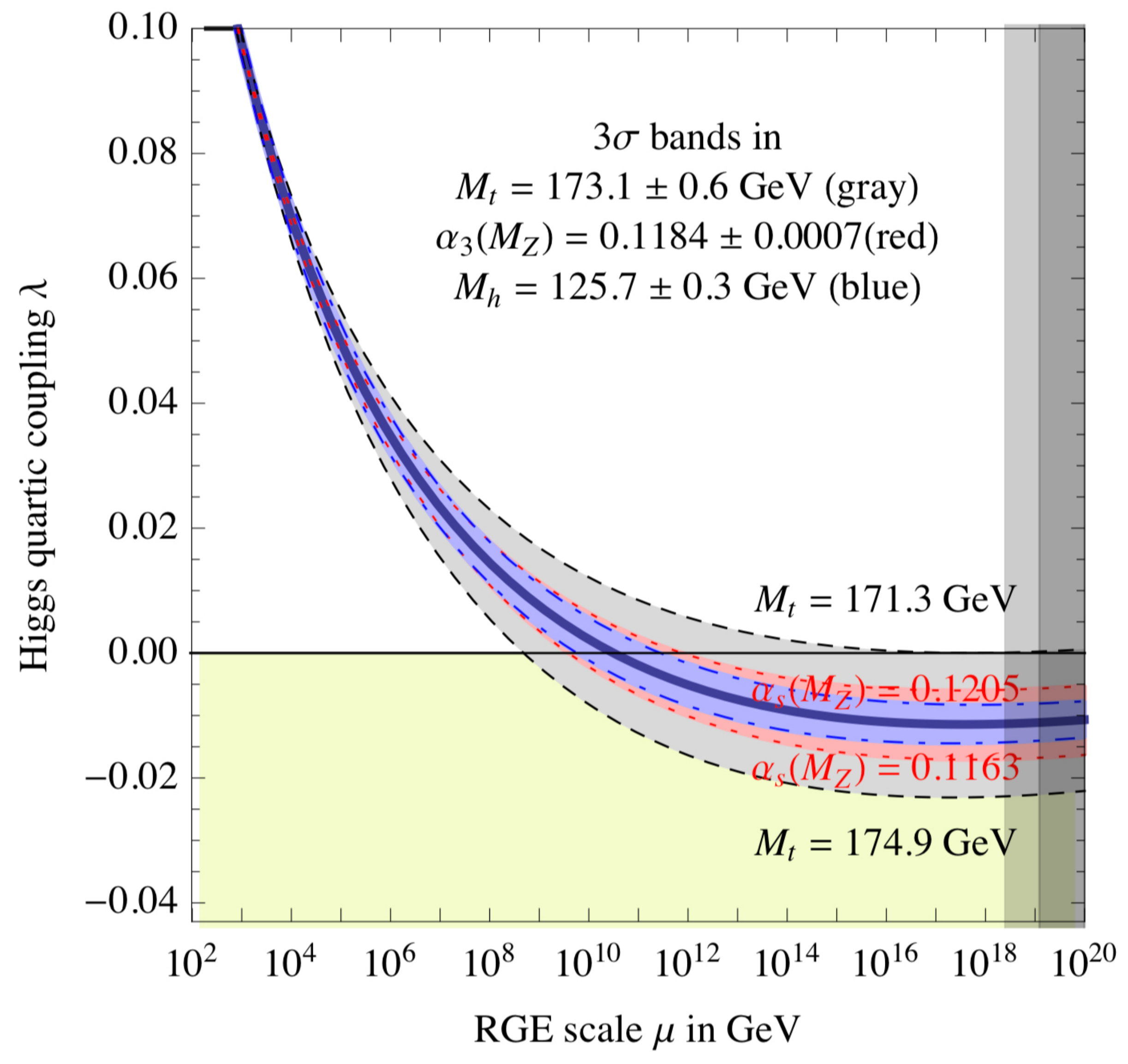}
\caption{Renormalization group running of the Higgs self-coupling
  $\lambda$, and its dependence on the top and Higgs masses, and on
  the strong coupling constant $\alpha_s(M_Z)$. From Ref.~\cite{Degrassi:2012ry}.
}
\label{fig:lambdaH}
\end{figure}
The Higgs quartic coupling $\lambda$ is modified only logarithmically
by loop corrections, but the effect of running to high energy can be
dramatic. At leading order, and neglecting small contribtions from the
gauge couplings, the renormalization group running of $\lambda$ is
given by the following expression:
\begin{equation}
  \beta(\lambda) \, = \,
  \frac{d\lambda}{d \log Q^2} \, = \, \frac{3}{4\pi^2} \left( \lambda^2 +
  \lambda y_t^2 - y_t^4 \right ) \; ,
\end{equation}
where $y_t=m_t/v \sim 0.7$ is the top Yukawa coupling, and
$\lambda=\lambda(Q^2)$ is the running Higgs selfcoupling. It is
straightforward to verify that, for the actual values of the top and Higgs
masses,
$\beta(\lambda)<0$, and $\lambda(Q^2)$ is driven to smaller values at
large $Q^2$. A
complete analysis, including higher order terms (see
e.g.~\cite{Degrassi:2012ry}),
indicates that $\lambda$
turns negative at energies in the range of $10^{10}$~GeV, as shown in
Fig.~\ref{fig:lambdaH}.
A negative
$\lambda$ would make the potential unstable, and short-distance
quantum fluctuations could therefore potentially
destabilize the SM Higgs vacuum. The timescale for ``our'' vacuum to
run away, calculated with the given values of top and Higgs mass,
is much longer than the age of the universe, making the
vacuum metastable and consistent with
observation~\cite{Espinosa:2015qea}. But it is disturbing, once more,
that the dynamics of the Higgs field be influenced so much
by physics taking place at scales much higher than the weak scale!

Both the puzzle of the hierarchy problem and the issue of the
metastability of the Higgs vacuum point to the existence of a more
fundamental layer behind EW symmetry breaking, and galvanise the need
to understand the deeper origin of the Higgs potential.

\subsection{More exploration targets for future colliders}
Even setting to the side the key issue of the origin of the Higgs,
there are other very concrete reasons why the Higgs deserves further
study, and may provide a window to undiscovered phenomena. As it
carries no spin and is electrically neutral, the Higgs may have
so-called `relevant' (i.e. dimension-4) interactions
(e.g. $\vert\Phi\vert^2 \vert S\vert^2$) with a scalar particle, S,
living in sectors of particle physics that are otherwise totally
decoupled from the SM interactions.  These interactions, even if they
only take place at very high energies, remain relevant at low energies
-- contrary to interactions between new neutral scalars and the other
SM particles.  The possibility of new hidden sectors already has
strong experimental support: there is overwhelming evidence from astrophysical
observations that a large fraction of the observed matter density in
the universe is invisible. This so-called dark matter (DM) makes up
26\% of the total energy density in the universe and more than 80\% of
the total matter~\cite{Ade:2015xua}. Despite numerous observations of
the astrophysical properties of DM, not much is known about its
particle nature.  This makes the discovery and identification of DM
one of the most pressing questions in science, a question whose answer
may hinge on the role of the Higgs boson.

The current main constraints on a particle DM candidate $\chi$ are
that it: a) should gravitate like ordinary matter, b) should not carry
colour or electromagnetic charge, c) is massive and non-relativistic
at the time the CMB forms, d) is long lived enough to be present in
the universe today ($\tau \gg \tau_{\rm universe}$), and e) does not
have too strong self-interactions ($\sigma/M_{\rm DM} \lesssim
100~{\rm GeV}^{-3}$). 
While no SM particles satisfy these criteria, they do not pose very
strong constraints on the properties of new particles to play the role
of DM. In particular the allowed range of masses spans almost 80
orders of magnitude. Particles with mass below $10^{-22}$\,eV would
have a wave length so large that they wipe out structures on the kPc
(kilo-Parsec) scale and larger~\cite{Hu:2000ke}, disagreeing with
observations, while on the other end of the scale micro-lensing and
MACHO (Massive Astrophysical Compact Halo Objects) searches put an
upper bound of $2\times 10^{-9}$ solar masses or $10^{48}$\,GeV on the
mass of the dominant DM
component~\cite{Griest:2013aaa,Alcock:1998fx,Yoo:2003fr}.
We shall discuss later on how future colliders can attack this pressing question,
providing comprehensive exploration of the class of `thermal
freeze-out' DM, which picks out a particular broad mass range as a
well-motivated experimental target, as well as unique probes of weakly
coupled dark sectors. 

Returning to the matter which is observable in the Universe, the SM
alone cannot explain baryogenesis, namely the origin of the dominance
of matter over antimatter that we observe today. Since the
matter-antimatter asymmetry was created in the early universe when
temperatures and energies were high, higher energies must be explored
to uncover the new particles responsible for it, and the LHC can only
start this search.  In particular, a well-motivated class of
scenarios, known as EW baryogenesis theories, can explain the
matter-antimatter asymmetry by modifying how the transition from the
high-temperature EW-symmetric phase to the low-temperature
symmetry-broken phase occurred. Independently of the problem of the
matter-antimatter asymmetry, there is the question of the nature of
the EW phase transition (EWPT): was it a smooth cross-over, as predicted by
the SM, or a first-order one, as possible in BSM scenarios (and as
necessary to enable EW baryogenesis)?  Since this phase transition
occurred at temperatures near the weak scale, the new states required
to modify the transition would likely have mass not too far above the weak
scale, singling out future 100~TeV colliders as the leading
experimental facility to explore the nature of this foundational epoch
of the early Universe.

Another outstanding question lies in the origin of the neutrino
masses, which the SM alone cannot account for. As with DM, there are
numerous models for neutrino masses that are within the discovery
reach of future lepton and hadron colliders, as discussed in
Ref.~\cite{Mangano:2651294}.

These and other outstanding questions might also imply the existence
of further spatial dimensions, or larger symmetries that unify leptons
and quarks or the known forces. The LHC's findings notwithstanding,
higher energy and larger statistics will be needed to explore these
fundamental mysteries more deeply and possibly reveal new paradigm
shifts.

\section{The way forward with future colliders}
Since the mid 70's, the path to establish experimentally the SM was
clear: discover the gauge bosons and complete the fermion sector
(e.g. determine the number of SM-like neutrino species and eventually
discover the top quark), test strong and EW interactions at the
level of quantum corrections (comparing precise measurements and
accurate theoretical predictions), test the CKM framework of flavor
phenomena, and discover the Higgs boson. Having accomplished all this,
the situation today is less well defined. In spite of the fact that the
formulation of the open problems, as reviewed in the previous Section,
is rather clear, there is however no experimental approach known today
that can guarantee conclusive answers. This is underscored by the fact
that our prejudices on where to look have not given results.  One of
the main questions we face in planning our future is therefore ``why
don't we see as yet any sign of the new physics that we confidently
expected to be present around the TeV scale?''. The question admits
two possible answers: (i) the mass scale of the new physics lies
beyond the LHC reach, or, (ii) while being within LHC's reach, its
manifestations are elusive and escaped so far the direct search. These
two scenarios are a priori equally likely, but they clearly impact in
different ways the future of our field, and thus the assessment of the
physics potential of possible future facilities. Our safest hedge is
therefore the readiness to cover both scenarios, via an experimental
programme relying on higher precision and sensitivity (to address
possible elusive signatures), and on an extended energy and mass reach
relative to the LHC.

A possible way to assess the value of a future collider facility
is to consider the following three criteria:
\begin{itemize}
\item {\bf The guaranteed deliverables.} This criterion is what I refer to as
  the ``value of measurements'': the new information that we can collect
  to probe the SM to a deeper level, pushing further the exploration of
  particles and processes that are still poorly known. The main targets of this
  component of the programme include of course the Higgs boson, the
  gauge bosons and EW interactions at energies above the EWSB scale,
  the flavor phenomena, in particular those related to the least known
  fermions, such as the top quark or the tau lepton.
\item {\bf The discovery potential.}  While the emergence of phenomena
  beyond the SM cannot be guaranteed, a future facility must promise a
  significant extension of today's sensitivity to new physics,
  addressing the most relevant and compelling BSM scenarios under
  consideration, and with sufficient flexibility to accommodate new
  ideas. The increase in the reach for direct discovery at the highest
  masses should be accompanied by the increased sensitivity throughout
  the whole mass range, thanks to higher precision and statistics. The
  mass reach for direct discovery should ideally match the sensitivity
  reach obtained indirectly via precision measurements.
  \item {\bf Conclusive answers.} Unless an actual discovery is made,
    no experiment  can provide conclusive answers to general
    questions such as ``what is DM?'', ``do supersymmetry or new Z'
    bosons exist?''. Lack of evidence can be evaded by pushing the
    relevant mass spectrum beyond reach. But there exist important,
    less generic, questions, for which it is reasonable to expect that
    a conclusive answer can be found below well-defined mass
    scales. Some examples were given before: did EWSB induce a first
    order phase transition? Is DM made of particles coupled to the SM via the weak
    interaction? Do neutrino masses arise from the weak scale?  Even
    negative answers, if firm, would be of great value, since they
    would force us to focus the searches elsewhere. While current
    experiments (LHC and others) could find partial answers,
    conclusive statements are expected to require higher energy and
    sensitivity, setting performance targets for the evaluation
    of future experiments.
\end{itemize}
These lectures will present an overview of the
physics potential of the various elements of the FCC programme,
in the light of those three criteria.

\subsection{The Role of FCC-ee}  
The capabilities of circular $\rm e^+e^-$ colliders are well
illustrated by LEP, which occupied the LHC tunnel from 1989 to
2000. Its point-like collisions between electrons and positrons and
precisely known beam energy allowed the four LEP experiments to test
the SM to new levels of precision, particularly regarding the
properties of the W and Z bosons. Putting such a machine in a 100 km
tunnel and taking advantage of advances in accelerator technology such
as superconducting radio-frequency cavities would offer even greater
levels of precision on a greater number of processes. For example, it
would be possible to adapt the collision energy during about 15 years
of operation, to examine physics at the Z pole, at the WW production
threshold, at the peak of ZH production, and above the $\rm
t\overline{t}$ threshold. Controlling the beam energy at the 100 keV
level would allow exquisite measurements of the Z and W boson masses,
whilst collecting samples of up to $10^{13}$ Z and $10^8$ W bosons,
not to mention several million Higgs bosons and top quark pairs. The
experimental precision would surpass any previous experiment and
challenge cutting edge theory calculations.

FCC-ee would quite literally provide a quantum leap in our
understanding of the Higgs. Like the W and Z gauge bosons, the Higgs
receives quantum EW corrections typically measuring a few per cent in
magnitude due to fluctuations of massive particles such as the top
quark. This aspect of the gauge bosons was successfully explored at
LEP, but now it is the turn of the Higgs -- the keystone in the EW
sector of the SM. The millions of Higgs bosons produced by FCC-ee,
with its clinically precise environment, would push the accuracy of
the measurements to the per mille level, accessing the quantum
underpinnings of the Higgs and probing deep into this hitherto
unexplored frontier. In the process $\rm e^+e^- \to HZ$, the mass
recoiling against the Z has a sharp peak that allows a unique and
absolute determination of the Higgs decay-width and production cross
section. This will provide an absolute normalisation for all Higgs
measurements performed at the FCC, enabling exotic Higgs decays to be
measured in a model independent manner.

The high statistics promised by the FCC-ee programme goes far beyond
precision Higgs measurements. Other signals of new physics could arise
from the observation of flavour changing neutral currents or
lepton-flavour-violating decays, by the precise measurements of the Z
and H invisible decay widths, or by direct observation of particles
with extremely weak couplings, such as right-handed neutrinos and
other exotic particles. The precision of the FCC-ee programme on EW
measurements would allow new physics effects to be probed at scales as
high as 100 TeV, anticipating what the FCC-hh must focus on.

\subsection{The Role of FCC-hh and FCC-eh}
The FCC-hh would operate at seven times the LHC energy, and collect
about 10 times more luminosity. The discovery reach for high-mass particles
-- such as Z$^\prime$ or W$^\prime$ gauge bosons corresponding to new
fundamental forces, or gluinos and squarks in supersymmetric theories
-- will increase by a factor five or more, depending on the final
statistics. The production rate of particles already within the LHC
reach, such as top quarks or Higgs bosons, will increase by even
larger factors. During the planned 25 years of data taking, a total of
more than $10^{10}$ Higgs bosons will be created, several thousand
times more than collected by the LHC through Run 2 and 200 times more
than will be available by the end of its operation. These additional
statistics will enable the FCC-hh experiments to improve the
separation of Higgs signals from the huge backgrounds that afflict
most LHC studies, overcoming some of the dominant systematics that
limit the precision attainable at the LHC. While the ultimate
precision of most Higgs properties can only be achieved with FCC-ee,
several demand complementary information from FCC-hh. For example, the
direct measurement of the coupling between the Higgs and the top quark
requires that they be produced together, requiring an energy beyond
the reach of the FCC-ee. At 100 TeV, almost $10^9$ out of the
$10^{12}$ top quarks produced will radiate a Higgs boson, allowing the
top-Higgs interaction to be measured at the 1\% level -- several times
better than at the HL-LHC and probing deep into the quantum structure
of this interaction. Similar precision can be reached for Higgs decays
that are too rare to be studied in detail at FCC-ee, such as those to
muon pairs or to a Z and a photon. All of these measurements will be
complementary to those obtained with FCC-ee and will use them as
reference inputs to precisely correlate the strength of the signals
obtained through various production and decay modes.

One respect in which a 100~TeV proton-proton collider would really
come to the fore is in revealing how the Higgs behaves in private.
The rate of Higgs pair production events, which in some part occur
through Higgs self-interactions, would grow by a factor of 40 at
FCC-hh, with respect to 14\,TeV, and enable this unique property of
the Higgs to be measured with an accuracy reaching 5\%. Among many
other uses, such a measurement would comprehensively explore classes
of models that rely on modifying the Higgs potential to drive a strong
first order phase transition at the time of EW symmetry breaking, a
necessary condition to induce baryogenesis.

FCC-hh would also allow an exhaustive exploration of new TeV-scale
phenomena. Indirect evidence for new physics can emerge from the
scattering of W bosons at high energy -- where the Higgs boson plays a
key role in controlling the rate growth -- from the production of
Higgs bosons at very large transverse momentum, or by testing the far
`off-shell' nature of the Z boson via the measurement of lepton pairs
with invariant masses in the multi-TeV region. The plethora of new
particles predicted by most models of symmetry-breaking alternatives
to the SM can be searched for directly, thanks to the immense mass
reach of 100~TeV collisions. The search for DM, for example, will
cover the possible space of parameters of many theories relying on
weakly interacting massive particles, guaranteeing a discovery or
ruling them out. Several theories that address the hierarchy problem
will also be conclusively tested. For supersymmetry, the mass reach of
FCC-hh pushes beyond the regions motivated by the hierarchy problem
alone. For composite Higgs theories, the precision Higgs coupling
measurements and searches for new heavy resonances will fully cover
the motivated territory. A 100~TeV proton collider will even confront
exotic scenarios such as the twin Higgs, which are extremely difficult
to test. These theories predict very rare or exotic Higgs decays,
possibly visible at FCC-hh thanks to its enormous Higgs production
rates.

The FCC-eh collider could operate in synchronous, symbiotic operation
alongside the pp collider.  The facility would serve as the most
powerful, high-resolution microscope onto the substructure of matter
ever built. High-energy ep collisions would provide precise
information on the quark and gluon structure of the proton, and how
they interact.  FCC-eh would complement and enhance the study of the
Higgs, and broaden the new physics searches also performed at FCC-hh
and FCC-ee, with a specific focus on phenomena such as quark
substructure, leptoquarks, heavy sterile neutrinos and long-lived
particles.

While not discussed at all in these lectures, FCC-hh would also enable
the continuation of the LHC successful programme of heavy ion
collisions, extending studies of the thermodynamic behaviour of QCD of
crucial relevance to multiple topics, ranging from the fundamental
properties of quantum field theory, to cosmology and astrophysics.

\section{Higgs boson properties}
\label{sec:higgs}
Indirect information about the Higgs boson is accessible through
precision EW measurements, as proven by the global fits to the LEP and
SLC data, which set very tight constraints on the Higgs mass well
before its discovery. But, following the Higgs discovery, the most direct way to test the
Higgs properties is to produce it and observe its decay features. With
the knowledge of the Higgs mass, the SM predicts uniquely its
couplings to each SM particle, and therefore all production and decay
rates are fixed. Since our target is to explore the origin of
EWSB, and possibly identify the underlying BSM phenomena that trigger
it, we must be open however to all sorts of deviations from the
SM. For example, while the couplings of the Higgs to the gauge bosons
are determined by the Higgs quantum numbers (an SU(2) doublet), the
existence of an additional Higgs scalar, acquiring its own
expectation value, could lead to a mixing in the Higgs sector, and the
mass eigenstate at 125~GeV could couple to the W and the Z with a
slightly reduced strength. The existence of additional Higgses opens
the door to the possibility that different fermions couple to
different Higgses, modifying the direct relation between fermion mass
and Yukawa coupling to the 125~GeV state (as in the case in
supersymmetry).
The study of Higgs couplings, threfore, requires as much as possible a
model independent approach.

Establishing the gauge couplings of the known fermions was relatively
straightforward, since they are quantized and the fermion assignment
to a gauge group representation ranges over a discrete set of
possibilities. Deviations are possible of course, but only in presence
of additional BSM interactions, that appear at low energy as operators
of dimension higher than 4. The basic, leading-order and
renormalizable interactions of SM fermions are therefore easily
established experimentally. That the top quark is an $SU(3)$ color
triplet, for example, can follow from the analysis of its production
rate and decay patterns\footnote{This is so straightforward in
  principle, that I am not even sure there has ever been an explicit
  experimental analysis to confirm that the top quark is a triplet. I
  leave it to you as an exercise to list the data and signatures that
  could be used to confirm it.}. On the contrary, the leading-order Higgs
couplings are a priori a generic real (or complex) number, and the
confidence on whether they agree or not with the SM will always only
be conditional to the precision of the available data.

\begin{figure}[th!]
\centering
\includegraphics[width=0.75\textwidth]{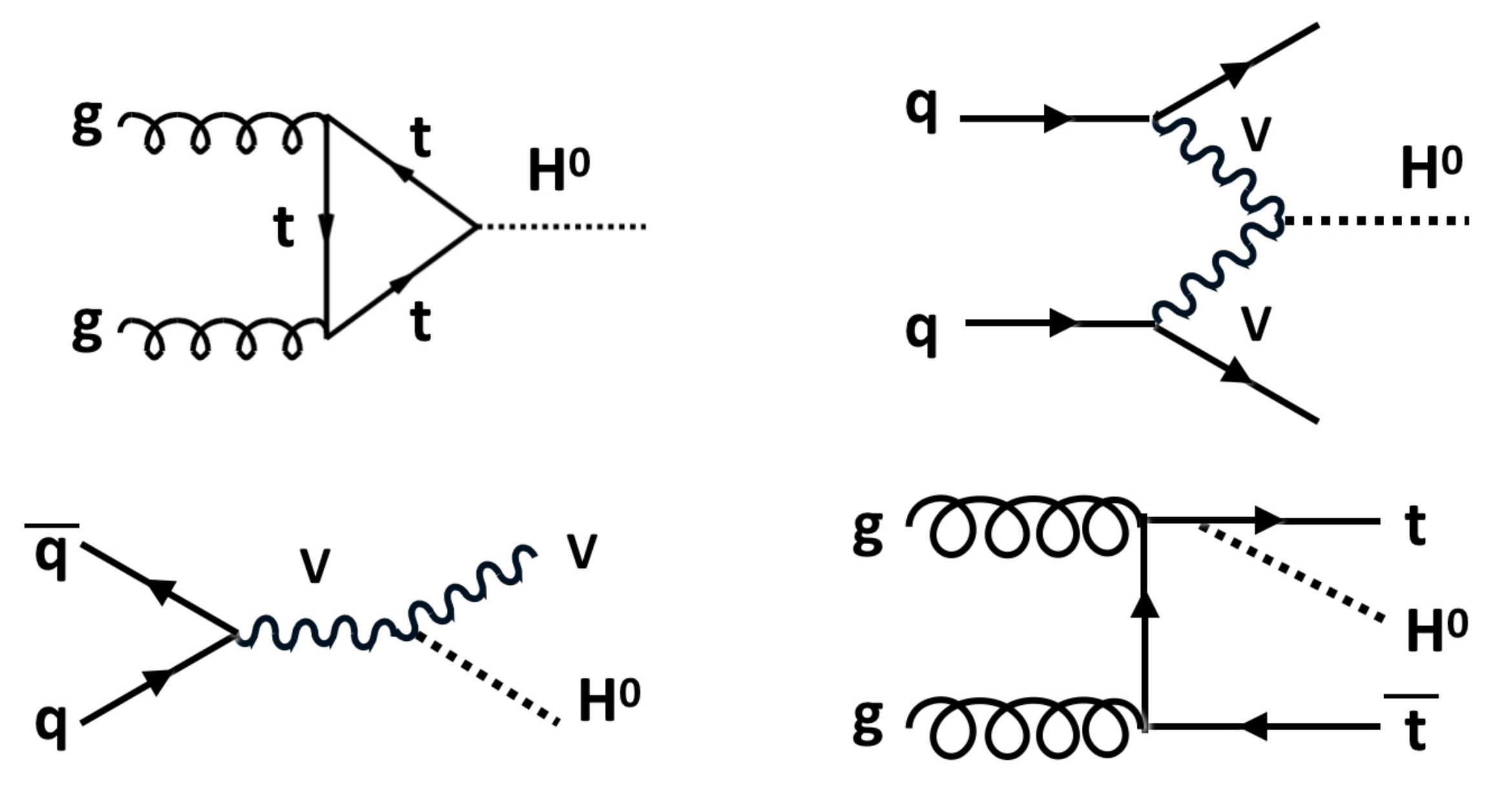}
\caption{Basic Higgs production processes in hadronic collisions. From
  the top left, clockwise: gluon-gluon fusion (ggH), vector boson fusion
  (VBF), associate production with a gauge boson (VH), and with a $\rm
  t\bar{t}$ pair (ttH). V=$\rm W^\pm$ or Z$^0$, throughout.}
\label{fig:ppH}
\end{figure}

The dominant Higgs production channels, in hadronic collisions like at
the LHC, are shown in Fig.~\ref{fig:ppH}. In these examples, the
production rates are proportional to the coulings to the gauge bosons,
or to the top quark. In the ideal world in which the strong coupling
$\alpha_s$, the partonic densities (PDFs) and the QCD matrix elements
were perfectly known, counting events in a given $\rm H\to Y$ decay
mode would provide a measurement of $g_X^2 \, \Gamma_Y/\Gamma_H$,
where $g_{X,Y}$ are the Higgs couplings to initial and final state
state, and $\Gamma_Y\propto g_Y^2$ ($\Gamma_H$) is the partial (total)
decay width.  If we could observe every possible Higgs decay, summing
over all Y states for a given production channel X would allow the
measurement of $g_X$, since $\sum_Y \Gamma_Y=\Gamma_H$. At the LHC and
in general in hadronic collisions, this is hardly possible: several SM
decay modes with a substantial branching ratio (BR), like $\rm H\to c\bar{c}$,
are very difficult to measure, and possible exotic Higgs decays are
also likely to escape detection. A completely model-independent
extraction of Higgs couplings in hadronic collisions can therefore
only reach a limited precision, independently of the theoretical
challenge of properly calculating the QCD part of the reactions.

As we show in the next sections, the measurements at an electron
collider can provide the needed input of $\Gamma_H$, and open the way
for a powerful synergetic programme of precision measurements with the
next generation of hadron colliders.

A complete compilation and critical review of the Higgs coupling measurement prospects covering all proposed future colliders can be found in Ref.~\cite{deBlas:2019rxi}.
\subsection{Higgs coupling measurements at FCC-ee}
\begin{figure}[th!]
\centering
\includegraphics[width=0.45\textwidth]{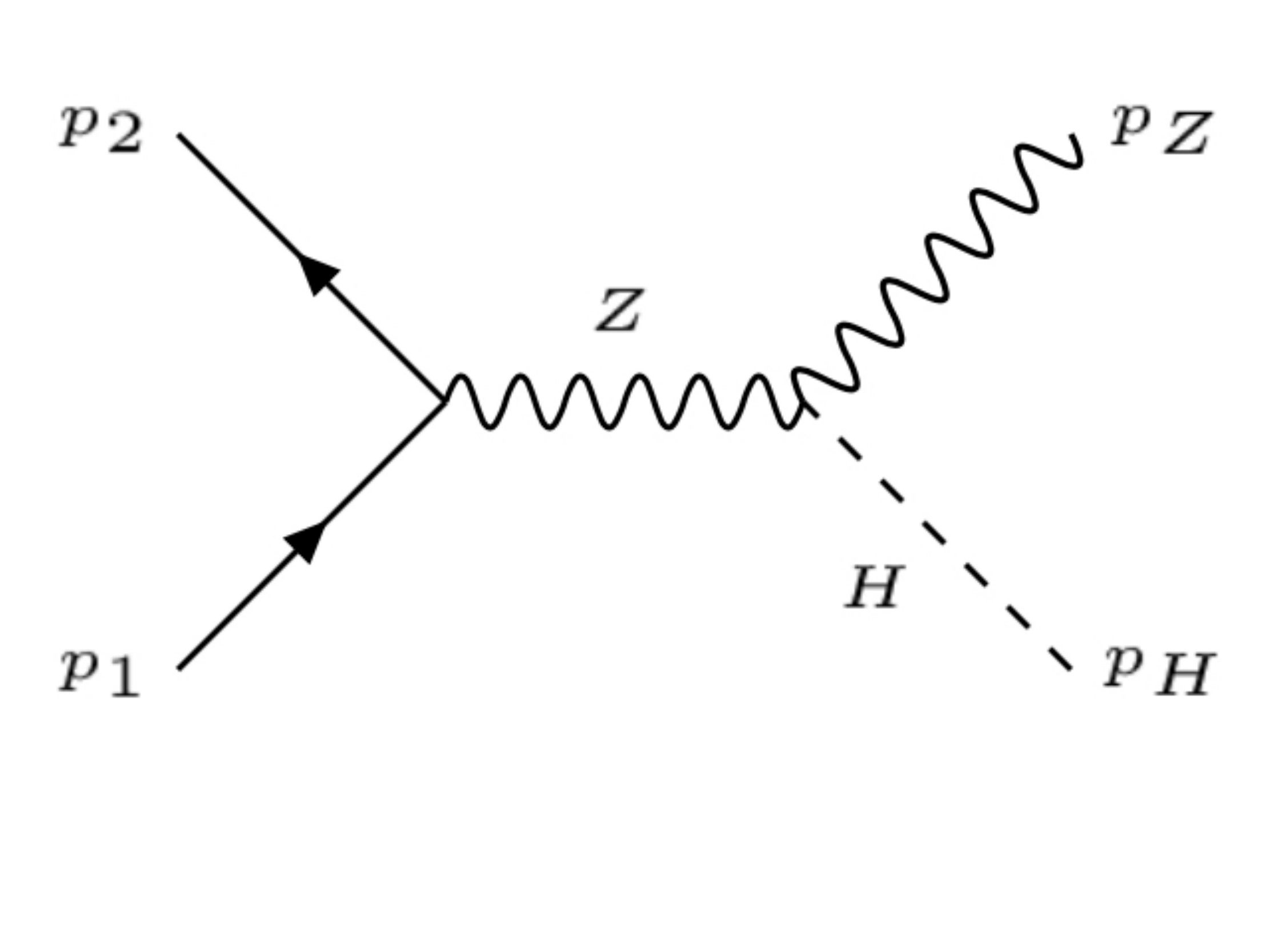}
\hfil
\includegraphics[width=0.45\textwidth]{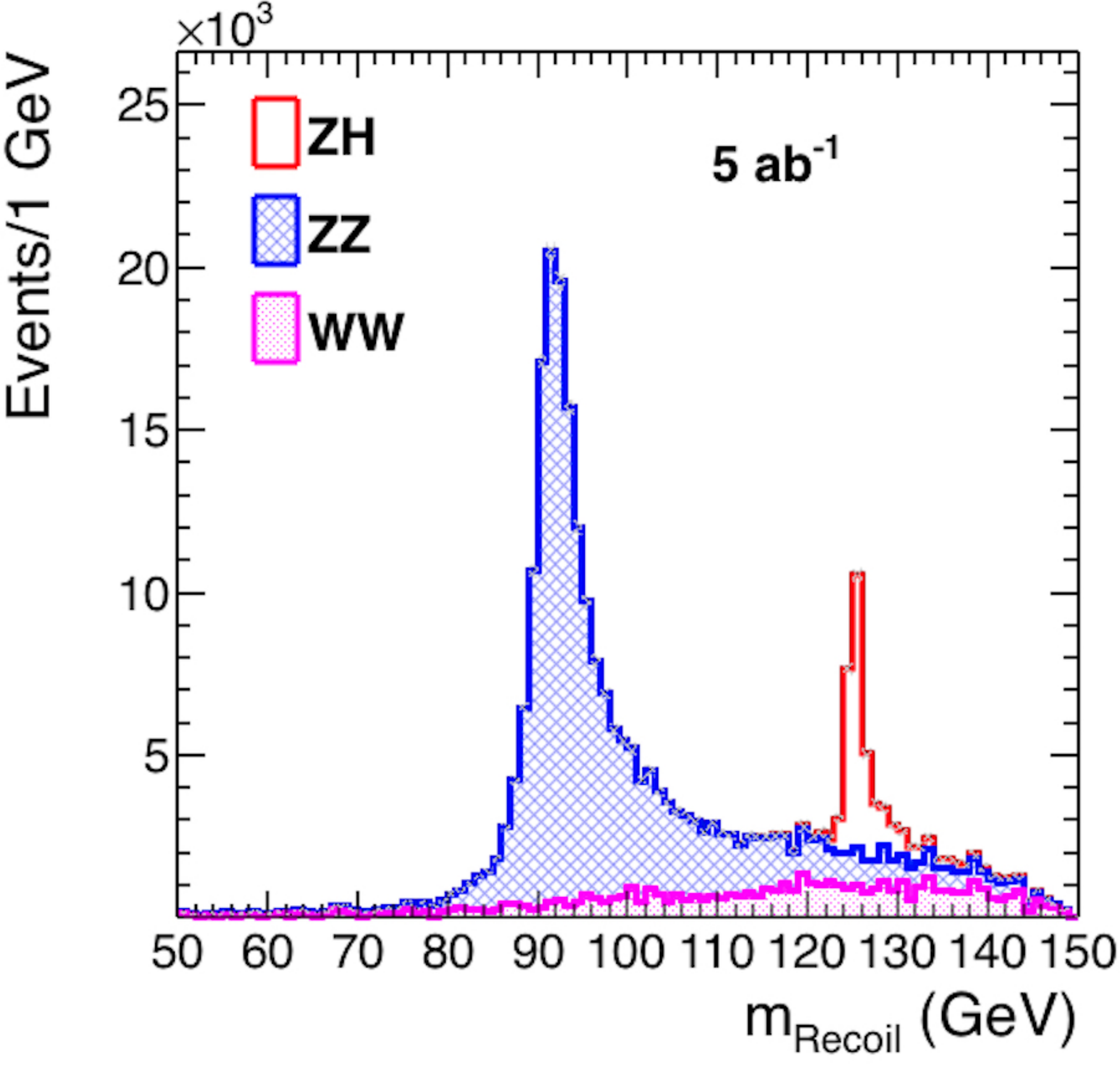}
\vskip -3mm
\caption{Left: the $\rm e^+e^- \to HZ$ production process. Right:
  Higgs signal and background contributions to the recoil mass in $\rm
  e^+e^- \to \mu^+\mu^- X$ events.}
\label{fig:ZH}
\end{figure}
The determination of $\Gamma_H$ is however possible at future $\rm
e^+e^-$ colliders, operating above the ZH threshold.  Here, the
production of a Higgs boson can be reconstructed, in a
model-independent way, with the so-called recoil-mass technique. One
considers $\rm e^+e^-\to \mu^+\mu^- +X$ final states, and for each
event defines the recoil mass as $m^2_{Recoil}=P_{X}^2 = (p_{e^+} +p_{e^-}
- p_{\mu^+} - p_{\mu^-})^2$. Most $\rm \mu^+\mu^-+X$ final states arise
from ZZ production ($\rm Z[\to \mu^+\mu^-]\, Z[\to X]$), in which case
$P_{X}$ is the momentum of the second Z, and the recoil mass equals
(up to finite-width and experimental resolution effects) the Z mass. A
further contribution comes from WW production ($\rm W^+[\to
  \mu^+\nu]\, W^-[\to \mu^-\bar{\nu}]$), in which case $P_X$ represents
the missing momentum, and the recoil mass is a broad continuum. In the
case of $\rm Z[\to\mu^+\mu^-]\, H[\to X]$ production (see the left image
of Fig.~\ref{fig:ZH}), the recoil mass coincides with the Higgs mass,
regardless of the H decay mode. These three contributions are shown,
for the simulation of an FCC-ee experiment, in Fig.~\ref{fig:ZH}. A
global fit of the recoil mass spectrum returns the total number of
Higgs produced in $\rm e^+e^- \to HZ[\to\mu^+\mu^-]$, and a direct
measurement of the HZZ coupling, $g_{ZZ}$. If we now focus on events
with the $\rm H\to ZZ^* \to 4\ell$ decay, and consider that their rate
is proportional to $g_{ZZ}^2 \times \Gamma_{ZZ}/\Gamma_H \propto
g_{ZZ}^4/\Gamma_H$, the knowledge of $g_{ZZ}$ allows to extract $\Gamma_H$ in a model-independent
way.

\noindent\rule{5cm}{1pt} \\
{\bf Exercise}: discuss, in a qualitative way, to which extent EW
radiative corrections or BSM effects influence this line of reasoning, and whether
they affect the ``model-independent'' argument.\\
{\bf Exercise}: discuss, in a qualitative way, the backgrounds under
the H peak in the recoil mass spectrum, and how they can be estimated,
and subtracted, precisely.\\
{\bf Exercise}: discuss how the recoil mass observable can be used to
determine the presence of exotic (in particular, invisible) H decays.\\
\rule{5cm}{1pt}

\begin{figure}[h!]
\centering
\includegraphics[width=0.4\textwidth]{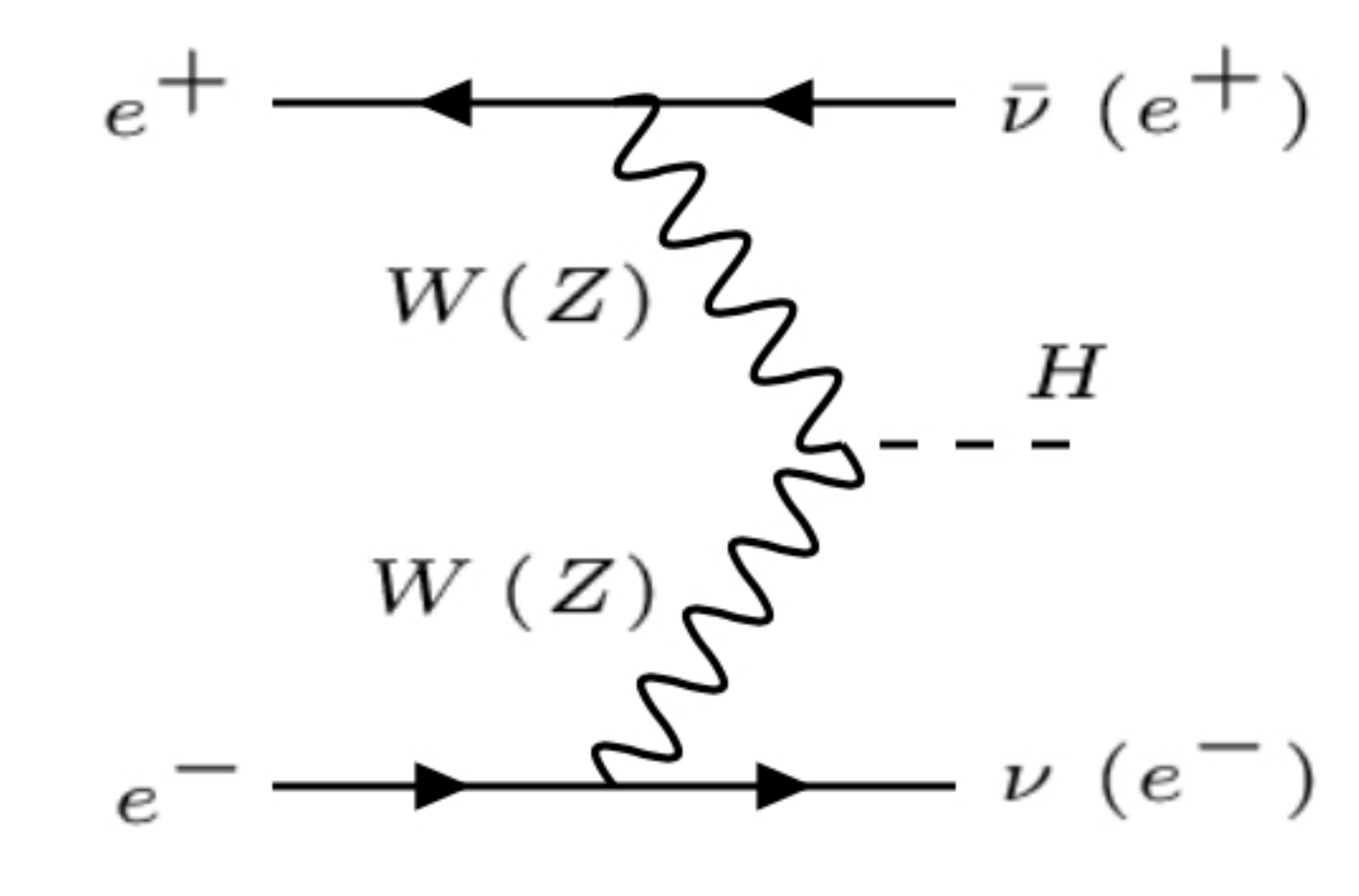}
\hfil
\includegraphics[width=0.5\textwidth]{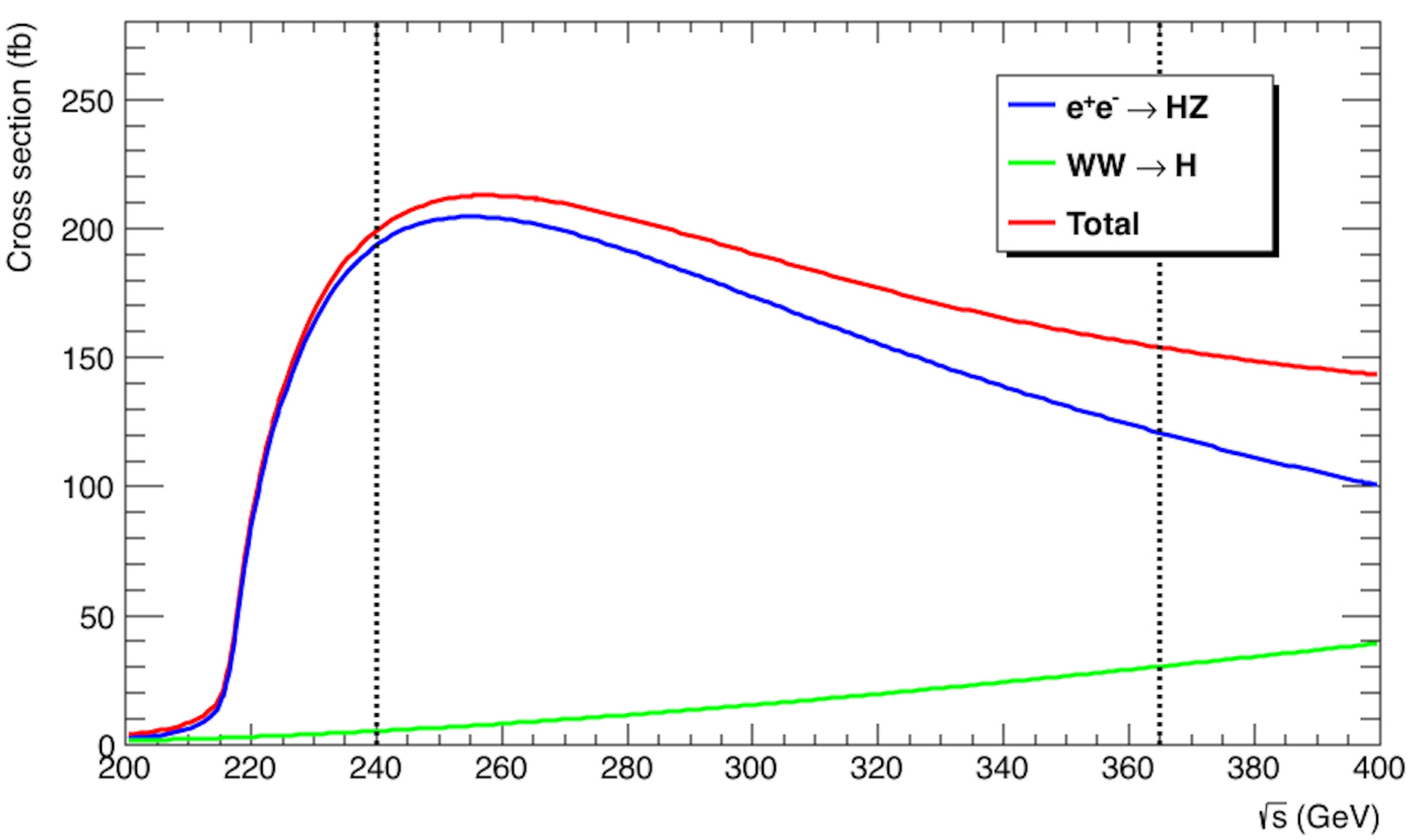}
\vskip -3mm
\caption{Left: the $\rm e^+e^- \to H \nu\bar{\nu} (e^+e^-) $ production process. Right:
  Higgs production rates for the leading $\rm e^+e^-$ processes in the
  FCC-ee energy range.}
\label{fig:eeVBF}
\end{figure}

Having established the value of $\Gamma_H$, further dedicated
measurements allow to determine the absolute value of the Higgs
couplings to all particles accessible via decay modes or production
channels. Assuming SM couplings, the statistical precision that can be achieved
for several BRs measurable at FCC-ee is summarized in
Table~\ref{tab:eeH} (for the details of the reconstruction of
individual final states, see e.g.~\cite{Benedikt:2651299}). These
include the results obtained from the run just above the Higgs
threhsold, at 240~GeV, and the runs above the $t\bar{t}$ thresholds,
where the VBF process $e^+e^-\to \nu\bar{\nu}H$, shown in
Fig.~\ref{fig:eeVBF}, becomes relevant.

\begin{table}[ht!]
\begin{center}
\caption{Relative statistical uncertainty on the measurements of event
  rates, providing $\sigma_{\rm HZ} \times {\rm BR}({\rm H \to XX})$
  and $\sigma_{\rm \nu\bar\nu H} \times {\rm BR}({\rm H \to XX})$, as
  expected from the FCC-ee data~\cite{Benedikt:2651299}.  This is
  obtained from a fast simulation of the reference FCC-ee detector and consolidated
  with extrapolations from full simulations of similar linear-collider
  detectors (SiD and CLIC). All numbers indicate 68\% C.L. intervals,
  except for the 95\% C.L. sensitivity in the last line. The
  accuracies expected with $5~{\rm ab}^{-1}$ at 240 GeV are given in
  the middle columns, and those expected with $1.5~{\rm ab}^{-1}$ at
  $\sqrt{s} = 365$~GeV are displayed in the last
  columns.\label{tab:eeH}}
\begin{tabular}{|l|r r|r r|}
\hline \hline
$\sqrt{s}$ (GeV) & \multicolumn{2}{|c|}{$240$} & \multicolumn{2}{|c|}{$365$} \\ \hline
Luminosity (${\rm ab}^{-1}$) & \multicolumn{2}{|c|}{5} & \multicolumn{2}{|c|}{$1.5$} \\ \hline
$\delta (\sigma {\rm BR}) / \sigma {\rm BR}$ (\%) & HZ &
\textnu$\overline{\mbox{\textnu}}$ H &  HZ &
\textnu$\overline{\mbox{\textnu}}$ H \\  \hline 
${\rm H \to any}$& $\pm 0.5$ & & $\pm 0.9$    &   \\ 
${\rm H \to b \bar b}$& $\pm 0.3$ & $\pm 3.1$ & $\pm 0.5$    & $\pm 0.9$   \\ 
${\rm H \to c \bar c}$& $\pm 2.2$ & & $\pm 6.5$    & $\pm 10$    \\ 
${\rm H \to gg}$ & $\pm 1.9$ & & $\pm 3.5$    & $\pm 4.5$   \\ 
${\rm H \to W^+W^-}$  & $\pm 1.2$ & & $\pm 2.6$    & $\pm 3.0$   \\ 
${\rm H \to ZZ}$ & $\pm 4.4$ & & $\pm 12$& $\pm 10$    \\ 
${\rm H \to}$ \texttau\texttau& $\pm 0.9$ & & $\pm 1.8$    & $\pm 8$\\ 
${\rm H \to}$ \textgamma\textgamma & $\pm 9.0$ & & $\pm 18$& $\pm 22$    \\ 
${\rm H \to}$ \textmu$^+$\textmu$^-$   & $\pm 19$  & & $\pm 40$&   \\ 
${\rm H \to invis.}$  & $<0.3$    & & $<0.6$  &   \\ \hline \hline
\end{tabular} 
\end{center}
\end{table}

\begin{table}
\centering
\caption{Precision determined in the $\kappa$ framework of the Higgs
  boson couplings and total decay width, as expected from the FCC-ee
  data, and compared to the projections for
  HL-LHC~\cite{Cepeda:2019klc}. All numbers indicate 68\% CL
  sensitivities, except for the last line which gives the 95\% CL
  sensitivity on the "exotic" branching fraction, accounting for final
  states that cannot be tagged as SM decays. The FCC-ee accuracies are
  subdivided in three categories: the first sub-column give the
  results of the model-independent fit expected with $5~{\rm ab}^{-1}$
  at 240 GeV, the second sub-column in bold -- directly comparable to
  the other collider fits -- includes the additional $1.5~{\rm
    ab}^{-1}$ at $\sqrt{s} = 365$~GeV, and the last sub-column shows
  the result of the combined fit with HL-LHC. The fit to the HL-LHC
  projections alone (first column) requires two additional assumptions
  to be made: here, the branching ratios into ${\rm c\bar c}$ and into
  exotic particles are set to their SM values. From~\cite{Mangano:2651294}.
\label{tab:FitResults}
}
\begin{tabular}{|l|r|r|r|r|}
\hline \hline Collider & { HL-LHC} & \multicolumn{3}{|c|}{FCC-ee$_{240+365}$} \\ \hline
Lumi (${\rm ab}^{-1}$) & { 3} & $5_{240}$ & $+1.5_{365}$ & { $+$ HL-LHC} \\ \hline
Years & { 25} & 3 & $+$4 &  \\ \hline
${} \delta\Gamma_{\rm H}/\Gamma_{\rm H}$ { (\%)} &
{ SM} & 2.7 & {\bf 1.3} & 1.1 \\ \hline 
${} \delta g_{\rm HZZ}/g_{\rm HZZ}$ { (\%)} & {
  1.5} & 0.2 & {\bf 0.17}  & 0.16 \\  
${} \delta g_{\rm HWW}/g_{\rm HWW}$ { (\%)} & {
  1.7} & 1.3& {\bf 0.43}  &  0.40 \\ 
${} \delta g_{\rm Hbb}/g_{\rm Hbb}$ { (\%)} & {
  3.7} & 1.3 & {\bf 0.61}  &  0.56 \\
${} \delta g_{\rm Hcc}/g_{\rm Hcc}$ { (\%)} & {
  SM} & 1.7 & {\bf 1.21}  &  1.18 \\  
${} \delta g_{\rm Hgg}/g_{\rm Hgg}$ { (\%)} & {
  2.5} & 1.6& {\bf 1.01}  &  0.90 \\ 
${} \delta g_{\rm H\mbox{\texttau\texttau}}/g_{\rm
  H\mbox{\texttau\texttau}}$ { (\%)} & { 1.9} & 1.4 & {\bf
  0.74}  &  0.67\\ 
${} \delta g_{\rm H\mbox{\textmu\textmu}}/g_{\rm
  H\mbox{\textmu\textmu}}$ { (\%)} & { 4.3} & 10.1 &
{\bf 9.0}  & 3.8 \\ 
${} \delta g_{\rm H\mbox{\textgamma\textgamma}}/g_{\rm
  H\mbox{\textgamma\textgamma}}$ { (\%)} & { 1.8} & 4.8 &
{\bf 3.9}  & 1.3 \\ 
${} \delta g_{\rm Htt}/g_{\rm Htt}$ { (\%)} & {
  3.4} & -- & -- &  3.1 \\ 
{ BR}$_{\rm  EXO}$ { (\%)} & { SM} &
$ < 1.2 $ & ${\bf < 1.0}$  & ${\bf < 1.0}$ \\  \hline \hline
\end{tabular} 
\end{table}
In practice, the width and the couplings are determined with
a global fit, which closely follows the logic of
Ref.~\cite{Peskin:2012we}. The results of this fit are summarised in
Table~\ref{tab:FitResults} and are compared to the same fit applied to
HL-LHC projections~\cite{Cepeda:2019klc}.
Table~\ref{tab:FitResults} also shows that the extractions of
$\Gamma_{\rm H}$ and of $g_{\rm HWW}$ from the global fit are
significantly improved by the addition of the WW-fusion process at
$\sqrt{s} = 365$~GeV, as a result of the correlation between the HZ
and \textnu$\overline{\mbox{\textnu}}$ H processes.
In particular the Higgs EW couplings have a permille-level
precision, and the couplings to the tau, the bottom and charm
quarks and the effective couplng to the gluon reach the percent level or better. 

Several SM couplings are left out of these projections: to the
lightest quarks (u, d, s), to the electron, to the top quark, to the
$\rm Z\gamma$ pair, and the Higgs self-coupling. To access the light
quarks, several ideas have been proposed: exclusive decays to hadronic
resonances, such as $\rm H\to Vh$ ($\rm h=\phi,\; \rho)$,
V=W/Z/\textgamma)~\cite{Bodwin:2013gca,Isidori:2013cla,Kagan:2014ila,Koenig:2015pha},
light-jet tagging techniques~\cite{Duarte-Campderros:2018ouv}, or
kinematical distributions of the Higgs boson in hadronic
collisions~\cite{Soreq:2016rae,Bishara:2016jga}. Experimental searches
for exclusive radiative hadronic decays have started already at the
LHC~\cite{Aaboud:2017xnb}, to at least establish upper limits, even
though well beyond the SM expectations. Given the small BRs, an
electron collider will barely have sufficient statistics to gain the
required SM sensitivity. At the FCC-ee, the most promising channel is
H$\to$\textgamma\textrho[$\to$\textpi\textpi], with about 40 events
expected~\cite{dEnterria:2017dac}.  A future hadron collider will have
much more events to play with, but backgrounds and experimental
conditions will be extremely challenging, and only detailed
simulations will be able to establish their true potential.

To probe the Hee coupling, the best hope appears to be the direct
resonant production in $\rm e^+e^- \to H$. The low rate demands high
luminosity, and a tuning of the beam energy to exactly match
$m_H/2$. Preliminary studies~\cite{dEnterria:2017dac} indicate that a
3$\sigma$ observation requires an integrated luminosity of 90\iab,
namely several years of dedicated running at 125~GeV.

While the direct access to the Htt coupling in an $\rm e^+e^-$
collider requires a center-of-mass energy of 500~GeV and more, FCC-ee
will expose an indirect sensitivity to it, through its effect at
quantum level on the ${\rm t\bar t}$ cross section just above
production threshold, $\sqrt{s} = 350$~GeV. The precise measurement of
$\alpha_s$ from the runs at the Z pole will allow the QCD effects to
be disentangled from those of the top Yukawa coupling at the ${\rm
  t\bar t}$ vertex, to achieve a precision of $\pm 10\%$~\cite{Benedikt:2651299}.

To access in a direct way the top Yukawa coupling, and to improve to
the percent level the measurement of small BR decays such as
H$\to$\textgamma\textgamma, Z\textgamma{} and \textmu$^+$\textmu$^-$,
we can then appeal to the huge statistics available to a hadron
collider.

\subsection{Higgs couplings measurements at FCC-hh}
\begin{figure}[t]
\centering
\includegraphics[width=0.85\textwidth]{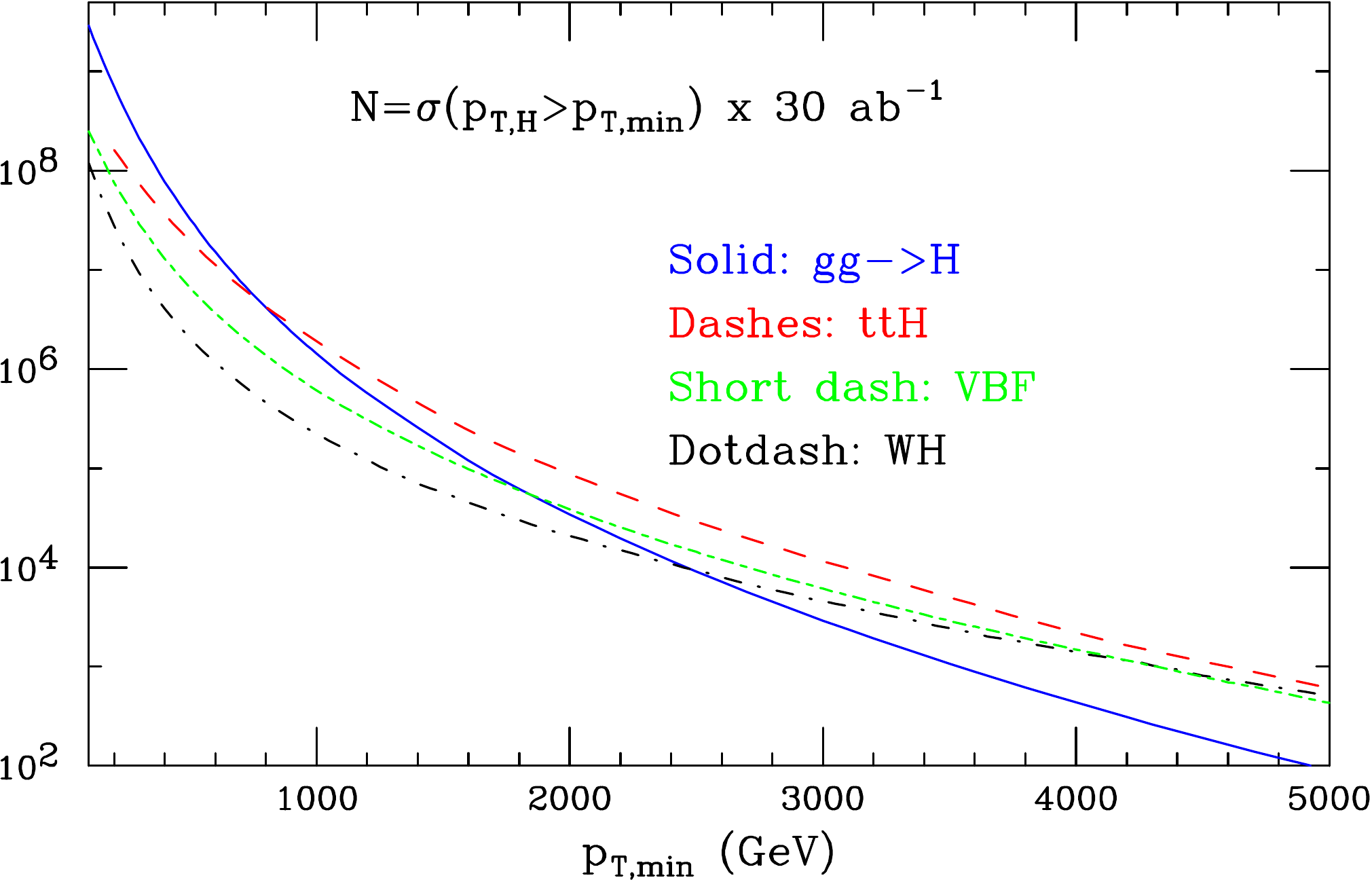}
\caption{Production rates of Higgs bosons at high $p_T$, for various
  production channels at 100~TeV and 30~\iab.} 
\label{fig:Hhighpt}
\end{figure}

Two elements characterise Higgs production at the FCC-hh: the large
statistics (see Table~\ref{tab:Hrates}), and the large kinematic
range, which, for several production channels, probes $p_T$ in the
multi-TeV region (see Fig.~\ref{fig:Hhighpt}).
\begin{table}[th]
 \begin{center}
   \caption{\label{tab:Hrates}
 Higgs production event rates for selected processes
     at 100~TeV ($N_{100}$) and 27~TeV ($N_{27}$), and
  statistical increase with 
  respect to the statistics of the HL-LHC
  ($N_{100/27}=\sigma_{100/27~\mathrm{TeV}} \times 30/15$~\iab,  
  $N_{14}=\sigma_{14~\mathrm{TeV}} \times 3$~\iab).} 
\begin{tabular}{|l|c|c|c|c|c|c|} 
  \hline\hline 
 &  gg$\to$H   & 
  VBF  & 
  WH  &
  ZH  &
  t\={t}H &
  HH 
  \\
  \hline 
$N_{100}$  & $24\times 10^9$ & $2.1\times 10^9$ & $4.6\times 10^8$ & $3.3\times
  10^8$ &  $9.6\times 10^8$ & $3.6 \times 10^7$  \\
$N_{100}/N_{14}$ &  180 & 170 & 100 & 110 & 530 & 390 
\\ 
  \hline 
$N_{27}$  
& $2.2\times 10^9$ 
& $1.8\times 10^8$ 
& $5.1 \times 10^7$ 
& $3.7 \times 10^7$ 
& $4.4 \times 10^7$ 
& $2.1 \times 10^6$  \\
$N_{27}/N_{14}$ &  16 & 15 & 11 & 12 & 24 & 19 
\\ 
\hline \hline
\end{tabular}
\end{center}
\end{table}
These factors lead to an extended and diverse sensitivity to possible
deviations of the Higgs properties from their SM predictions: the
large rates enable precise measurements of branching ratios for rare
decay channels such as \textgamma\textgamma\ or \textmu\textmu, and
push the sensitivity to otherwise forbidden channels such as
\texttau\textmu. The large kinematic range can be used to define cuts
improving the signal-to-background ratios and the modelling or
experimental systematics, but it can also amplify the presence of
modified Higgs couplings, described by higher-dimension operators,
whose impact grows with $Q^2$. Overall, the Higgs physics programme of
FCC-hh is a fundamental complement to what can be measured at FCC-ee,
and the two Higgs programmes greatly enrich each other. This section
contains some examples of these facts, and documents the current
status of the precision projections for Higgs measurements. A more
extensive discussion of Higgs production properties at 100~TeV and of
possible measurements is given in Ref.~\cite{Contino:2016spe}.

Figure~\ref{fig:Hhighpt} shows the Higgs rates above a given $p_T$
threshold, for various production channels. It should be noted that
these rates remain above the level of one million up to $p_T\sim
1$~TeV, and there is statistics for final states like H$\to$b\={b} or
H$\to$\texttau\texttau\ extending up to several TeV. Furthermore, for
$p_T(\mathrm{H})\gsim$1~TeV, the leading production channel becomes
t\={t}H, followed by vector boson fusion when
$p_T(\mathrm{H})\gsim$2~TeV. The analysis strategies to separate
various production and decay modes in these regimes will therefore be
different to what is used at the LHC. Higgs measurements at 100~TeV
will offer many new options and precision opportunities with respect
to the LHC, as it happened with the top quark moving from the
statistics-hungry Tevatron to the rich LHC.

\noindent
\rule{5cm}{1pt}\\
{\bf Exercise:} discuss possible strategies to separate the different
Higgs production processes in the various ranges of $p_T(H)$ shown in
Fig.~\ref{fig:Hhighpt}.
\\
\rule{5cm}{1pt}

\begin{figure}
  \begin{center}
    \includegraphics[height=0.45\textwidth]{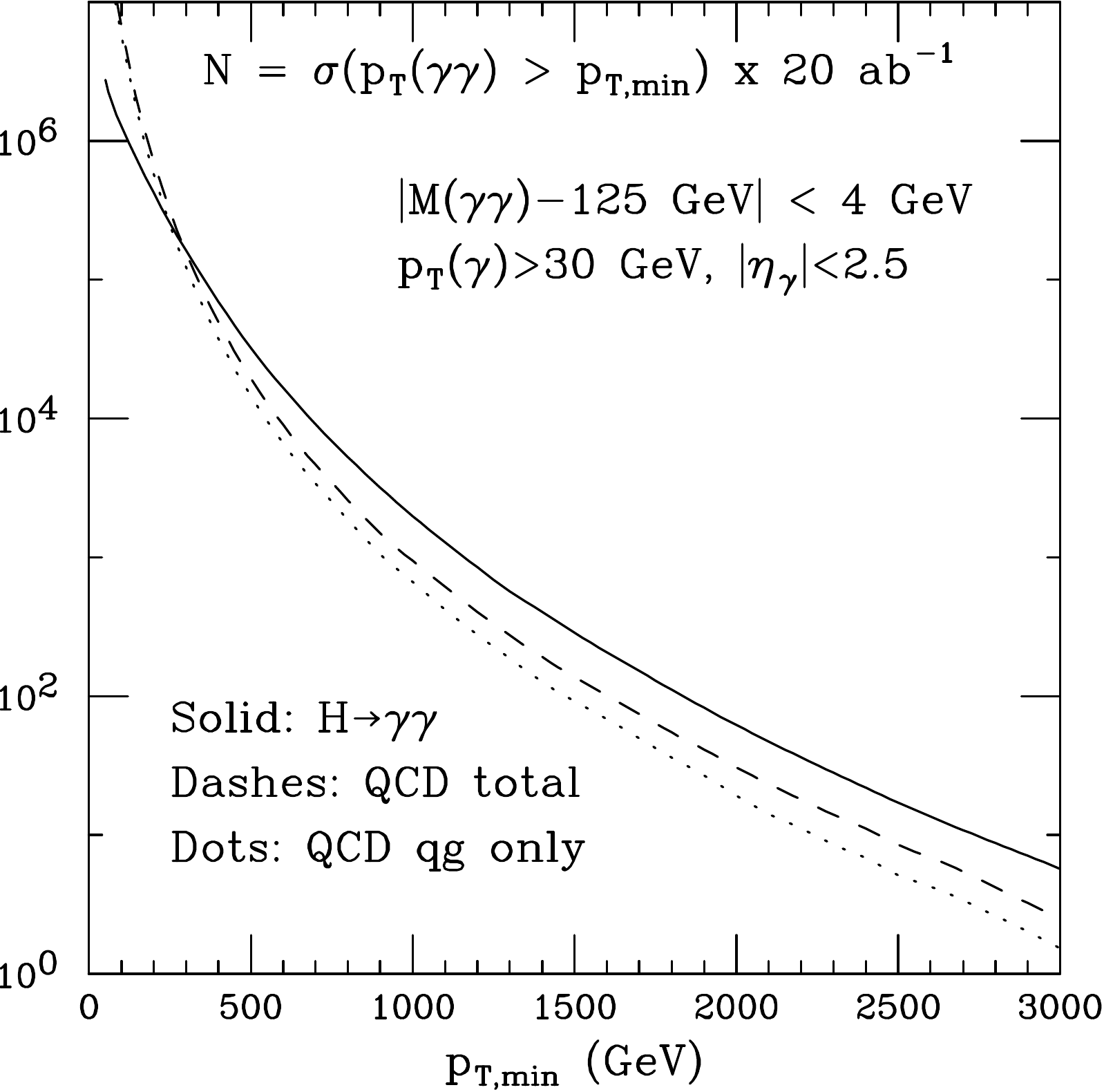}
    \hfill 
    \includegraphics[height=0.45\textwidth]{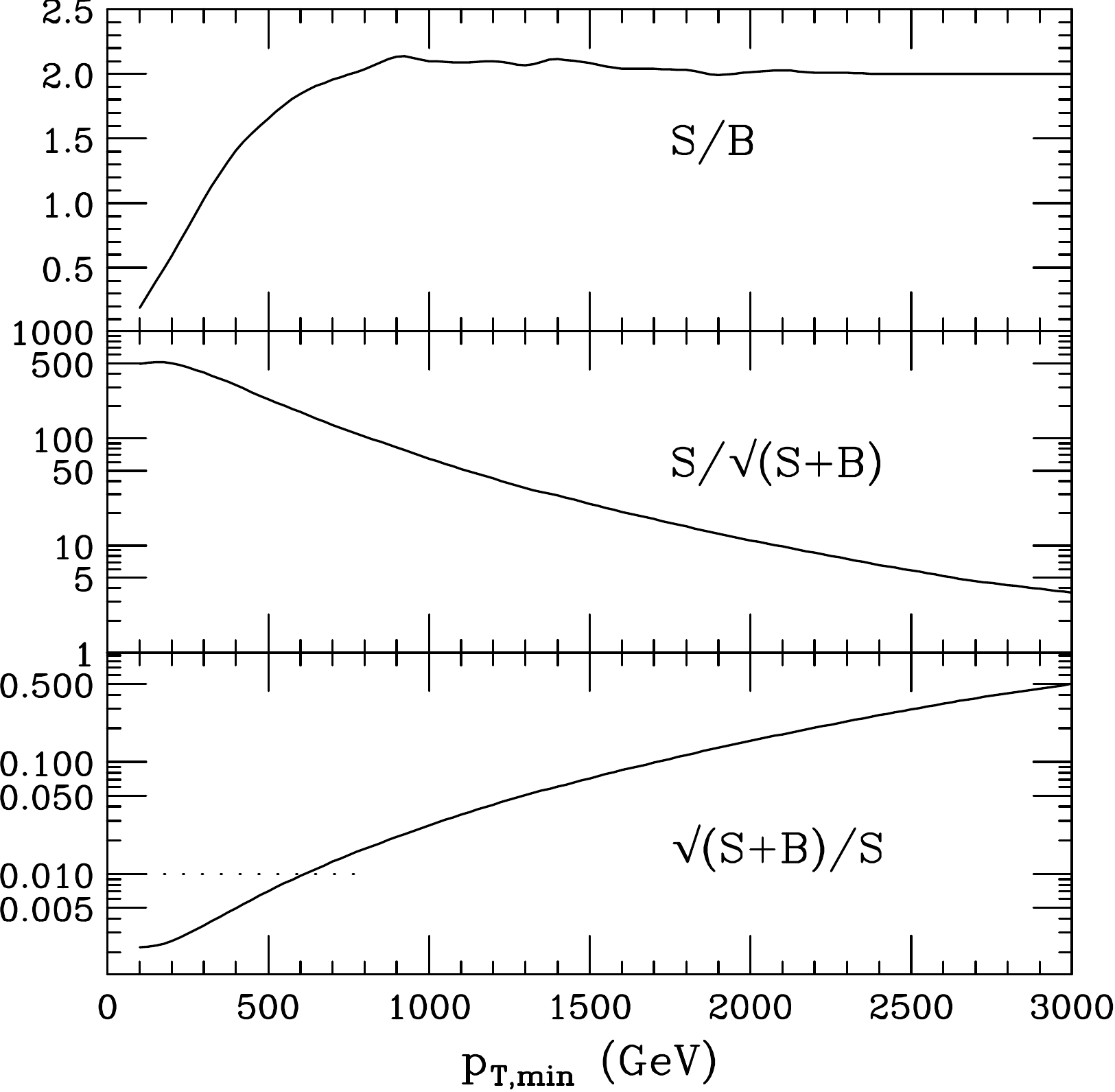}
  \end{center}
  \caption{Left: Integrated transverse momentum rates (20~\iab) for a
    photon pair with mass close to the Higgs mass: signal and QCD
    background. Right: $S/B$, significance of the signal, and potential 
    statistical accuracy of the sample. From Ref.~\cite{Contino:2016spe}}
\label{fig:Hgg}
\end{figure}

For example, as shown in Ref.~\cite{Contino:2016spe}, $S/B$
improves for several final states at large $p_T$. In the case of the
important \textgamma\textgamma\ final state, Fig.~\ref{fig:Hgg} shows
that $S/B$ increases from $\sim3\%$ at low $p_T$ (a value similar to
what observed at the LHC), to $\gsim 1$ at $p_T\gsim 300$~GeV. In this
range of few hundred GeV, some experimental systematics will also
improve, from the determination of the energies (relevant e.g. for the
mass resolution of H$\to$\textgamma\textgamma\ or b\={b}) to the
mitigation of pile-up effects.

\noindent
\rule{5cm}{1pt}\\
     {\bf Exercise:} why do you think S/B improves at
large $p_T(\rm H)$ for a process like $\rm gg\to
H[\to$\textgamma\textgamma]+jet?
  \\ \rule{5cm}{1pt}

\begin{figure}
  \centering
  \includegraphics[width=0.50\columnwidth]{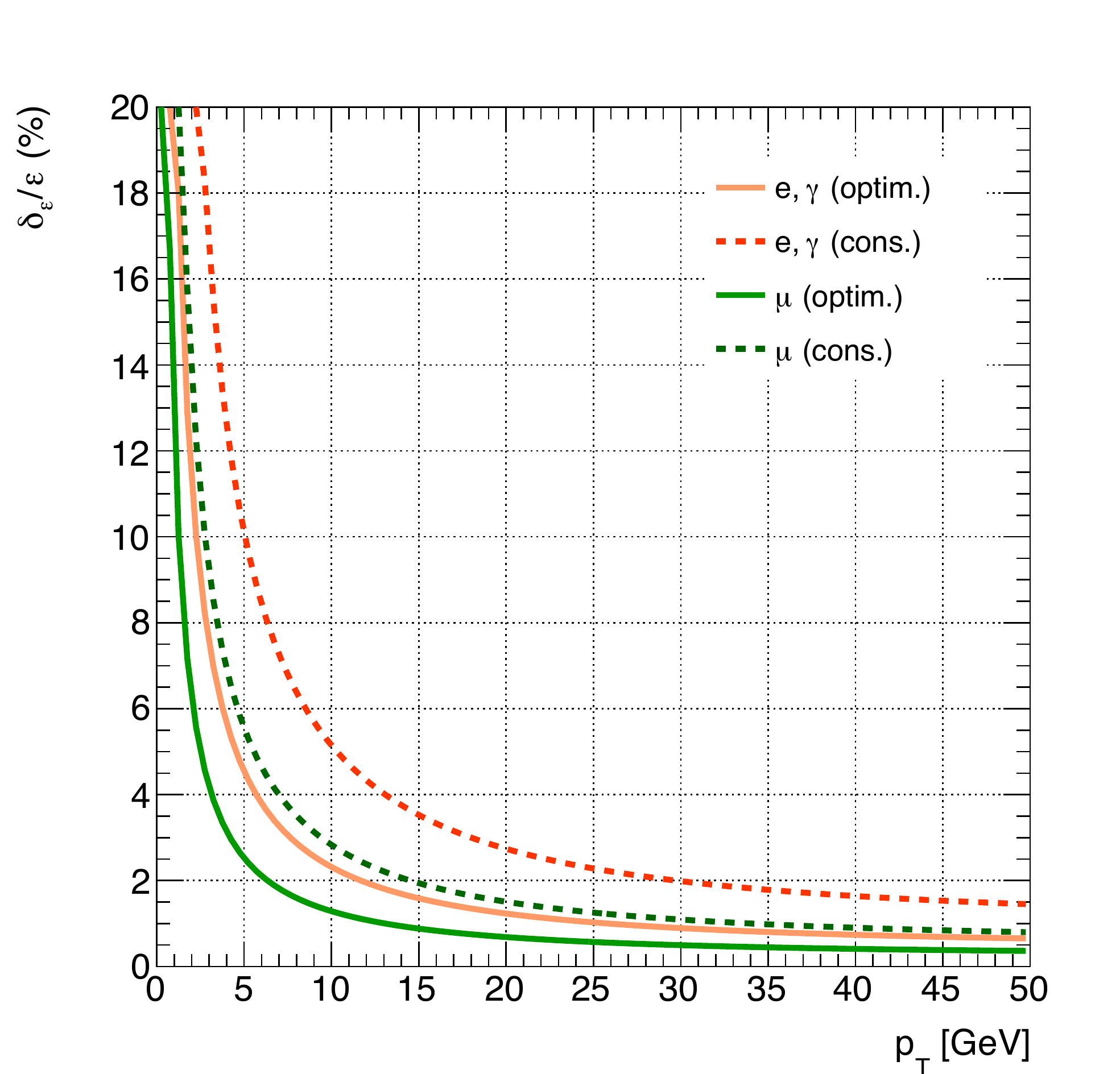}
  \caption{The uncertainty on the reconstruction efficiency of
    electrons, photons and muons as a function of transverse
    momentum. An optimistic (solid) and a conservative (dashed)
    scenario are considered.} 
  \label{fig:effunc}
\end{figure}

\begin{figure}
\centering
\includegraphics[width=0.45\textwidth]{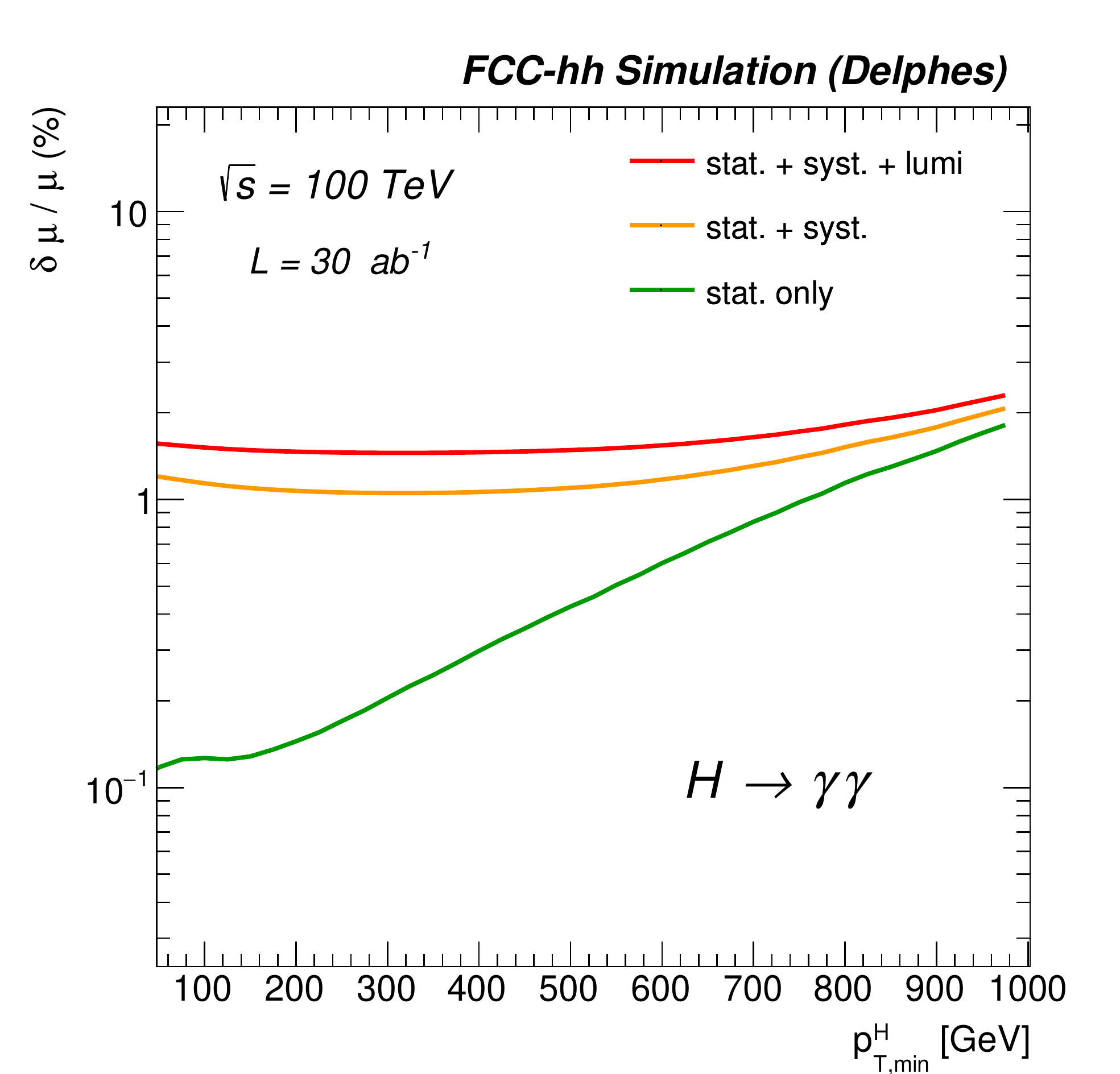}
\includegraphics[width=0.45\textwidth]{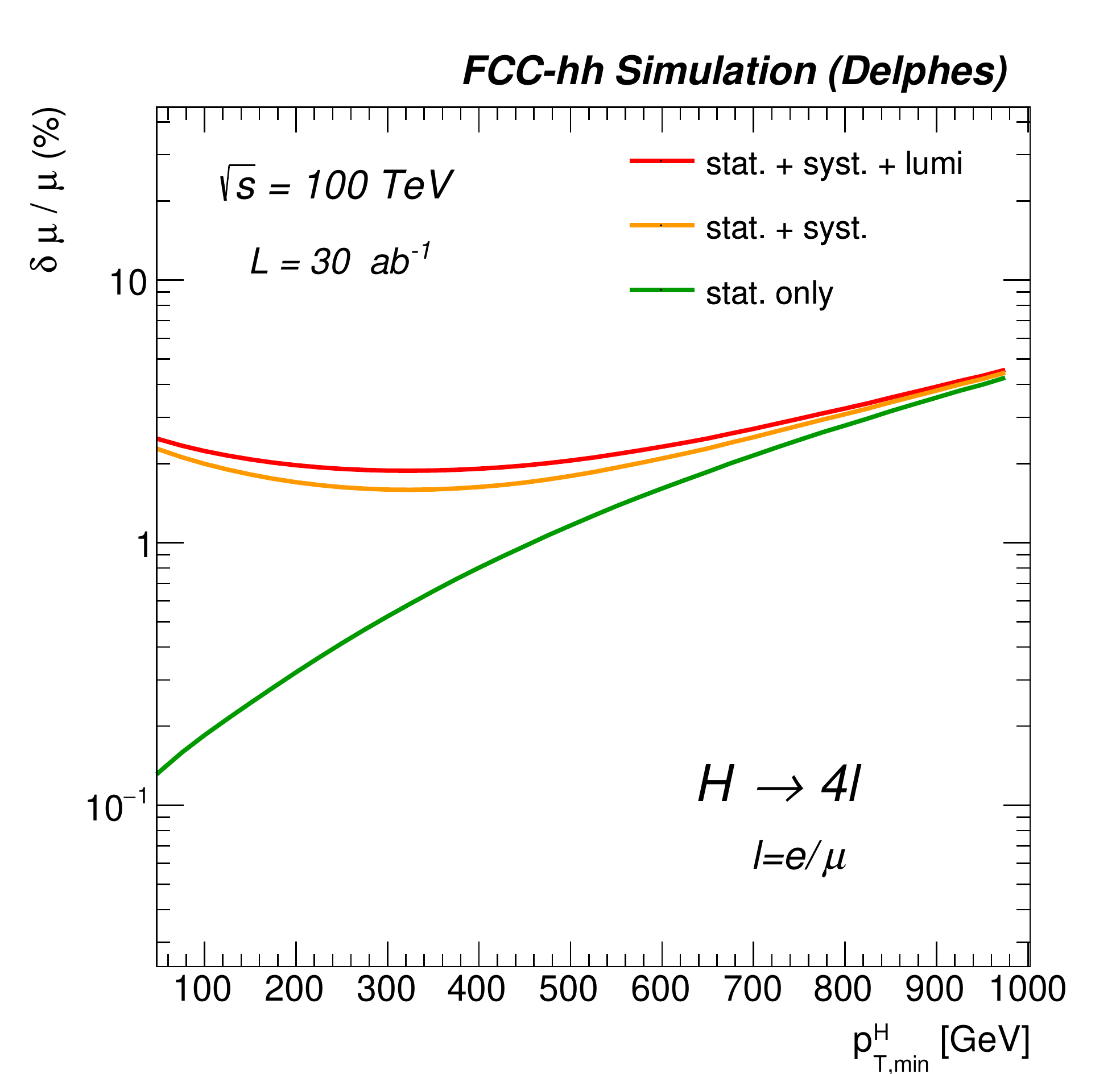}
\\
\includegraphics[width=0.45\textwidth]{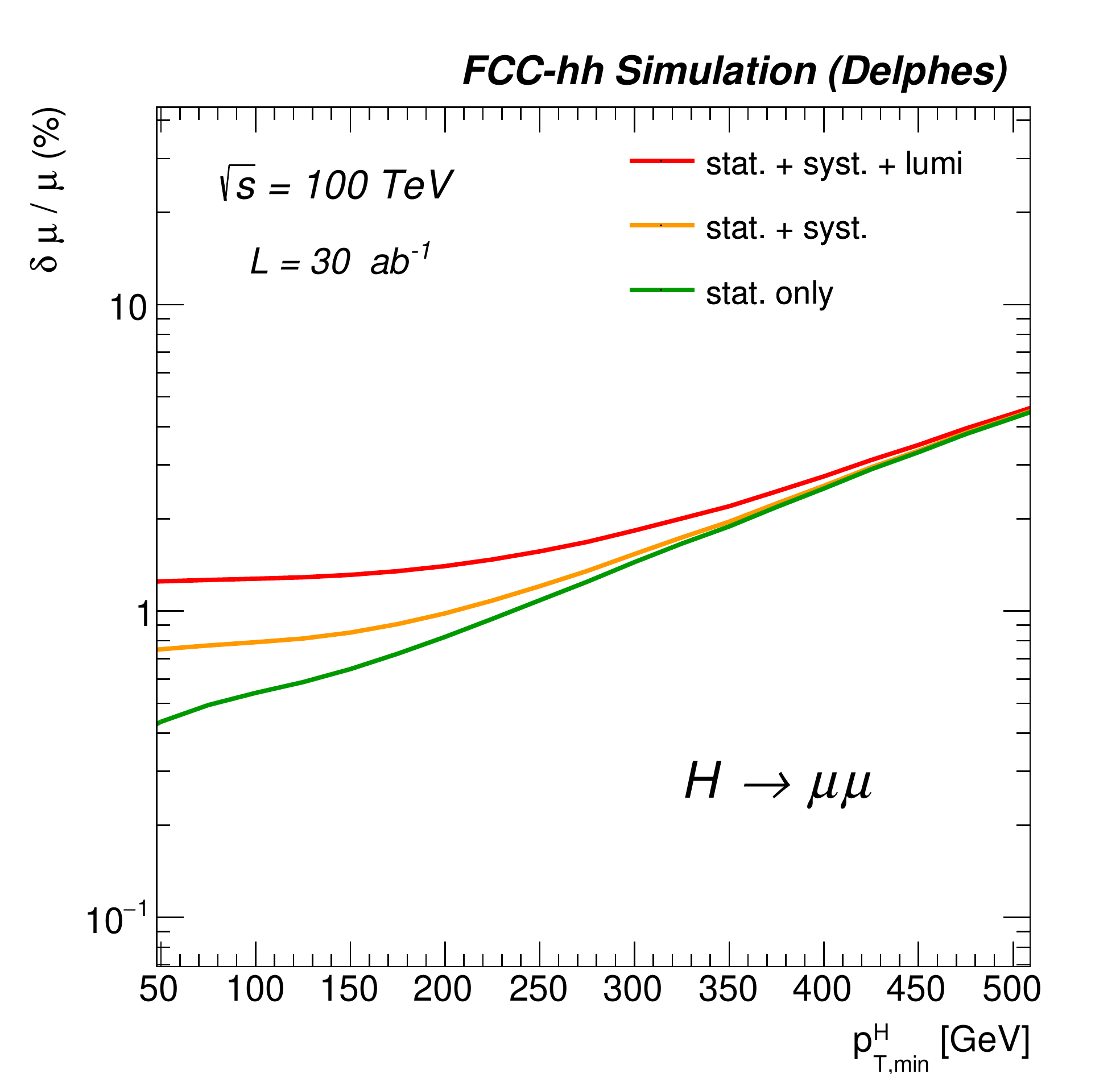}
\includegraphics[width=0.45\textwidth]{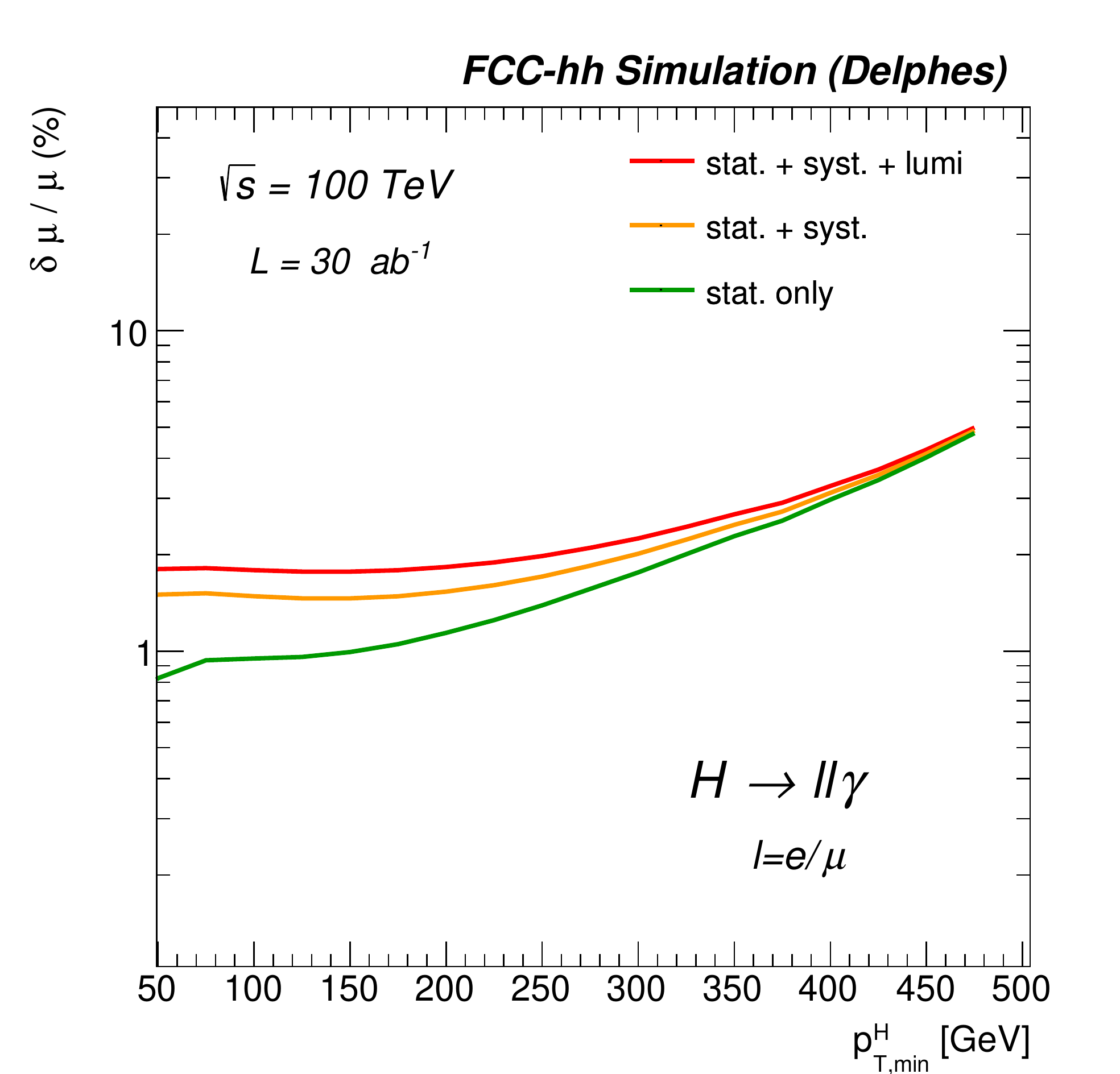}
\caption{Projected precision for the rate measurement of various Higgs final states, in the gg$\to$H production channel. The label ``lumi'' indicates the inclusion of a 1\% overall uncertainty. The systematic uncertainty ``syst'' is defined in the text.}
\label{fig:higgs_rates}
\end{figure}

The analyses carried out so far for FCC-hh are still rather crude when
compared to the LHC standards, but help to define useful targets for
the ultimate attainable precision and the overall detector
performance. The details of the present detector simulations for Higgs
physics at FCC-hh are contained in Ref.~\cite{Borgonovi:2642471}.

The target uncertainties considered include statistics (taking into
account analysis cuts, expected efficiencies, and the possible
irreducible backgrounds) and systematics (limited here to the
identification efficiencies for the relevant final states, and an
overall 1\% to account for luminosity and modelling
uncertainties). While these estimates do not reflect the full
complexity of the experimental analyses in the huge pile-up
environment of FCC-hh, the systematics assumptions that were used are
rather conservative. Significant improvements in the precision of
reconstruction efficiencies would arise, for example, by applying
tag-and-probe methods to large-statistics control samples. Modelling
uncertainties will likewise improve through better calculations, and
broad campaigns of validation against data. By choosing here to work
with Higgs bosons produced at large $p_T$, the challenges met by
triggers and reconstruction in the high pile-up environment are
eased. The projections given here are therefore considered to be
reasonable targets for the ultimate precision, and useful benchmarks
to define the goals of the detector performance.  

The consideration of the reconstruction efficiency of leptons and
photons is relevant in this context since, to obtain the highest
precision by removing global uncertainties such as luminosity and
production modelling, ratios of different decay channels can be
exploited. The reconstruction efficiencies are shown in
Fig.~\ref{fig:effunc} as a function of $p_T$. The uncertainties on the
electron and photon efficiencies are assumed to be fully correlated,
but totally uncorrelated from the muon one. The curves in
Fig.~\ref{fig:effunc} reflect what is achievable today at the LHC, and
it is reasonable to expect that smaller uncertainties will be
available at the FCC-hh, due to the higher statistics that will allow
statistically more powerful data-driven fine tuning. For example,
imposing the identity of the Z boson rate in the ee and
\textmu\textmu\ decay channels will strongly correlate the e and
\textmu\ efficiencies.

The absolute uncertainty expected in the measurement of the production
and decay rates for several final states is shown in
Fig.~\ref{fig:higgs_rates}, as a function of the minimum $p_T(\rm
H)$.
The curves labeled by ``stat+syst'' include the optimal
reconstruction efficiency uncertainties shown in
Fig.~\ref{fig:effunc}. The curves labeled by ``stat+syst+lumi''
include a further 1\%, to account for the overall uncertainty related
to luminosity and production systematics.  The luminosity itself could
be known even better than that by using a standard candle process such
as Z production, where both the partonic cross section and the PDF
luminosity will be pinned down by future theoretical calculations, and
by the FCC-eh, respectively. Notice that the gg luminosity in the mass
range between $m_H$ and several TeV will be measured by FCC-eh at the
few per mille level.

Several comments on these figures are in order. First of all, it
should be noted that the inclusion of the systematic uncertainty leads
to a minimum in the overall uncertainty for $p_T$ values in the range
of few hundred GeV. The very large FCC-hh statistics make it possible
to fully benefit from this region, where experimental systematics are
getting smaller. The second remark is that the measurements of the
Higgs $p_T$ spectrum can be performed with a precision better than
10\%, using very clean final states such as \textgamma\textgamma\ and
4$\ell$, up to $p_T$ values well in excess of 1~TeV, allowing the
possible existence of higher-dimension operators affecting Higgs
dynamics to be probed up to scales of several TeV.

\begin{figure}[t]
\centering
\includegraphics[width=0.45\textwidth]{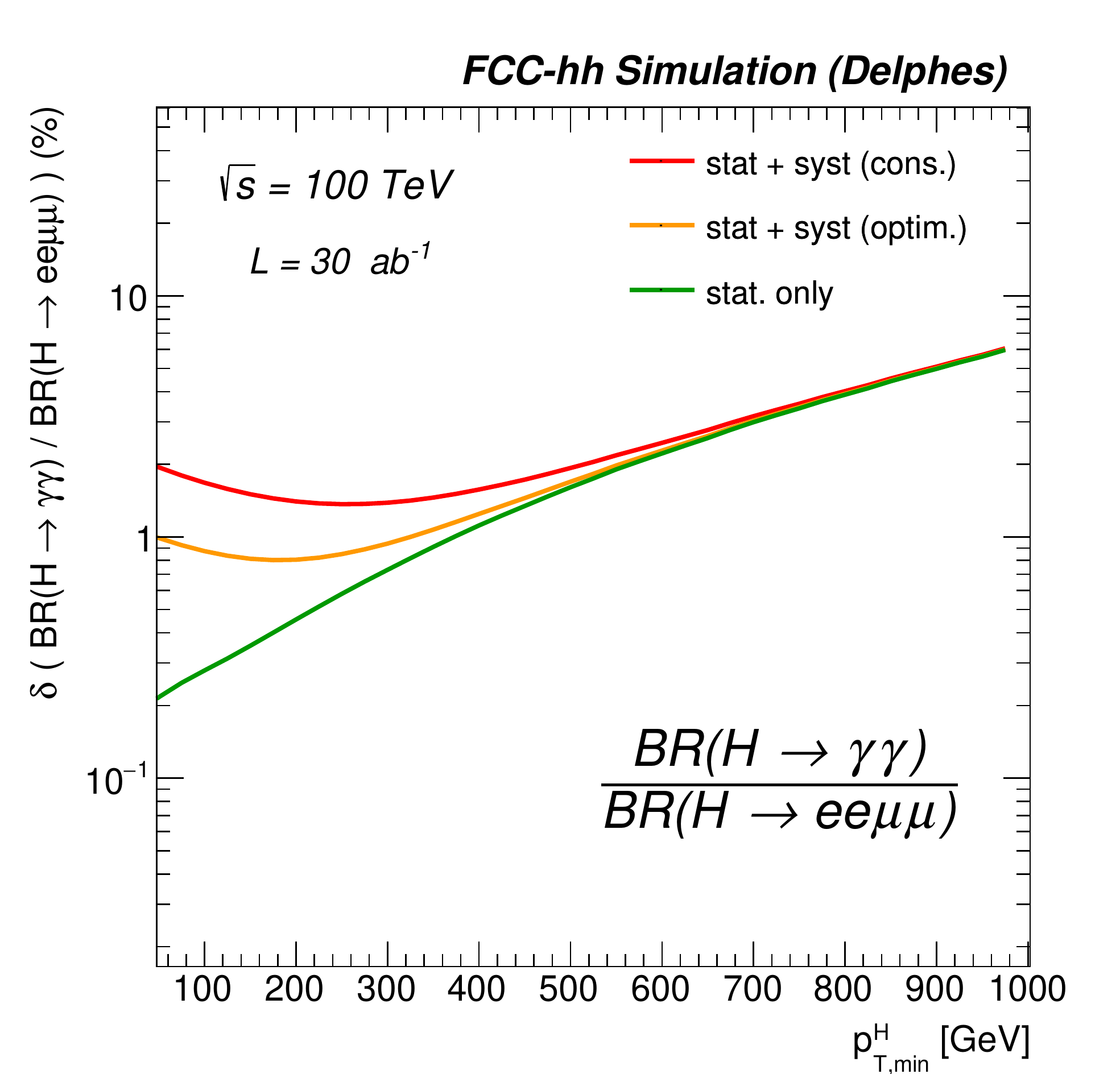}
\includegraphics[width=0.45\textwidth]{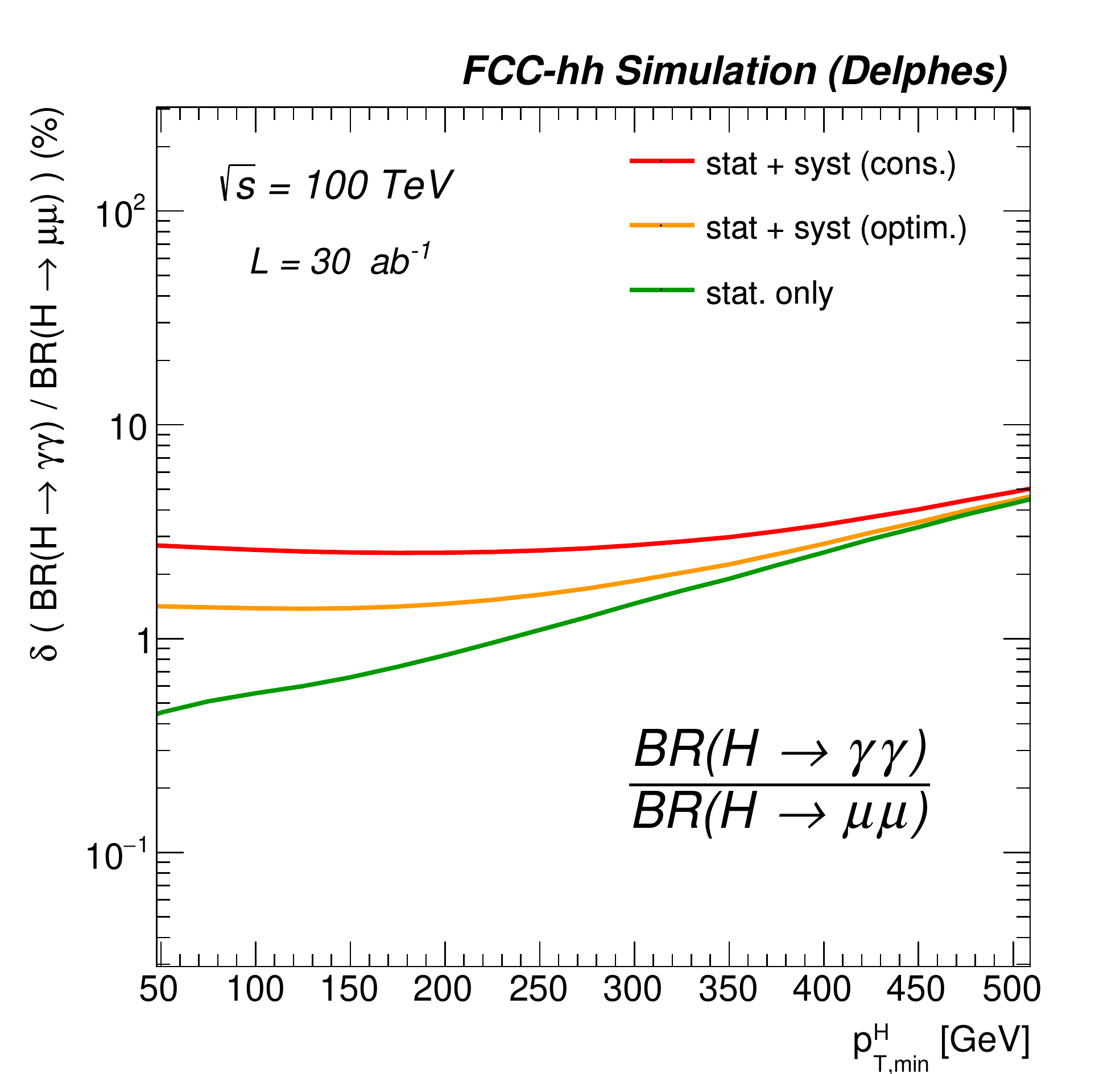}
\\
\includegraphics[width=0.45\textwidth]{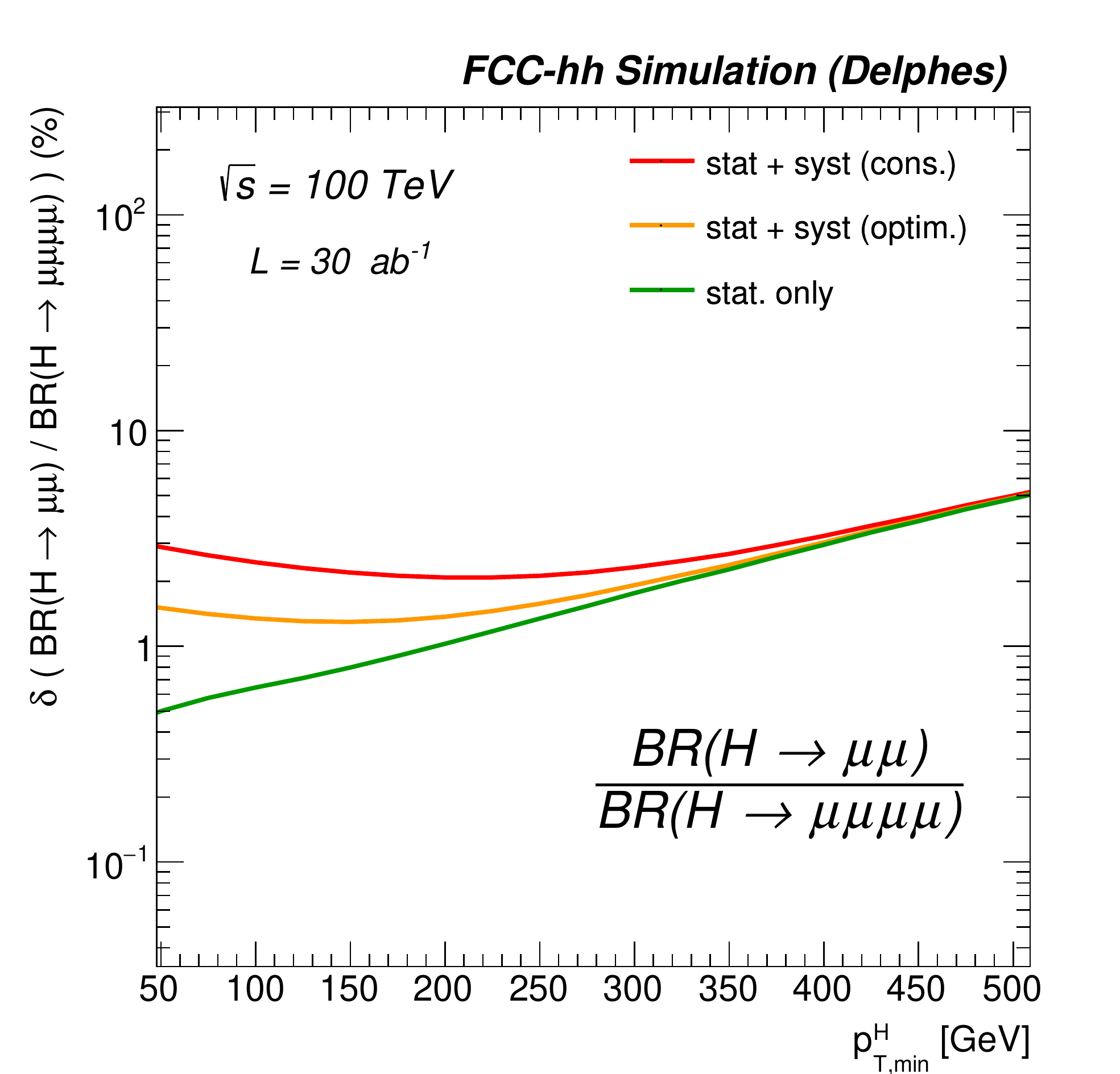}
\includegraphics[width=0.45\textwidth]{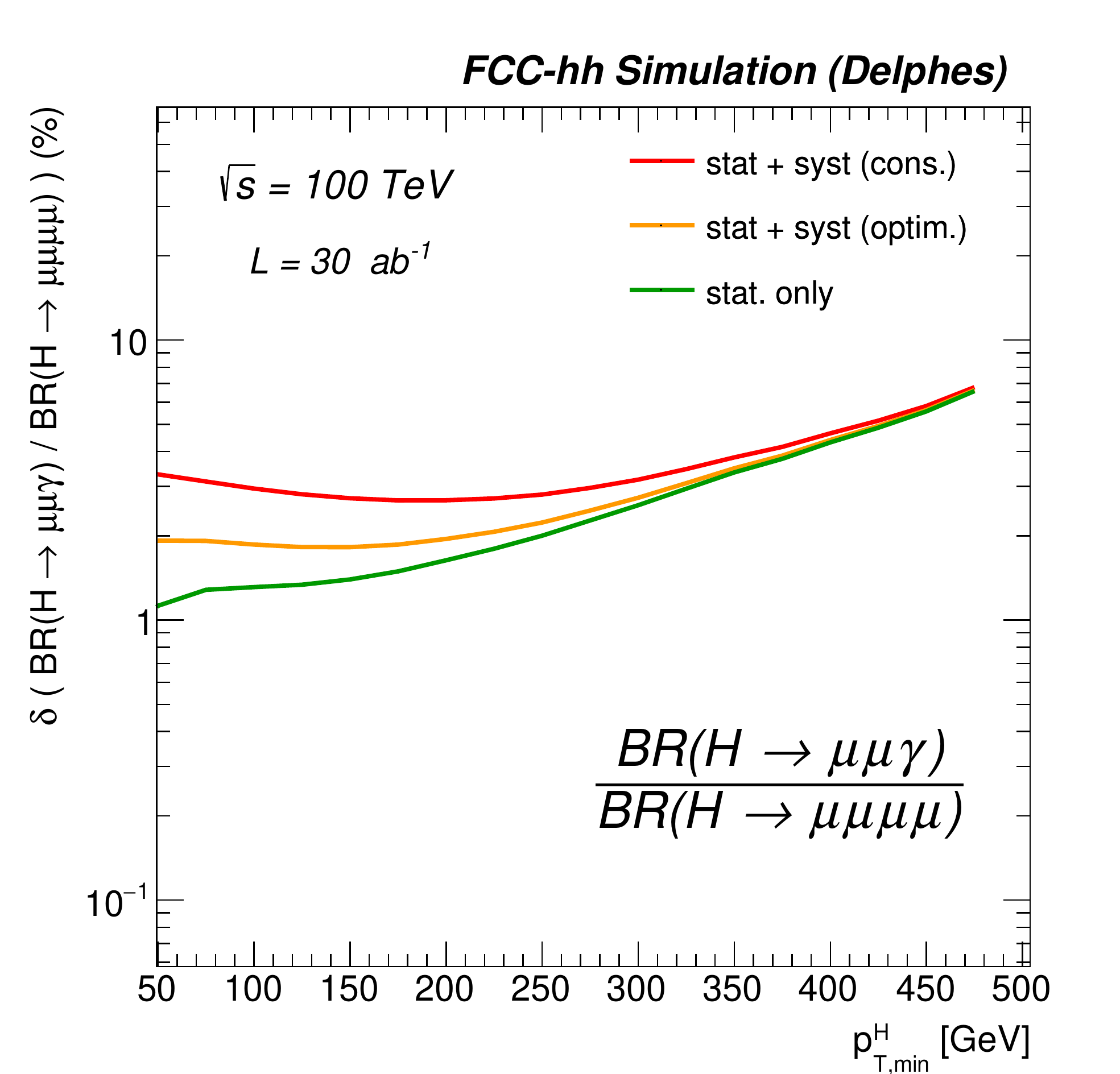}
\caption{Projected precision for the measurement of ratios of rates of
  different Higgs final states, in the gg$\to$H production
  channel. The  systematic uncertainty labels are defined in the
  text.} 
\label{fig:higgs_ratios}
\end{figure}

Independently of future progress, the systematics related to
production modelling and to luminosity cancel entirely by taking the
ratio of different decay modes, provided selection cuts corresponding
to identical fiducial kinematic domains for the Higgs boson are
used. This can be done for the final states considered in
Fig.~\ref{fig:higgs_rates}. Ratios of production rates for these
channels provide absolute determinations of ratios of branching
ratios, with uncertainties dominated by the statistics, and by the
uncorrelated systematics such as reconstruction efficiencies for the
different final state particles. These ratios are shown in
Fig.~\ref{fig:higgs_ratios}. The curves with the systematics labeled
as ``cons'' use the conservative reconstruction uncertainties plotted
in Fig.~\ref{fig:effunc}.
\begin{table}[th]
 \begin{center}
   \caption{\label{tab:Hprecision}
Target
precision for the parameters relative to
the measurement of various Higgs decays, ratios thereof, and of the Higgs
  self-coupling $\lambda$. Notice that Lagrangian couplings have
   a precision that is typically half that of what is shown here, since
   all rates and branching ratios depend quadratically on the couplings.} 
\begin{tabular}{|l|c|c|c|} 
  \hline\hline 
   Observable & Parameter & Precision & Precision \\
  &  & (stat) & (stat+syst+lumi)
   \\ \hline 
  $\mu=\sigma$(H)$\times$B(H$\to$ \textgamma\textgamma)   & $\delta\mu/\mu$ & 0.1\% & 1.5\%
  \\
  $\mu=\sigma$(H)$\times$ B(H$\to$\textmu\textmu)   & $\delta\mu/\mu$ & 0.28\% & 1.2\%
  \\
  $\mu=\sigma$(H)$\times$B(H$\to$ 4\textmu)   &  $\delta\mu/\mu$ & 0.18\% & 1.9\%
  \\
  $\mu=\sigma$(H)$\times$B(H$\to$ \textgamma\textmu\textmu)   &  $\delta\mu/\mu$ & 0.55\% & 1.6\%
  \\
  $\mu=\sigma$(HH)$\times$B(H$\to$\textgamma\textgamma)B(H$\to$b\={b})
  & $\delta\lambda/\lambda$ & 5\% & 7.0\%
  \\
  $R=$ B(H$\to$\textmu\textmu)/B(H$\to$4\textmu)   & $\delta R/R$ & 0.33\% & 1.3\%
  \\
  $R=$ B(H$\to$\textgamma\textgamma)/B(H$\to$ 2e2\textmu)   & $\delta R/R$ & 0.17\% & 0.8\%
  \\
  $R=$ B(H$\to$\textgamma\textgamma)/B(H$\to$ 2\textmu) & $\delta R/R$ & 0.29\% & 1.4\%
  \\
  $R=$ B(H$\to$\textmu\textmu\textgamma)/B(H$\to$\textmu\textmu)   & $\delta R/R$ & 0.58\% & 1.8\%
  \\
$R=\sigma$(t\={t}H)$\times$ B(H$\to$ b\={b})/$\sigma$(t\={t}Z)$\times$B(Z$\to$ b\={b}) & $\delta R/R$ & 1.05\% & 1.9\%
   \\
  $B$(H$\to$ invisible) & $B@$95\%CL & $1\times 10^{-4}$ & $2.5\times
  10^{-4}$
\\
\hline\hline 
\end{tabular}
\end{center}
\end{table}

\begin{figure}[th]
\centering
\includegraphics[width=0.7\textwidth]{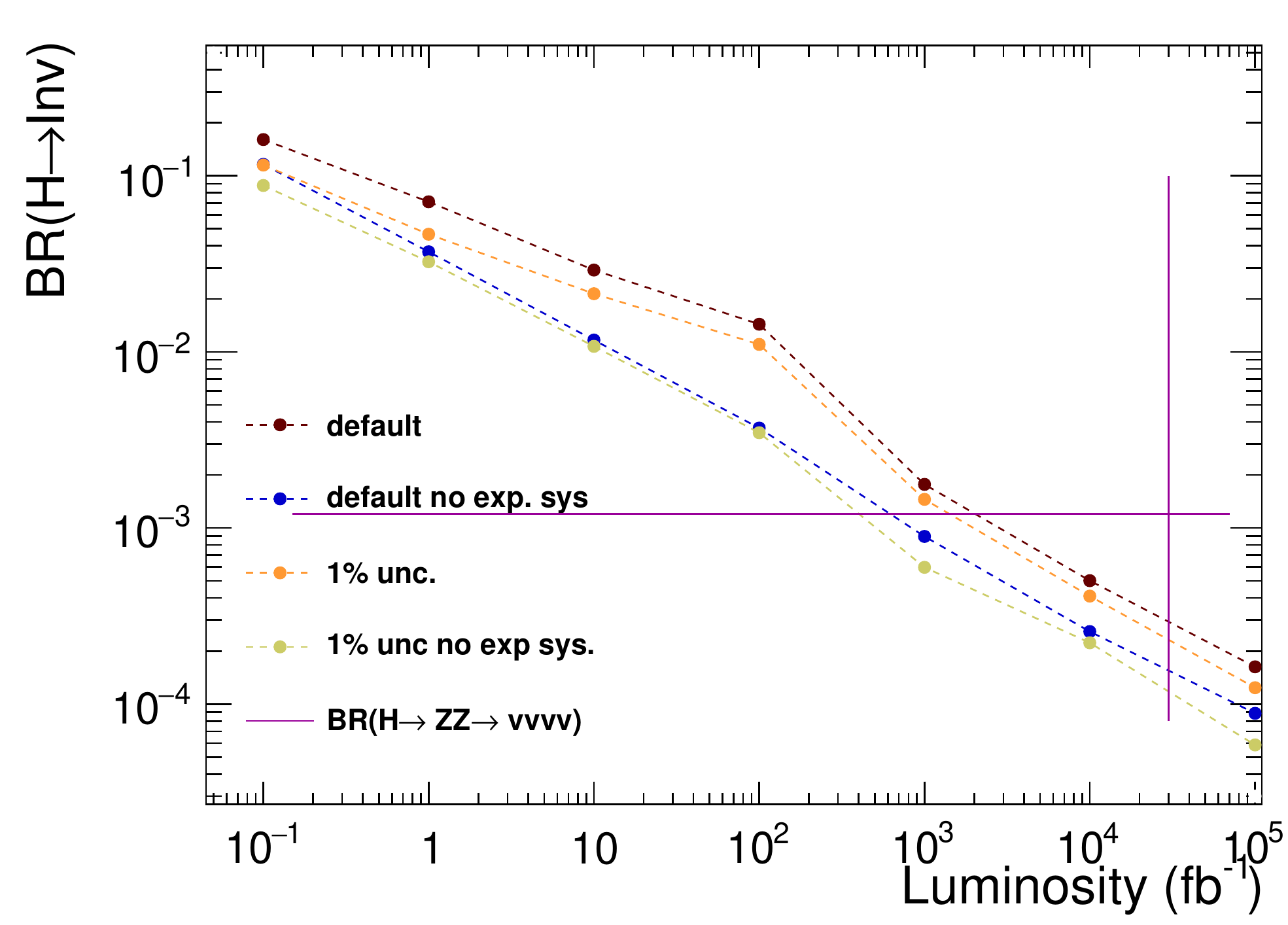}
\caption{Integrated luminosity evolution of the H$\to$invisible branching ratio, under various systematics assumptions.}
\label{fig:ppHinv}
\end{figure}

These results are summarised in Table~\ref{tab:Hprecision}, separately
showing the statistical and systematic uncertainties obtained in our
studies. As remarked above, there is in principle room for further
progress, by fully exploiting data-driven techniques to reduce the
experimental systematics. At the least, one can expect that these
potential improvements will compensate for the current neglect of
other experimental complexity, such as pile-up. The most robust
measurements will involve the ratios of branching ratios. Taking as a
given the value of the HZZ coupling (and therefore $B$(H$\to 4\ell$)),
which will be measured to the few per-mille level by FCC-ee, from the
FCC-hh ratios it could be possible to extract the absolute couplings
of the Higgs to \textgamma\textgamma\ (0.4\%),
\textmu\textmu\ (0.7\%), and Z\textgamma (0.9\%).

\noindent
\rule{5cm}{1pt}\\
{\bf Exercise:} discuss the possible role of precise measurememts of
ratios of BRs in exploring the microscopic origin of potential
deviations from the SM expectations. Which type of models can give
rise to deviations in the ratios considered here? Which models would
leave no signatures in these ratios?
\\
\rule{5cm}{1pt}

The ratio with the t\={t}Z process is considered for the t\={t}H
process, as proposed in Ref.~\cite{Plehn:2015cta}. This allows the
removal of the luminosity uncertainty, and reducing the theoretical
systematics on the production modelling below 1\%. An updated study of
this process, including the FCC-hh detector simulation, is presented
in Ref.~\cite{Borgonovi:2642471}. Assuming FCC-ee will deliver the
expected precise knowledge of $B$(H$\to$b\={b}), and the confirmation
of the SM predictions for the Zt\={t} vertex, the t\={t}H/t\={t}Z
ratio should therefore allow a determination of the top Yukawa
coupling to 1\%.

The limit quoted in Table~\ref{tab:Hprecision} on the decay rate of
the Higgs boson to new invisible particles is obtained from a study of
large missing-$E_T$ signatures. The analysis, discussed in detail in
Ref.~\cite{Borgonovi:2642471}, relies on the data-driven
determination of the leading SM backgrounds from W/Z+jets. The
integrated luminosity evolution of the sensitivity to invisible H
decays is shown in Fig.~\ref{fig:ppHinv}.  The SM decay
H$\to$4\textnu, with branching ratio of about $1.1\times 10^{-3}$,
will be seen after $\sim$1~\iab, and the full FCC-hh statistics will
push the sensitivity to $2\times 10^{-4}$.

\noindent
\rule{5cm}{1pt}\\
{\bf Exercise:} if the Higgs admits an important decay rate to non-SM
particles, $\Gamma_H$ will increase. A larger width will reduce all BRs by a
common factor. Assuming that these decay signatures are elusive in a
hadron collider, discuss how ratios of BRs could still be used to learn
more about their origin.
\\
\rule{5cm}{1pt}

Last but not least, Table~\ref{tab:Hprecision} reports a 7\% expected
precision in the extraction of the Higgs self-coupling $\lambda$. This
result is discussed in more detail in a later Section, with other probes of the Higgs
self-interaction.
 
\begin{figure}[th!]
\centering
\includegraphics[width=0.57\textwidth]{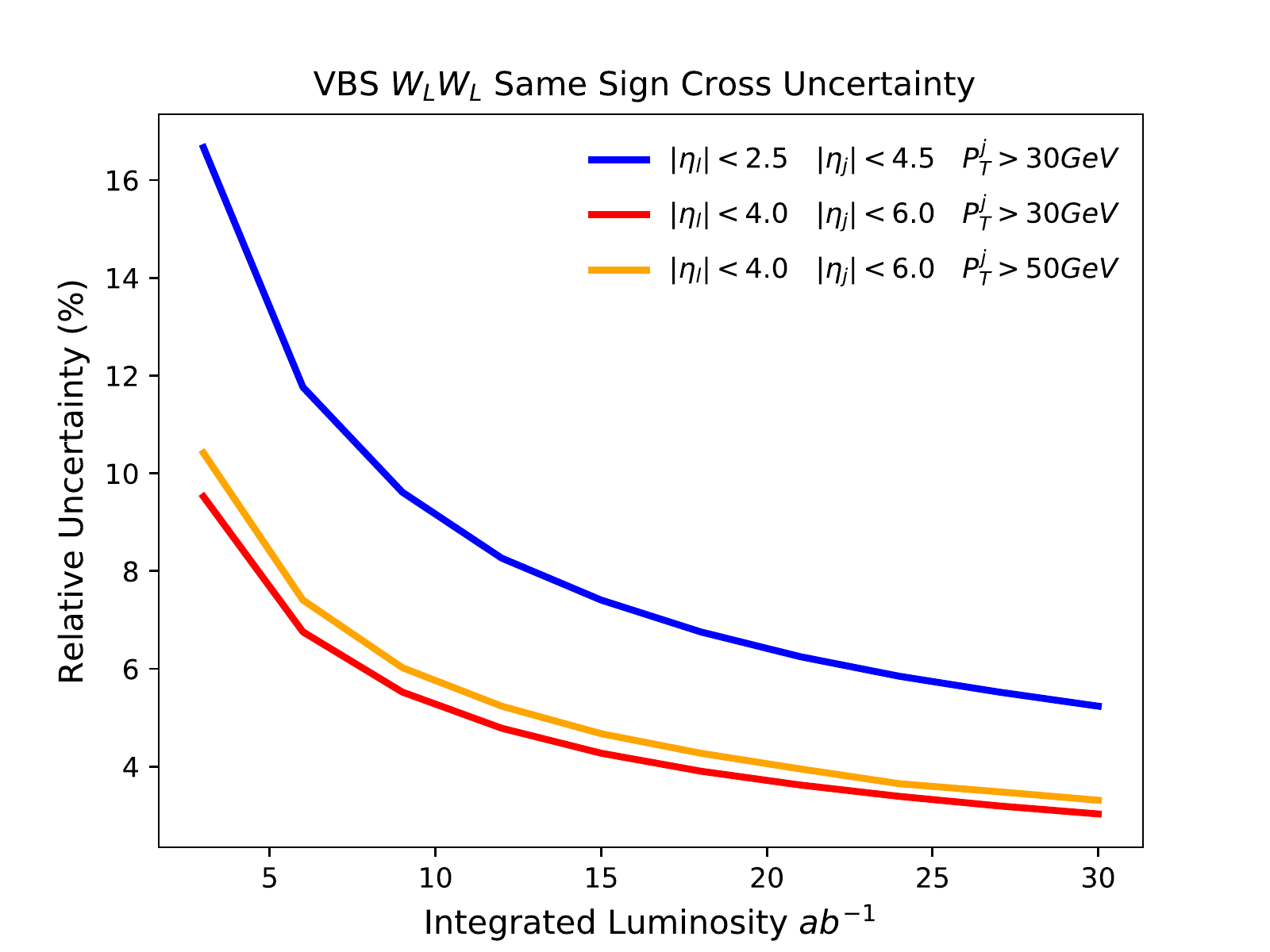}
\hspace{-0.5cm}
\includegraphics[width=0.42\textwidth]{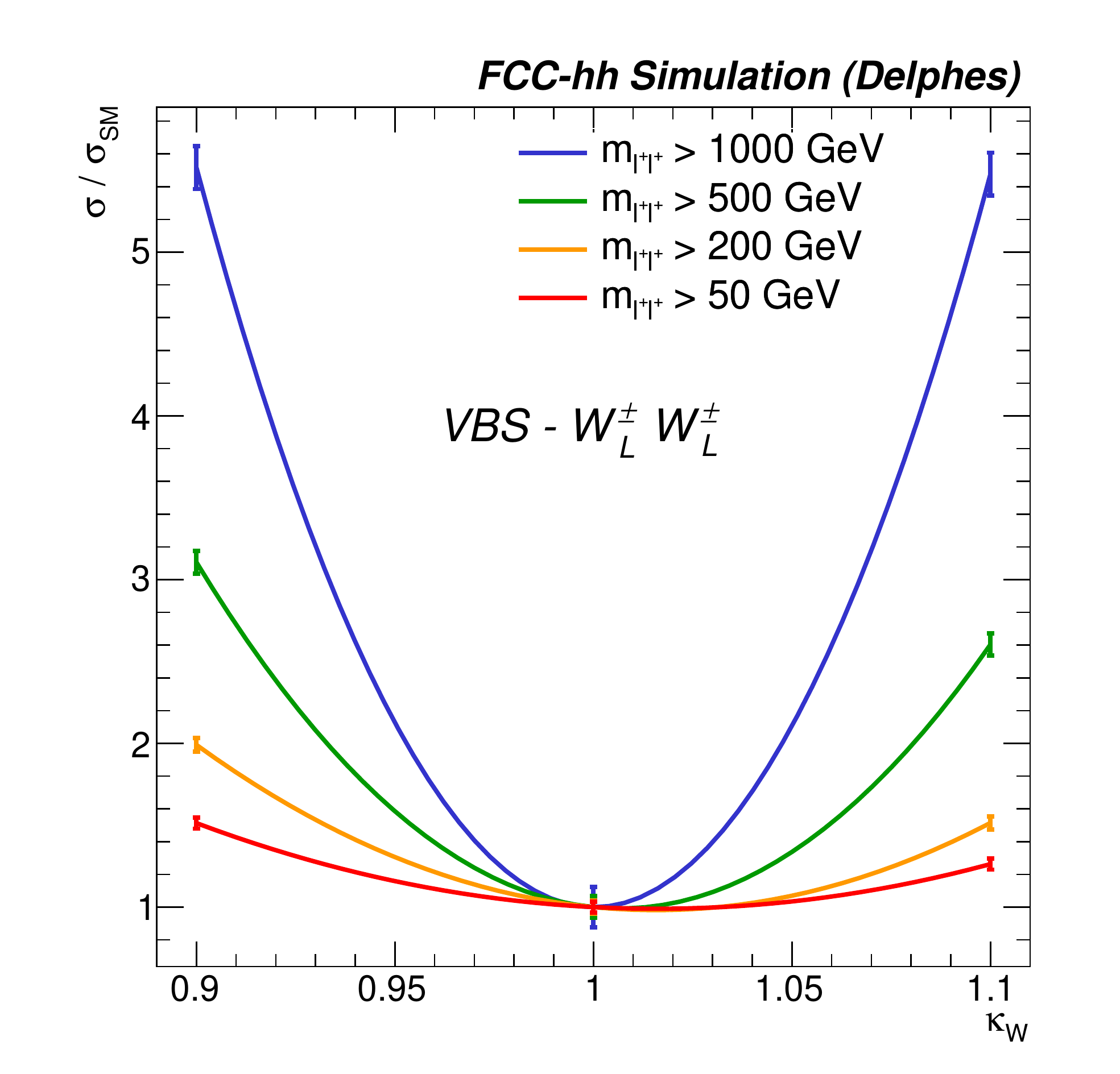}
\caption{Left: precision in the determination of the scattering of same-sign longitudinal W bosons, as function of luminosity, for various kinematic cuts. Right: sensitivity of the longitudinal boson scattering cross section w.r.t. deviations of the WWH coupling from its SM value ($\kappa_W=1$), for various selection cuts on the final-state dilepton invariant mass. The vertical bars represent the precision of the measurement, for 30\,\iab.}
\label{fig:WLWL}
\end{figure}

\begin{table}[ht!]
\begin{center}
\caption{Constraints on the HWW coupling modifier $\kappa_W$ at 68\% CL, obtained for various cuts on the di-lepton pair invariant mass in the $\rm W_LW_L\to HH$ process.}
\label{tab:WLWL}
\begin{tabular}{|c|c|c|c|c|}
\hline \hline
$m_{l^{+}l^{+}}$ cut &  $>50$~GeV &  $>200$~GeV &  $>500$~GeV &  $>1000$~GeV \\
$\kappa_W \in$ & [0.98,1.05] & [0.99,1.04] & [0.99,1.03] & [0.98,1.02] \\
\hline \hline
\end{tabular}
\end{center}
\end{table}

\subsection{Longitudinal Vector Boson Scattering}
\label{sec:WLWL}
The scattering of the longitudinal components of vector bosons is
particularly sensitive to the relation between gauge couplings and the
VVH coupling. A thorough analysis of same-sign $\rm W_L W_L$
scattering, in the context of the FCC-hh detector performance studies,
is documented in Ref.~\cite{Borgonovi:2642471}. The extraction of the
$\rm W^\pm_L W^\pm_L$ signal requires the removal of large QCD
backgrounds ($\rm W^\pm W^\pm$+jets, WZ+jets) and the separation of
large EW background of transverse-boson scattering. The former is
suppressed by requiring a large dilepton invariant mass and the
presence of two jets at large forward and backward rapidities. The
longitudinal component is then extracted from the scattering of
transverse states by exploiting the different azimuthal correlations
between the two leptons. The precision obtained for the measurement of
the $\rm W_L W_L$ cross section as a function of integrated
luminosity, is shown in Fig.~\ref{fig:WLWL} (left). The three curves
correspond to different assumptions about the rapidity acceptance of
the detector and drive the choice of the detector design, setting a
lepton (jet) acceptance out to $\vert\eta\vert=4(6)$. The small change
in precision when increasing the jet cut from $p_T>30$ to
$p_T>50$\,GeV indicates a strong resilience of the results against the
presence of large pile-up. The quoted precision, reaching the value of
3\% at 30\,\iab, accounts for the systematic uncertainties of
luminosity (1\%), lepton efficiency (0.5\%), PDF (1\%) and the shape
of the distributions used in the fit (10\%).  The right plot in
Fig.~\ref{fig:WLWL} shows the impact of rescaling the WWH coupling by
a factor $\kappa_W$. The effect is largest at the highest dilepton
invariant masses, as expected. The measurement precision, represented
by the small vertical bars, indicates a sensitivity to
$\delta\kappa_W$ at the percent level, as shown also in
Table~\ref{tab:WLWL}.

For a review of the Higgs measurements at HE-LHC, see
Refs.~\cite{Mangano:2651294,Cepeda:2019klc}.

\section{Precision EW measurements}
\label{sec:EW}
The Higgs boson forms an integral part of the EW sector, and its properties
are deeply intertwined with those of EW phenomena. A thorough program
of EW measurements goes hand in hand with the study of Higgs
properties, and is an essential complement to it. At the FCC, EW
interactions can be studied from multiple perspectives, extending by
large factors all previous targets of precision and energy reach. We
summarize here the main results from the existing studies, documented
in more detail in Ref.~\cite{Mangano:2651294,Benedikt:2651299}.

The FCC-ee run at the $\rm Z^0$ peak will deliver about $2\times 10^5$
times the LEP statistics, with about $10^{11}$ $\rm \mu^+\mu^-$ or
$\rm \tau^+\tau^-$ final states, and $3\times 10^{12}$ hadronic
decays. The larger statistics w.r.t. LEP will be accompanied by
significant efforts to minimize the systematic uncertainties. For
example, the beam energies will be measured more precisely, and better
detectors will improve the efficiency of b-tagging or the precision in
the absolute luminosity determination. Significant theoretical
improvements in the calculation of higher-loop EW and QCD corrections
are also foreseen, and necessary to fully exploit the potential
improvement by over two orders of magnitude in the statistical
precision. 

$\alpha_{\rm EM}(M_Z)$ is a crucial input parameter to interpret SM
precision observables. The EM coupling at the scale of the electron
mass is the best known fundamental constant of nature, but its
renormalization group evolution to the scale of weak interactions is
subject to important uncertainties, due to non-perturbative hadronic
physics, which enters the photon self-energy corrections as  $Q^2$
evolves through the region of the hadronic resonances. The systematic
uncertainties of the experimental data on $\sigma(\rm e^+e^-
\to$hadrons$)$ end up dominating the precision of the extrapolation to
$Q^2=M_Z^2$, which is limited to the level of $10^{-4}$. Dedicated
runs at $\sqrt{s}$=87.7 and 93.9~GeV will extract $\alpha_{\rm
  EM}(M_Z)$ directly (namely without an extrapolation in $Q$) from the
energy-dependence of the $\rm \mu^+\mu^-$ forward-backward asymmetry,
improving the current uncertainty by a factor of 4.

Forward-backward and polarization asymmetries will also allow to
reduce by a factor of 30-50 the uncertainty in $\sin^2\theta_W$. I
recall that today's determination of the weak mixing angle,
$\sin^2\theta^{\rm lept}_{\rm eff} = 0.23153 \pm
0.00015$~\cite{ALEPH:2005ab}, is dominated by the combination of two
precise measurements (the b-quark forward-backward asymmetry from LEP
and the left-right polarization asymmetry from SLD), which differ
among themselves by 3.2 standard deviations. While future HL-LHC
data~\cite{Azzi:2019yne} will provide an independent determination of
the weak mixing angle with a precision approaching the current LEP/SLD
one, using the lepton charge asymmetry in Z boson decays, it is only
with the future FCC-ee data that this puzzling result will be
clarified.

\begin{figure}
\begin{center}
\includegraphics*[width=0.85\textwidth]{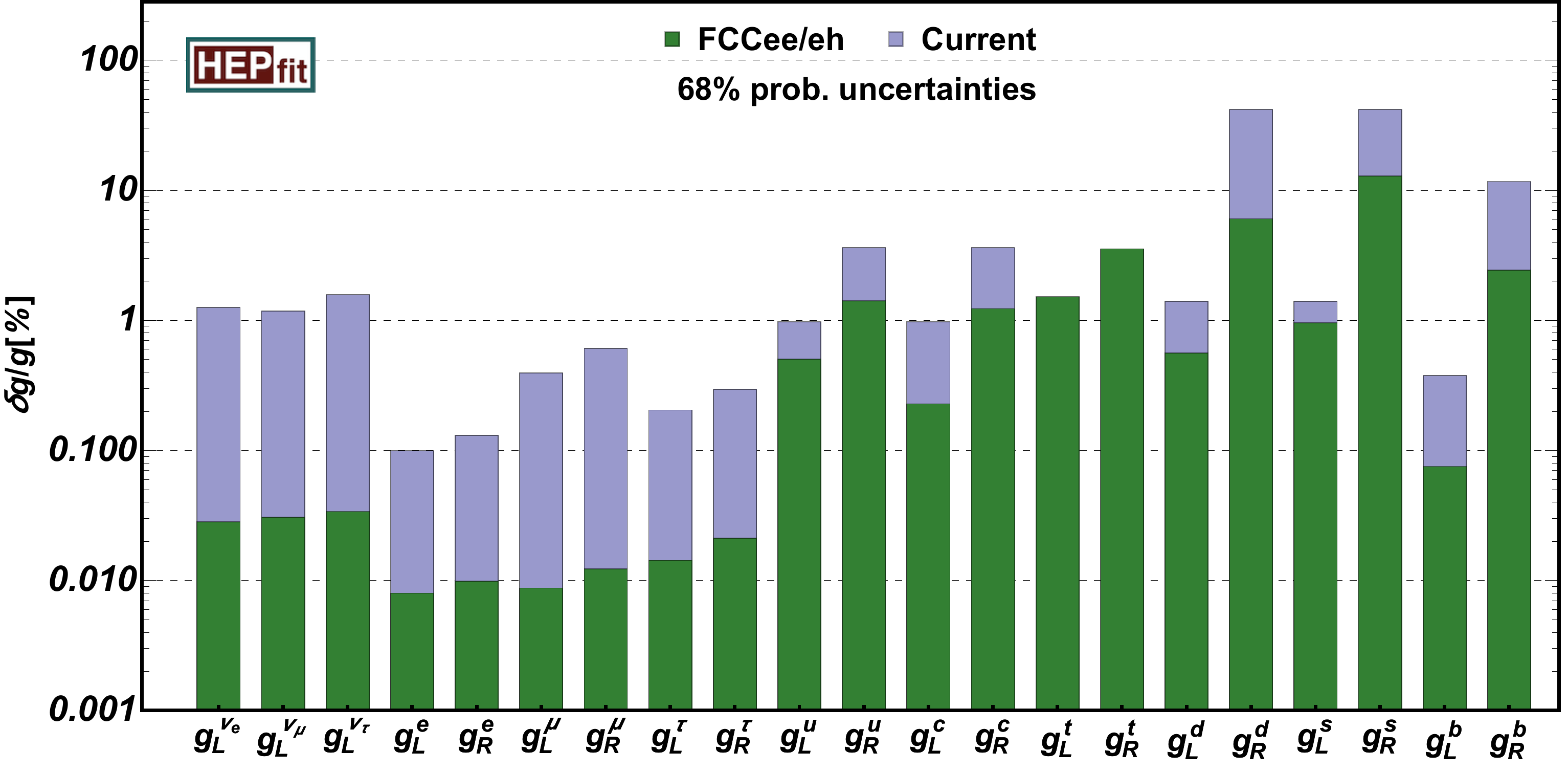}
\end{center}
\caption{Sensitivity, at the 1-$\sigma$ level, to deviations of the
  neutral current couplings resulting from a global EFT fit at the
  dimension-6 level to EW precision measurements at FCC-ee and
  FCC-eh. From~\cite{Mangano:2651294}.} 
\label{fig:Zff} 
\end{figure}

The Z-decay asymmetries will also help improving the measurement of
vector and axial couplings of the leptons and of the charm and bottom
quarks. In absence of a reliable technique to distinguish Z decays to
the different lighter quarks (u, d and s), the most precise
determination of their couplings to the neutral current will come from
FCC-eh. There, a simultaneous fit to the light quarks EW couplings and
to the PDFs, using both charged and neutral current data, will
disentangle the individual quarks and allow the measurement of their
respective vector and axial couplings. The projection in in
Fig.~\ref{fig:Zff} for the precision of all fermionic couplings, from
a global fit~\cite{deBlas:2016ojx,hepfitsite} to both FCC-ee and
FCC-eh data treating each lepton and quark flavour as independent,
shows the improvement expected with respect to today's knowledge.

The measurement of the total Z width $\Gamma_Z$, and of its visible
fraction, will allow to extract the invisible component of
$\Gamma_Z$. Today, the number of neutrino species obtained from the
LEP data is $N_\nu=2.984\pm 0.008$, which is low by two standard
deviations. A deficit in the neutrino counting from Z decays could be
attributed to a violation of unitarity in the neutrino mixing matrix,
or to the presence of right-handed
neutrinos~\cite{Jarlskog:1990kt}. FCC-ee will improve the precision on
$N_\nu$ by almost a factor of 10, down to 0.001.

The $10^8$ pairs of W bosons produced at the two energies of
$\sqrt{s}=$157.5 and 162.5~GeV will reduce the uncertainty on the W
mass, $m_W$, to 0.5~MeV, and of its width to 1.2 MeV. The limited
statistics of W bosons from LEP2 left us with a puzzling $\sim 3\sigma$ discrepancy
between the decay branching ratio of the W to the tau lepton,
$(11.38\pm0.21)\%$, and to the e and \textmu ($(10.71\pm 0.16)\%$ and
$(10.63\pm 0.15)\%$, resp.). FCC-ee can reduce these uncertainties by
almost two orders of magnitude, greatly increasing the sensitivity to
possible violations of lepton flavour universality, a topic that is
receiving great attention nowadays. For comparison, the $10^{11}$ Z decays to individual
leptons will allow to test neutral-current lepton universality at the
level of of $10^{-5}$.  \texttau{} semileptonic decays, and the
\texttau{} lifetime, could achieve a sensitivity to deviations from lepton flavour
universality in weak charged currents at a similar level.

\begin{figure}[th]
\begin{center}
\includegraphics*[width=0.8\textwidth]{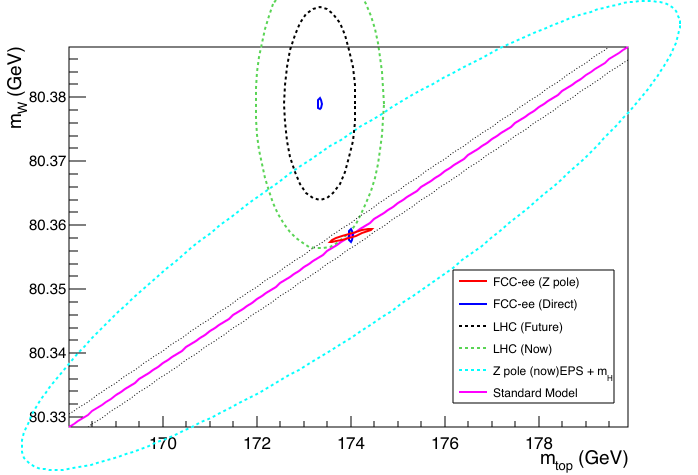}
\end{center}
\caption{\label{fig:GFitterCDR} Contours of 68\% confidence level in
  the $(m_{\rm top}, m_{\rm W})$ plane obtained from fits of the SM to
  the EW precision measurements offered by the FCC-ee, under the
  assumption that all relevant theory uncertainties can be reduced to
  match the experimental uncertainties: the red ellipse is obtained
  from the FCC-ee measurements at the Z pole, while the blue ellipses
  arise from the FCC-ee direct measurements of the W and top masses.
  The two dotted lines around the SM prediction illustrate the
  uncertainty from the Z mass measurement if it were not improved at
  the FCC-ee. The green ellipse corresponds to the current W and top
  mass uncertainties from the Tevatron and the LHC. The potential
  future improvements from the LHC are illustrated by the black dashed
  ellipse. Plot from Ref.~\cite{Benedikt:2651299}, obtained using the GFitter code
  introduced in~\cite{Baak:2014ora}.}
\end{figure}

A collection of the various EW precision measurements possible at the
FCC-ee, including those relative to the top quark properties, is shown
in Table~\ref{tab:FCCEW}. Their overall impact  in testing the
SM relation between the top and W masses is shown in Fig.~\ref{fig:GFitterCDR}.

\setlength{\tabcolsep}{2pt}
\renewcommand{\arraystretch}{0.8}
\begin{table}[htbp!]
\centering
\caption{Measurement of selected EW quantities at FCC-ee, compared
  with the present precisions~\cite{Benedikt:2651299}   \label{tab:FCCEW}}
\begin{tabular}{|l|rcl|c|c|r|}
\hline \hline 
Observable  & present &  &          &  FCC-ee  &  FCC-ee  &  Comment and   \\ 
            & value  &$\pm$& error  &  Stat. &   Syst.     &  dominant exp. error \\ 
\hline \hline 
$ \mathrm{ m_Z  ~(keV) } $  &  91186700   & $\pm$ &  2200    & 5  & 100  & From Z line shape scan  \\ 
$  $  &  & &    &   &  &  Beam energy calibration  \\
\hline
$ \mathrm{  \Gamma_Z  ~(keV) } $  & 2495200   & $\pm$ &  2300    & 8  & 100  & From Z line shape scan  \\
$  $  &  & &    &   &   &  Beam energy calibration  \\
\hline
$ \mathrm{  R_{\ell}^{Z}} ~(\times 10^3) $  & 20767 & $\pm$ &  25   & 0.06   & 0.2-1.0   &  ratio of hadrons to leptons \\
$  $  &  & &    &   &   &  acceptance for leptons  \\
\hline
$ \mathrm{ \alpha_{s} (m_Z) } ~(\times 10^4) $  & 
 1196 & $\pm$ &  30  &  0.1  &  0.4-1.6  &   
from $\mathrm{  R_{\ell}^{Z}}$ above~\cite{dEnterria:2015kmd}\\
\hline
$ \mathrm{  R_b} ~(\times 10^6) $  & 216290 & $\pm$ &  660   & 0.3   &  <60  &  ratio of $\rm{ b\bar{b}}$  to hadrons  \\
$  $  &  & &    &   &   &  stat. extrapol. from SLD
~\cite{Abe:2005nqa}\\
\hline
$ \mathrm{\sigma_{had}^0} ~(\times 10^3)$ (nb) & 41541 & $\pm$ &  37   & 0.1  &  4  &  peak hadronic cross-section  \\
$  $  &  & &    &   &   &  luminosity measurement  \\
$ \mathrm{  N_{\nu}}  (\times 10^3) $  & 2991  & $\pm$ &  7   & 0.005   &  1  &  Z peak cross sections \\
$  $  &  & &    &   &  &   Luminosity measurement \\
\hline
$ \mathrm{ sin^2{\theta_{W}^{\rm eff}}} (\times 10^6) $  & 231480   & $\pm$ &  160   & 3   &  2 - 5  &   
from $ \mathrm{ A_{FB}^{{\mu} {\mu}}}$  at Z peak\\
$  $  &  & &    &   &   &  Beam energy calibration  \\
\hline
$ \mathrm{ 1/\alpha_{QED} (m_Z) } (\times10^3) $  & 128952 
  & $\pm$ &  14   & 4   &  small  &   
from $ \mathrm{ A_{FB}^{{\mu} {\mu}}}$ off peak~\cite{Janot:2015gjr}\\
\hline
$ \mathrm{  A_{FB}^b,0} ~(\times 10^4) $  & 992 & $\pm$ &  16   & 0.02   &  1-3  &  b-quark asymmetry at Z pole  \\
$  $  &  & &    &   &   &  from jet charge \\
\hline

$ \mathrm{A_{FB}^{pol,\tau} ~(\times 10^4)} $  & 1498 & $\pm$ &  49   & 0.15   &  <2  &  \texttau{} polarisation and charge asymmetry  \\
$  $  &  & &    &   &   &  \texttau{} decay physics \\
\hline \hline 
$ \mathrm{ m_W  ~(MeV) } $  &  80350   & $\pm$ &  15    & 0.5  & 0.3  & From WW threshold scan \\ 
$  $  &  & &    &   &  &  Beam energy calibration  \\
\hline
$ \mathrm{  \Gamma_W  ~(MeV) } $  & 2085   & $\pm$ &  42    & 1.2  & 0.3  & From WW threshold scan \\
$  $  &  & &    &   &   &  Beam energy calibration  \\
\hline
$ \mathrm{ \alpha_{s} (m_W) }  (\times 10^4)$  & 
 1170   & $\pm$ & 420 &  3  & small  &   
  from $ \mathrm{R_{\ell}^{W} }$~\cite{dEnterria:2016rbf}\\
\hline
$ \mathrm{  N_{\nu}}  (\times 10^3) $  & 2920 & $\pm$ &  50   & 0.8   & small   &   ratio of invis. to leptonic \\
$  $  &  & &    &   &  & in radiative Z returns  \\
\hline
\hline \hline
$ \mathrm{ m_{top}  ~(MeV) } $  &  172740   & $\pm$ &  500    & 17  & small  & From $\mathrm {t\bar{t}}$ threshold scan \\ 
$  $  &  & &    &   &  &  QCD errors dominate  \\
\hline
$ \mathrm{ \Gamma_{top}  ~(MeV) } $  &  1410   & $\pm$ &  190    & 45  & small  & From $\mathrm {t\bar{t}}$ threshold scan \\ 
$  $  &  & &    &   &  &  QCD errors dominate \\
\hline
$ \mathrm{ \lambda_{top}/\lambda_{top}^{SM}   } $  &   1.2 
   & $\pm$ &  0.3    & 0.1  & small  & From $\mathrm {t\bar{t}}$ threshold scan \\ 
$  $  &  & &    &   &  &  QCD errors dominate \\
\hline
$ \mathrm{ ttZ ~couplings   } $  &   
   & $\pm$ &  30\%   & $0.5 - 1.5$\%  & small  & From $ \mathrm{E_{CM}=365 GeV}$ run  \\ 
\hline \hline
\end{tabular} 
\end{table}
\setlength{\tabcolsep}{6pt}

\subsection{Complementarity of EW and Higgs measurements}
The best framework to expose the complementarity between Higgs and EW
probes of new physics at large scales, is that of the SM effective
field theory (SMEFT~\cite{Buchmuller:1985jz,
  Grzadkowski:2010es,Deflorian:2016spz}). Here one assumes that, as in
the SM, the Higgs boson transforms as doublet of $SU(2)_L$, and
considers all $SU(2)_L\times U(1)_Y$ invariant operators, classified
according to their dimension:
\begin{equation}
{\cal L}_{\mathrm{Eff}}=\sum_{d=4}^\infty \frac{1}{\Lambda^{d-4}}{\cal
  L}_d={\cal L}_{\mathrm{SM}}+\frac{1}{\Lambda} {\cal L}_5
+\frac{1}{\Lambda^2} {\cal L}_6 + \ldots,~~~~~~{\cal L}_d=\sum_i C_i
{\cal O}_i.
\label{eq:LEff}
\end{equation}
At dimension 4, one finds the SM
itself. At dimension 5 appear the operators that generate Majorana neutrino
masses. At dimension 6 one first finds operators that, parameterizing
the low-energy behaviour of new interactions beyond the EW scale, lead
to modifications of EW and Higgs observables. Focusing on lepton and
baryon-number conserving operators, and assuming flavour universality,
leaves a basis of 59 dim-6 operators~\cite{Grzadkowski:2010es}.

\begin{figure}[th]
\begin{center}
\includegraphics*[width=0.8\textwidth]{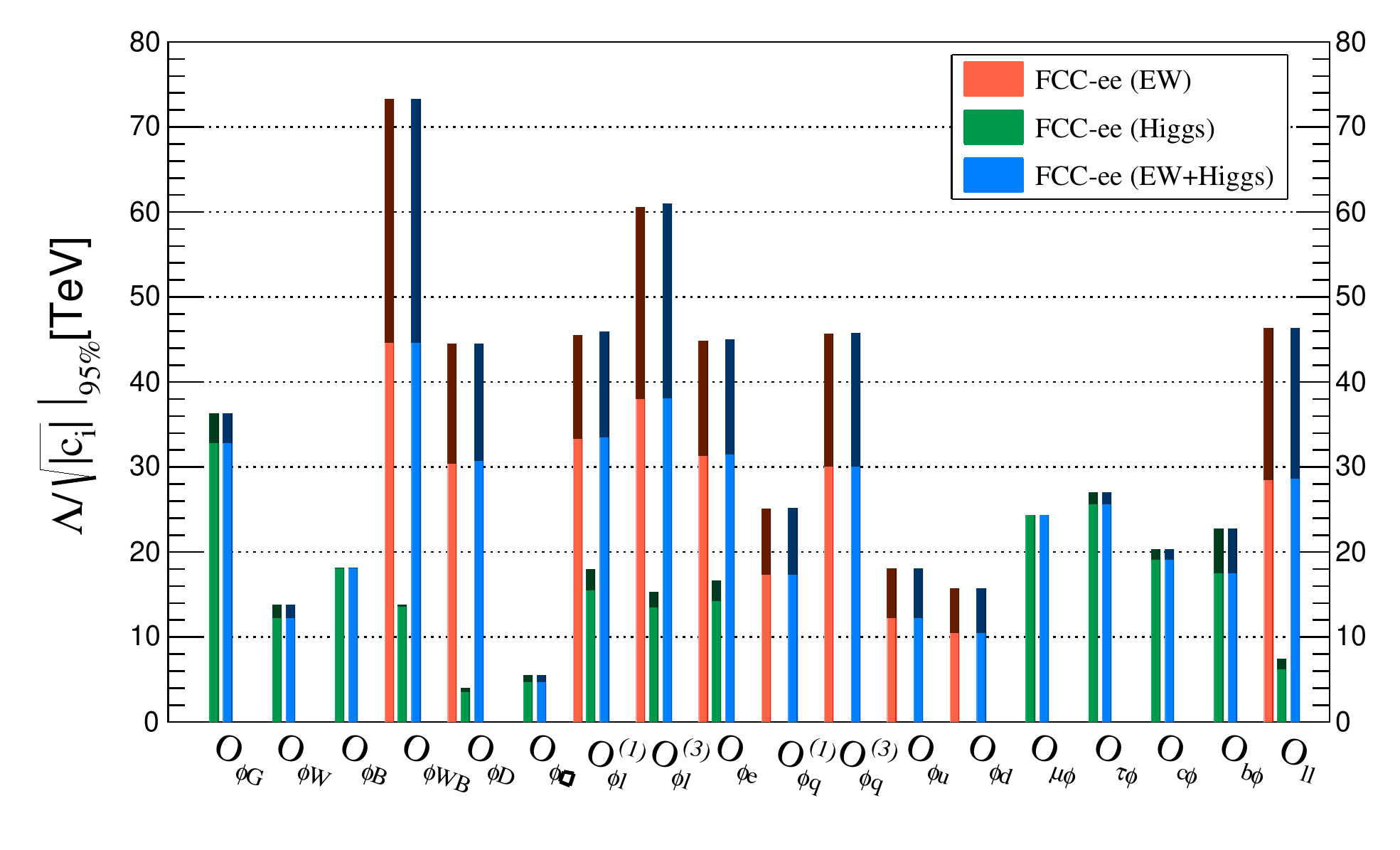}
\end{center}
\caption{\label{fig:ZHfit} FCC-ee constraints on various EFT
  operators, using EW and Higgs observables, separately and combined.
  Darker shades of each color indicate the results neglecting all SM
  theory uncertainties. Plot from Ref.~\cite{Benedikt:2651299},
  obtained using the HEPfit framework~\cite{hepfitsite}.}
\end{figure}

\newcommand{\lrD}{~\!\stackrel{\leftrightarrow}{\hspace{-0.1cm}D}\!}
\newcommand{\lD}{~\!\stackrel{\leftarrow}{\hspace{-0.1cm}D}\!}
\newcommand{\rD}{~\!\stackrel{\rightarrow}{\hspace{-0.1cm}D}\!}
\newcommand{\lrDa}{~\!\stackrel{\leftrightarrow}{\hspace{-0.1cm}D}\!^{\!~a}}
Figure~\ref{fig:ZHfit} shows the constraints that future FCC-ee data
can impose on the coefficients of the subset of operators that play a
role in the EW and Higgs observables discussed so far. On the EW side,
one has the following 10  operators:
\begin{eqnarray}
  && {\cal O}_{\phi D}=\left|\phi^\dagger D^\mu \phi\right|^2, \quad
  {\cal O}_{\phi WB}=\left(\phi^\dagger \sigma_a \phi\right)
  W_{\mu\nu}^a B^{\mu\nu}, \nonumber\\
  &&
  {\cal O}_{\phi \psi}^{(1)}=(\phi^\dagger \lrD_\mu \phi)
  (\overline{\psi}^i\gamma^\mu \psi^i),  \quad
  {\cal O}_{\phi  F}^{(3)}=(\phi^\dagger \lrDa_\mu \phi)
  (\overline{F}^i\gamma^\mu \sigma_a F^i ), \quad
  {\cal O}_{ll}=\left(\overline{l}\gamma_\mu l\right) \left(\overline{l}\gamma^\mu l\right),
\label{eq:EWops}
\end{eqnarray}
where $\phi$ is the Higgs scalar doublet, $\psi$ runs over all the 5 types
of SM fermion multiplets, while $F$ only refers to the 2 types of SM
left-handed fermion doublets. The $\phi$ field can induce effects in
processes where an explicit Higgs particle is present, or can influence
EW observables indirectly, when $\phi$ is set to its vacuum
expectation value. The other 8 operators shown in Fig~\ref{fig:ZHfit}
are mostly affecting Higgs observables:
\begin{eqnarray}
&& {\cal O}_{\phi G}=\phi^\dagger \phi~\! G_{\mu\nu}^A G^{A~\!\mu\nu}, \quad
   {\cal O}_{\phi W}=\phi^\dagger \phi W_{\mu\nu}^a W^{a~\!\mu\nu}, \quad
   {\cal O}_{\phi B}=\phi^\dagger \phi B_{\mu\nu} B^{\mu\nu}, \quad
   {\cal O}_{\phi \Box}= (\phi^\dagger \phi) \Box (\phi^\dagger \phi),\nonumber\\
&&
    {\cal O}_{\mu \phi}= (\phi^\dagger \phi) (\bar{l}^2 \phi~\!\mu), \quad
    {\cal O}_{\tau \phi}= (\phi^\dagger \phi) (\bar{l}^3 \phi~\!\tau),  \quad
    {\cal O}_{b \phi}= (\phi^\dagger \phi) (\bar{q}^3 \phi\!b), \quad
    {\cal O}_{c \phi}= (\phi^\dagger \phi) (\bar{q}^2 
    \tilde{\phi}~\!c).\label{eq:Hops}
\end{eqnarray}

The sensitivities to the ratios $C_i/\Lambda^2$ are reported in
Fig.~\ref{fig:ZHfit} as  95$\%$ probability bounds on the {\it
  interaction} scale, $\Lambda/\sqrt{C_i}$, associated to each
operator. Notice that, in the same way that the energy scale
associated to the Fermi constant $G_{\rm F}^{-{1}/{2}}$ differs from
the W boson mass by a factor related to the weak coupling, the
interaction scale defined by $\Lambda/\sqrt{C_i}$ does not correspond
exactly to the mass of new particles.

Few remarks are in order. First
of all, comparison~\cite{Benedikt:2651299} with the current constraints shows that
the FCC-ee data will increase by a factor of 4-5 the
sensitivity to new energy scales. This matches well
the factor of 7 increase in energy of the FCC-hh w.r.t. the LHC, which
will lead to a comparable incease in the direct sensitivity to new physics. In
other words, a 100~TeV pp collider is typically properly scaled to search for
the microscopic orgin of possible deviations observed by the FCC-ee
precision measurements. In absolute terms, the scales that can be
probed in the case of weakly-interacting new physics ($C_I\lsim 1$)
range between a few and few tens of TeV, while they can go up to
O(100)~TeV in the case of strongly interacting forces. 
The second remark is that Fig.~\ref{fig:ZHfit} underscores the
complementarity between EW and Higgs observables: even though
operators considered here (except ${\cal O}_{ll}$) include both Higgs
and gauge bosons or fermions, their constraints come primarily from either EW
or Higgs measurements. Both types of measurements are therefore
necessary for a systematic exploration of BSM contributions induced
via SMEFT operators. The respective scale sensitivity observed for
both types of operators is rather consistent, with most limits on interaction
scales in the range of 20-30~TeV, showing that the precision
targets set by the Higgs and EW programs are coherent. 

\section{Precision versus sensitivity}
In the previous sections, we focused on the very precise measurements
that the FCC enables, both in the Higgs and in the EW sector of the
SM. These measurements have a value per se, independently of whether
we eventually find discrepancies w.r.t. the SM expectations. We need
these precise measurements to test up to which point our understanding
of fundamental EW phenomena is controlled by the SM. But we also need
precise measurements because, the day new physics were found some
place, we shall need as much data as possible to evaluate and
constrain the many models that will be proposed to explain it. In this
context, data that agree with the SM can be as useful as data that may
not agree with it. Precise data in agreement with the SM, for example,
can help us rule out claims of observations of new physics that were
to openly clash with established measurements.

So, the success of a precision physics program is not necessarily tied
to the discovery of discrepancies with the SM, but builds on reliable,
unbiased and ever more accurate measurements of how nature
behaves. Treating the achievable precision as a gauge of the
sensitivity to new physics phenomena, however, is critical to examine
and characterize the potential relevance of a given measurement in a
specific BSM context. It is also a useful way to evaluate the reach of
different facilities, which might not measure exactly the same
observable: the interpretation of their measurements in terms of
constraints on some common BSM phenomenon can be seen as a useful
standard candle for their comparison, to assess synergies and
complementarities.

\begin{figure}[!ht]
\begin{center}
\includegraphics[width=0.8\textwidth]{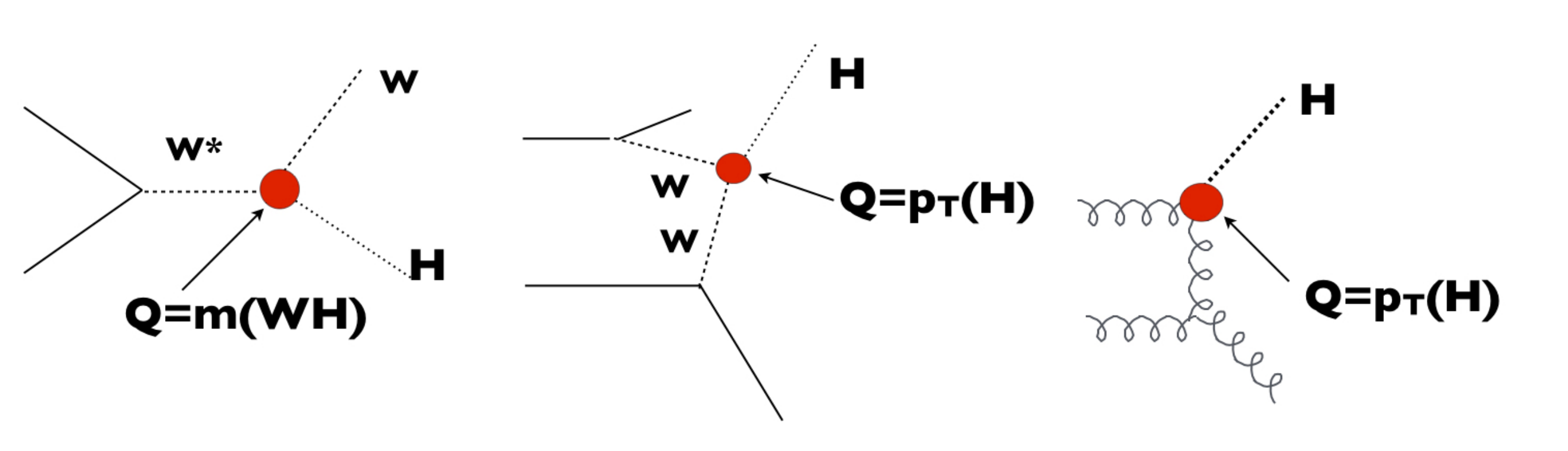} 
\end{center}
\vspace{-3mm}
\caption{\label{fig:HQ} Example of high-$Q$ Higgs production
  processes.}
\end{figure}
In the context of the FCC, there is great richness of synergies and
complementarities. The same new physics model can manifest itself via
departures in indirect precision measurements at FCC-ee, and directly
via the production of new particles at FCC-hh. But low- and
high-energy observables can also both participate in building indirect
evidence for new phenomena living at scales that are beyond the direct
discovery reach. This interplay can be shown with a general example,
using the EFT language. Let us consider a dimension-6 operator $O$,
parameterizing at energies $E<\Lambda$ the effects of some new physics
present at a scale $\Lambda$. This induces a contribution to the
Lagrangian given by $O/\Lambda^2$, and leads to corrections
$\delta_{if} = \langle f \vert O \vert i \rangle / \Lambda^2$ to the
transition between the initial and final states $\vert i \rangle$ and
$\vert f \rangle$. By dimensional analysis, $\delta_{if}$ will scale,
relative to a dimension-4 SM contribution, like $\delta_{if} \propto
(\mu/\Lambda)^2$, where $\mu$ is a mass scale characteristic of the
$i\to f$ transition. In the case of a Higgs decay, or of Higgs
production at threshold as in $gg\to H$, the only possible scales are
$m_H$ or $v$. Taking e.g. $\mu=m_H$, gives $\delta_{if} \propto
m_H^2/\Lambda^2 \sim 1.5\% \times ({\rm 1~TeV}/\Lambda)^2$. If the
transition involves another large kinematical scale $Q$, with $m_H < Q
< \Lambda$, as in the case of Higgs production at large $p_T$ (see
Fig.~\ref{fig:HQ}), the deviation could scale like $Q$, $\delta_{if}
\propto (Q/{\rm 1\;TeV})^2 \times ({\rm 1\;TeV}/\Lambda)^2$. The
impact of the EFT operator would therefore be greatly enhanced at
large $Q$. Assuming $\Lambda=1$~TeV, for example, a measurements at a
scale $Q\sim 400$~GeV would lead to an effect of approximatey 15\%,
ten times larger than for the decay or production at threshold. This
means that, at large $Q$, one can detect the presence of new physics
effects with lesser precision than is required at smaller
$Q$\footnote{Needless to say the EFT formalism looses validity once
  the hard scale becomes large enough to induce direct manifestation
  of new physics.} . A hadron collider cannot provide precision
comparable to an electron collider, but its higher kinematical reach
can compensate for this!

Several studies of this interplay between precision and sensitivity
from high-energy indirect measurements have appeared in the literature (see
e.g. \cite{Grojean:2013nya,Franceschini:2017xkh,Azatov:2017kzw,Grojean:2018dqj,Banerjee:2018bio,DiLuzio:2018jwd,Lee:2018fxj,Banfi:2018pki})
covering both Higgs and EW observables. Several of these papers test
these ideas in the context of the LHC, and the sensitivity
comparison is therefore drawn w.r.t. the LEP precision
measurements. We shall give here a couple of examples specific to the
100~TeV collider, where the kinematic reach and the large Higgs
production rates make this approach even more powerful.

\subsection{Example: the VVHH coupling}
\begin{figure}[htbp]
\begin{center}
\includegraphics*[width=0.75\textwidth]{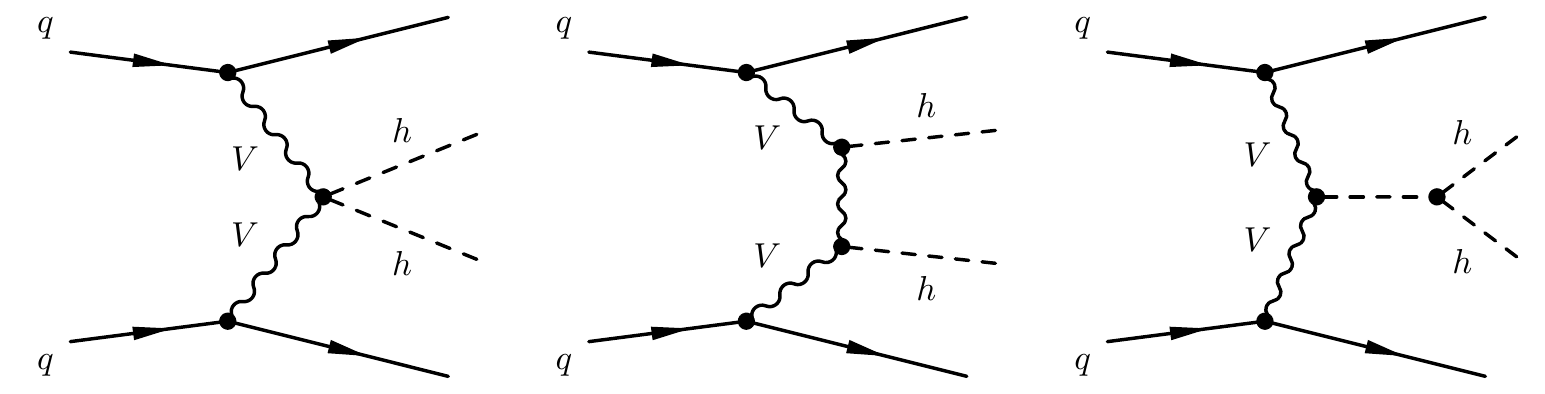}
\end{center}
\caption{\label{fig:vbfHH} Leading order Feynman diagrams for Higgs-pair production via vector boson fusion.}
\end{figure}

Figure~\ref{fig:vbfHH} shows the diagrams for Higgs-pair production
via vector boson fusion, studied in Ref.~\cite{Bishara:2016kjn}.
At large invariant mass of the Higgs pair,
$m_{HH}$, the triple-Higgs coupling contribution is suppressed by the
$1/s$ off-shell Higgs propagator,   
and the amplitude is controlled by the behaviour of the longitudinal-longitudinal
component of the amplitude, characterised by the destructive
interference between the first two diagrams:
\begin{equation} 
A({\rm V_LV_L\to HH}) \sim \frac{\hat{s}}{v^2}(c_{2V}-c_V^2) + {\cal
  O}(m_W^2/\hat{s}) \; .
\label{eq:c2v}
\end{equation}
Here, $c_{2V}$ and $c_V$ represent, respectively, the coefficients of
the VVHH and VVH couplings, normalised to their SM values. $\delta_c =
c_{2V}^2-c_V$ vanishes in the SM and in models where the $SU(2)$
symmetry is linearly realized, and the growth of the amplitude with
energy is suppressed. In composite Higgs models based e.g. on an
$SO(5)/SO(4)$ symmetry~\cite{Contino:2006qr}, on the other hand, $c_V=\sqrt{1-\xi}$ and
$c_{2V}=1-2\xi$ ($\xi \sim v^2/\Lambda^2$), so the scattering
amplitude grows with energy as $A\sim \hat{s}/\Lambda^2$.
\begin{figure}[htbp]
\begin{center}
\includegraphics*[width=0.48\textwidth]{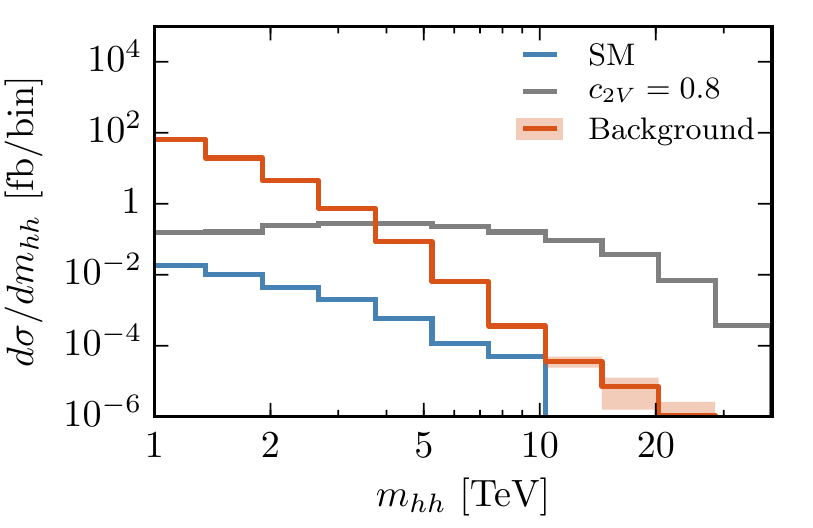}
\includegraphics*[width=0.48\textwidth]{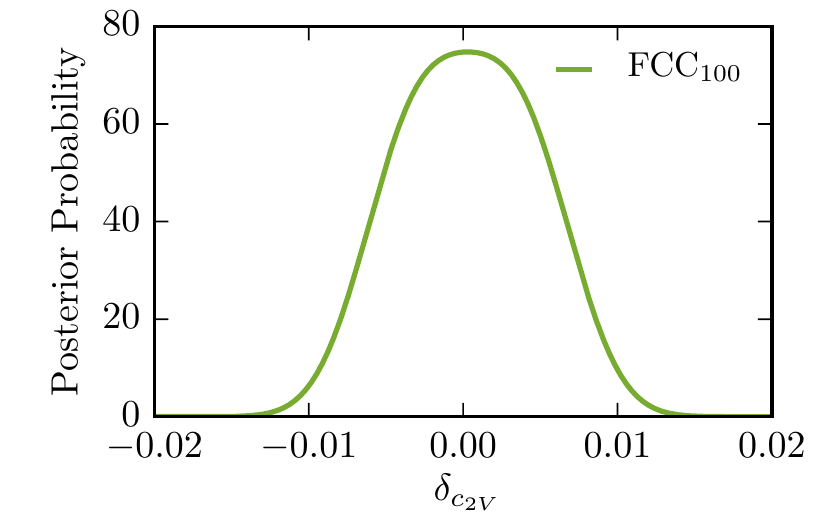}
\end{center}
\caption{\label{fig:vbfHH2} Left: HH invariant mass spectrum in vector
  boson scattering. Right: probability density distribution for
  $\delta c_{2V}$, assuming $c_{V}=1$ in Eq.~\protect\ref{eq:c2v}. {} From
  Ref.~\cite{Bishara:2016kjn}.
}
\end{figure}
The study of Ref.~\cite{Bishara:2016kjn} considered the HH$\to$4b final state, applying
boosted-jet tagging techniques, given the high $p_T$ of the
Higgs bosons at high $m_{HH}$. The
impact of $\delta_c\ne 0$ is visible in Fig.~\ref{fig:vbfHH2}, which shows the
$m_{HH}$ distribution in the SM in a $c_V=1$, $c_{2V}=0.8$
scenario, and the expected backgrounds. $c_V$ will
be measured with a few per-mille precision at FCC-ee (independently of
whether it agrees or not with the SM), and the constraints on $\delta_c$ at
FCC-hh will translate directly into a constraint on $c_{2V}$. The
detailed study of Ref.~\cite{Bishara:2016kjn} projects a sensitivity
to deviations of $c_{2V}$ from its SM value of better
than $\pm 1\%$, in spite of the much coarser precision in the
measurement of the HH rates! See Ref.~\cite{Bishara:2016kjn} for the 
discussion of the validity range of the EFT approximation.

\subsection{Example: Drell-Yan at large mass}
\begin{figure}[ht!]
\centering
\includegraphics[width=0.47\textwidth]{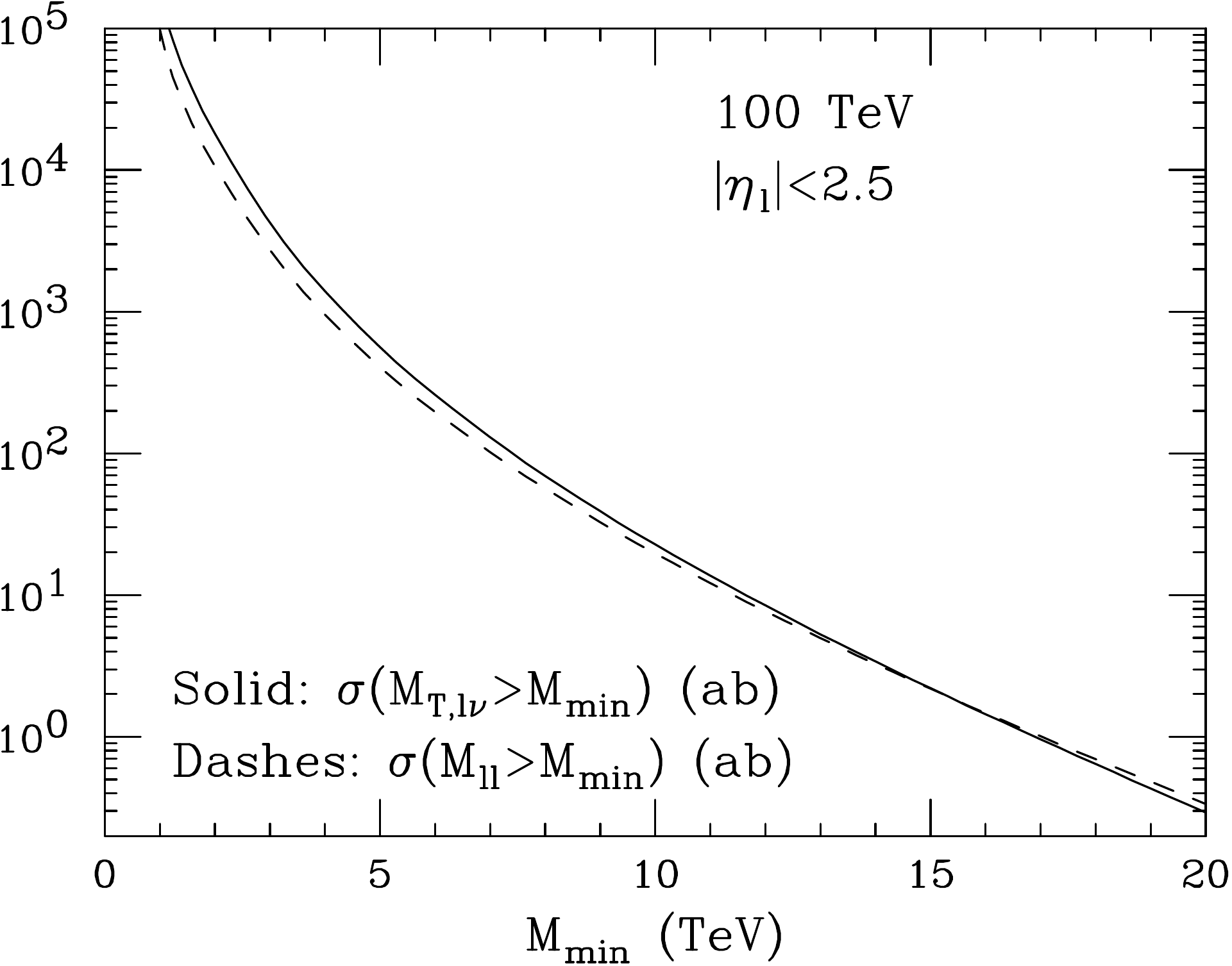}
\hfill
\includegraphics[width=0.47\textwidth]{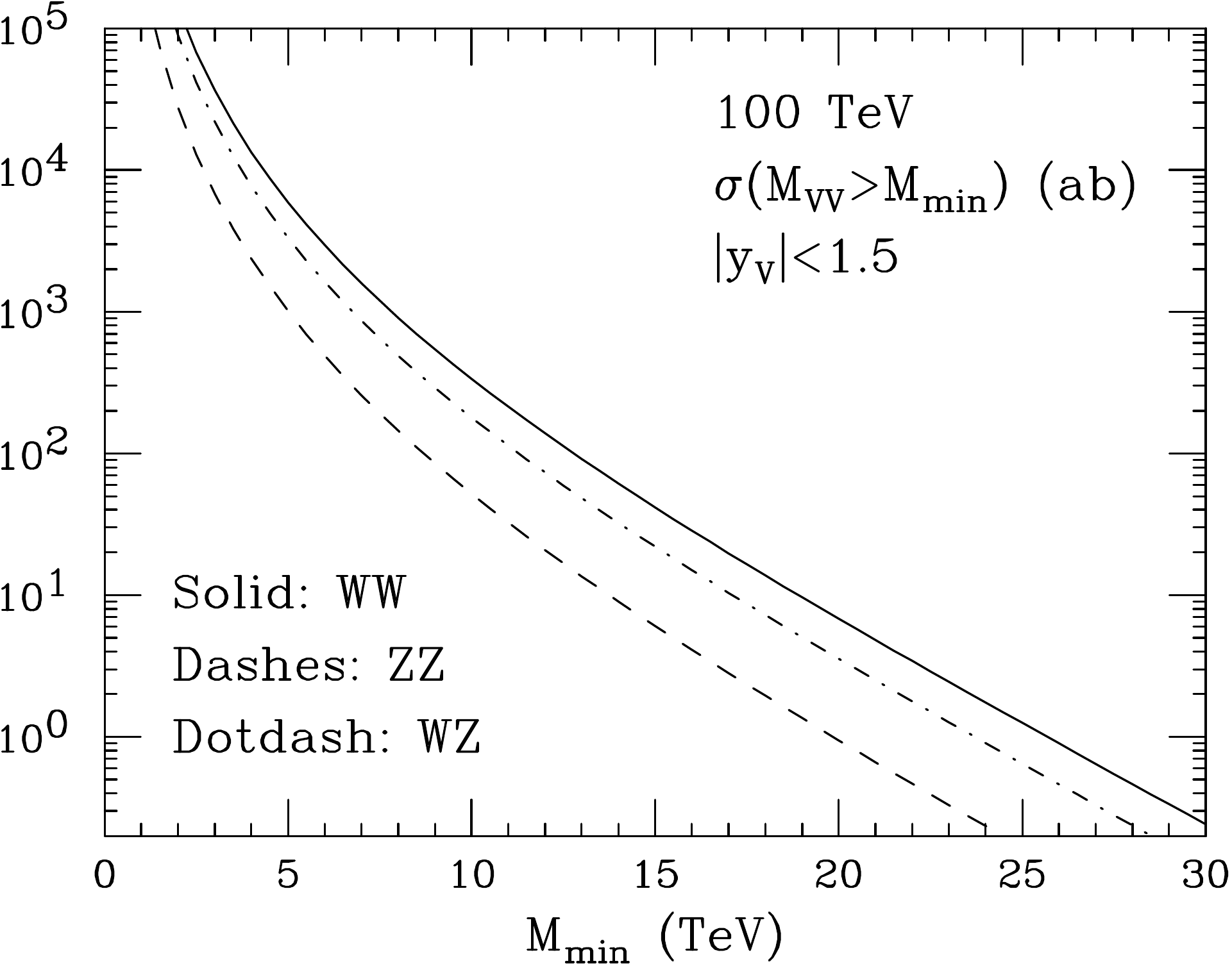}
\caption{Left: integrated lepton transverse (dilepton) mass
  distribution in $\rm pp\to W^*\to\ell$\textnu\ ($\rm pp\to
  Z^*/$\textgamma$^*\to \ell^+\ell^-$). One lepton family is included,
  with $\vert \eta_\ell \vert<2.5$. Right: Integrated invariant mass
  spectrum for the production of gauge boson pairs in the central
  kinematic range $\vert y \vert<1.5$. No branching
  ratios included. }
\label{fig:DYmass}
\end{figure}

Drell-Yan (DY) production is another example where energy can
complement precision, achieving sensitivity to new high-scale
phenomena~\cite{Farina:2016rws}.  The total production rates of
W$^\pm$ and Z$^0$ bosons at 100~TeV are about 1.3 and 0.4~\textmu{}b,
respectively, i.e. samples of ${\cal O}(10^{10-11})$ leptonic decays
per \iab! A large fraction of these events probe very high
energies. Figure~\ref{fig:DYmass}, left panel, shows the integrated
spectra of the W boson transverse mass
($M^2_T=2p_{T,\ell}p_{T,\nu}(1-\cos\theta_{\ell\nu})$) and of the
\textgamma/Z dilepton mass.

\noindent
\rule{5cm}{1pt}\\ {\bf Exercise:} The right panel of
Fig.~\ref{fig:DYmass} shows the invariant mass spectrum of gauge boson
pairs. Compare these rates to those for lepton-pair production (left
panel), and discuss the reasons for the large differences.
\\ \rule{5cm}{1pt}

In presence of new physics, large corrections to the SM prediction can
arise from the $W$ and $Y$ oblique parameters defined
by the following dim-6 EFT operators ~\cite{Barbieri:2004qk}:
\begin{equation}
\hat{W}=-\frac{W}{4m_W^2}(D_\rho W^a_{\mu\nu})^2 \quad, \quad
\hat{Y}=-\frac{Y}{4m_W^2}(\partial_\rho B_{\mu\nu})^2 \; .
\end{equation}
These parameters capture the universal
modifications of the EW gauge boson propagators, and are already
constrained, at the per mille level, from LEP-2 precision measurements
and from the W mass measurements at Tevatron and LHC (as well as other
precision measurements at the Z pole at LEP/SLD).  The FCC-hh DY
statistics at very high mass will contribute, with the precision
measurements at lower energy by FCC-ee, to improve the
current constraints by two orders of magnitude, as shown in
Table~\ref{tab:WY}. In terms of a new physics scale $\Lambda$ defined by
$g_*^2/\Lambda^2=(W,Y)/4m_W^2$, the FCC-hh reach corresponds to
$\Lambda\gsim g_*\times 80$~TeV, a sensitivity that could be matched
by a multi-TeV lepton collider such as CLIC.
We should remark that the low- and high-energy approaches should
not be seen as alternative, but as synergetic: should deviations be
observed in either of them, the other measurement would serve as
an independent probe to confirm the SM departure, and to help pinning
down its origin.
\begin{table*}[t]
\centering
  \caption{ \label{tab:WY} Reach at 95\%~CL on \textsc{W} and
    \textsc{Y}~from different experiments. The bounds from neutral DY
    are obtained by setting the unconstrained parameter to zero. From
    Ref.~\cite{Farina:2016rws}.}  {\small
\begin{tabular}{|c|c|c|c|c|c|c|c|c|} 
\hline \hline
\multicolumn{2}{|c|}{  }
 & LEP& ATLAS\,8 & CMS\,8 & \multicolumn{2}{c |}{ LHC\,13}  & FCC-hh  & FCC-ee   \\ \hline
\multicolumn{2}{|r|}{ luminosity \hspace{4cm}}  &
~~$2\times10^7\,Z$~~& 19.7\,\ifb & 20.3\,\ifb & ~ 0.3\,\iab & ~ 3\,\iab &
10\,\iab &  ~$10^{12}\,Z$~ \\ \hline
~NC~  & ~\sc{W} $\times 10^{4}$~  & $[-19, 3]$ & $[-3,15]$  & $[-5,22]$  & $\pm1.5$ & $\pm0.8$  & $\pm0.04$  &  $\pm 1.2$ \\ 
 \cline{2-9}
  & ~\sc{Y} $\times 10^{4}$~ & $[-17, 4]$  & $[-4,24]$  & $[-7,41]$  & $\pm2.3$ & $\pm1.2$  & $\pm0.06$  &  $\pm1.5$ \\ \hline
~CC~  & \sc{W}  $\times 10^{4}$~& ---  &\multicolumn{2}{ c |}{ $\pm 3.9$}  & $\pm0.7$ & $\pm0.45$  & $\pm0.02$   &   ---   \\ 
\hline \hline 
\end{tabular}
}
\end{table*}
Further applications of the huge DY lever arm in $Q^2$ are (i) the determination of the
running of EW couplings, by measuring the transverse (invariant) mass
spectrum of (di)leptons produced by far off-shell W (Z)
bosons~\cite{Alves:2014cda}, and (ii) the indirect search for new heavy EW
particles (like gauginos in supersymmetry) using the distortion of the
DY shape near their production threshold~\cite{Chigusa:2018vxz}.

\section{The Higgs potential}
\newcommand*{\kl}{\ensuremath{\kappa_{\lambda}}}
\newcommand*{\effg}{\ensuremath{\epsilon_{\gamma}}}
\newcommand*{\misg}{\ensuremath{\epsilon_{j \rightarrow \gamma}}}
\newcommand*{\effb}{\ensuremath{\epsilon_{b}}}
\newcommand*{\misc}{\ensuremath{\epsilon_{c \rightarrow b}}}
\newcommand*{\mislc}{\ensuremath{\epsilon_{l(c) \rightarrow b}}}
\newcommand*{\maa}{\ensuremath{m_{\gamma\gamma}}}
\newcommand*{\mbb}{\ensuremath{m_{bb}}}
As we discussed at the beginning of these lectures, understanding the
origin of the Higgs potential is among the most, if not {\em the} most, outstanding target
of future colliders. As an essential part of this understanding, we
must start measuring it. Let us briefly rediscuss the relations between
the parameters of the Higgs potential and the physical observables, in
a context slightly more general than the SM. To
simplify the notation, let us consider a single real field, and consider
the following simple generalization of the SM potential (you could consider repeating
the exercise with a more general functional form):
\begin{equation} \label{eq:Hpotn}
  V(\phi) = -\frac{\mu^2}{2} \phi^2 + \frac{\lambda}{n}\phi^n \; ,
\end{equation}
$n$ must be even, and of course $n=4$ in the SM. The two key relations obtained by setting
$\langle \phi \rangle=v$ at the minimum of the potential, and defining
$m_\phi^2=\partial^2 V(\phi)/\partial \phi^2 \vert_{\phi=v}$ give:
\begin{equation} \label{eq:mh}
  v^{n-2}=\frac{\mu^2}{\lambda} \quad , \quad m_\phi^2 = (n-2) \mu^2
  \; .
\end{equation}
We stress that, while dimensional analysis makes the Higgs mass
proportional to $\mu^2$, its precise value is not defined by the
dynamics near the origin $\phi\sim 0$, where the quadratic term
dominates, but it is defined by the dynamics around the minimum
$\phi=v$, which could be far away from the origin. This is reflected
by the coefficient $(n-2)$ in Eq.~\ref{eq:mh}, whose specific value
depends on the shape of the potential at large $\phi$. In other words,
the quadratic term of the potential only provides the overall scale of
the Higgs mass, but the specific value is controlled by the Higgs
dynamics at the minimum, where the higher-order terms of the potential
are important: in the same way that the masses of SM particles are
related to the strength of their interaction with the Higgs, it is
reasonable that the mass of the Higgs be related to its own
self-interaction. Therefore studying the structure of the Higgs
potential is also a way to address the question of ``what gives mass
to the Higgs''.

Now, the potential above has three parameters, $\mu^2$, $\lambda$ and
the power $n$. However we only have two measurements, the Higgs mass
and its expectation value $v$. The two are absolute numbers,
independent of the shape of the potential: the mass is what we measure
in the experiments, while $v$ is given by the relation $M_W=g v$,
where $g$ is the weak gauge coupling (recall also that
$v=(\sqrt{2}G_F)^{-1/2}$). For each $n$, we can extract a value
 for $\lambda$ and $\mu^2$, but we won't be able to determine
what the shape of the potential is. In other words, the only thing we
know experimentally about the Higgs potential is that its second derivative at
the origin is negative, to drive $v$ away from 0. We have no
experimental evidence,
as of today, that $n=4$ rather than 6. To
make progress in learning about the structure of $V(\phi)$, we need a
further measurement, sensitive to the power $n$ in
Eq.~\ref{eq:Hpotn}. Expanding the potential around its minimum, the
cubic term controls the cubic self-interaction, and its strength is
given by:
\begin{equation} \label{eq:lambda3}
  \lambda_{\phi\phi\phi} = \frac{\partial^3 V}{\partial \phi^3}
  \vert_{\phi=v} = (n-1) \frac{m_\phi^2}{v} \; .
\end{equation}
Assuming a Higgs potential given by Eq.~\ref{eq:Hpotn}, and given that
$m_\phi$ and $v$ are known, the measurement of the Higgs cubic
self-interaction would be directly a measurement of $n$. More in
general, it is clear that the Higgs self-coupling is a mandatory
measurement to start learning about the Higgs potential. Notice one
point: the change from $n=4$ (as in the SM), to something like $n=6$,
has a big impact on the self-coupling, namely it increases it by a
factor of 5/3. With this modification of the Higgs potential, there is
no continuous knob that allows to smoothly change
$\lambda_{\phi\phi\phi}$ from its SM value.  However large, this
change would have still failed to be detected experimentally, so for
all we know this is still an open option. So we should be open to the
possibility that $\lambda_{\phi\phi\phi}$ differs from its SM value by
${\cal O}(1)$, and a 20\% uncertainty in its determination might
already explore possible deviations at the 5$\sigma$ level.

Needless to say, the most likely scenario assumes the presence of the
$\lambda \vert H \vert^4$ SM quartic coupling\footnote{The presence of
  a quartic coupling is unavoidable, even if, for some odd reason, the
  underlying fundamental theory did not have such a coupling: the
  quartic would in fact be generated via radiative corrections at the
  one-loop order, starting from the $\vert H \vert^6$ term.}, supplemented by other
higher-order terms, like $1/\Lambda^2 \vert H \vert^6$. In this case,
$v^2/\Lambda^2$ becomes a continuous tunable parameter, which can
alter the SM Higgs self-coupling $\lambda$ by arbitrarily small
values:
\begin{equation} \label{eq:HiggsPot6}
\delta \kappa_\lambda \equiv \frac{\delta \lambda}{\lambda} \sim
\frac{2 v^4}{m_H^2 \Lambda^2} \; .
\end{equation}

One more small remark\footnote{Notice that obviously
  $\lambda_{\phi\phi\phi}$ has mass dimension 1, independently of the
  form of the potential, since it's the parameter of a 3-boson
  coupling.}: we can rewrite Eq.~\ref{eq:lambda3} as $m_\phi^2=v
\lambda_{\phi\phi\phi}/(n-1)$. This highlights the fact that, contrary
to the case of all other SM particles, where the strength $y$ of their
interaction with the Higgs uniquely defines their mass as $m=yv$, the
mass vs $v$-times-coupling relation for the Higgs has an additional
dependence on $n$. So, until we measure the Higgs self-coupling, we
cannot claim to know ``how'' the Higgs gets its mass ....

\subsection{Higgs Self-Coupling Probes at FCC}
At the FCC two approaches to measure the Higgs self-coupling, known as
`direct' and `indirect', can be followed. In the former the Higgs
self-coupling enters at tree-level and in the latter it enters via
processes at the loop level.  In general, to unambiguously probe the
coefficient of any dimension-6 operator one should perform a global
fit of all available observations, including the effects of all
operators in all physical observables. In the case that the heavy
particles contribute dominantly to some subset of dimension-6
operators, it is well-motivated to include only those operators in the
analysis, even though the result now becomes model-dependent, owing to
the assumption made that other operator coefficients are small.

\begin{figure}[htbp]
\begin{center}
\includegraphics[width=0.30\textwidth]{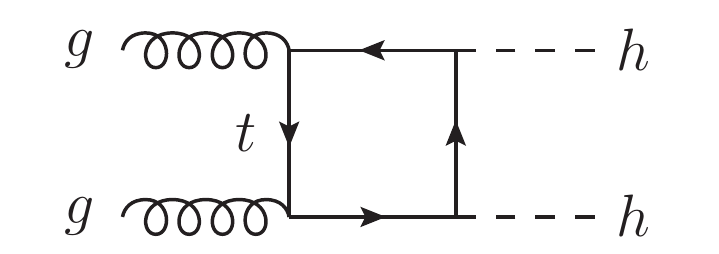}
\hspace{2.5em}
\includegraphics[width=0.335\textwidth]{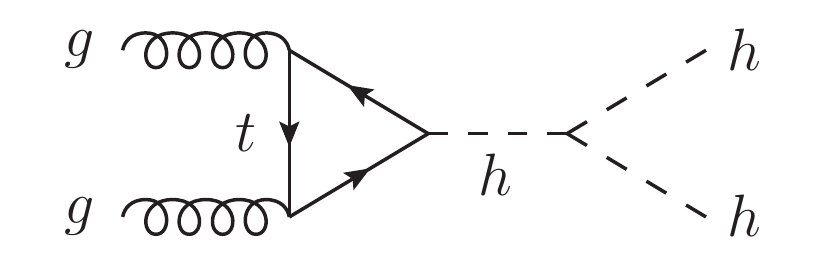}
\\
\vspace{1.2em}
\includegraphics[width=0.28\textwidth]{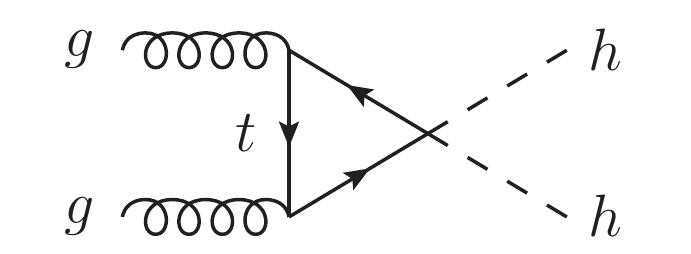}
\hspace{0.8em}
\includegraphics[width=0.30\textwidth]{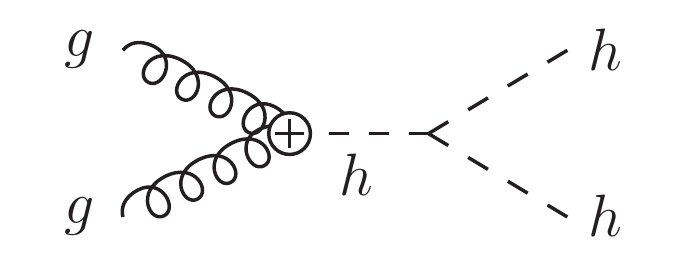}
\hspace{0.8em}
\includegraphics[width=0.233\textwidth]{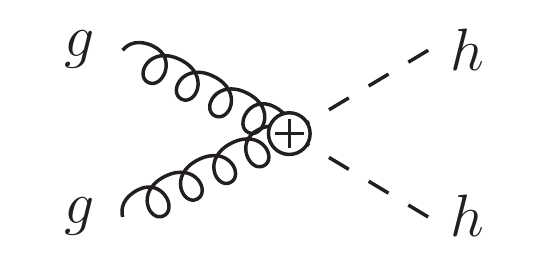}
\end{center}
\caption{\label{fig:ggHH} Feynman diagrams contributing to double Higgs production via gluon fusion: in the SM (upper set) and with higher-dimension operators affecting the tthh, ggh and gghh couplings, respectively (lower set).}
\end{figure}

In the case of Higgs pair production at FCC-hh, for example via the
gluon-fusion process gg$\to$HH, the Higgs self-coupling enters at tree
level. The `direct' constraint on $\delta \kappa_\lambda$ from a
global analysis~\cite{Azatov:2015oxa} would require the inclusion of
other dimension-6 operators that may contribute, for example those
involving gluons or top quarks that give rise to the diagrams in the
second row of Fig.~\ref{fig:ggHH}. But assuming that the deviations
induced by these additional operators are already constrained to be
small, as justified by the high precision achievable at the FCC in the
measurement of ggh and ggt couplings, one may directly extract a
constraint on $\delta \kappa_\lambda$ from the study of Higgs-pair
production.

For the case of indirect constraints the situation is somewhat
different.  For a given single-Higgs production process one has the
tree-level process but also higher loop processes in which the Higgs
self-coupling may enter.
The first discussion of such effects was in \cite{vanderBij:1985ww},
however, it was more recently emphasised for current and future
experiments in~\cite{McCullough:2013rea}.  Using these effects one may
search for the influence of a modified Higgs self-coupling on
experimental observables involving a single Higgs boson.  Such probes
are known as `indirect'.  For indirect probes, if additional
dimension-6 operators contribute to this process at tree-level then,
for a comparable magnitude of coefficient, they would lead to
deviations that are a loop factor greater than from the self-coupling
effects.  Thus to assume that the self-coupling modification is the
leading source of deviations in an observable would necessarily imply
the assumption that the other operators that contribute at tree-level
have coefficients that are smaller than the self-coupling by more than
a loop factor.  This assumption is too strong to cover a wide range of
scenarios for new physics beyond the SM that might modify the
self-coupling, thus for an indirect probe it is only realistic to
perform a global analysis, allowing all dimension-6 operators to enter
the scattering process whilst considering all available measurements
to over constrain the system of unknown coefficients.

\subsection{FCC-ee: Indirect Probe}
\label{sec:FCCee_indirect_probe}

\begin{figure}[htbp]
\begin{center}
\includegraphics*[width=0.55\textwidth]{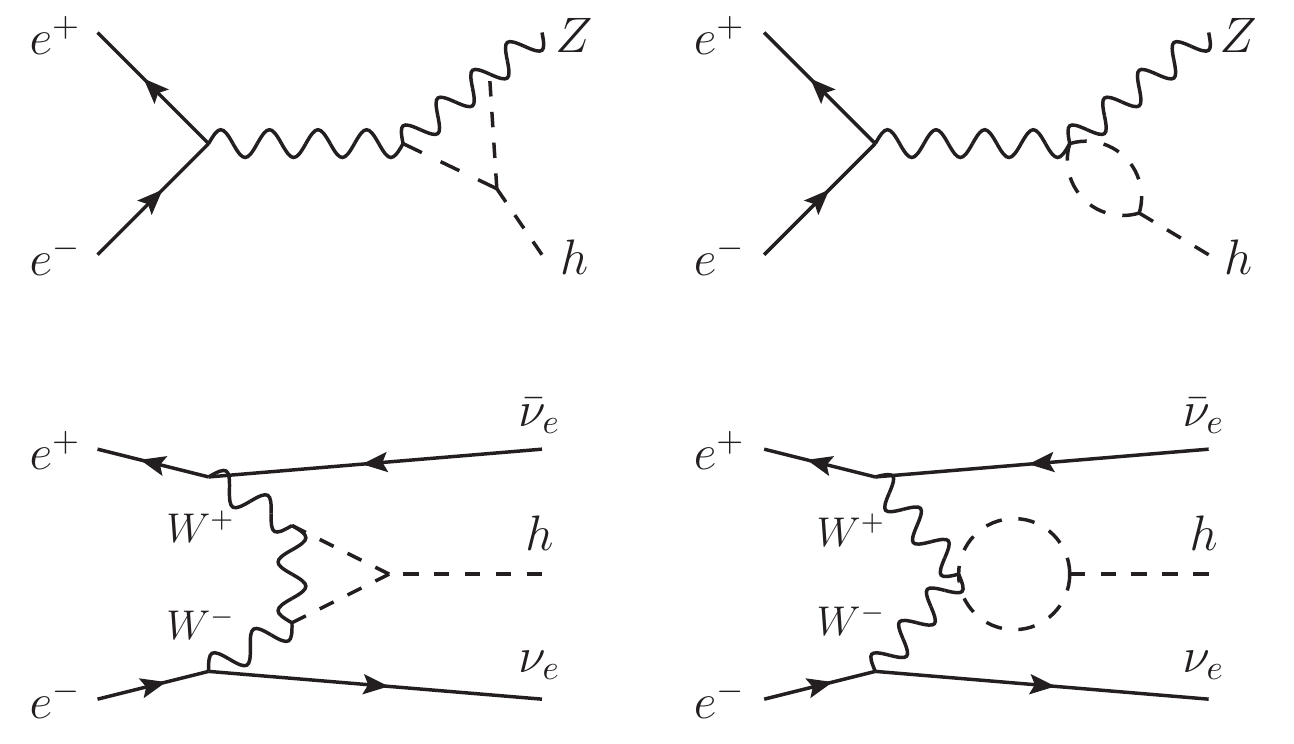}
\end{center}
\caption{\label{fig:h3-TGC-atEE} From Ref.~\cite{DiVita:2017vrr},
  sample Feynman diagrams illustrating the effects of the Higgs
  trilinear self-coupling on single Higgs process at next-to-leading
  order.} 
\end{figure}

With the large luminosity delivered at 240 and 365~GeV, the FCC-ee has
privileged sensitivity to the Higgs self-coupling by measuring its
centre-of-mass-energy-dependent effects at the quantum level on single
Higgs observables~\cite{McCullough:2013rea}, such as the HZ and the
\textnu\textnu H production cross sections, representative diagrams of
which are displayed in Fig.~\ref{fig:h3-TGC-atEE}.

\begin{figure}[ht!]
\begin{center}
\includegraphics[width=0.55\textwidth]{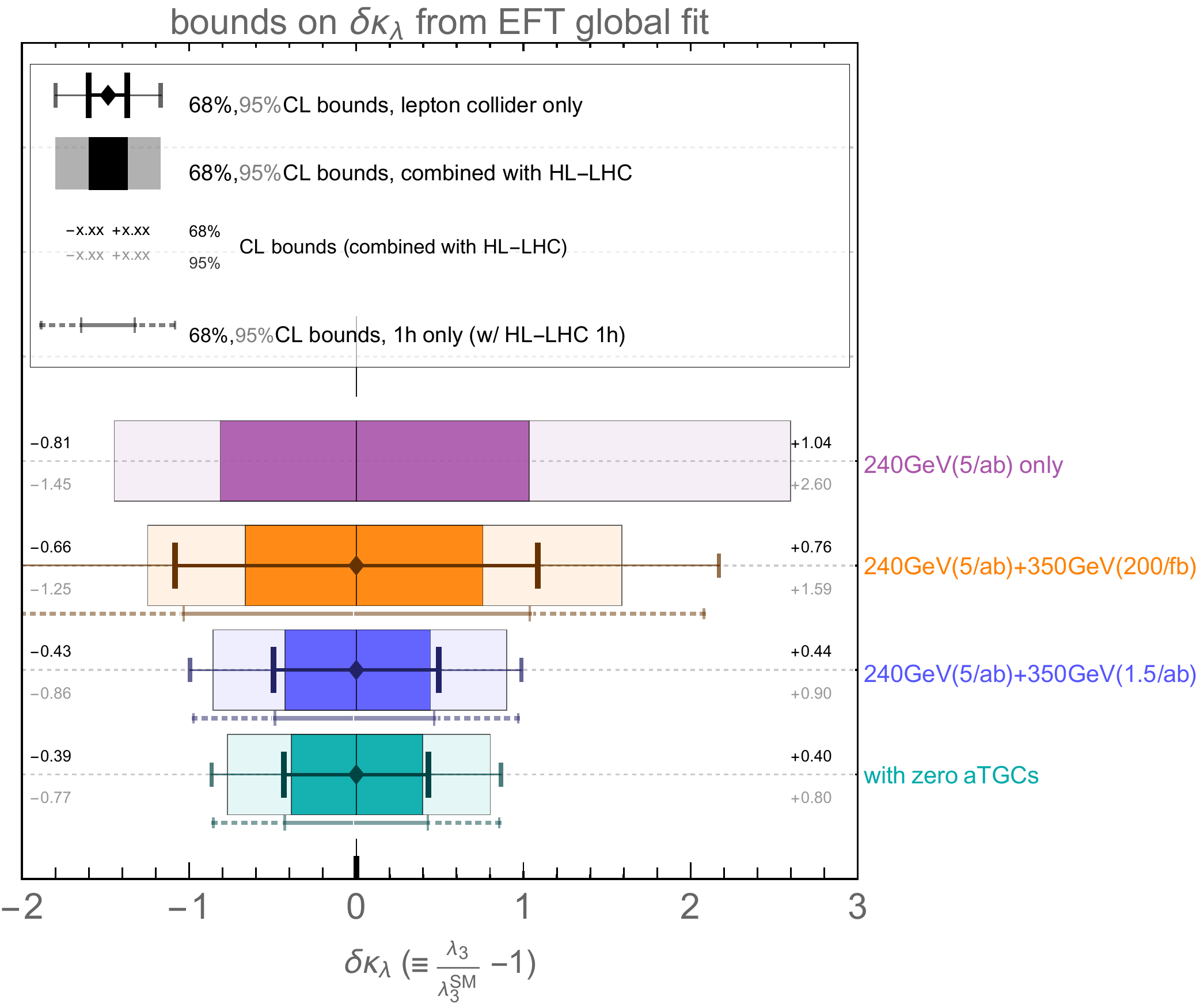} 
\end{center}
\vspace{-3mm}
\caption{\label{fig:selfFCCee} Indirect measurements of the Higgs
  self-coupling at FCC-ee combining runs at different energies.}
\end{figure}

In Fig.~\ref{fig:selfFCCee} the results of a global analysis for
FCC-ee are shown.  This figure is taken from \cite{DiVita:2017vrr},
which the reader is referred to for details.  Notably, through a global
analysis which allows for the presence of all dimension-6 operators
with arbitrary coefficients, at FCC-ee a robust and model-independent
measurement of the Higgs self-coupling can be made with a precision of
$44\%$, to be compared with the HL-LHC projection, in which the
precision is around $50\%$.  This would establish, at better than
$95\%$ confidence level, a non-zero value for the Higgs self-coupling
and would probe a scale of $\Lambda \sim 1$ TeV in
Eq.~\ref{eq:HiggsPot6}.  In this analysis the FCC-ee precision
electroweak measurements at lower energies are equally important to
fix extra parameters that would otherwise enter the global Higgs fit
and open flat directions that cannot be resolved.

\subsection{FCC-hh: Direct Probes}
At FCC-hh, the Higgs self-coupling can be probed directly via
Higgs-pair production. The cross sections for several production
channels are given~\cite{hhxswg} in Table~\ref{tab:HHrates}, where the
quoted systematics reflect today's state of the art, and are therefore
bound to be significantly improved by the time of FCC-hh operations.
\begin{table}[th]
\renewcommand{\arraystretch}{1.2}
 \begin{center}
   \caption{\label{tab:HHrates} Higgs-pair cross sections rates for
     various production processes~\cite{hhxswg}. The first uncertainty
     corresponds to the scale choice, the second combines $\alpha_S$
     and PDF systematics (PDF4LHC15NNLO), the third estimates
     finite-$m_{top}$ effects in the NNLO contribution to the gg
     channel.}
\begin{tabular}{|l|c|c|} 
  \hline\hline 
 &  $\sigma$[100 TeV](fb)  &$\sigma$[27 TeV](fb)  \\ \hline
gg$\to$HH 
& $1.22\times 10^3 {+0.9\%\atop -3.2\%} \pm 2.4\% \pm 4.5\%_{m_{t}}$ 
& $140             {+1.3\%\atop -3.9\%} \pm 2.5\% \pm 3.4\%_{m_{t}}$ 
\\
HHjj  
& $80.5 \pm 0.5 \% \pm 1.8\%$ 
& $1.95 \pm 2\% \pm 2.4\%$ 
\\  
W$^+$HH  
& $4.7 \pm 1\% \pm 1.8\%$  
& $0.37 \pm0.4\% \pm 2.1\% $ 
\\
W$^-$HH  
& $3.3 \pm 4\% \pm 1.9\%$  
& $0.20 \pm 1.3\% \pm 2.7\% $ 
\\
ZHH
& $8.2 \pm 5\% \pm 1.7\% $  
& $0.41 \pm 3\% \pm 1.8\%$
\\
t\={t}HH 
& $ 82.1 \pm 8\% \pm 1.6\%$ 
& $0.95 {+1.7\% \atop -4.5\%} \pm 3.1\%$ 
\\
\hline \hline
\end{tabular}
\end{center}
\end{table}
The most studied channel, in view of its large rate, is gluon fusion
(see Fig.~\ref{fig:ggHH}).  In the SM there is a large destructive
interference between the diagram with the top-quark loop and that with
the self-coupling.  While this interference suppresses the SM rate, it
makes the rate more sensitive to possible deviations from the SM
couplings, the sensitivity being enhanced after NLO corrections are
included, as shown in the case of gg$\to$HH in
Ref.~\cite{Borowka:2016ypz}, where the first NLO calculation of
$\sigma$(gg$\to$HH) inclusive of top-mass effects was performed.  For
values of $\kappa_\lambda$ close to 1, $1/\sigma_{\rm HH} d\sigma_{\rm
  HH}/d\kappa_\lambda\sim -1$, and a measurement of $\kappa_\lambda$
at the few percent level requires the measurement and
theoretical interpretation of the Higgs-pair rate at a similar level
of precision. Table~\ref{tab:HHrates} shows that the current
theoretical systematics on the signal is at the 5\% level (for a
complete discussion see~\cite{Grazzini:2018bsd}), which is already
competitive with the statistical and experimental systematics, to be
presented shortly. It is reasonable to predict a further reduction to
the percent level.

\begin{figure}
  \centering
\includegraphics[width=0.45\columnwidth]{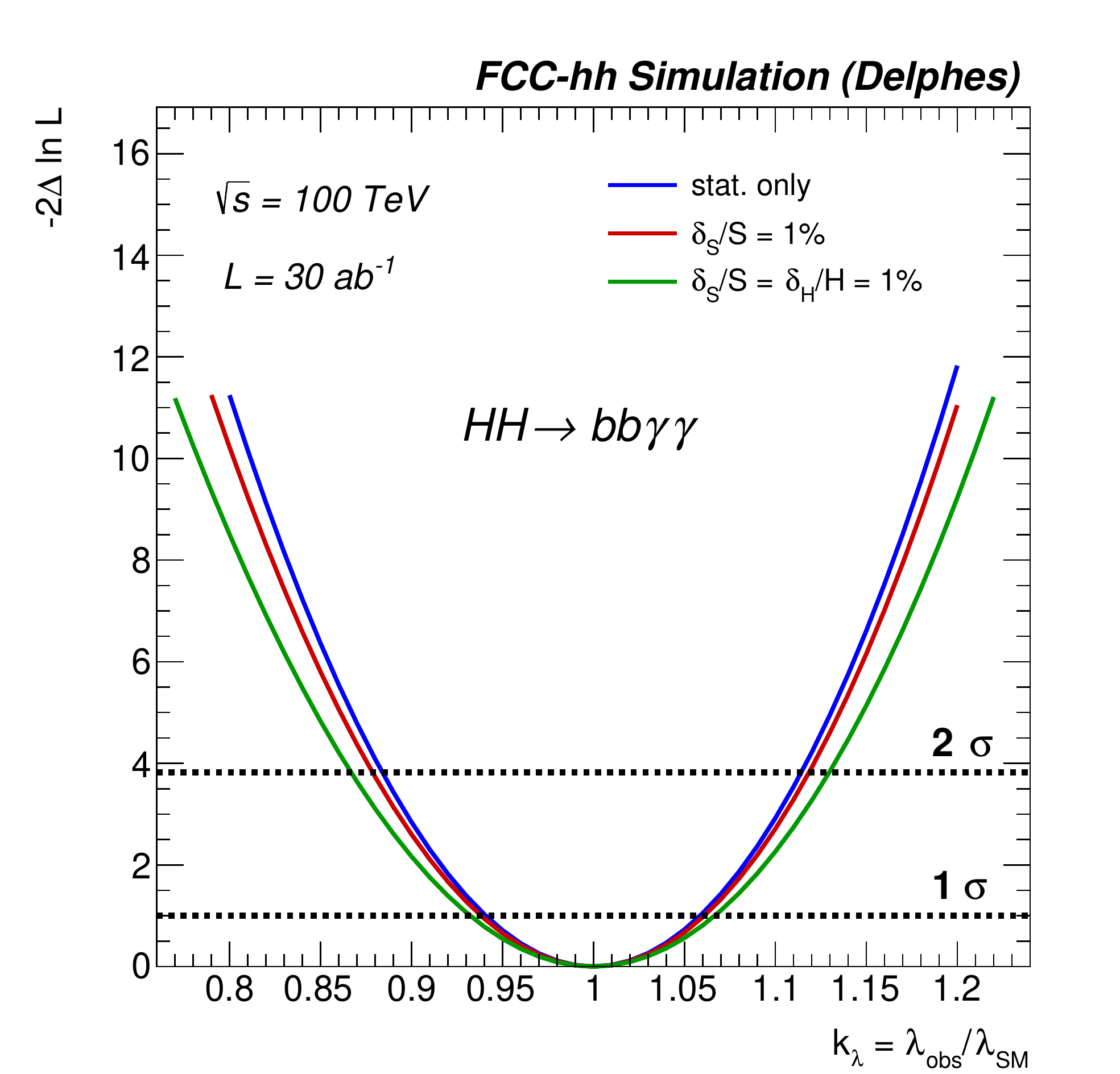}
\includegraphics[width=0.45\columnwidth]{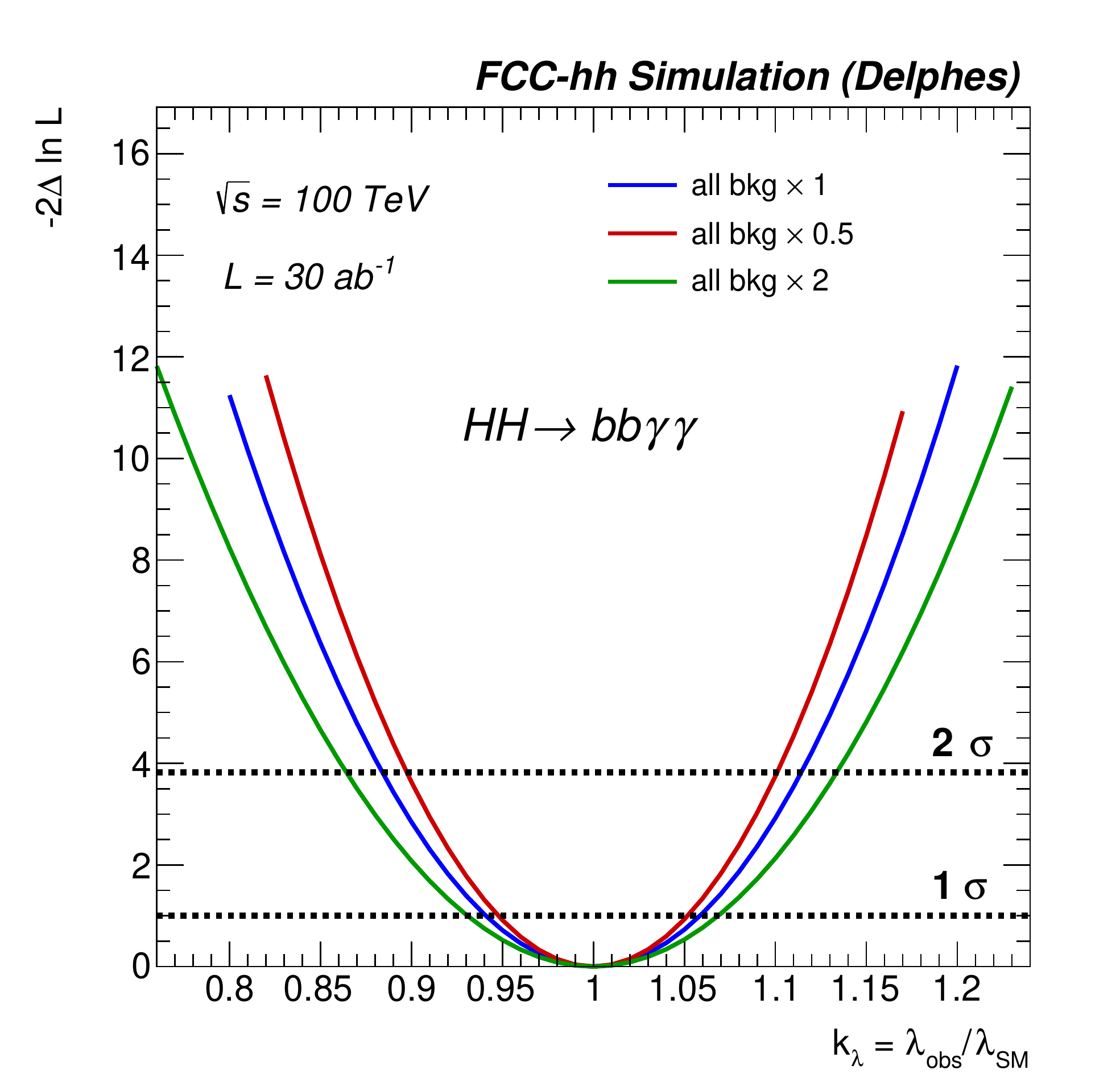}
  \caption{Expected precision on the Higgs self-coupling modifier
    $\kl$ with no systematic uncertainties (only statistical), 1\%
    signal uncertainty, 1\% signal uncertainty together with 1\%
    uncertainty on the Higgs backgrounds (left) and assuming
    respectively $\times 1$, $\times 2$, $\times 0.5$ background
    yields (right).)} 
   \label{hhbbaa_kl_main}
\end{figure}

The full analysis of the Higgs pair observation potential and its
interpretation in terms of the Higgs self-coupling is rather complex,
due to the presence of significant backgrounds and experimental
challenges, ranging from the tagging of b quarks to the optimization
of the mass resolution in the reconstruction of the Higgs boson
peaks. All the details can be found in
Refs.~\cite{Mangano:2651294,Borgonovi:2642471}.  I limit myself here
to showing the bottom line, Fig.~\ref{hhbbaa_kl_main}, with the
results relative to the most significant channel, $\rm HH\to
b\bar{b}$\textgamma\textgamma.

Decay modes other than HH$\to$b\={b}\textgamma\textgamma\ have also
been considered in the detector performance studies. These include
b\={b}ZZ*[$\to$4$\ell$] ($\ell=$e,\textmu),
b\={b}WW*[$\to$2j$\ell$\textnu], and 4b+jet. A summary of the target
precision in the measurement of $\kappa_\lambda$ is given in
Table~\ref{tab:kappalambda}, where the results were obtained with the
baseline detector performance parameters, and a 1\% systematics on the
rates of the signals and of the leading backgrounds.
\begin{table}[th]
\renewcommand{\arraystretch}{1.2}
 \begin{center}
   \caption{\label{tab:kappalambda} Precision of the direct Higgs
     self-coupling measurement in gg$\to$HH production, for various
     decay modes, from the FCC-hh detector performance studies.}
\begin{tabular}{|l|c|c|c|c|} 
  \hline\hline 
 & b\={b}\textgamma\textgamma 
 & b\={b}ZZ*[$\to$4$\ell$]
 & b\={b}WW*[$\to$2j$\ell$\textnu] 
 & 4b+jet   \\ \hline
 $\delta\kappa_\lambda$ 
 & 6.5\% 
 & 14\%
 & 40\%
 & 30\%
 \\
\hline \hline
\end{tabular}
\end{center}
\end{table}

Additional studies, of a more phenomenological nature, have appeared
in the literature, see e.g. Refs.~\cite{Goncalves:2018qas,Banerjee:2018yxy}.
Their results are consistent with those presented above.
To summarise, within the stated assumptions on the expected
performance of the FCC-hh detector, a precision target on the Higgs
self-coupling of $\delta \kl =$~5\% in the gg$\to$HH channel appears
achievable, by exploiting several techniques and decay modes, and
assuming the future theoretical progress in modelling signals and
backgrounds.

\subsection{The EW phase transition}
Aside from the need to clarify the deep origin of the Higgs potential,
there is a second very interesting issue raised by EWSB: what was the
nature of the transition that, during the big bang, led to the broken
symmetry phase in which we live today? This is more than a matter of
curiosity: the EW phase transition (EWPT) could have been the seed for
the out-of-equilibrium state needed to freeze into a matter-antimatter
asymmetric world some primordial baryon number and CP violating
interactions~\cite{Sakharov:1967dj}.  This process of EW baryogenesis
(EWBG~\cite{Cohen:1993nk,Morrissey:2012db}), relying on the SM baryon
number violation induced at high temperatures by
sphalerons~\cite{Kuzmin:1985mm}, requires a strong first-order EWPT,
and a sufficent source of CP asymmetry.

These conditions, unfortunately, are not met by the SM with a 125~GeV
Higgs. On one side the CKM source of CP violation is too
weak~\cite{Gavela:1993ts}. On the other~\cite{Kajantie:1995kf}, the SM
Higgs potential for $m_H\gsim 70$~GeV causes the transition from the
high-temperature vacuum at $H=0$ to smoothly cross over to the $H=v$
vacuum, without creating the potential barrier needed for a
first-order transition.  The required conditions, however, can be met
in a variety of BSM scenarios.  CP violation relevant to the
matter-antimatter asymmetry can arise from new interactions over a
broad range of mass scales, possibly well above 100~TeV (for a recent
review, see e.g. Ref.~\cite{Servant:2018xcs}).  Exhaustively testing
these scenarios may therefore go beyond the scope of the FCC. For the
phase transition to be sufficiently strong, on the other hand, one
expects new particles to exist with masses typically below one TeV,
whose interactions with the Higgs boson modify the Higgs potential.
Should they exist, these particles and interactions could appear at
FCC, setting a key scientific opportunity and priority for the FCC.

\begin{figure}[ht]
\centering
\includegraphics[width=0.62\textwidth]{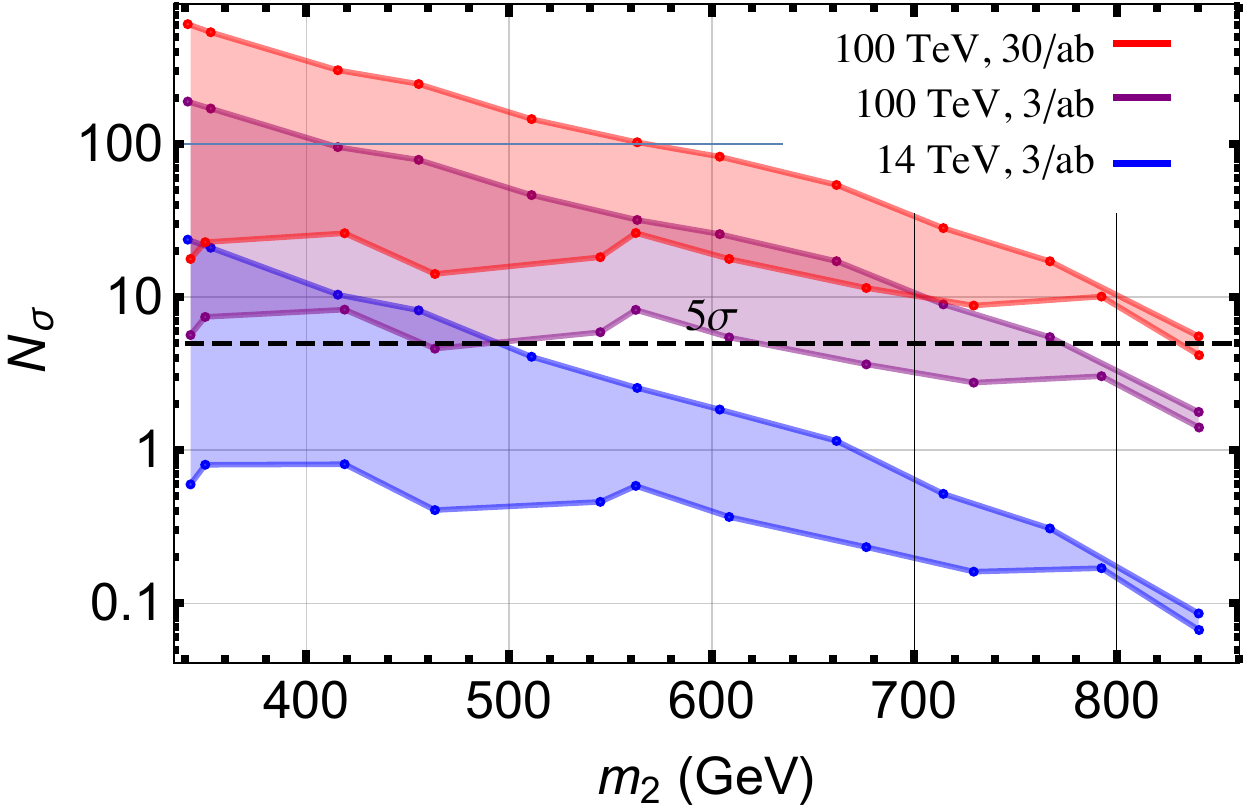}
\caption{Discovery potential for the singlet-induced strong first
  order EWPT using resonant di-Higgs production combining
  4\texttau\ and b\={b}\textgamma\textgamma\ final
  states~\cite{Kotwal:2016tex}. Vertical axis gives significance as a
  function of the singlet-like scalar mass $m_2$ for the HL-LHC (blue
  band) and the FCC-hh with 3 ab$^{-1}$ and 30 ab$^{-1}$ (purple and
  red bands, respectively). }
\label{fig:ewpt}
\end{figure}

An important task to be addressed by a future collider, therefore, is
to establish in a conclusive way whether or not the EWPT has been of
strong first order. This was first introduced as a possible no-lose
target for future colliders in Ref.~\cite{Curtin:2014jma}, which 
considered in detail ``nightmare scenarios'', namely models
particularly challenging to assess at colliders.  The challenge can be
met by probing the existence of suitable particles and interactions,
either through a direct search, or indirectly through the
modifications induced on the Higgs interactions. A more complete
overview of the studies carried out so far in the context of future
colliders in given in Ref.~\cite{Mangano:2651294}. Here we present
just one example, given by a theory with an extra real scalar S,
coupled as follows:
\begin{eqnarray}
  V(H,S)&=& -\mu^2(H^\dagger H) + \lambda(H^\dagger H)^2 +
  \frac{a_1}{2} S (H^\dagger H) \nonumber \\
  && +  \frac{a_2}{2} S^2 (H^\dagger H)
  +\frac{b_2}{2} S^2 +\frac{b_3}{3} S^3 +\frac{b_4}{4} S^4 \; .
\label{eq:HS}
\end{eqnarray}
As a singlet under the SM, S can only be produced at colliders in
association with, or in the decay of, Higgs bosons, which makes its
rate highly suppressed. This type of model is therefore among those
experimentally most challenging. Depending on the values of the $a_i$
and $b_i$ parameters, several scenarios are possible, where S itself
may or may not acquire an expectation value, at temperatures higher or
lower than H. Some of these scenarios will lead to acceptable
phenomenology and to a strong first-order EWPT, others will not. In
general, the 125~GeV Higgs boson will be a mixture of H and S, which
we call here h$_1$, and h$_2$ will be the heavier state. The mixing
between h$_1$ and h$_2$ will make it possible for h$_2$ to be produced
in gluon fusion, and, when sufficiently heavy, to decay to h$_1$
pairs. Figure~\ref{fig:ewpt}, from Ref.~\cite{Kotwal:2016tex}, shows
the significance of the searches for $\rm h_2\to h_1h_1 \to$
b\={b}\textgamma\textgamma, at the HL-LHC and FCC-hh, for various
luminosities. The shaded areas correspond to the envelope of results
obtained for combination of parameters in Eq.~\ref{eq:HS} satisfying
the strong first-order EWPT conditions.  The energy and luminosity of
FCC-hh are necessary, but sufficient, to meet the 5$\sigma$ goal for
the full spectrum of viable models.

\begin{figure}[!htb]
\centering
\includegraphics[width=0.7\textwidth]{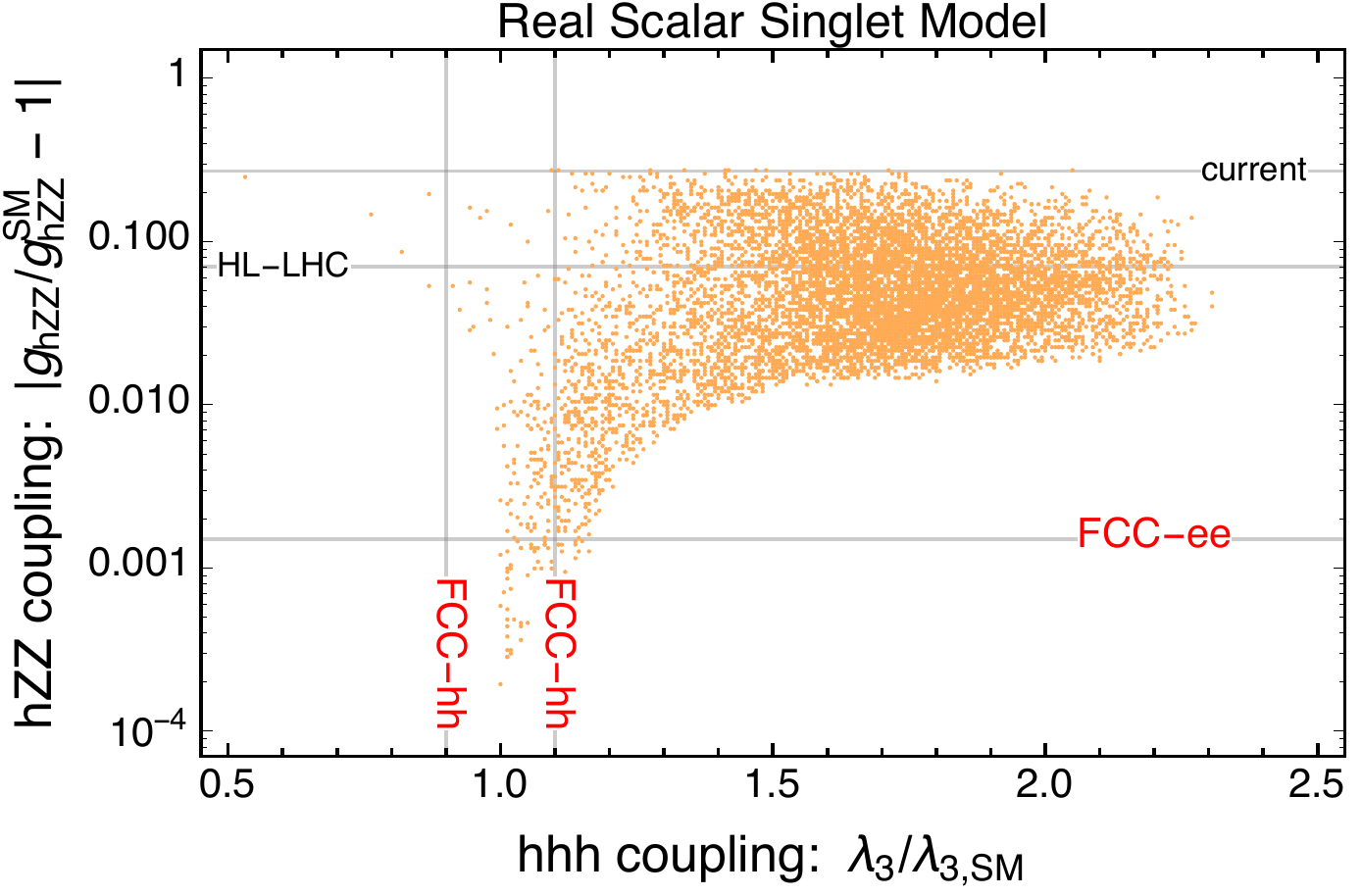}
\caption{Correlation between changes in the HZZ coupling (vertical
  axis) and the HHH coupling scaled to its SM value (horizontal axis),
  for a space scan of the S singlet model parameters leading to a
  first-order phase transition. Adapted from
  Refs.~\cite{Huang:2016cjm}, where similar plots, for a broad set of
  BSM models, are also shown.}
\label{fig:HuangLongWang}
\end{figure}

The observation of these signatures could then be correlated to
precise measurements of the Higgs couplings, in order to learn more
about the underlying origin of these signals, and break possible
degeneracies among various
interpretations. Figure~\ref{fig:HuangLongWang}, from
Ref.~\cite{Huang:2016cjm}, shows the size of deviations, from their SM
value, of the Higgs triple self-coupling and its coupling to a pair of
Z$^0$ bosons. The points in the scatter
plot correspond to parameters of the S scalar model fulfilling the
condition of a strong first-order EWPT, and detectable directly as in
Fig.~\ref{fig:ewpt}. The projected precision at FCC-ee (for the ZHH
coupling) and at FCC-hh (using a conservative $\pm 10\%$ uncertainty),
can cover most of the parameter range.

Independently of the issue of EWBG, collisions among bubbles created
by a strong phase transition during the early universe, can
lead~\cite{Hogan:1986qda,Kamionkowski:1993fg} to the generation of a
stochastic background of primordial gravitational waves (GWs), adding
a further possible remarkable signature for a strong first-order
EWPT. Reference~\cite{Huang:2016cjm} studied the models discussed
above, in the context of the observation potential of future satellite
GW interferometers, like eLISA~\cite{Caprini:2015zlo}.

Establishing the nature of the EWPT is therefore a target with a broad
range of phenomenological ramifications, which makes the FCC a
facility of unique value!

\section{Direct searches for new physics}
The LHC experiments have already published many more than 1000 papers
reporting the results of searches for new physics. While there is
certainly some duplication (the same search was repeated at different
energies, and by different experiments), this number gives a
sense of the immense variety of models that can be probed by a
collider. The FCC facility will allow to extend all of these searches,
and to probe new parameter ranges, both at the high- and at the
lower-end of the mass spectrum. Many more models will emerge in the
future, adding to the immense discovery potential enabled by the
energies and rates available at the FCC. In this lecture we give a
brief overview of some of the new features that characterize this
potential. In the case of FCC-hh, a more extensive review can be found
in Ref.~\cite{Golling:2016gvc}.

\subsection{The reach at high mass}
\begin{figure}[!ht]
\begin{center}
\includegraphics[width=0.47\textwidth]{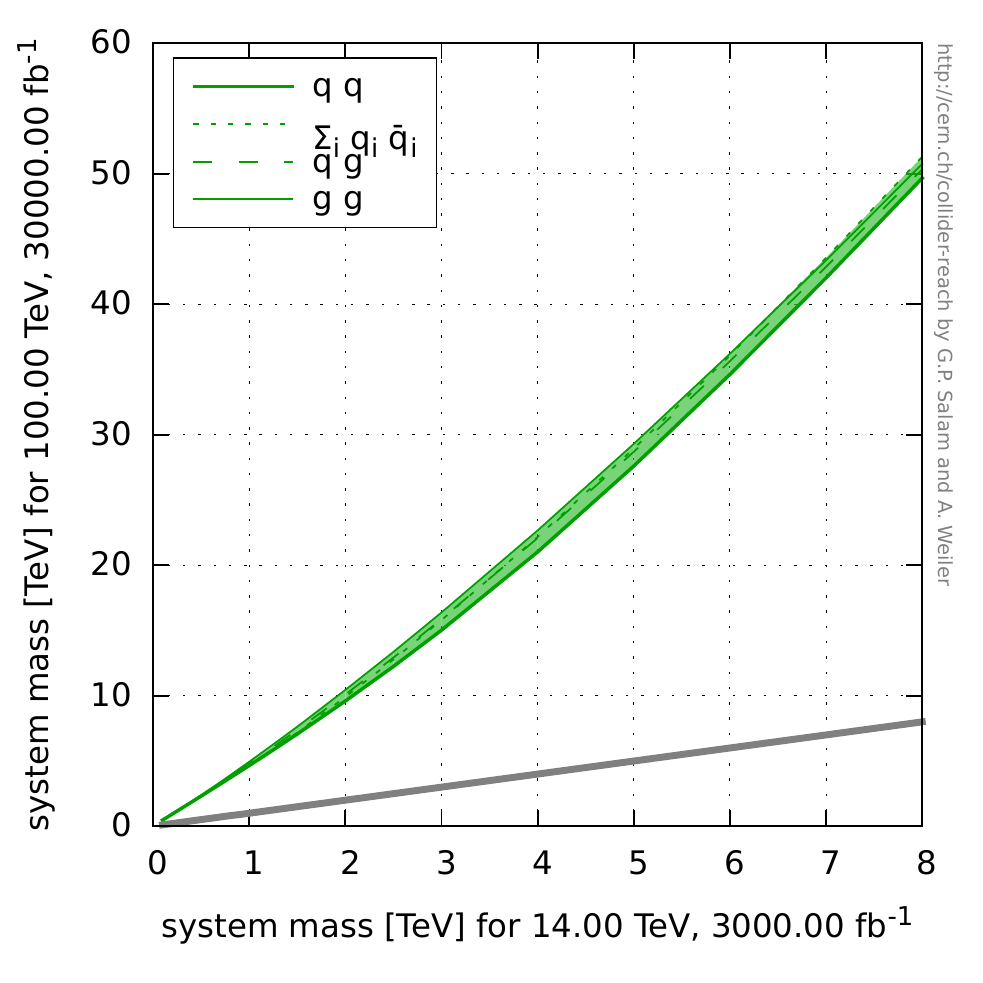} 
\hfill
\includegraphics[width=0.47\textwidth]{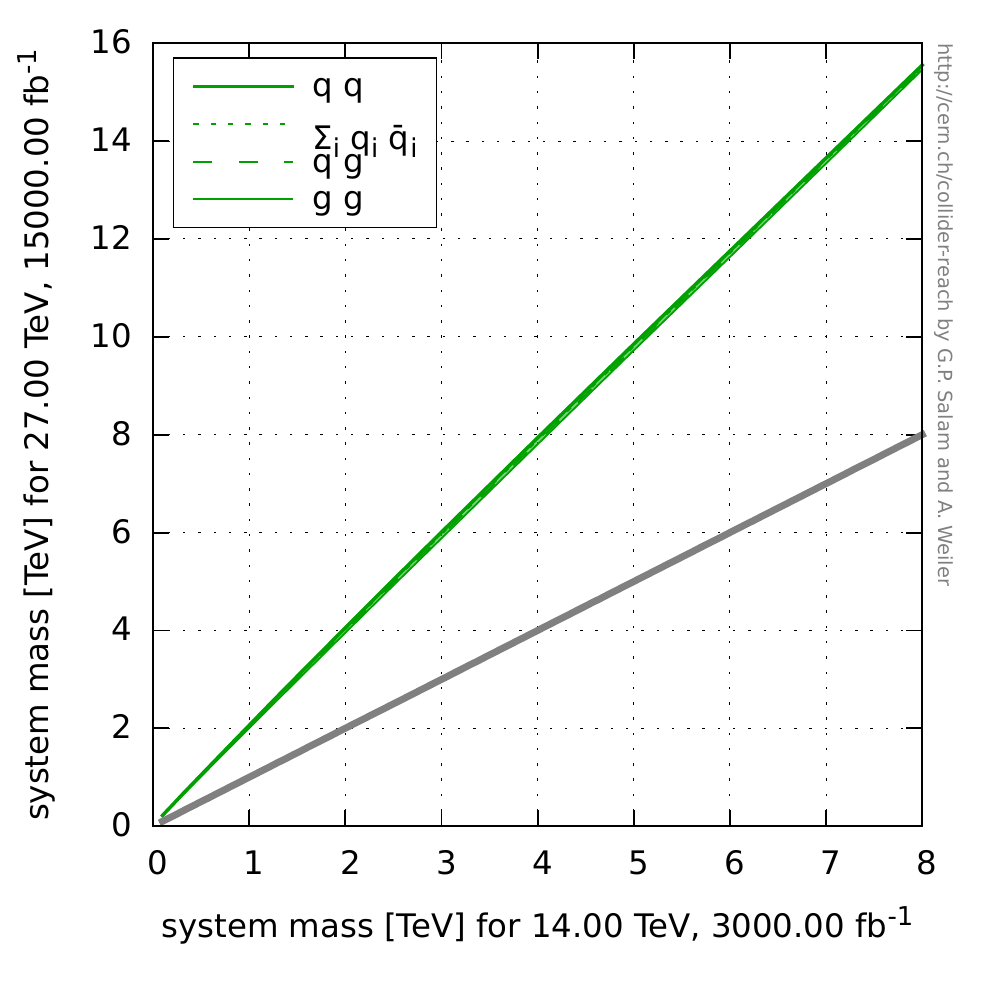} 
\end{center}
\vspace{-3mm}
\caption{\label{fig:pp-massreach} Left (right): the
  discovery/exclusion mass reach for resonance searches at FCC-hh
  (HE-LHC), as a function of the equivalent target for the
  HL-LHC~\cite{colliderreach}.}
\end{figure}

To first approximation, if the beam energy of a pp collider goes up by
a factor $z$, we expect the mass reach to go up by a similar factor,
provided the integrated luminosity is properly increased by a factor
$z^2$. This is because the partonic cross section to produce an object
of mass M scales typically like $\sigma(M) \sim 1/M^2$, and therefore
$\sigma(zM) \sim \sigma(M) / z^2$. This is confirmed by
Fig.~\ref{fig:pp-massreach}, which shows the projections for the
discovery or exclusion reach at the FCC-hh (or at the HE-LHC), as a
function of the equivalent HL-LHC targets, and assuming backgrounds
scale with energy like the signals.

However things can be more subtle. For example, when
potential signals are subject to large backgrounds, it is
important to consider whether the signal and the backgrounds evolve
with energy in the same way or not. Or detector properties can force
changes of analysis strategies (for better or for worse) when the mass
scale of final states is increased, and this can modify the discovery
opportunities at larger energies and mass. Finally, in specific cases
one may not be interested in extending the discovery reach in mass,
but in sensitivity. This is the case of the search for rare or
forbidden decay BRs of a given particle, say of the Higgs, or the
search for superweakly coupled resonances at low masses. In all of
these cases, the energy scaling of the discovery reach does not admit
a simple universal rule, and should be considered on a case by case
basis. We shall return to these issues in Sect.~\ref{sec:Escaling},
and we start now with an overview of the general search potential at
the highest masses.

\subsection{$s$-channel resonances}
\label{sec:Zprime}
Searching for heavy resonances produced in the $s$ channel probes a
large variety of BSM scenarios, from the existence of new gauge
interactions, to the existence of excited quarks, a signal of quark's
compositness. The signatures are typically straightforward, with mass
peaks in various two-body final states. The measurement of the
heaviest objects potentially within the reach of a 100~TeV collider,
however, poses new challenges relative to the LHC. For example, in
this multi-TeV domain, hadronic decays of final-state top quarks or
W/Z gauge bosons appear as collimated jets, and require enhanced
calorimeter granularity to apply substructure discriminators tagging
the object and reducing the potentially overwhelming QCD dijet
backgrounds. Reference~\cite{Jamin:2019mqx} describes in more detail
the detector features that are necessary to fully exploit the
discovery potential at the highest masses, and provides concrete
examples of analyses for many BSM models. We summarize here some of
the main results. 

\begin{figure}[!htb]
\centering
 \includegraphics[width=0.32\textwidth]{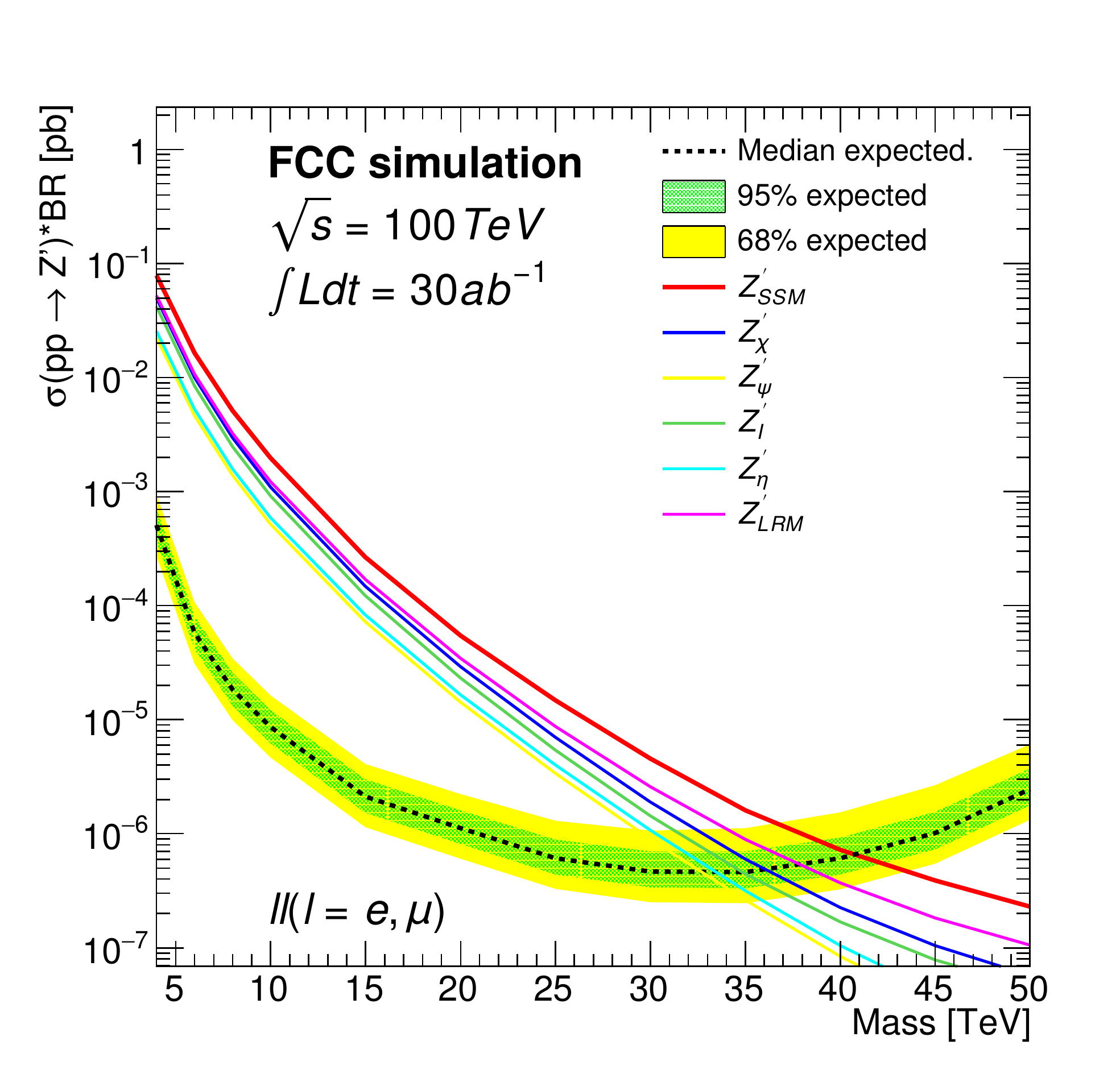} 
\includegraphics[width=0.32\textwidth]{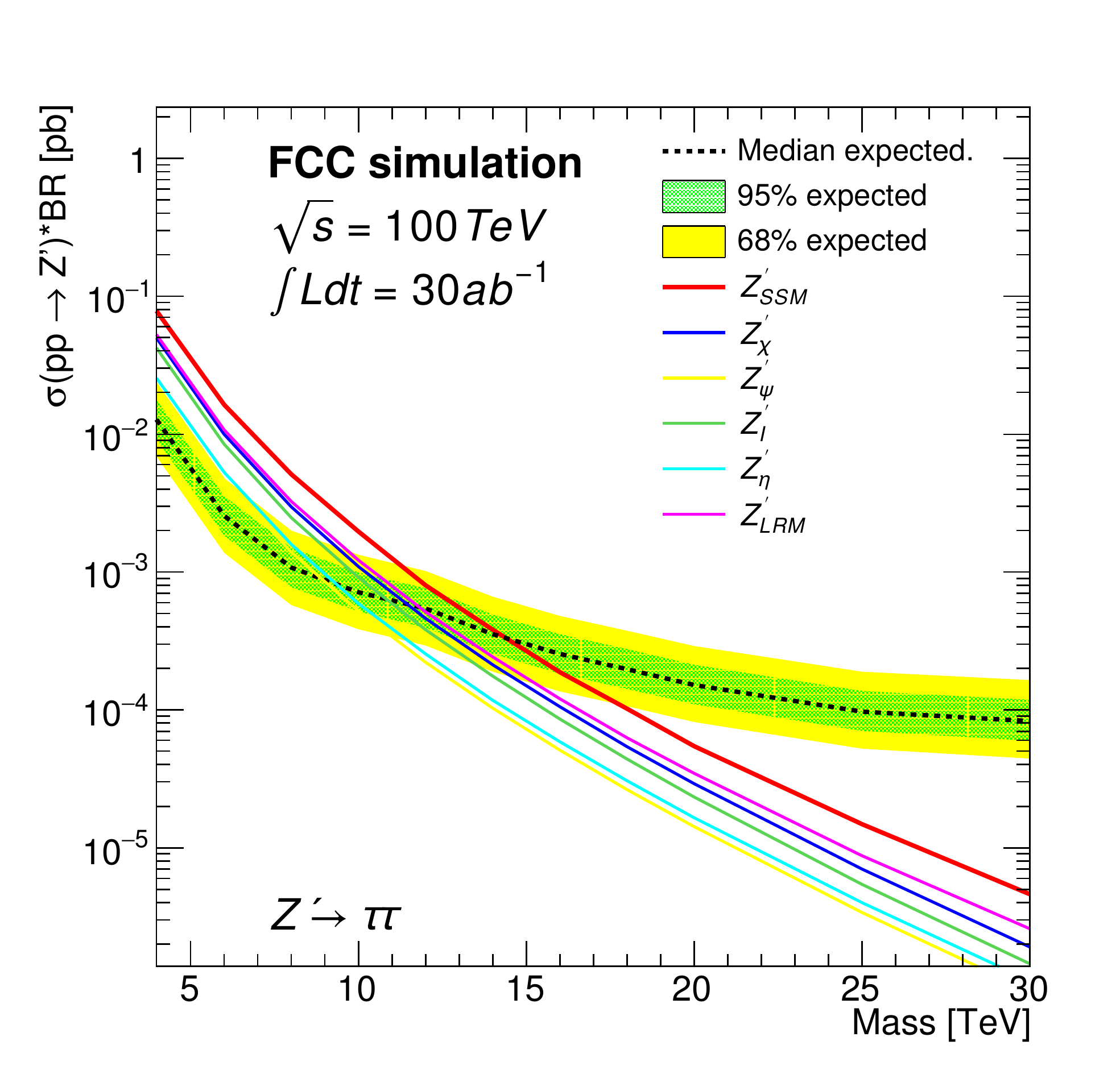}
\includegraphics[width=0.32\textwidth]{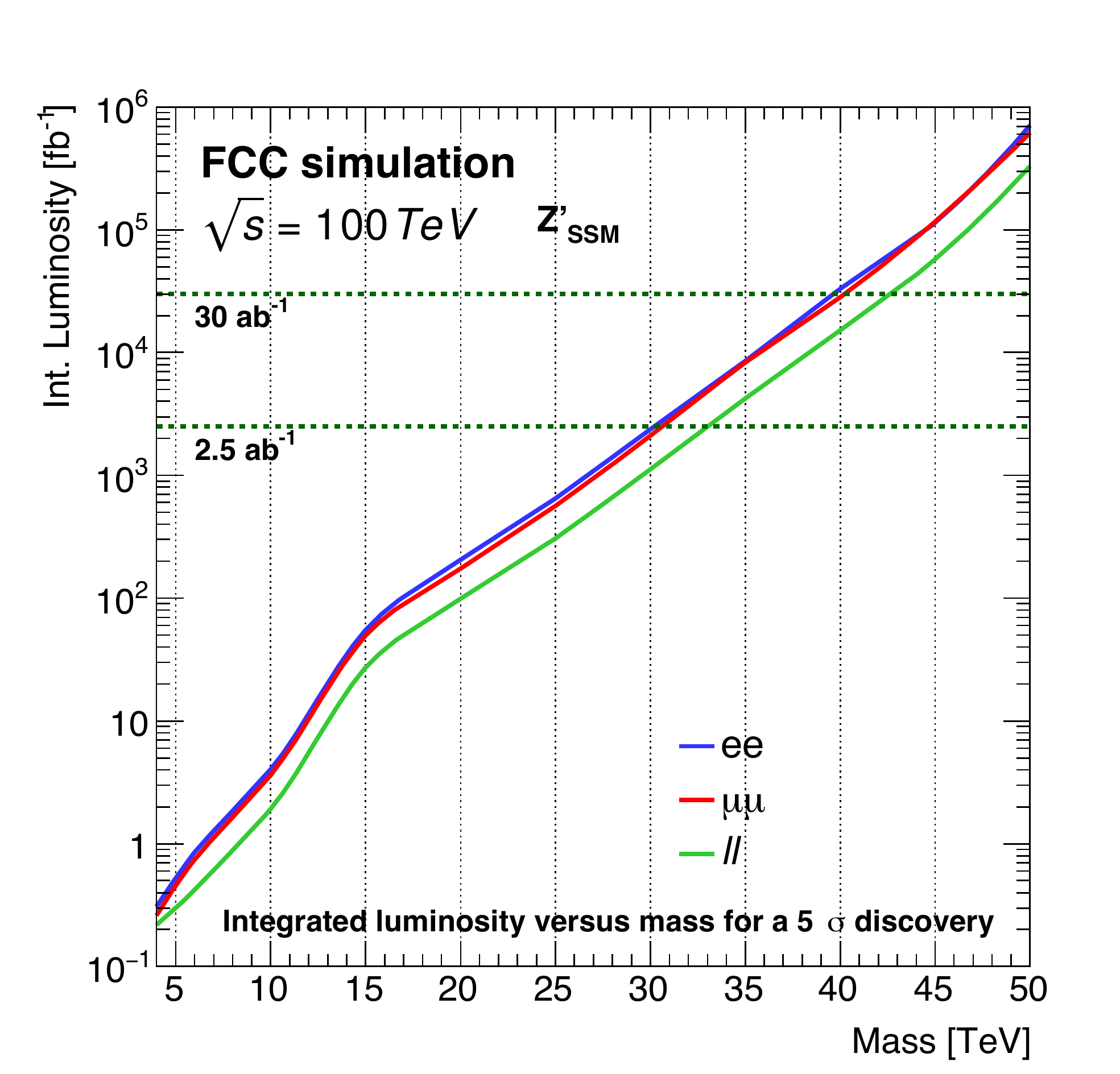}
  \caption{Left (centre): Cross section times branching ratio 95\%CL
    limits versus mass, for the ee+\textmu\textmu (\texttau\texttau)
    final states, compared to the expectations of several Z$^\prime$
    models. Right: luminosity required for the $5\sigma$ discovery of
    a SSM Z$^\prime$ in the dilepton channel, versus the hypothetical
    resonance mass. From Ref.~\cite{Jamin:2019mqx}.}
  \label{fig:Ztoleptons}
\end{figure}

The sensitivity to new Z$^\prime$ gauge bosons decaying to leptons is
shown in Fig.~\ref{fig:Ztoleptons}, for a set of extra-U(1) models
considered in the literature~\cite{Rizzo:2014xma,Han:2013mra}. The
right plot of this figure shows the luminosity required for the
5$\sigma$ discovery of a sequential SM (SSM) Z$^\prime$ (namely a
gauge boson with couplings to the SM particles identical to those of
the Z), as a function of its mass.  

\begin{figure}[!htb]
\centering
 \includegraphics[width=0.32\textwidth]{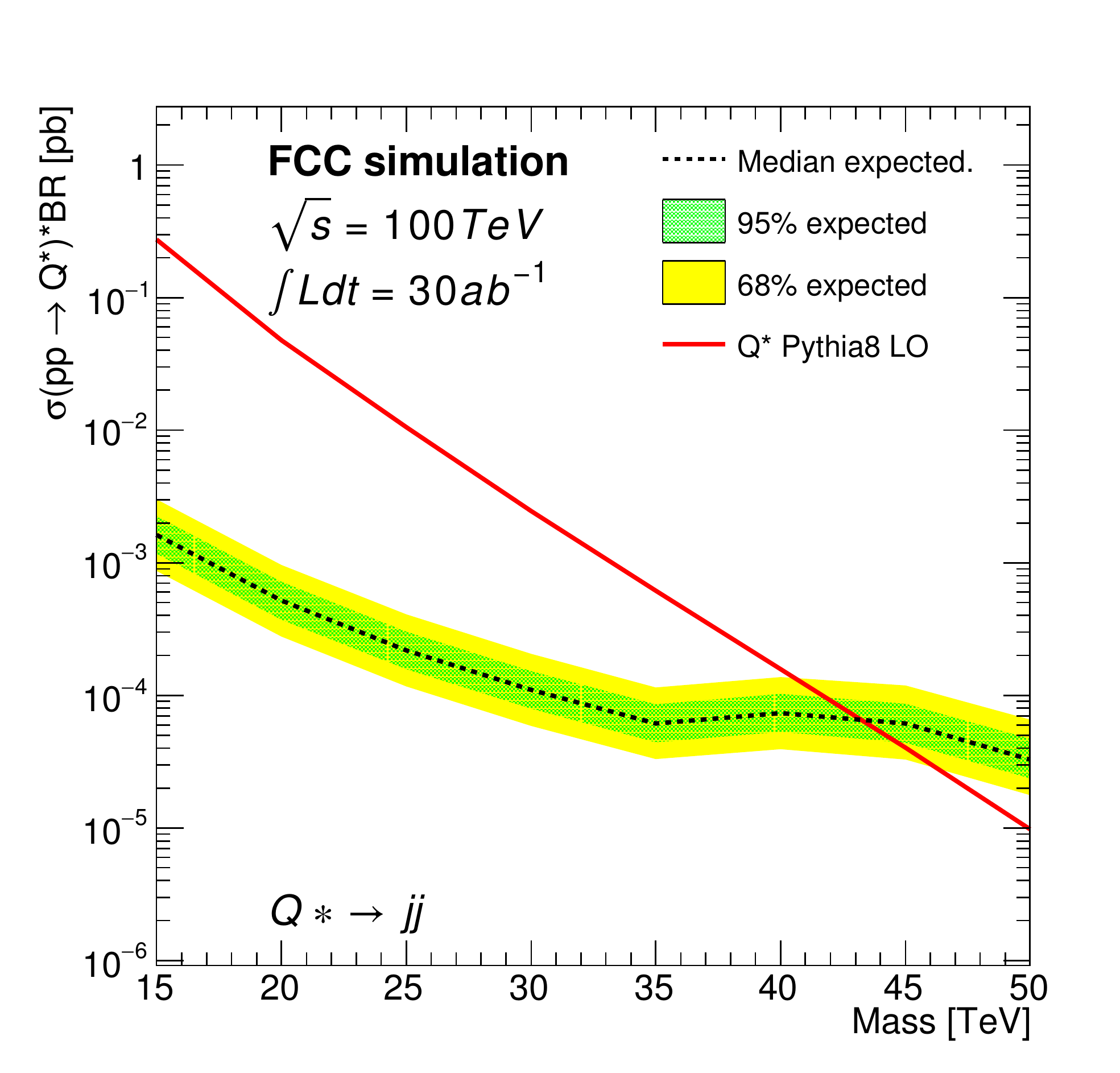} 
\includegraphics[width=0.32\textwidth]{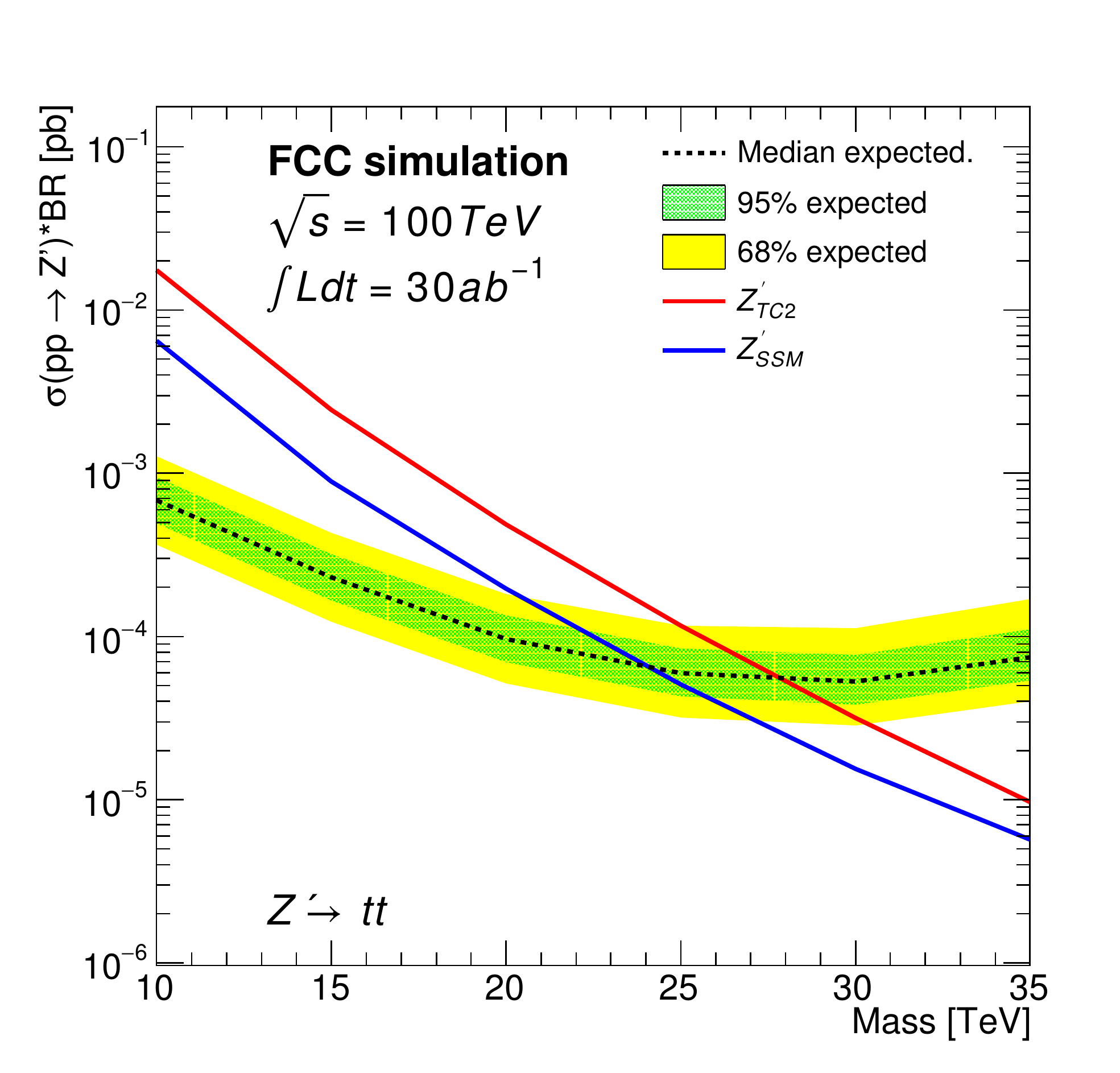}
\includegraphics[width=0.32\textwidth]{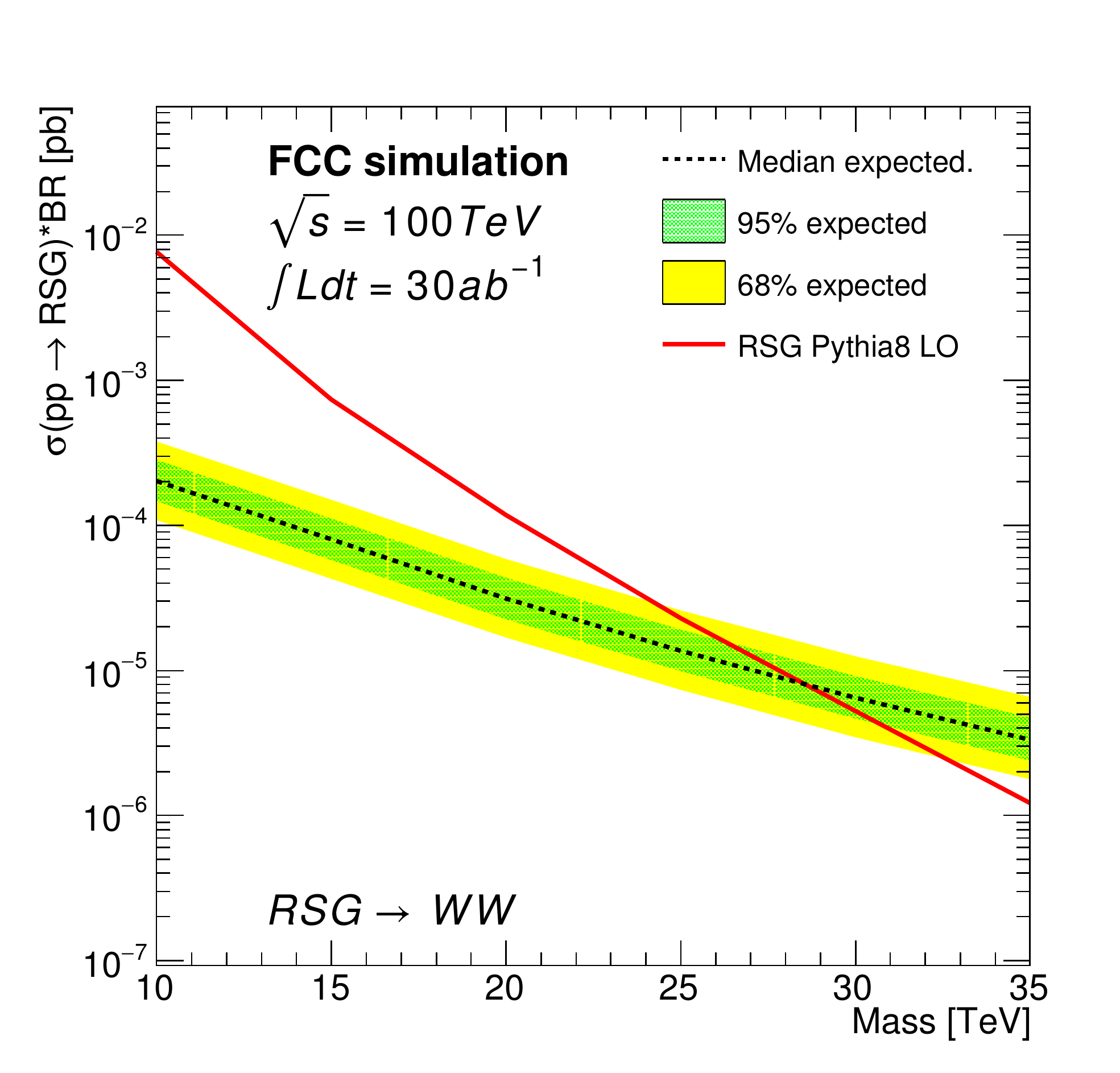}
  \caption{Sensitivity to various resonance models using fully
    hadronic final states. For left to right: excited quarks Q$^*$ in
    the dijet channel; Z$^\prime$ in the $\rm t\bar{t}$ hadronic
    channel (SSM and top-assisted technicolour~\cite{Hill:1994hp},
    TC2, models); $\rm G_RS$ in the WW fully hadronic channel. From
    Ref.~\cite{Jamin:2019mqx}.} 
  \label{fig:Ztojets}
\end{figure}

\begin{figure}[!htb]
  \centering
  \includegraphics[width=0.70\textwidth]{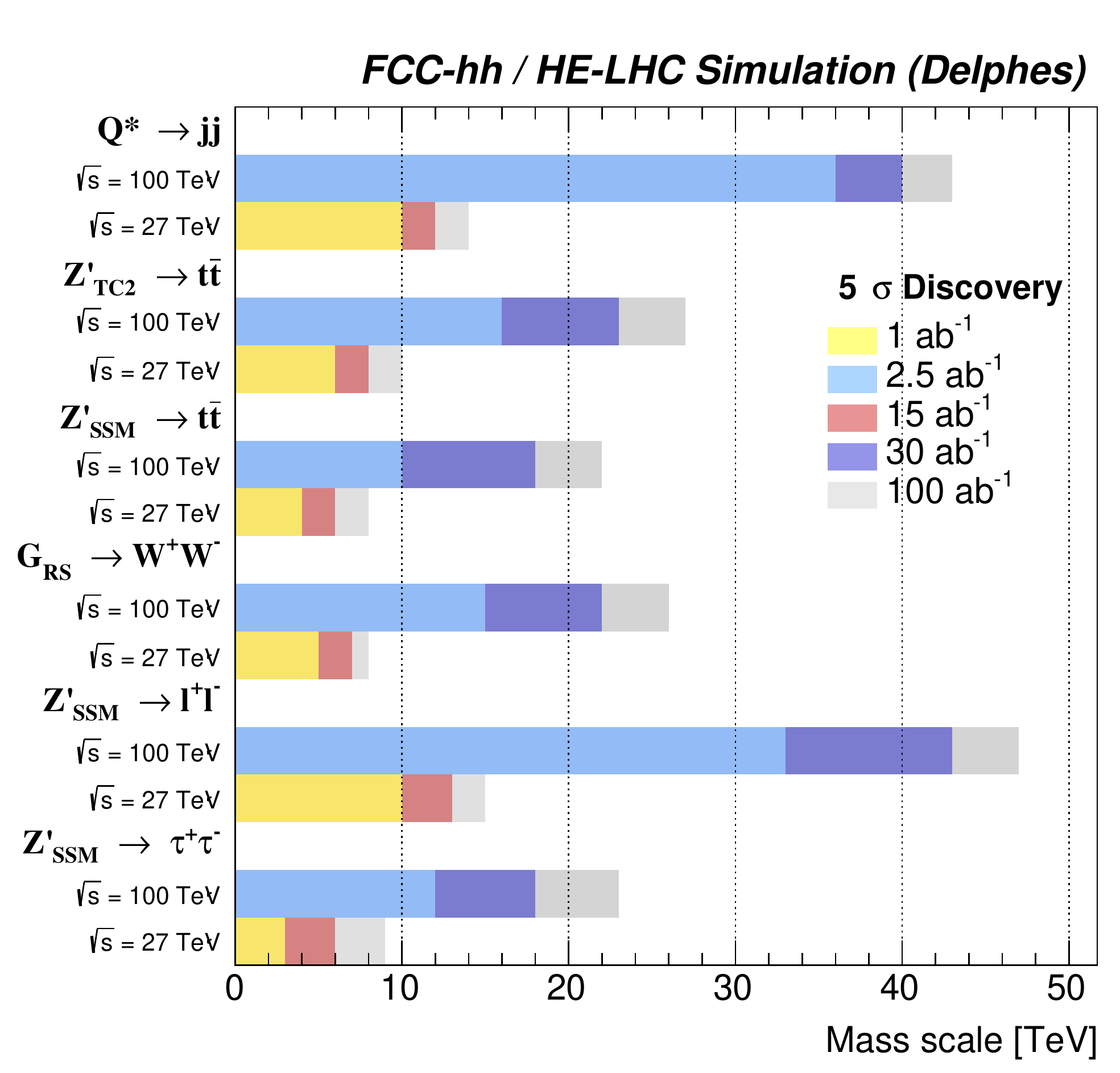}
  \caption{Summary of a $5\sigma$ discovery reach as a function of the resonance mass for different luminosity scenario of FCC-hh and HE-LHC.}
  \label{fig:Zsummary}
\end{figure}

To explore the sensitivity to hadronic final states, three scenarios
were considered in~\cite{Jamin:2019mqx}: a Z$^\prime$ in $\rm q\bar{q} \to$ Z$^\prime \to
\rm t\bar{t}$, a Randall-Sundrum graviton~\cite{Randall:1999ee} in
$\rm gg/q\bar{q} \to G_{RS}\to W^+W^-\to$ jets, and an excited quark
resonance~\cite{Baur:1987ga,Baur:1989kv} in $\rm qg\to Q^*\to$
dijet. The 5$\sigma$ discovery range reaches 18~TeV (24) for $\rm
Z_{SSM}'\to$t\={t} ($\rm Z_{TC2}'$), 22~TeV for $\rm G_{RS}\to WW$ and
40~TeV for the excited quark. The corresponding exclusion limits are
shown in Fig.~\ref{fig:Ztojets}.
A summary of the discovery reach for various models, for various
integrated luminosities at 27 and
100~TeV, is shown in Fig.~\ref{fig:Zsummary}. You can verify that the
gain in reach, from 27 to 100~TeV, is indeed the expected approximate factor of
4. We also notice that, with the assumed target integrated
luminosities for HE-LHC and FCC-hh, 15 and 30\iab{} respectively, the
discovery reach is close to saturation, as the gain obtained by
further increasing the luminosity to 100\iab{} is rather marginal. 

\subsection{Pair production of new particles}
\begin{figure}[!ht]
\begin{center}
\includegraphics[width=0.65\textwidth]{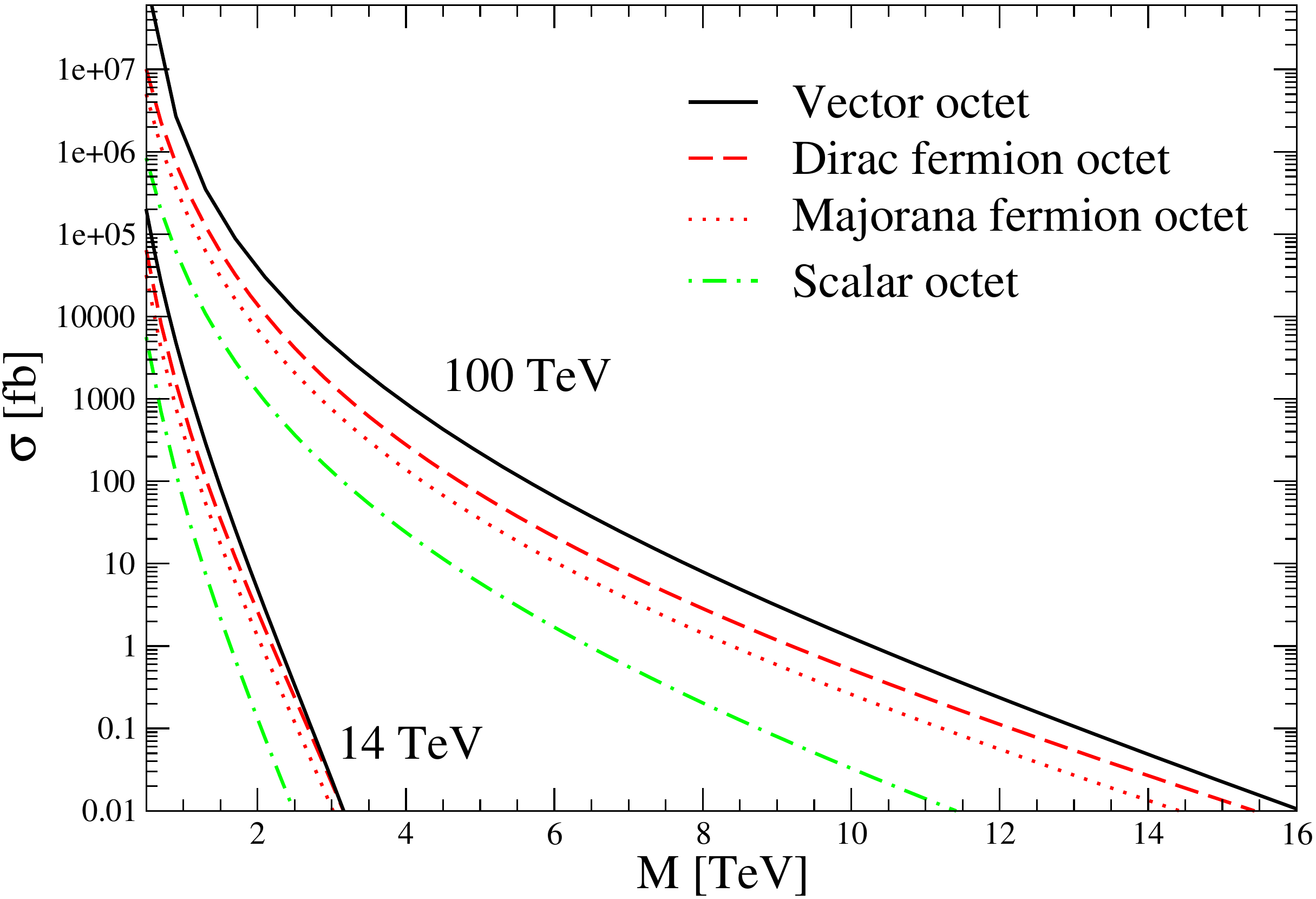} 
\end{center}
\vspace{-3mm}
\caption{\label{fig:octets} The pair production cross section for colour octets of various spins.}
\end{figure}

Looking beyond resonances, heavy particles with SM gauge charges
feature in many scenarios beyond the SM. Typically, the production
cross section may be calculated for a given particle mass and gauge
representation.  As a hadron collider, the FCC-hh discovery reach for
new coloured particles is extensive.  As a simple example for
illustration, the pair production cross section for colour octet
particles of various spins at the LHC versus FCC-hh is shown in
Fig.~\ref{fig:octets}.  At a given mass value the cross section at 100
TeV is orders of magnitude greater than at 14 TeV.  Not only does this
demonstrate a significant increase in the discovery potential at
FCC-hh, but also implies that if a new particle were discovered in the
HL-LHC runs, it would be possible to study this particle in
significantly greater qualitative detail at FCC-hh.

\begin{figure}[th]
\begin{center}
\includegraphics[width=0.54\textwidth]{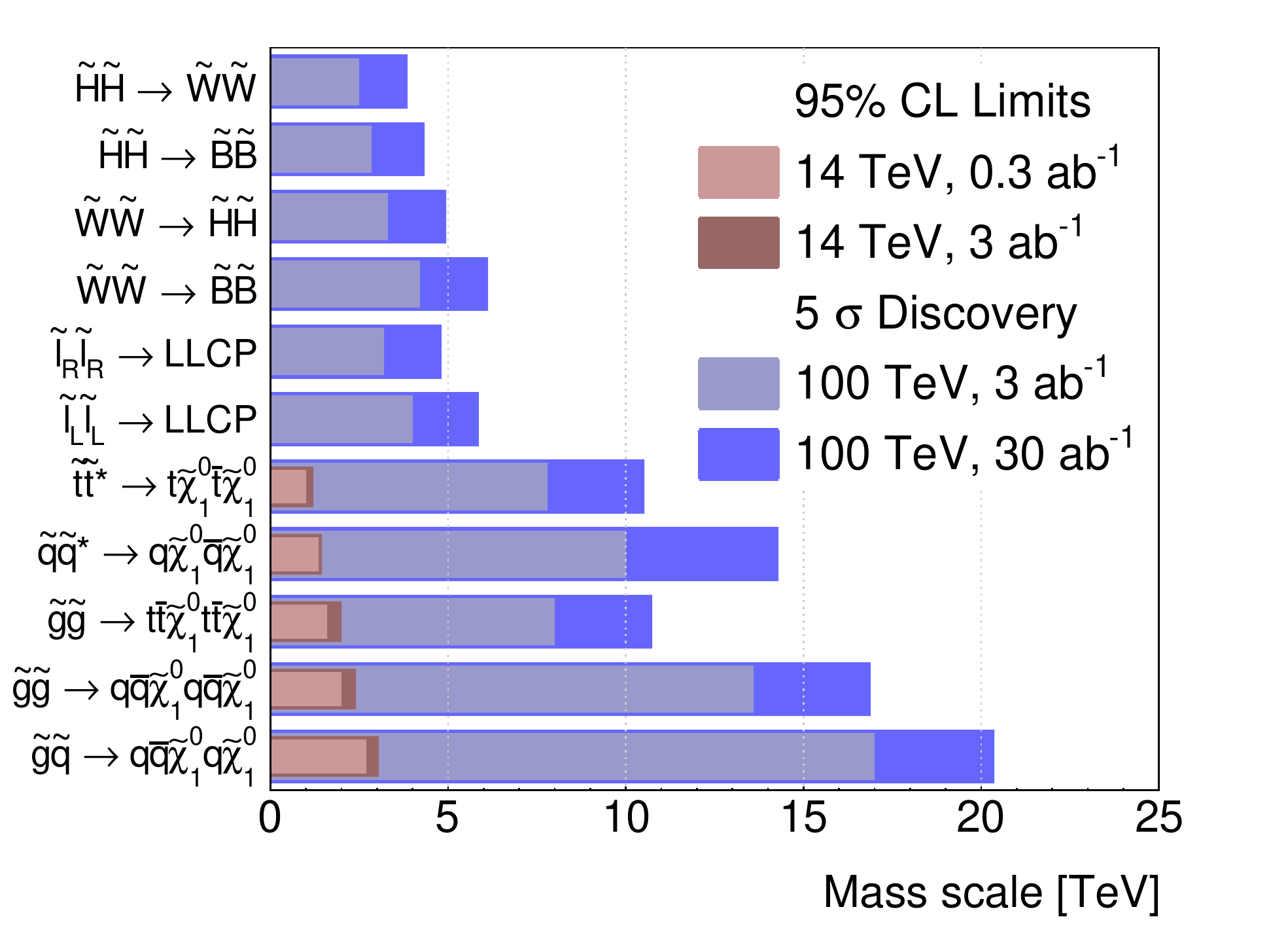}
\hfil
\includegraphics[width=0.42\textwidth]{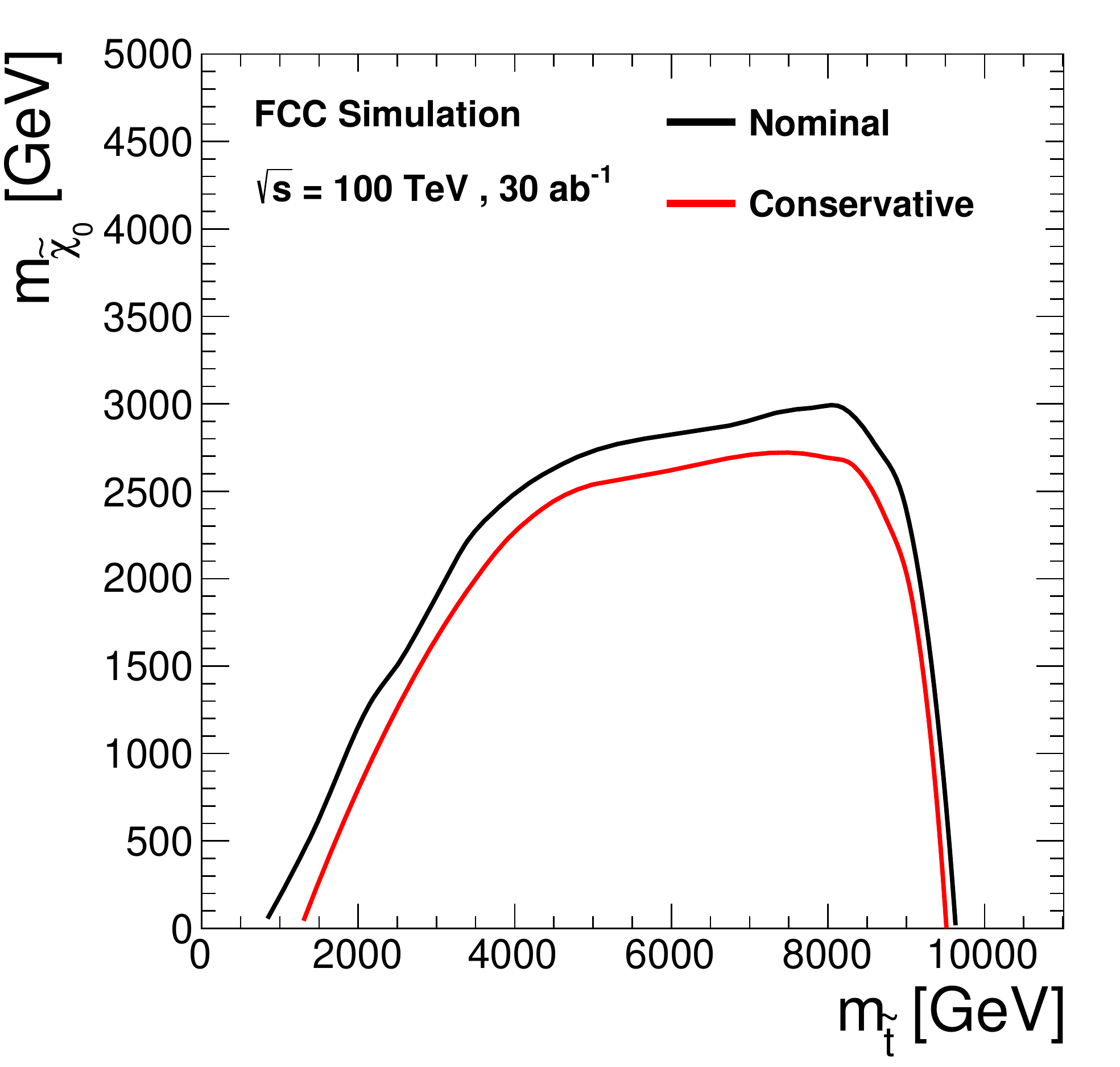} 
\end{center}
\vspace{-8mm}
\caption{\label{fig:SUSY} Left: Projected direct FCC-hh $2 \sigma$ and
  $5 \sigma$ discovery reach for supersymmetric Higgsinos, Winos,
  sleptons, stops, squarks, and gluinos (see
  Ref.~\cite{Golling:2016gvc} for details). Right: 5$\sigma$ discovery
  reach for pair-produced stops at FCC-hh. The area below the solid red (black)
  curve represents the expected exclusion and the $\pm 1\sigma$
  contours for the nominal (conservative) scenario of associated
  systematic uncertainties~\cite{Gouskos:2642475}.}
\end{figure}

Dedicated studies have been carried out in the context of specific
models, taking into proper account backgrounds and analysis
opportunities. As an example, Fig.~\ref{fig:SUSY} shows the reach for pair production
of supersymmetric particles. The reach for the degenerate light-generation
squarks extends up to 15~TeV, for pair-produced gluinos up to 10-15~TeV,
depending on the decay mode, and goes up to 20~TeV for the associated production
of a squark and a gluino, assuming equal mass. More details on the
discovery reach for stop squarks, decaying to a top quark and a
neutralino, are shown in the right plot of
Fig.~\ref{fig:SUSY}, for different values of the neutralino mass (see
\cite{Gouskos:2642475} for the discussion of the analysis). 

\subsection{Considerations on the energy scaling of production rates at hadron colliders}
\label{sec:Escaling}
\begin{figure}[h]
\centering
\includegraphics[width=0.75\textwidth]{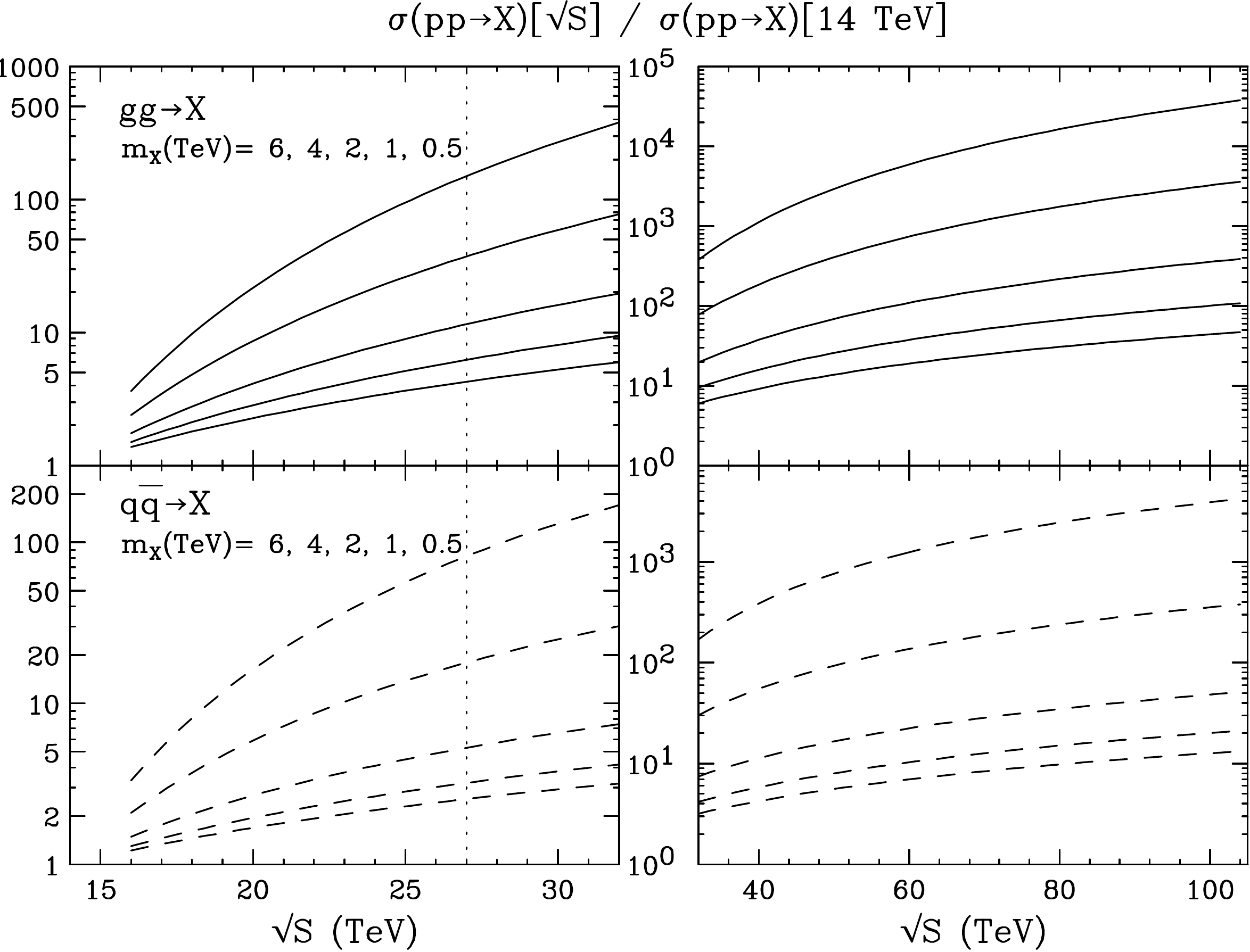}
\caption{Scaling with beam energy of the production cross section for
  final states of various masses, in the gg and $\rm q \bar{q}$
  channels. The final state X is characterized here by its mass: it
  can be a single resonance, as in $\rm q\bar{q}to Z'$, with
  $m_X=m_{Z'}$, or it could be a pair production, as in $\rm gg\to
  Q\bar{Q}$, in which case $m_X \gsim 2 m_Q$.}
\label{fig:sigvsE}
\end{figure}

There are circumstances in which it might make sense to
optimize the beam energy choice, as a function of a specific physics
goal, or when considerations of cost and technology may limit our
ambition to reach the highest possible energy. For example, if a
particle were discovered at the LHC, we might want to optimize the
design of the next collider in order to collect as soon as possible a
statistics large enough to study its properties.
Appendix~\ref{app:Escaling} presents, through simple analytic
approximations, some general qualitative features of production rates,
which help understanding the main properties of their dependence on
the pp center-of-mass energy\footnote{The results shown in this
  Section use parton luminosities calculated from real proton PDFs,
  not from the approximated forms used in
  Section~\ref{app:Escaling}.}. We learn there that what drives the
energy dependence of cross sections is just the partonic luminosity:
once the initial state is assigned, there is no need to specify the
details of the produced final state, and it is possible therefore to
make general statements valid for a large class of models.  We apply
this concept to the case of particles that could be discovered at the
HL-LHC, produced by gg or $\rm q\bar{q}$ initial
states. Figure~\ref{fig:sigvsE} shows the growth of the production
rates as a function of beam energy, relative to the rates at 14
TeV, for systems of mass $m_X$ between 0.5 and 6~TeV. These could be
$s$-channel resonances (e.g. $\rm q\bar{q}\to Z'$, with $m_{Z'}=m_X$) or new particles
pair-produced (e.g. gluinos $\rm \tilde{g}$ in $\rm gg \to
\tilde{g}\tilde{g}$, with $m_{\tilde{g}}\sim m_X/2$). Few comments
are in order.

To start with, the figure shows clearly that the gain with energy is
most significant for the heaviest final states.  Furthermore, objects
produced in the $\rm gg$ channel receive a much larger boost in rate
than objects produced in the $\rm q\bar{q}$ channel. 

\noindent
\rule{5cm}{1pt}\\
{\bf Exercise:} discuss and explain the origin of these behaviors.
\\
\rule{5cm}{1pt}

 In practice, a 6~TeV object produced in the $\rm gg$ channel can see
 a factor of 40,000 increase in going from 14 to 100~TeV, while for a
 0.5~TeV object produced in the $\rm q\bar{q}$ channel (this could be,
 for example, a very weakly coupled 500~GeV $\rm Z'$, or a pair of 250~GeV
 electroweakinos from supersymmetry), the rate goes up only by a factor
 of 10! In the former case, it is clear that one would like to go to
 100~TeV to study in real detail the new discovery. In the latter
 case, one might argue that it is more economical to just sit at the
 HL-LHC for a bit longer, or maybe to move to a TeV-scale $\rm e^+e^-$
 collider, where the study of a weakly interacting sub-TeV particle is
 likely cleaner than in a hadron collider. You can play with
 Fig.~\ref{fig:sigvsE} and with your favourite new physics scenarios
 accessible to the LHC, and make up your mind on what is the best way
 to go after discovery!

 In my mind, when dealing with the further exploration of possible LHC
 discoveries, nothing beats the additional power provided by the
 higher energy of a 100~TeV collider.  The higher energy can turn out
 to be crucial not just in studying this object with greater
 statistics (even if only by a factor of 10 as in the 500~GeV case
 discussed above), but it allows us to explore the other new particles
 that will most likely accompany any new BSM scenario, of which the
 LHC discovery can just be a first signal of. For example, the new
 resonance could be the first state in a tower of Kaluza-Klein (KK)
 states, and we may want to search for further KK recurrences. If
 these states were EW gauginos, we would want to search for the
 heavier supersymmetric particles. And so on...

Another important consideration is that, as the LHC has shown, a
greater kinematical reach can lead to greater benefits even when
dealing with lighter objects, which are produced well below the
kinematic limit of the collider. Let me discuss a couple of examples:
$\rm H\to b\bar{b}$ and the search for resonances in the 100~GeV
mass range.

It is well known that the huge QCD background makes it almost
impossible to pull out the $\rm H\to b\bar{b}$ signal from the
inclusive $\rm gg\to H$ sample, in spite of the fact that we are well
below the kinematic threshold and that the inclusive Higgs production
rate is huge. At high energy, however, one can consider the $\rm VH$
($\rm V=W,Z$) associated production, in a configuration of large
invariant mass of the $\rm VH$ pair, where the Higgs is highly
boosted.  The large momentum of the b-pair gives a higher b-tagging
efficiency, allows the use of jet-substructure techniques, where the
b-pair is confined within a broad jet with a two-prong substructure,
and gives a more precise measurement of the invariant mass of $\rm
b\bar{b}$ jet.  This makes the peak structure of the signal at
$m_{bb}\sim m_H$ more pronounced, and better visible over the
continuum background from $\rm q\bar{q} \to V b\bar{b}$ and $\rm
q\bar{q} \to V Z[\to b\bar{b}]$. This technique has allowed
ATLAS~\cite{Aaboud:2018zhk} and CMS~\cite{Sirunyan:2018kst} to
discover the $\rm H\to b\bar{b}$ decay, with a signal enhanced by the
requirement of $p_T(H)\gsim 150$~GeV. For the use of these signals at
high transverse momentum, in the study of SM EFT operators, see
e.g.~\cite{Banerjee:2018bio}.

\begin{figure}[h]
\centering
\includegraphics[width=0.75\textwidth]{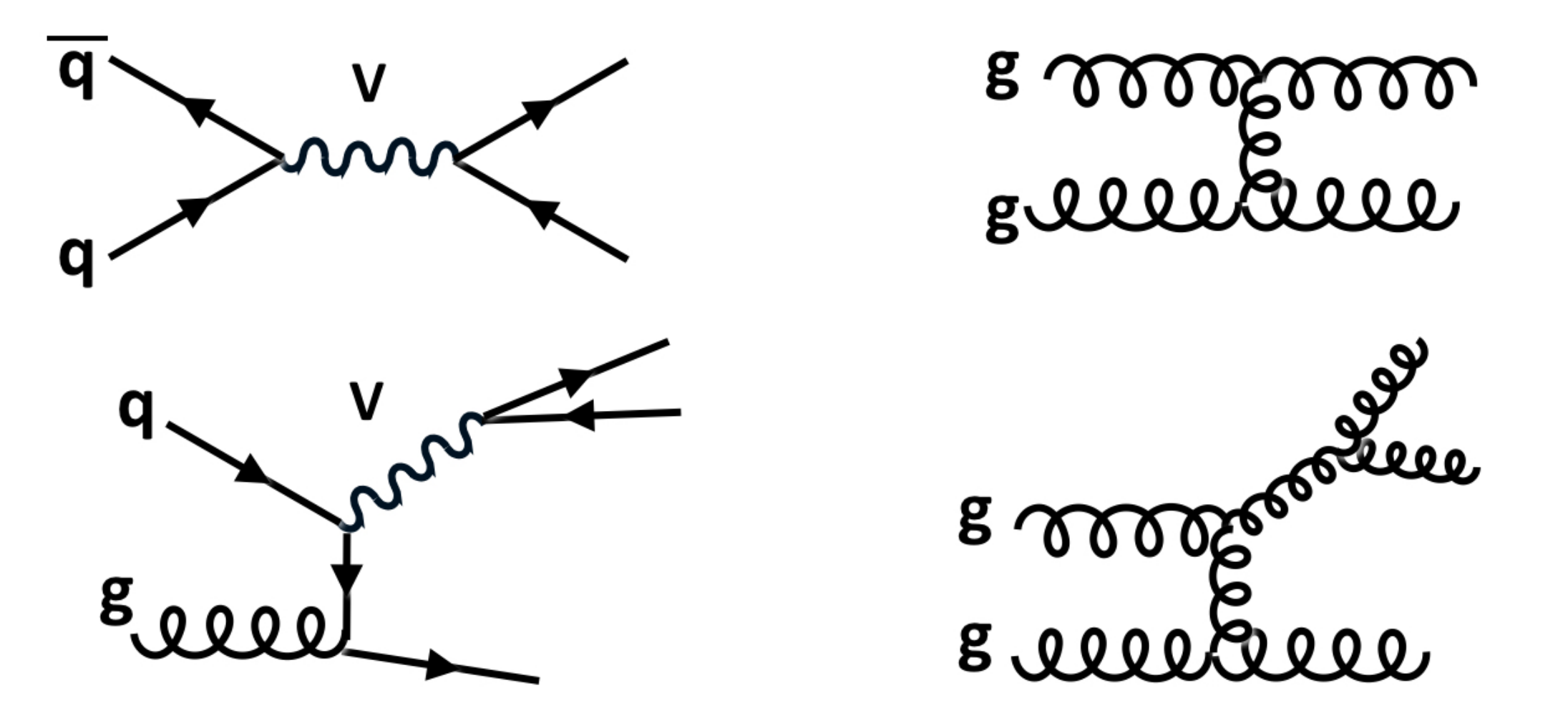}
\caption{Signal and background diagrams for the production of a
  hadronic resonance. Top row: inclusive production. Bottom row:
  production at large $p_T$ in association with a jet. Similar
  processes, with a $\rm q\bar{q}$ initial state, are possible with
  the emission of photons or W/Z bosons.}
\label{fig:Vj}
\end{figure}
 A second important example is the search for a weakly interacting
 hadronic resonance $V\to jj$ with mass of (few) hundred GeV. This is
 relevant in the search of potential candidates for mediators between
 the SM and a DM sector~\cite{Abercrombie:2015wmb}. The search for
 low-mass hadronic resonances is difficult, because of the huge QCD
 dijet background. On one side this background forces high thresholds
 on the minimum jet $E_T$ at the trigger level, strongly suppressing
 the signal. On the other, what's left of the resonant signal ends up
 being swamped by the large QCD continuum. Requiring the resonance to
 be highly boosted (see Fig.~\ref{fig:Vj}), will significantly improve
 the situation. The large $E_T$ of the jets from the V decay, or of
 the recoil jet, will maximize the trigger efficiency. The S/B ratio
 will also be strongly enhanced, since the signal initial state will
 be $\rm qg$, rather than $\rm q\bar{q}$, gaining a factor of $\rm
 g(x)/\bar{q}(x)$. Jet substructure techniques will also help, since
 the structure of a high-$E_T$ fat jet coming from the decay of a
 color-singlet resonance has distinctive features that can be used to
 separate it from a generic QCD high-$E_T$ jet. Selecting events with
 a jet of $E_T>500$~GeV, CMS~\cite{Sirunyan:2017dnz} could probe
 resonance masses down to 100~GeV, with a sensitivity orders of
 magnitude better than any previous measurement at previous hadron
 colliders, in spite of the fact that this mass region was well within
 their kinematic reach. In this powerful analysis, even the hadronic
 decays of W and Z bosons, which have always been extremely hard
 to detect using inclusive dijet samples, emerge very clearly from
 the QCD background. The analysis can be extended to the case in which
 the resonance recoils against a photon~\cite{Aaboud:2018zba}, or
 decays into a pair of b quarks~\cite{ATLAS:2018hzj}.

To reiterate the previous message: high energy is a blessing even when
we are interested in searching low-mass objects, well below the
kinematic reach, since it provides a multitude of handles required to
promote otherwise elusive signatures into detectable final states!

\subsection{Dark matter searches}
The search for DM particles is another top-priority target of
any future collider. Of the motivations, possible explanations, and
search strategies for DM, you already learned a lot from other
lectures at this school~\cite{Lin:2019uvt,Hooper:2018kfv}, so I give all of
this for granted.

I start with one general comment, to put in perspective the value of
DM searches at a future collider. For all we know, DM could be
anything from an ultralight particle~\cite{Hui:2016ltb} in the ${\cal
  O}(10^{-20})$\,eV mass range, with a Compton wavelength as large as
a galaxy, to a primordial black hole~\cite{Ali-Haimoud:2019khd}, with
a mass of several solar masses. There are models for almost every mass
window in between, from axions~\cite{Hook:2018dlk} to sterile
neutrinos~\cite{tasi6}, from dark photons to weakly interacting
massive particles (WIMPs~\cite{tasi4}). It is clear that there is no
single experiment, whether in space or in the laboratory, that can
guarantee the discovery of DM. Hopefully, at least one of those under
way or under planning will soon succeed. Its success will reward those who
built it, and those who proposed the relevant model. But this
discovery will not diminuish the value of all the other attempts,
theoretical and experimental, that have gone and will go into the
search across the board. As a community, we must endeavour to explore
all reasonable hypothesis.

In this context, colliders provide a powerful window on the broad, and
theoretically well justified, class of WIMP models.

If at any point in the history of the early universe the DM is in
thermal equilibrium with the SM particles, then its relic density
today can be estimated by studying how it decouples from the SM, a
process known as freeze-out. For particles which are held in
equilibrium by pair creation and annihilation processes, $(\chi \chi
\leftrightarrow {\rm SM})$ one finds the simple relation
that~\cite{Kolb:1990vq}
\begin{equation}
	\Omega_{\rm DM}h^2 \sim \frac{10^{9} {\mathrm GeV}^{-1}}{ M_{\rm pl}}
        \frac{1}{\langle \sigma v \rangle} \,, \label{eqn:relicD1}
\end{equation}
where $\langle \sigma v \rangle$ is the velocity averaged annihilation
cross section of the DM candidate $\chi$ into SM particles,
$\Omega_{\rm DM}h^2 \approx 0.12$ is the observed relic abundance of
DM~\cite{Ade:2015xua}, $M_{\rm pl}$ is the Planck scale and order one
factors have been neglected.

For a particle annihilating through processes which do not involve any
larger mass scales, the annihilation cross section scales as $\langle
\sigma v \rangle \sim g_{\rm eff}^4/M_{\rm DM}^2$, where $g_{\rm eff}$
is the effective coupling strength which parameterises the process. It
follows that
\begin{equation}
	\Omega_{\rm DM}h^2 \sim 0.12 \times \left(\frac{M_{\rm
            DM}}{2\;\mathrm{TeV}}\right)^2 \left(\frac{0.3}{g_{\rm
            eff}}\right)^4 .
    \label{eqn:relicD2}
\end{equation}
This approximate relation implies that a DM candidate with a mass at
or below the TeV~scale and which couples to the SM with a strength
similar to the weak interactions naturally has a relic density in
agreement with observations.  Furthermore, as the DM mass is reduced,
ever weaker couplings are required.  On one hand this is the main
reason why it is hoped to find evidence for DM at the LHC, but on the
other hand it already shows that a higher energy collider will be
necessary to fully probe the WIMP paradigm for DM.

Eq.~\ref{eqn:relicD2} shows that as the mass of DM increases, in order
to maintain the observed relic abundance, the annihilation cross
section also has to increase. This becomes inconsistent with unitarity
of the annihilation amplitudes at $M_{\rm DM} \lesssim 110$\,TeV, the
so called unitarity bound on the mass of
DM~\cite{Griest:1989wd,Blum:2014dca}. Most well motivated models of
WIMP DM do not saturate this bound, but rather have upper limits on
the DM mass in the TeV~range.

\begin{figure}[h]
  \centering
     \includegraphics[width=0.52\textwidth]{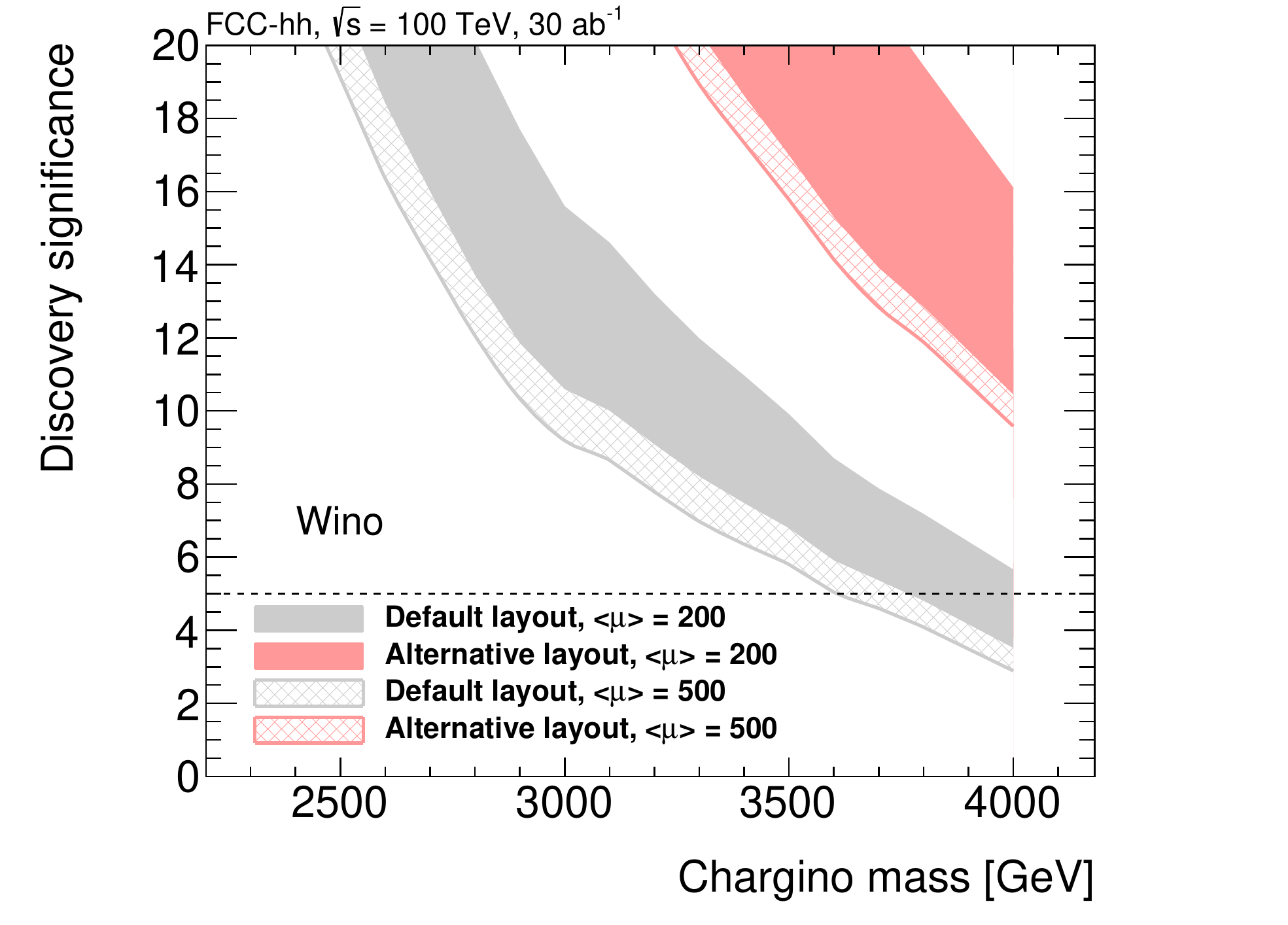}
     \hspace{-1cm}
     \includegraphics[width=0.52\textwidth]{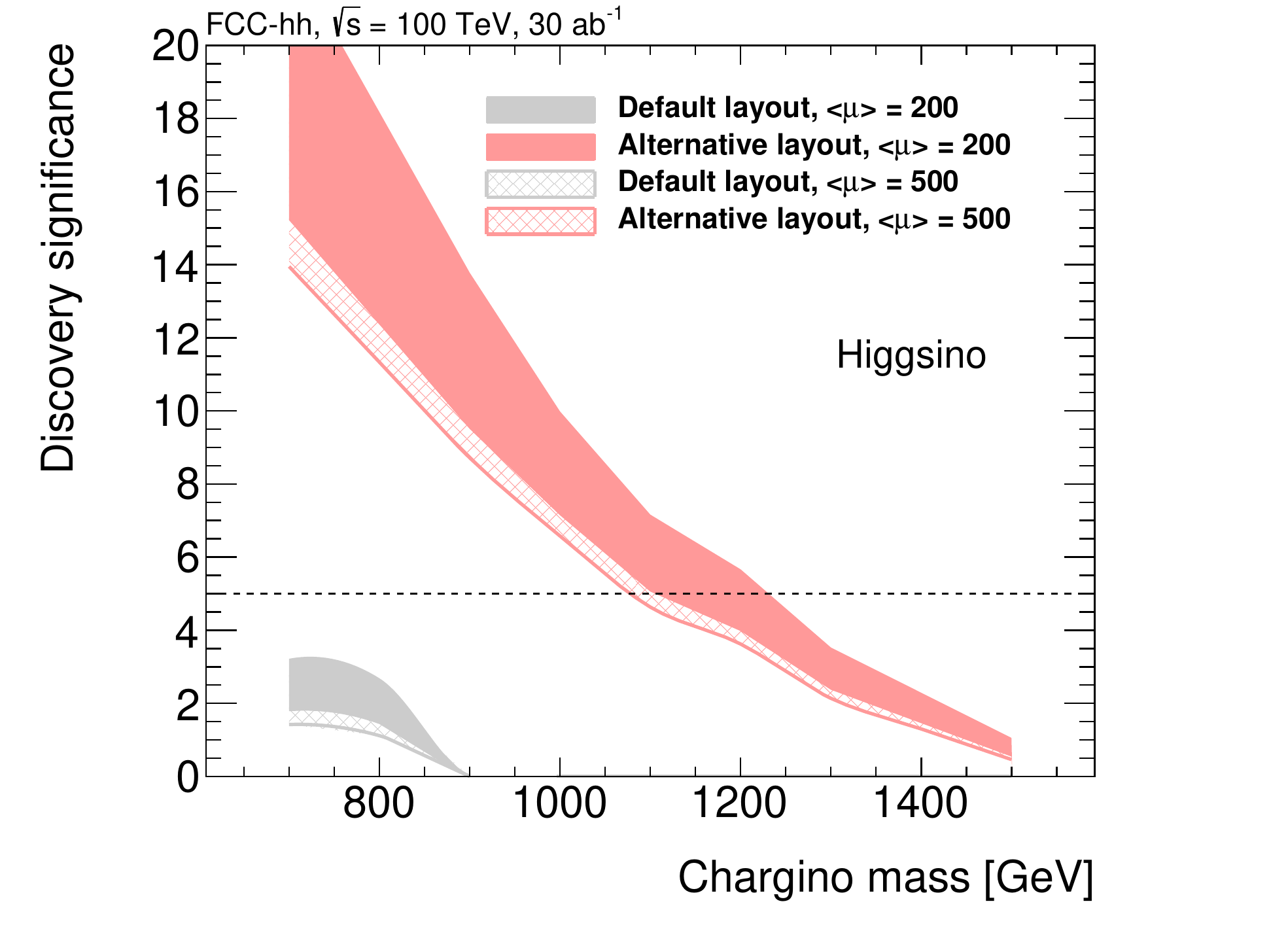}
  \caption{Expected discovery significance at 30 ab${}^{-1}$ with 500
    pile-up collisions.  The black (red) band shows the significance
    using two alternative tracker layouts, described in
    Ref.~\cite{Saito:2019rtg}. The band width represents the
    difference between two models of the soft QCD processes.}
  \label{fig:dis_track}
\end{figure}

For example, DM particles coupled to the SM via EW interactions, are
subject to mass upper limits of about 1~TeV (in the case of EW
doublets, such as supersymmetry higgsinos), and about 3~TeV (in the
case of EW triplets, such as the supersymmetric partners of the W/Z
bosons). These mass values are well beyond the reach of the HL-LHC,
because of the low production rates. The challenge to find them, or
rule them out, is therefore left to a future high-energy collider.  At
these high masses, EW symmetry leads to a further difficulty:
particles in EW multiplets tend to be degenerate in mass, with
electromagnetic mass splittings in the range of few hundred MeV, and
decay lifetimes in the range of several cm. The standard
production/decay signatures, such as $\rm pp
\to$\textchi$^+$\textchi$^{-}\to$\textchi$^0$\textchi$^{0}$+leptons or
jets, lead to very little visible energy, and are not even recorded by
the triggers. In these cases, a crucial viable signature is the
so-called ``disappearing track''~\cite{Mahbubani:2017gjh}: the charged
state travels for few cm through the tracking layers of the detector,
and suddenly disappears, vanishing into the non-interacting DM
particle and a small-momentum charged track, typically undetected.
The trigger is provided by a
recoiling jet. Gaining sensitivity to these
disappearing tracks is a crucial criterion in the design of future
trackers for the FCC-hh experiments~\cite{Benedikt:2651300}. Without
entering into the details of the analyses~\cite{Saito:2019rtg}, we
report here the results in Fig.~\ref{fig:dis_track}, which confirm
that the reach for discovery, or exclusion, of WIMPs at the upper mass
end is achieved.

For DM masses at the lower end of the WIMP spectrum, one typically
expects that annihilation proceeds through a mediator with $M_{\rm
  med} > 2 M_{\rm DM}$. Then the annihilation cross section is
suppressed by $(M_{\rm DM}^4/M_{\rm med}^4)$. Assuming that no
mediator particle exists with a mass below the Higgs mass, then this
puts a lower bound close to a GeV on the mass of the WIMP DM
candidate. In this region, and in general in the mass region which
allows for direct DM decays of the Higgs or Z bosons, the sensitivity
to their invisible decays discussed in Section~\ref{sec:higgs} and
\ref{sec:EW} provides an extensive
probe to probe dark sectors of relevance to the DM
issue~\cite{Mangano:2651294}. 

While there is no firm no-lose theorem yet, covering all potential
loopholes, the multiple studies carried out so far provide solid
evidence that the combination of FCC-ee and FCC-hh could indeed
exhaustively close the WIMP DM window.

\subsection{Other discovery opportunities}
As mentioned at the beginning of this section, there are countless
areas of opportunity for direct discovery, throughout the mass and
coupling spectrum. I briefly show here a few additional examples,
which emphasize once again the great potential coming not only from
the high-energy reach of the pp collider, but also from the increased
sensitivity in the low-mass region, thanks to the high statistics and
the flexibility of the overall program.

\begin{figure}[h]
\centering
\includegraphics[width=0.75\textwidth]{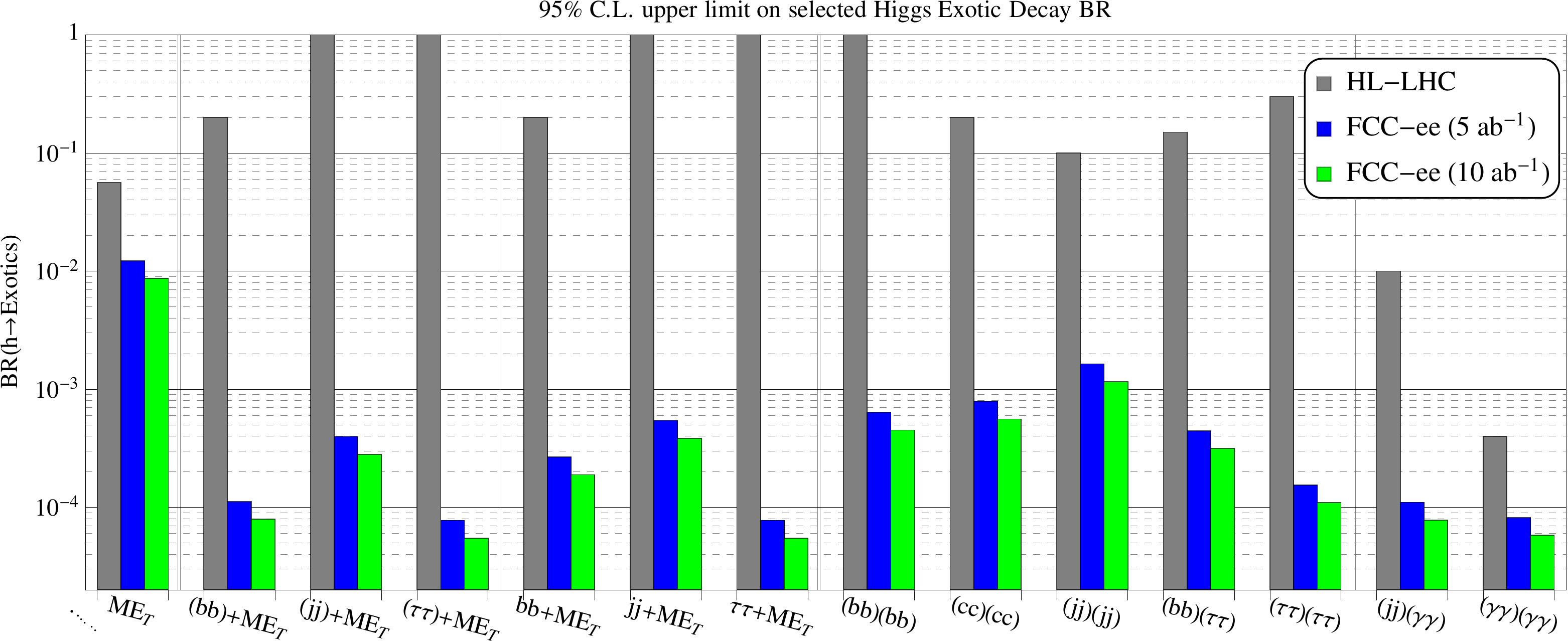}
\caption{Sensitivity to exotic Higgs decays at
  FCC-ee. From~\cite{Benedikt:2651299,Liu:2016zki,Liu:2017zdh}.  }
\label{fig:Hexodec}
\end{figure}

\begin{figure}[t]
\begin{center}
\includegraphics[width=0.85\textwidth]{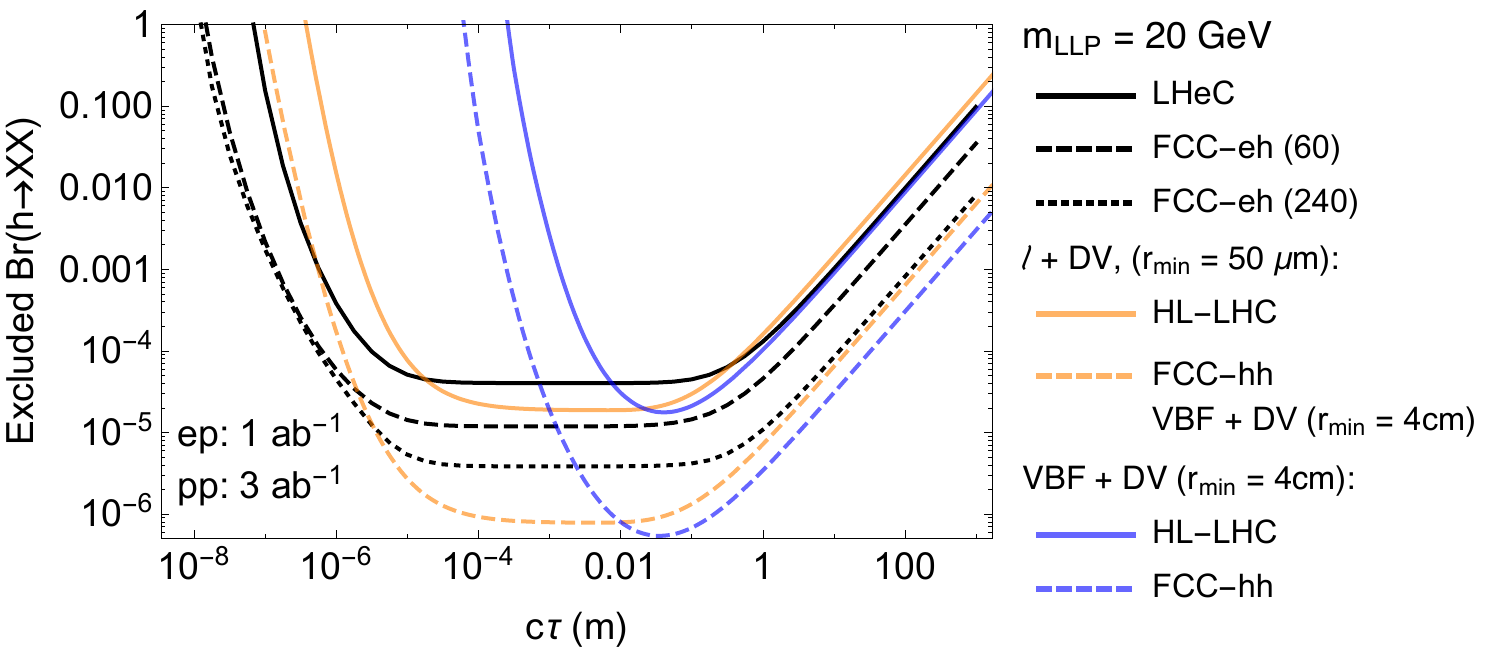} 
\end{center}
\vspace{-3mm}
\caption{\label{fig:displaced} Projected exclusion limits on exotic
  Higgs decay branching fractions to LLPs $X$ as a function of
  lifetime $c \tau$ for the LHeC, and for the FCC-eh at two different
  energies of the electron beam, 60 and 240 GeV. The
  excluded branching ratio scales linearly with luminosity under the
  assumption of no background.  The LLP mass in the plot is
  20\,GeV. For comparison, an estimate for the sensitivity of proton
  colliders without background is shown (blue), as well as a very
  optimistic estimate which assumes extremely short-lived LLP
  reconstruction (orange), from \cite{Curtin:2015fna}. This plot is
  taken from \cite{Curtin:2017bxr}.}
\end{figure}

To complete the exploration of the Higgs properties, it is important
to guarantee sensitivity to exotic decays, which could arise in a
variety of BSM models~\cite{Curtin:2013fra}, particularly in the
context of Higgs-portal scenarios, where the Higgs boson is the only
SM particle coupled to an otherwise dark
sector via dim-4 interactions like $\vert H \vert^2 \varphi^2$.
Figure~\ref{fig:Hexodec} shows the prospects for the detection
of decays covering a wide array of possible final states. The
sensitivity of the $\rm e^+e^-$ collider is crucial, as shown by the
significant improvement over the HL-LHC prospects, in spite of the much
higher Higgs statistics available. It is possible that the
FCC-hh could also contribute to this searches, at least in some
channels, although dedicated studies have not been carried out as
yet. 
A notable feature of the Higgs portal interaction $\vert H \vert^2
\varphi^2$ is that it respects a $\mathcal{Z}_2$ symmetry, $\varphi
\to - \varphi$.  Such a symmetry can make $\varphi$ stable or, if it
is explicitly broken by some small amount, $\varphi$ may be
long-lived.  This alone motivates searching for exotic decays of the
Higgs boson to particles that decay on a macroscopic time scale.  Such
scenarios arise frequently in theories beyond the SM, such as models
of neutral naturalness. Projections for sensitivity to such exotic
Higgs decays are shown in Fig.\ref{fig:displaced} for a variety of
collider scenarios, including the HL-LHC, FCC-hh, LHeC and FCC-eh.  At
hadron colliders two different triggering strategies are considered.
One involves VBF-tagging and displaced vertices and the other only a
single lepton and displaced vertices.
It is notable that at FCC-hh one can access exotic Higgs decays down
to branching ratios smaller than $10^{-6}$, demonstrating extreme
sensitivity to very exotic Higgs decays.  Furthermore, due to the
cleaner detection environment, the FCC-eh projections can push to
shorter lifetimes than are accessible at FCC-hh. 

A similar approach can be applied to Z decays. In this case, there is
no way any alternative collider could match the power of the
$10^{13}$ Z statistics of FCC-ee!
\begin{figure}[t]
\begin{center}
\includegraphics[width=0.85\textwidth]{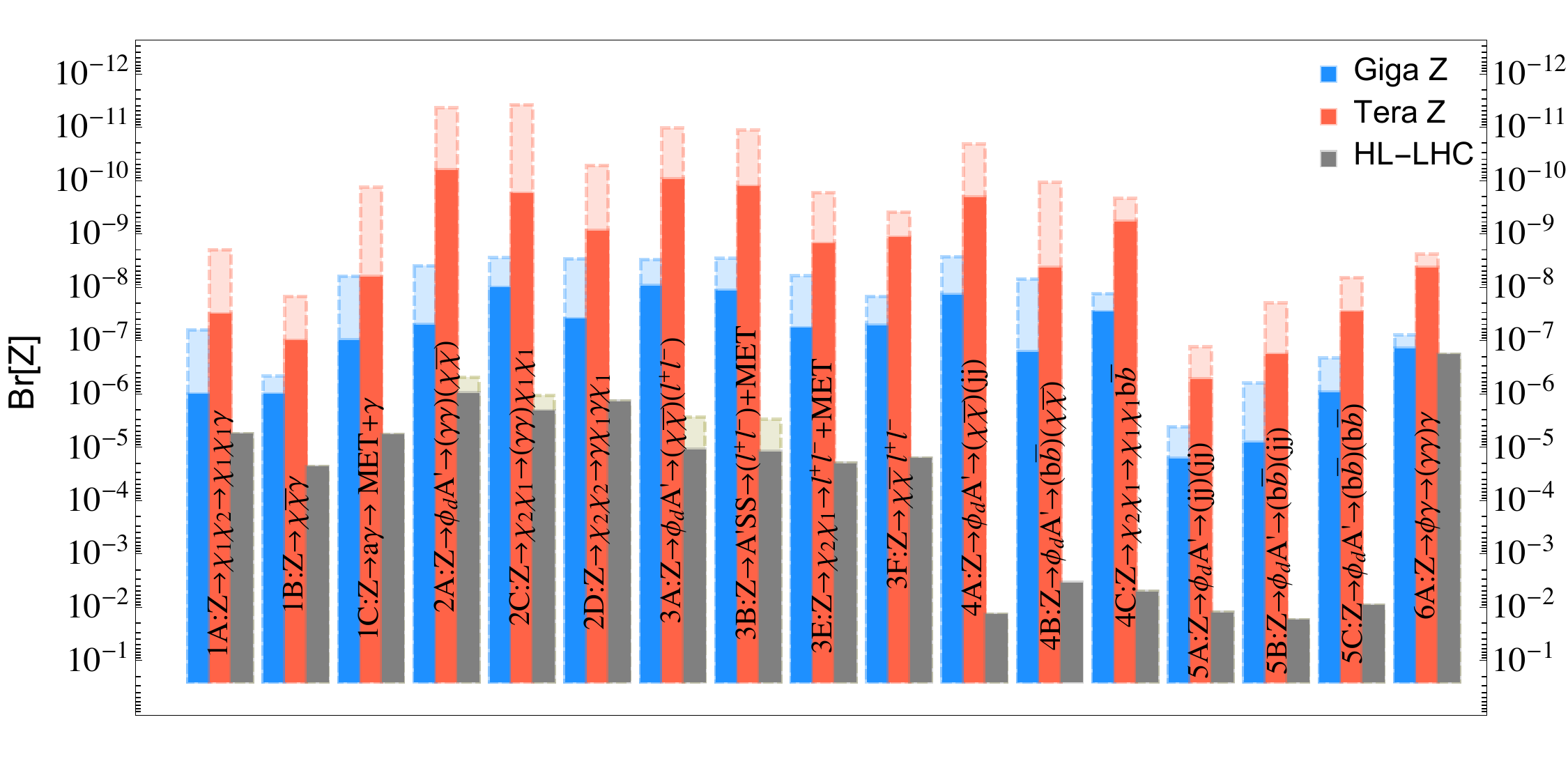} 
\end{center}
\vspace{-3mm}
\caption{\label{fig:exoticZ} The 95\% C.L. sensitivity reach for
  various exotic Z decay branching ratios at the future Z-factory,
  where the Tera Z benchmark corresponds to the FCC-ee Z-pole run.
  This figure is taken from \cite{Liu:2017zdh}, where further details
  on the various decay topologies considered, including any
  model-dependence, are presented.}
\end{figure}

\section{Final remarks}
The physics program of the FCC facility goes well beyond what we have
shown here. We focused in these lectures on the most visible elements,
which support the promise to fulfill an ambitious program of
measurements related to the EW symmetry breaking sector of the SM, and
of searches for new physics. Much more is discussed in the literature,
and in the more complete physics reports and CDRs. There it is shown
that the FCC can extend our knowledge over many fronts I did not
cover.  The mass and EW couplings of the top quark will be studied
with great precision at FCC-ee, and the huge statistics of FCC-hh will
push the search for rare or forbidden decays, and the sensitivity to
anomalous couplings, beyond what can be achieved elsewhere. Should
neutrino masses and mixings have their origin at the EW scale, the
combination of several observables at FCC will be able to probe the
underlying physics. Flavour phenomena, in the b and in the $\rm \tau$
sectors, will be explored with a sensitivity often exceeding what will
be known by the end of the HL-LHC and of the Belle~II programs, thanks
to the statistics and clean experimental environment of Z decays at
FCC-ee. The study of QCD dynamics will attain new levels of precision,
in the determination of $\alpha_s$, of the proton PDFs, in the
modeling of the structure of jets, and more. The program of collisions
with and between heavy ions will open new avenues in the exploration
of the high-density phase of QCD, for example acquiring new hard
probes (notably the top quark and its decay products) to monitor the
thermalization phase of the quark-gluon plasma, and extending to much
higher multiplicities the studies of collective effects.

When looking at the whole program, one feature emerges very clearly:
the overall coherence, completeness and far-sightedness. Each
component of the programme is part of a big picture, and contributes
in different ways to the overall complementarity and synergy. No stone
will be left unturned, not a byte of data will be redundant. The huge
statistics at the Z is not just a blessing for EW precision physics,
but provides also opportunities for discovery in otherwise elusive
areas, from exotic Higgs decays, to flavour physics of the tau lepton
and b quark, to neutrino physics. The run at the top threshold,
primarily focused on the study of the top mass and couplings, will
also push the precision of Higgs measurements, thanks to the emergence
of the WW$\to$H fusion channel, and thanks to the indirect access to
the ttH and HHH couplings. The FCC-ee results (e.g. the total H width
and the absolute H$\to$4$\ell$ branching ratio, or the vector/axial EW
top couplings) will provide the inputs necessary to fully exploit the
potential precision of systematics-dominated FCC-hh measurements, such
as the branching ratios for rare decay modes or the Higgs
self-coupling. FCC-eh will complement the set of inputs, providing the
most accurate determination of the proton PDFs; but it will also add
to the ee and hh results with its own Higgs studies, particularly
sensitive to the HWW and Hbb couplings, and with a complementary
sensitivity to new physics phenomena, from neutrino observables, to
long-lived particles or leptoquarks. The FCC-hh, finally, will
add  its own critical contributions to the
 precision measurements in the Higgs and EW symmetry
breaking sectors, and will push to the highest masses the direct
search for new particles, with a reach that promises to confirm or
exclude in a conclusive way the existence of WIMP DM candidates or the
strong 1$^{st}$ order nature of the EW phase transition.

All of these elements make the FCC the most comprehensive facility to
extend the exploration of high-energy physics throughout the XXI century.
\vskip 1cm
\noindent
{\bf Acknowledgements}: I am grateful to Tilman and Tracy for the kind
inivitation to lecture at TASI, and to the whole school team for their
support and assistance. The students were fantastic, it's been a real
pleasure and honour to spend time with them! The material for these
lectures is the result of several years of work by countless
colleagues, at the LHC, FCC, CEPC, CLIC and ILC: a world-wide community
sharing a challenging vision for the future of our field, extending
well through the XXI century. My thanks go therefore to all my
friends, who are working hard to turn this vision into a reality!
\newpage
\appendix

\section{Simple features of the energy scaling of hard cross sections
  in hadronic collisions}
\label{app:Escaling}
As well known, and recalled in the lectures by Frank Krauss, the
differential cross section for a process in hadronic collisions is
represented as follows: 
\begin{equation}
  d\sigma \; = \; \int \; \sum_{i,j} \; dx_1 dx_2 \, f_i(x_1,Q) f_j(x_2,Q)
  \ d\hat\sigma_{ij} 
\end{equation}
where $Q$ is the factorization scale, namely the characteristic hard
scale of the partonic process, $x_{1,2}$ are the momentum fractions of
the protons carried by the initial state partons, $f_i(x)$ are the
parton distribution functions (PDFs) for the parton of flavor $i$, and
the partonic cross section is given, in terms of the squared and
averaged matrix element, by:
\def\msquare{\overline{\vert M \vert^2}}
\begin{equation}
  d\hat\sigma \; = \; \frac{1}{2\hat s} \int \prod_i
  \frac{d^3p_i}{(2\pi)^3 2p_i^0} \; (2\pi)^4 \,
  \delta(P_{in}-P_{out}) \times \msquare
\end{equation}
It is an easy exercise to reexpress the integration measure $dx_1\,
dx_2$ as $dy \, d\tau$, where ($S=4E_{beam}^2$):
\begin{equation}
  \tau=x_1x_2=\frac{\hat{s}}{S} \quad, \quad y=\frac{1}{2}
  \log\frac{E^{in}+P_z^{in}}{E^{in}-P_z^{in}}=\frac{1}{2}\log\frac{x_1}{x_2}
  \quad \Rightarrow
  x_{1,2}=\sqrt{\tau}e^{\pm y} \; .
\end{equation}
The result can then be written as:
\begin{equation}
  d\sigma = \int \frac{d\tau}{\tau} \ \, \times \, \tau\frac{d{\cal
    L}}{d\tau} \, \times \, d\hat{\sigma}
\end{equation}
with
\begin{equation}
  \frac{d{\cal  L}}{d\tau} \; = \; \int_{y_{min}}^{y_{max}} \, dy \,
    f(\sqrt{\tau}e^y)f(\sqrt{\tau}e^{-y})  \; = \;
    \int_{\tau}^{1} \; \frac{dx}{x} f(x)f(\tau/x)
  \quad (y_{min,max}=\pm 0.5 \log\tau) \; .
\end{equation}
Notice that, for fixed partonic center-of-mass energy
$\sqrt{\hat{s}}$, the dependence of the cross sections on the beam
energy $E_{beam}$ is all built into the partonic luminosity factor
$\tau d{\cal L}/d\tau$, since both $\hat\sigma$ and $d\tau/\tau$ are
independent of $E_{beam}$ \footnote{Notice however that kinematic cuts
  to the final state particles will affect the integration ranges in
  an energy-dependent way.}.

Several useful results, which help getting a feeling for how cross
sections behave at high energies, can be easily obtained with simple assumptions
and parameterizations of the PDFs. Consider for example a $2\to 1$
process such as the production of a H boson in the $gg$ channel.
In this case it is easy to show that, for a final state of mass $m$ ($\tau=m^2/S$):
\begin{equation}
  \frac{d\sigma}{dy} = \frac{\pi}{m^4} \msquare \; \tau
  \, f(\sqrt\tau e^y)f(\sqrt\tau e^{-y})
  \end{equation}
If we parameterize $f(x) \sim 1/x^{1+\Delta}$, which as we shall see later
is a good approximation over a broad range of $x$ values, we can
obtain
\begin{eqnarray}
  \label{eq:sigtaudy}
  \frac{d\sigma}{dy} & = & \frac{\pi}{m^4} \msquare \; \frac{1}{\tau^{\Delta}}
  \\
  \sigma & = & \frac{\pi}{m^4} \msquare \; \frac{1}{\tau^{\Delta}} \;
  \log\frac{1}{\tau} \label{eq:sigtau}
\end{eqnarray}
The first equation says that the rapidity distribution $d\sigma/dy$ is
flat in $y$, in the $y$ range up to the kinematic limit, $\vert y\vert
= \log(E_{beam}/m)$. The total cross section will grow with energy at
least logarithmically, to reflect the increased rapidity range. The
growth is faster if $\Delta>0$~\footnote{Notice that $\Delta$ must be smaller
than 1, to ensure that the total momentum fraction carried by a parton
$i$, $\int_0^1 dx \, x f_i(x)$, be finite.}.

\begin{figure}[h]
\centering
\includegraphics[width=0.32\textwidth]{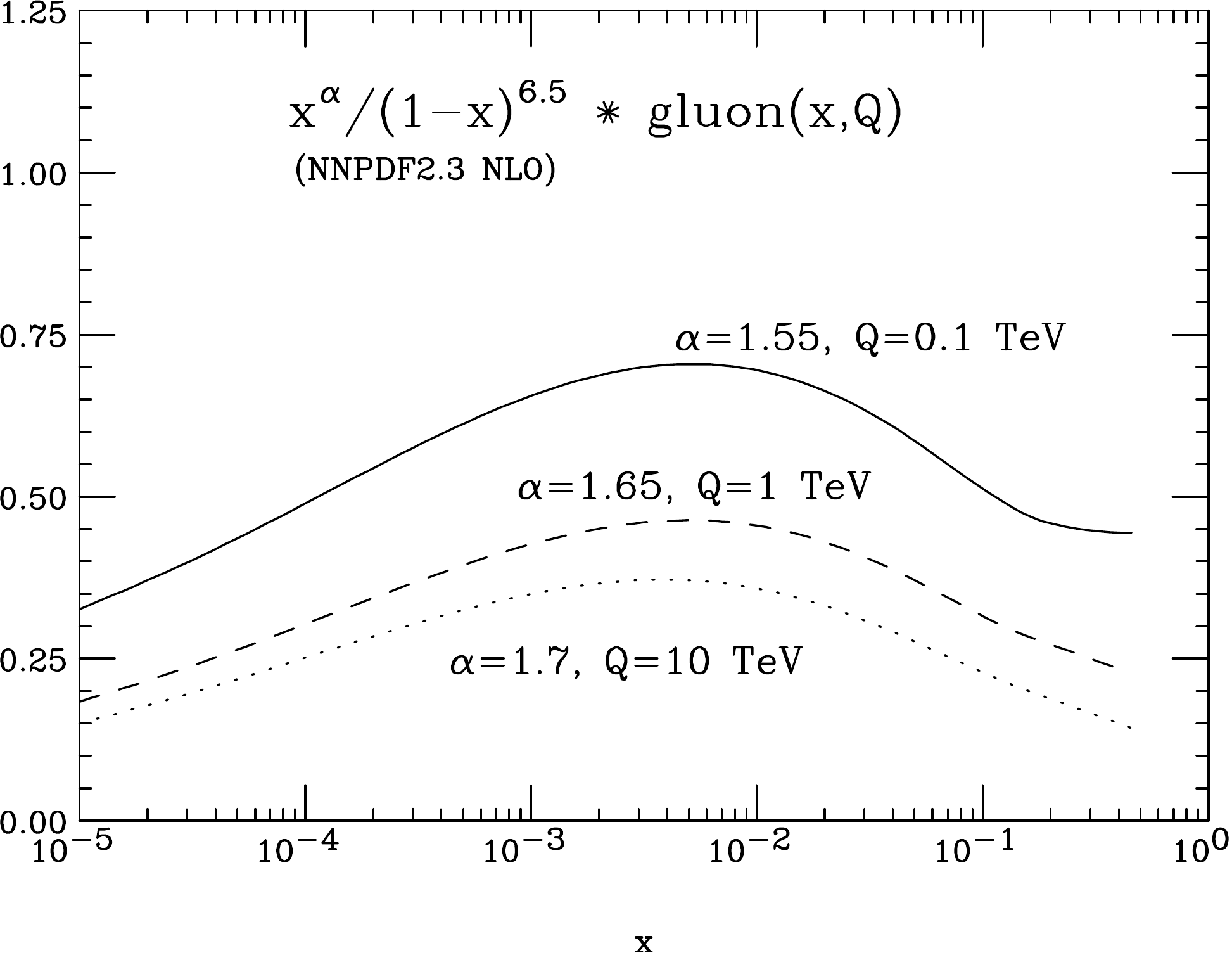}
\includegraphics[width=0.32\textwidth]{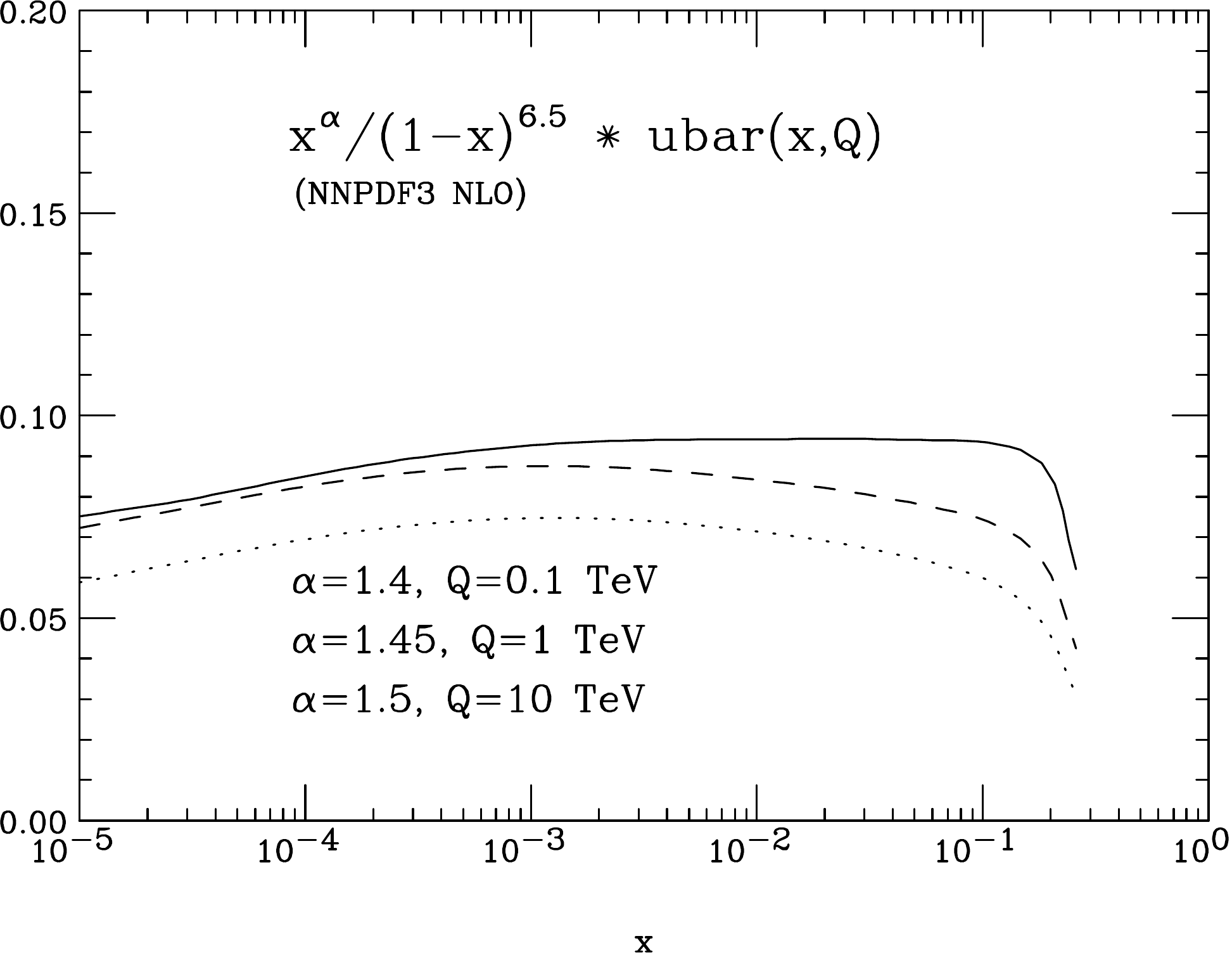}
\includegraphics[width=0.32\textwidth]{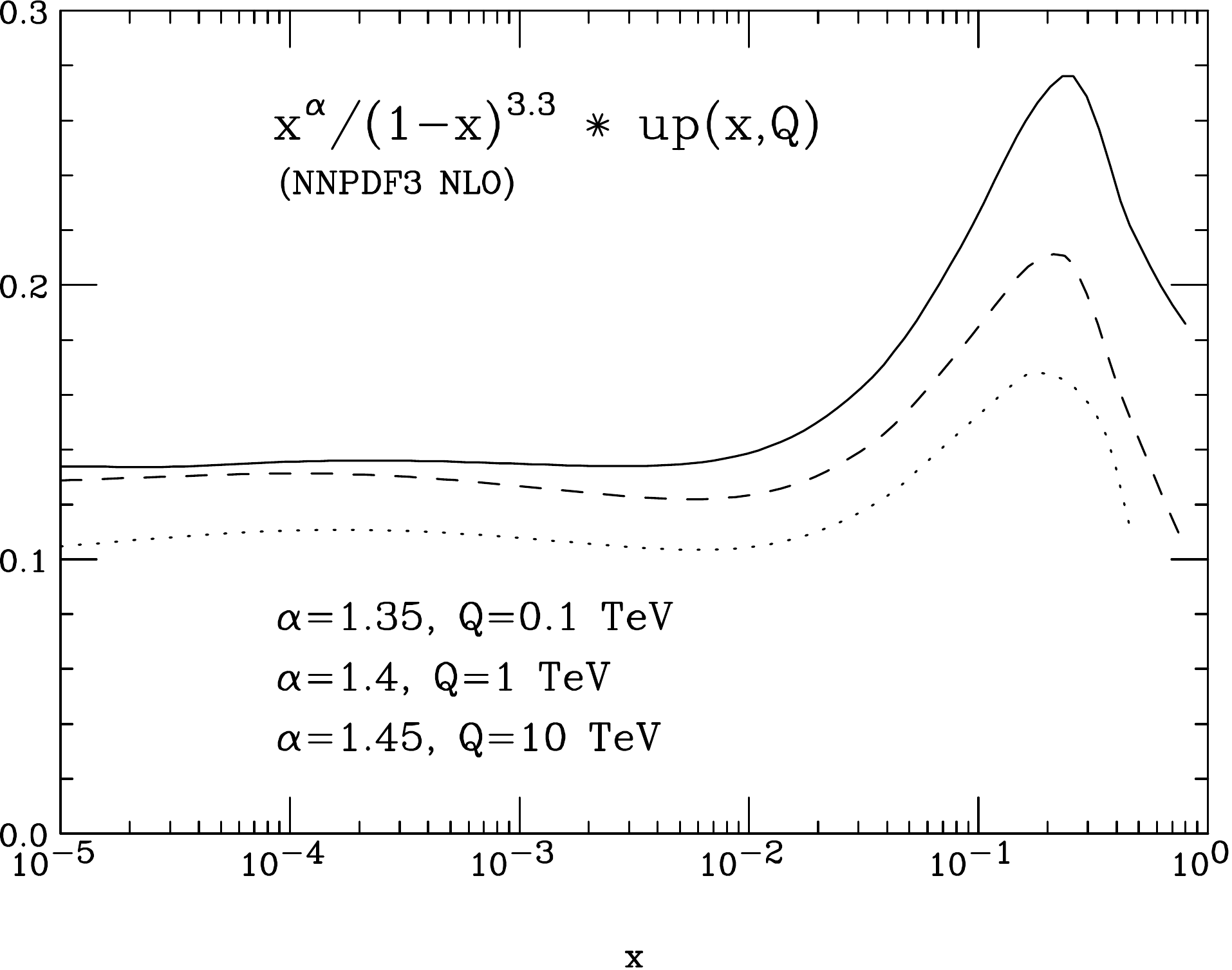}
\caption{Testing the simple approximations $f(x)\sim
  (1-x)^\beta/x^\alpha$ for the gluon, up and antiup quarks, at
  different values of the factorization scale $Q$. }
\label{fig:pdfpar}
\end{figure}

As a first simple application of these relations, let us estimate the
Higgs cross section growth between LHC ($\sqrt{S}=14$~TeV) and
FCC-hh ($\sqrt{S}=100$~TeV). Equation~\ref{eq:sigtau}, with $m_H=125$~GeV, gives:
\begin{equation}
  \frac{\sigma_{100}(gg\to H)}{\sigma_{14}(gg\to H)} \sim
  \left(\frac{100}{14}\right)^{2\Delta} \times \;
  \frac{\log(100\,\mathrm{TeV}/m_H)}{\log(14\,\mathrm{TeV}/m_H)} \sim
  15 \; ,
\end{equation}
where we used $\Delta \sim 0.6$, as suggested by the approximate fits
in Fig.~\ref{fig:pdfpar}. The actual result, as shown in
Table~\ref{tab:Hrates}, is 18, so not bad for a simple-minded
approach. In the case of W production, replacing $m_H$ with $m_W\sim
80$~GeV, and for $\Delta\sim 0.4$ (see Fig.~\ref{fig:pdfpar}), gives a
W cross section increase of about 7, compared to the factor of 6.5
increase of an actual complete calculation (1.3 vs 0.2 $\rm \mu b$).

\noindent
\rule{5cm}{1pt}\\
{\bf Exercise:} Eqs.~\ref{eq:sigtaudy} and \ref{eq:sigtau} assume $\Delta$ is the
same for both incoming partons. This is not a bad assumption for $u$
and $\bar{u}$, as shown in Fig.~\ref{fig:pdfpar}, but would not hold
for a $qg$ initial state. Discuss.
\\
\rule{5cm}{1pt}

\begin{figure}[h]
\centering
\includegraphics[width=0.45\textwidth]{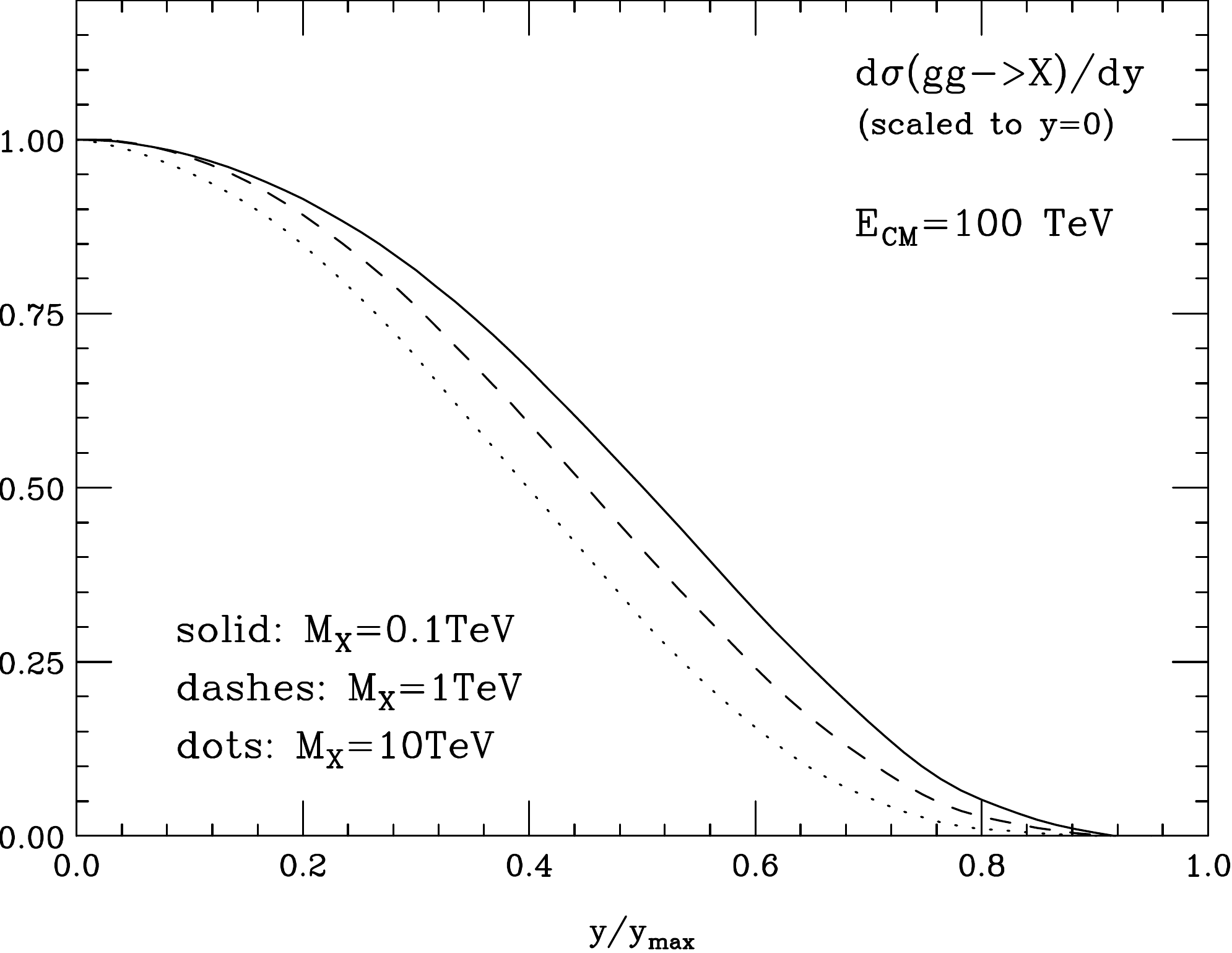}
\includegraphics[width=0.45\textwidth]{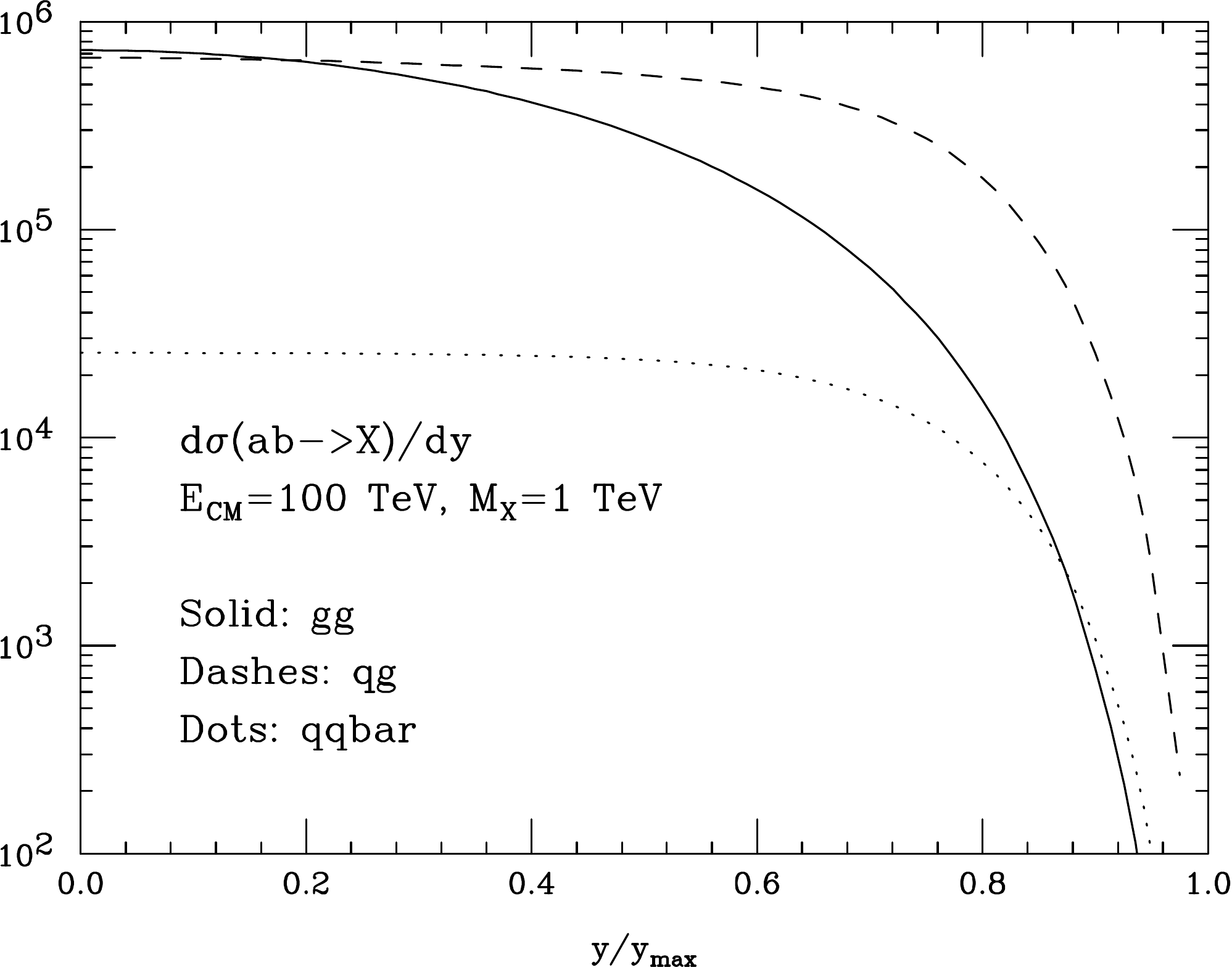}
\caption{Rapidity spectra of systems of mass $M_X$, produced
  in 100~TeV pp collisions by various initial-state channels.
}
\label{fig:dsigdy}
\end{figure}

In reality, the behaviour of the rapidity distribution
at $y\sim y_{min,max}$ is not a sharp edge,
since when $\vert y \vert \to y_{max}$ then $x\to 1$ for one of the
two initial partons, and the parameterization $f(x)\sim
1/x^{1+\Delta}$ is clearly inappropriate. To be more realistic, we can
replace this by
\begin{equation} \label{eq:xtoone}
  f(x) \sim \frac{(1-x)^{\beta}}{x^{1+\Delta}}  \; .
\end{equation}
Figure~\ref{fig:pdfpar} compares this approximation against the actual
PDF for the gluon, the u quark and the $\rm \bar{u}$ antiquark, for
various values of the factorization scale $Q$.  While the approximtion
is not perfect, it is remarkable that it does such a good job, over a
range of $x$ from $10^{-5}$ and $\sim 0.2-0.3$. Given the
$1/x^{1+\Delta}$ behaviour, these PDFs vary by 6 or 7 orders of
magnitude in this $x$ range, and, by choosing suitable parameters
$\Delta$ and $\beta$, the approximate functional form maps out the PDF
behaviour within a factor better than 2! Notice also the peak
structure of the $u(x)$ distribution at $x\sim 0.2$, due to the
valence contribution, which cannot be properly modeled by the simple
parameterization in Eq.~\ref{eq:xtoone}. The rapidity spectra of
systems of different mass produced with different initial states are
shown in Fig.~\ref{fig:dsigdy}. Notice that the $\rm q\bar{q}$ and qg
distributions are broader than the gg one, since the valence quark
distribution is much bigger than the gluon at large $x$.

\newpage
\section{FCC parameters}
For reference, we summarize in this Section the key parameters
established for the design of the various components of the FCC
facility. More details can be found in
Refs.~\cite{Benedikt:2651299,Benedikt:2651300,Zimmermann:2651305}. 
\subsection{FCC-ee}
The main design principles of FCC-ee are as follows.  
\begin{itemize}
\item It is a double ring collider with electrons and positrons
  circulating in separate vacuum chambers. This allows a large and
  variable number of bunches to be stored. The beam intensity can thus
  be increased in inverse proportion to the synchrotron radiation (SR)
  per particle per turn, to keep the total power constant to a set
  value of 100 MW for both beams, at all energies.
\item A common low emittance lattice for all energies, except for a
  small rearrangement in the RF section for the $\rm t\overline t$
  mode. The optics are optimised at each energy by changing the
  strengths of the magnets.
\item The length of the free area around the IP ($L^*$) and the
  strength of the detector solenoid are kept constant at 2.2~m and
  2~T, respectively, for all energies. 
\item A top-up injection scheme maintains the stored beam current and
  the luminosity at the highest level throughout the experimental
  run. This is achieved with a booster synchrotron situated in the
  collider tunnel itself.
\end{itemize}

As a requirement, the luminosity figures are very high, ranging from
$2\times 10^{36}$cm$^{-2}$s$^{-1}$ per IP at the Z pole, and
decreasing with the fourth power of the energy to $1.5\times
10^{34}$cm$^{-2}$s$^{-1}$ per IP at the top energies.  
The run plan spanning 15 years including commissioning is shown in
Table~\ref{tab:FCC-ee-runplan}. The number of Z bosons planned to be
produced by FCC-ee (up to $5\times 10^{12}$), for example, is  more
than five orders of magnitude larger than the number of Z bosons
collected at LEP ($2 \times 10^7$).

\begin{table}[b]
\begin{center}
\caption{Run plan for FCC-ee in its baseline configuration with two experiments. The number of WW events is given for the entirety of the FCC-ee running at and above the WW threshold. \label{tab:FCC-ee-runplan}}
\begin{tabular}{|l|c|c|c|r|} \hline \hline 
Phase & Run duration  & Centre-of-mass  &  Integrated   & Event   \\ 
& (years)  & Energies (GeV) & Luminosity (ab$^{-1}$) &  Statistics \\ 
\hline
FCC-ee-Z & 4  & 88-95   & 150   & $3\times 10^{12}$ visible Z decays   \\ 
FCC-ee-W & 2  & 158-162 &  12   & 10$^8$ WW events\\ 
FCC-ee-H & 3  & 240&  5    & 10$^6$ ZH events\\ 
FCC-ee-tt(1) & 1  & 340-350&  0.2    & ${\rm t\bar t} $ threshold scan\\ 
FCC-ee-tt(2) & 4  & 365 &  1.5 & 10$^6$ $\rm t\overline{t}$ events  \\
 \hline \hline
\end{tabular} 
\end{center}
\end{table}

\begin{table}[th!]
\centering
\caption{Reference parameters for operations at (HL-)LHC, HE-LHC and FCC-hh. More details on the structure of the minimum bias events at 100~TeV can be found in~\cite{Mangano:2016jyj}. }
\label{tab:FCCaccpar}
\begin{tabular}{|l|c|c|c|c|c|}
\hline
\hline
    Parameter & Unit & LHC  & HL-LHC & HE-LHC & FCC-hh \\
 \hline
  $E_{cm}$   & TeV &14 & 14 & 27 & 100 \\ 
  Circumference & km & 26.7 & 26.7 & 26.7 & 97.8 \\
  Peak ${\cal L}$, nominal (ultimate)  & $10^{34}$cm$^{-2}$s$^{-1}$ & 1 (2)  & 5 (7.5) & 16 & 30 \\
   Goal  $\int {\cal L}$ & ab$^{-1}$ & 0.3 & 3 &  10 & 30 \\  
   Bunch spacing  & ns & 25 & 25 & 25  & 25   \\
 Number of bunches & & 2808 & 2760 & 2808 & 10600 \\ 
   RMS luminous region $\sigma_z$ & mm & 45 & 57 & 57 & 49 \\
\hline
   $\sigma_{inel} $ \cite{Mangano:2016jyj} & mb & 80 & 80 & 86 & 103 \\
  $\sigma_{tot} $  \cite{Mangano:2016jyj}& mb & 108 & 108 & 120 & 150 \\
   Peak pp collision rate & GHz & 0.8 & 4 & 14 &  31 \\
   \hline
   $dN_{ch}/d \eta \vert_{\eta=0}   $\cite{Mangano:2016jyj} & & 6.0 & 6.0 & 7.2 & 10.2 \\
   Charged tracks per collision $N_{ch}$   \cite{Mangano:2016jyj}  &  & 70 & 70 & 85 & 122 \\ 
    Rate of charged tracks & GHz & 59 & 297 & 1234 & 3942 \\ 
   $ \langle p_T \rangle $  \cite{Mangano:2016jyj} & GeV/c & 0.56 & 0.56 & 0.6 & 0.7 \\ 
    $d E/d \eta \vert_{\eta=5} $   \cite{Mangano:2016jyj} & GeV & 316 & 316 & 427 & 765 \\ 
    $d P/d \eta \vert_{\eta=5} $  & kW & 0.04 & 0.2 & 1.0 & 4.0 \\ 
   \hline
\hline
\end{tabular}
\end{table}

\subsection{FCC-hh and HE-LHC}
The FCC-hh collider design, performance and operating conditions are
discussed in detail in Volume~3, Chapter~2, and summarized in
Table~\ref{tab:FCCaccpar}. The key parameters are the total
centre-of-mass energy, 100~TeV, and the peak initial (nominal)
luminosity of \lumi{5 (25)}{34}, with 25~ns bunch spacing.  At nominal
luminosity, the pile-up reaches 850 interactions per bunch
crossing. The feasibility and performance of alternative bunch
spacings of 12.5 and 5~ns are under study (see Volume~2,
Section~2.2.5). With two high-luminosity interaction points, and
taking into account the luminosity evolution during a fill and the
turn-around time, the optimum integrated luminosity per day is
estimated to be 2.3 (8.2)~\ifb.  The total integrated luminosity at
the end of the programme will obviously depend on its
duration. Assuming a 25 year life cycle, with 10+15 years at
initial/nominal parameters, allows a goal of 5+15=20~\iab to be set.
This has been shown to be adequate for the foreseeable
scenarios~\cite{Hinchliffe:2015qma}. A luminosity range up to 30~\iab
is considered for most of the physics studies. This allows the
ultimate physics potential to be assessed, considering that the two
experiments will probably combine their final results for the most
sensitive measurements.

The current design allows for two further interaction points (IPs),
where the pp luminosity can reach \lumi{2}{34}, with a free distance
between IP and the focusing triplets of 25~m. Apart from the case of
heavy ion collisions, no discussion of possible FCC-hh experiments
using these lower-luminosity IPs will be presented.   

For the HE-LHC the asumptions are for a collision energy
$\sqrt{S}=27$~TeV and a total integrated luminosity of 15~\iab, to be
collected during 20 years of operation. 

\begin{table}[!ht]
\begin{center}
\caption[Main beam and collider parameters.]{Beam and machine parameters.} 
\label{tab:IonParametersShortForPhysics}%
\begin{tabular}{|l|c|cc|cc|}
\hline \hline  
& Unit & \multicolumn{2}{c|}{Baseline} & \multicolumn{2}{c|}{Ultimate} \\
\hline
Operation mode&-& PbPb 	& pPb 	& PbPb & pPb \\
Number of Pb bunches & - & \multicolumn{2}{c|}{2760} & \multicolumn{2}{c|}{5400} \\
Bunch spacing & [ns] & \multicolumn{2}{c|}{100} & \multicolumn{2}{c|}{50}\\
Peak luminosity (1 experiment) &
[$10^{27}~$cm$^{-2}$s$^{-1}$]
& 80 & 13300 & 320 & 55500 \\
Integrated luminosity (1 experiment, 30 days) & [nb$^{-1}$] & 35 &
8000 & 110 & 29000\\
\hline \hline
\end{tabular}
\end{center}
\end{table}

\subsection{FCC-hh: Operations with Heavy Ions}
\label{sec:HI_machine}
It has been shown that the FCC-hh could operate very efficiently as a
nucleus-nucleus or proton-nucleus collider, analogously to the
LHC. Previous studies~\cite{Schaumann:2015fsa,Dainese:2016gch} have
revealed that it enters a new, highly-efficient operating regime, in
which a large fraction of the injected intensity can be converted to
useful integrated luminosity.
Table~\ref{tab:IonParametersShortForPhysics} summarises the key
parameters for PbPb and pPb operation.  Two beam parameter cases were
considered, \textit{baseline} and \textit{ultimate}, which differ in
the $\beta$-function at the interaction point, the optical function
$\beta^*$ at the interaction point, and the assumed bunch spacing,
defining the maximum number of circulating bunches.  The luminosity is
shown for one experiment but the case of two experiments was also
studied: this decreases the integrated luminosity per experiment by
40\%, but increases the total by 20\%.  The performance projections
assume the LHC to be the final injector synchrotron before the
FCC~\cite{Jowett2017}.  A performance efficiency factor was taken into
account to include set-up time, early beam aborts and other deviations
from the idealised running on top of the theoretical calculations.
Further details on the performance of the heavy-ion operation in
FCC-hh can be found in Section 2.6 of the FCC-hh CDR Volume.

\subsection{FCC-eh}
\label{sec:eh_acc}
The FCC-eh is designed to run concurrently with the FCC-hh.  The
electron-hadron interaction has a negligible effect on the multi TeV
energy hadron beams, protons or ions. The electron beam is provided by
an energy recovery linac (ERL) of $E_e=60$\,GeV energy which emerges
from a 3-turn racetrack arrangement of two linacs, located opposite to
each other. This ERL has been designed and studied in quite some
detail with the LHeC design. For FCC-eh, for geological reasons, the
ERL would be positioned at the inside of the FCC tunnel and tangential
to the hadron beam at point L. There will be one detector only, but
forming two data taking collaborations may be considered, for example,
to achieve cross check opportunities for this precision measurement
and exploratory programme.

The choice of $E_e=60$\,GeV is currently dictated by limiting
cost. Desirably one would increase it, to reduce the beam energy
uncertainty and access extended kinematics, but that would increase
the cost and effort in a non-linear way. This could happen,
nevertheless, if one expected, for example, leptoquarks with a mass of
$4$\,TeV which the FCC-eh would miss with a $60$\,GeV beam. Currently,
the energy chosen, taken from the LHeC design, is ample and adequate
for a huge, novel programme in deep inelastic physics as has been
sketched above.

In concurrent operation, the FCC-eh would operate for 25 years, with
the FCC-hh. This provides an integrated luminosity of
$\cal{O}$(2)\,ab$^{-1}$, at a nominal peak luminosity above
$10^{34}$\,cm$^{-2}$\,s$^{-1}$, at which the whole result of HERA's 15
year programme could be reproduced in about a day or two, with
kinematic boundaries extended by a factor of 100. The pile-up at
FCC-eh is estimated to be just 1. The forward detector has to cope
with multi-TeV electron and hadron final state energies, while the
backward detector (in the direction of the e beam) would only see
energies up to $E_e=60$\,GeV. The size of the detector corresponds to
about that of CMS at the LHC.

Special runs are possible at much lower yet still sizeable luminosity,
such as with reduced beam energies. There is also the important
programme of electron-ion scattering which extended the kinematic
range of the previous lepton-nucleus experiments by 4 orders of
magnitude. This is bound to revolutionise the understanding of parton
dynamics and substructure of nuclei and it will shed light on the
understanding of the formation and development of the Quark-Gluon
Plasma.

\newpage
\bibliographystyle{report}
\bibliography{mangano}

\providecommand{\href}[2]{#2}\begingroup\raggedright\begin{thebibliography}{100}

\bibitem{Behnke:2013xla}
T.~Behnke, J.~E. Brau, B.~Foster, J.~Fuster, M.~Harrison, J.~M. Paterson,
  M.~Peskin, M.~Stanitzki, N.~Walker, and H.~Yamamoto, {\em {The International
  Linear Collider Technical Design Report - Volume 1: Executive Summary}\/},
\href{http://arxiv.org/abs/1306.6327}{{\tt arXiv:1306.6327 [physics.acc-ph]}}.

\bibitem{Evans:2017rvt}
{Linear Collide Collaboration}, L.~Evans and S.~Michizono, {\em {The
  International Linear Collider Machine Staging Report 2017}\/},
\href{http://arxiv.org/abs/1711.00568}{{\tt arXiv:1711.00568
  [physics.acc-ph]}}.

\bibitem{Barklow:2017suo}
T.~Barklow, K.~Fujii, S.~Jung, R.~Karl, J.~List, T.~Ogawa, M.~E. Peskin, and
  J.~Tian, {\em {Improved Formalism for Precision Higgs Coupling Fits}\/},
  \href{http://dx.doi.org/10.1103/PhysRevD.97.053003}{Phys. Rev. {\bf D97}
  (2018) no.~5, 053003},
\href{http://arxiv.org/abs/1708.08912}{{\tt arXiv:1708.08912 [hep-ph]}}.

\bibitem{Fujii:2017vwa}
K.~Fujii et al., {\em {Physics Case for the 250 GeV Stage of the International
  Linear Collider}\/},
\href{http://arxiv.org/abs/1710.07621}{{\tt arXiv:1710.07621 [hep-ex]}}.

\bibitem{Aicheler:1500095}
M.~Aicheler, P.~Burrows, M.~Draper, T.~Garvey, P.~Lebrun, K.~Peach, N.~Phinney,
  H.~Schmickler, D.~Schulte, and N.~Toge, {\em {A Multi-TeV Linear Collider
  Based on CLIC Technology: CLIC Conceptual Design Report}}.
\newblock CERN Yellow Reports: Monographs. CERN, Geneva, 2012.
\newblock \url{https://cds.cern.ch/record/1500095}.

\bibitem{CLIC:2016zwp}
{CLICdp, CLIC Collaboration}, M.~J. Boland et al., {\em {Updated baseline for a
  staged Compact Linear Collider}\/},
\href{http://arxiv.org/abs/1608.07537}{{\tt arXiv:1608.07537
  [physics.acc-ph]}}.

\bibitem{Charles:2018vfv}
{CLICdp, CLIC Collaboration}, T.~K. Charles et al., {\em {The Compact Linear
  $e^+e^-$ Collider (CLIC) - 2018 Summary Report}\/},
  \href{http://dx.doi.org/10.23731/CYRM-2018-002}{CERN Yellow Rep. Monogr. {\bf
  1802} (2018)  1--98},
\href{http://arxiv.org/abs/1812.06018}{{\tt arXiv:1812.06018
  [physics.acc-ph]}}.

\bibitem{deBlas:2018mhx}
J.~de~Blas et al., {\em {The CLIC Potential for New Physics}\/},
\href{http://arxiv.org/abs/1812.02093}{{\tt arXiv:1812.02093 [hep-ph]}}.

\bibitem{Mangano:2651294}
M.~Mangano, P.~Azzi, M.~Benedikt, A.~Blondel, D.~A. Britzger, A.~Dainese,
  M.~Dam, J.~de~Blas, D.~Enterria, O.~Fischer, C.~Grojean, J.~Gutleber,
  C.~Gwenlan, C.~Helsens, P.~Janot, M.~Klein, U.~Klein, M.~P. Mccullough,
  S.~Monteil, J.~Poole, M.~Ramsey-Musolf, C.~Schwanenberger, M.~Selvaggi,
  F.~Zimmermann, and T.~You, {\em {Future Circular Collider Study. Volume 1:
  Physics Opportunities}\/},   CERN-ACC-2018-0056, CERN, Geneva, Dec, 2018.
\newblock \url{https://cds.cern.ch/record/2651294}.
\newblock To appear in Eur. Phys. J. C.

\bibitem{Benedikt:2651299}
M.~Benedikt, A.~Blondel, O.~Brunner, M.~Capeans~Garrido, F.~Cerutti,
  J.~Gutleber, P.~Janot, J.~M. Jimenez, V.~Mertens, A.~Milanese, K.~Oide, J.~A.
  Osborne, T.~Otto, Y.~Papaphilippou, J.~Poole, L.~J. Tavian, and
  F.~Zimmermann, {\em {Future Circular Collider Study. Volume 2: The Lepton
  Collider (FCC-ee) }\/},   CERN-ACC-2018-0057, CERN, Geneva, Dec, 2018.
\newblock \url{https://cds.cern.ch/record/2651299}.
\newblock To appear in Eur. Phys. J. ST.

\bibitem{Benedikt:2651300}
M.~Benedikt, M.~Capeans~Garrido, F.~Cerutti, B.~Goddard, J.~Gutleber, J.~M.
  Jimenez, M.~Mangano, V.~Mertens, J.~A. Osborne, T.~Otto, J.~Poole,
  W.~Riegler, D.~Schulte, L.~J. Tavian, D.~Tommasini, and F.~Zimmermann, {\em
  {Future Circular Collider Study. Volume 3: The Hadron Collider (FCC-hh)}\/},
   CERN-ACC-2018-0058, CERN, Geneva, Dec, 2018.
\newblock \url{https://cds.cern.ch/record/2651300}.
\newblock To appear in Eur. Phys. J. ST.

\bibitem{Zimmermann:2651305}
F.~Zimmermann, M.~Benedikt, M.~Capeans~Garrido, F.~Cerutti, B.~Goddard,
  J.~Gutleber, J.~M. Jimenez, M.~Mangano, V.~Mertens, J.~A. Osborne, T.~Otto,
  J.~Poole, W.~Riegler, L.~J. Tavian, and D.~Tommasini, {\em {Future Circular
  Collider Study. Volume 4: The High Energy LHC (HE-LHC)}\/},
  CERN-ACC-2018-0059, CERN, Geneva, Dec, 2018.
\newblock \url{https://cds.cern.ch/record/2651305}.
\newblock To appear in Eur. Phys. J. ST.

\bibitem{CEPCStudyGroup:2018rmc}
{CEPC Study Group Collaboration}, {\em {CEPC Conceptual Design Report: Volume 1
  - Accelerator}\/},
\href{http://arxiv.org/abs/1809.00285}{{\tt arXiv:1809.00285
  [physics.acc-ph]}}.

\bibitem{CEPCStudyGroup:2018ghi}
{CEPC Study Group Collaboration}, {\em {CEPC Conceptual Design Report: Volume 2
  - Physics \& Detector}\/},
\href{http://arxiv.org/abs/1811.10545}{{\tt arXiv:1811.10545 [hep-ex]}}.

\bibitem{ALEGRO:2019alc}
{ALEGRO Collaboration}, {\em {Towards an Advanced Linear International
  Collider}\/},
\href{http://arxiv.org/abs/1901.10370}{{\tt arXiv:1901.10370
  [physics.acc-ph]}}.

\bibitem{MuonCollider}
{\em Muon Accelerator Program home page\/},
\newblock \url{https://map.fnal.gov}.

\bibitem{Antonelli:2015nla}
M.~Antonelli, M.~Boscolo, R.~Di~Nardo, and P.~Raimondi, {\em {Novel proposal
  for a low emittance muon beam using positron beam on target}\/},
  \href{http://dx.doi.org/10.1016/j.nima.2015.10.097}{Nucl. Instrum. Meth. {\bf
  A807} (2016)  101--107},
\href{http://arxiv.org/abs/1509.04454}{{\tt arXiv:1509.04454
  [physics.acc-ph]}}.

\bibitem{tasi7}
{\em {{\em C. Englert, 2018 TASI lectures on} "The Higgs boson"}\/},
  \url{https://sites.google.com/a/colorado.edu/tasi-2018-wiki/}.

\bibitem{tasi4}
{\em {{\em P. Fox, 2018 TASI lectures on} "WIMPS and Supersymmetry"}\/},
  \url{https://sites.google.com/a/colorado.edu/tasi-2018-wiki/}.

\bibitem{tasi3}
{\em {{\em F. Krauss, 2018 TASI lectures on} "Dark Matter Models and Direct
  Searches"}\/},
  \url{https://sites.google.com/a/colorado.edu/tasi-2018-wiki/}.

\bibitem{Cohen:2019wxr}
T.~Cohen, {\em {As Scales Become Separated: Lectures on Effective Field
  Theory}\/},
\href{http://arxiv.org/abs/1903.03622}{{\tt arXiv:1903.03622 [hep-ph]}}.

\bibitem{tasi8}
{\em {{\em S. Gori, 2018 TASI lectures on} "Flavour physics"}\/},
  \url{https://sites.google.com/a/colorado.edu/tasi-2018-wiki/}.

\bibitem{Plehn:2015dqa}
T.~Plehn, \href{http://dx.doi.org/10.1007/978-3-319-05942-6}{{\em {Lectures on
  LHC Physics}}}, vol.~886.
\newblock
2015.
\newblock

\bibitem{Ginzburg:1950sr}
V.~L. Ginzburg and L.~D. Landau, {\em {On the Theory of superconductivity}\/},
Zh. Eksp. Teor. Fiz. {\bf 20} (1950)  1064--1082.

\bibitem{Bardeen:1957mv}
J.~Bardeen, L.~N. Cooper, and J.~R. Schrieffer, {\em {Theory of
  superconductivity}\/},
\href{http://dx.doi.org/10.1103/PhysRev.108.1175}{Phys. Rev. {\bf 108} (1957)
  1175--1204}.

\bibitem{Degrassi:2012ry}
G.~Degrassi, S.~Di~Vita, J.~Elias-Miro, J.~R. Espinosa, G.~F. Giudice,
  G.~Isidori, and A.~Strumia, {\em {Higgs mass and vacuum stability in the
  Standard Model at NNLO}\/},
  \href{http://dx.doi.org/10.1007/JHEP08(2012)098}{JHEP {\bf 08} (2012)  098},
\href{http://arxiv.org/abs/1205.6497}{{\tt arXiv:1205.6497 [hep-ph]}}.

\bibitem{Espinosa:2015qea}
J.~R. Espinosa, G.~F. Giudice, E.~Morgante, A.~Riotto, L.~Senatore, A.~Strumia,
  and N.~Tetradis, {\em {The cosmological Higgstory of the vacuum
  instability}\/},  \href{http://dx.doi.org/10.1007/JHEP09(2015)174}{JHEP {\bf
  09} (2015)  174},
\href{http://arxiv.org/abs/1505.04825}{{\tt arXiv:1505.04825 [hep-ph]}}.

\bibitem{Ade:2015xua}
{Planck Collaboration}, P.~A.~R. Ade et al., {\em {Planck 2015 results. XIII.
  Cosmological parameters}\/},
  \href{http://dx.doi.org/10.1051/0004-6361/201525830}{Astron. Astrophys. {\bf
  594} (2016)  A13},
\href{http://arxiv.org/abs/1502.01589}{{\tt arXiv:1502.01589 [astro-ph.CO]}}.

\bibitem{Hu:2000ke}
W.~Hu, R.~Barkana, and A.~Gruzinov, {\em {Cold and fuzzy dark matter}\/},
  \href{http://dx.doi.org/10.1103/PhysRevLett.85.1158}{Phys. Rev. Lett. {\bf
  85} (2000)  1158--1161},
\href{http://arxiv.org/abs/astro-ph/0003365}{{\tt arXiv:astro-ph/0003365
  [astro-ph]}}.

\bibitem{Griest:2013aaa}
K.~Griest, A.~M. Cieplak, and M.~J. Lehner, {\em {Experimental Limits on
  Primordial Black Hole Dark Matter from the First 2 yr of Kepler Data}\/},
  \href{http://dx.doi.org/10.1088/0004-637X/786/2/158}{Astrophys. J. {\bf 786}
  (2014) no.~2, 158},
\href{http://arxiv.org/abs/1307.5798}{{\tt arXiv:1307.5798 [astro-ph.CO]}}.

\bibitem{Alcock:1998fx}
{MACHO, EROS Collaboration}, C.~Alcock et al., {\em {EROS and MACHO combined
  limits on planetary mass dark matter in the galactic halo}\/},
  \href{http://dx.doi.org/10.1086/311355}{Astrophys. J. {\bf 499} (1998)  L9},
\href{http://arxiv.org/abs/astro-ph/9803082}{{\tt arXiv:astro-ph/9803082
  [astro-ph]}}.

\bibitem{Yoo:2003fr}
J.~Yoo, J.~Chaname, and A.~Gould, {\em {The end of the MACHO era: limits on
  halo dark matter from stellar halo wide binaries}\/},
  \href{http://dx.doi.org/10.1086/380562}{Astrophys. J. {\bf 601} (2004)
  311--318},
\href{http://arxiv.org/abs/astro-ph/0307437}{{\tt arXiv:astro-ph/0307437
  [astro-ph]}}.

\bibitem{deBlas:2019rxi}
J.~de~Blas et al., {\em {Higgs Boson Studies at Future Particle Colliders}\/},
\href{http://arxiv.org/abs/1905.03764}{{\tt arXiv:1905.03764 [hep-ph]}}.

\bibitem{Cepeda:2019klc}
{Physics of the HL-LHC Working Group Collaboration}, M.~Cepeda et al., {\em
  {Higgs Physics at the HL-LHC and HE-LHC}\/},
\href{http://arxiv.org/abs/1902.00134}{{\tt arXiv:1902.00134 [hep-ph]}}.

\bibitem{Peskin:2012we}
M.~E. Peskin, {\em {Comparison of LHC and ILC Capabilities for Higgs Boson
  Coupling Measurements}\/},
\href{http://arxiv.org/abs/1207.2516}{{\tt arXiv:1207.2516 [hep-ph]}}.

\bibitem{Bodwin:2013gca}
G.~T. Bodwin, F.~Petriello, S.~Stoynev, and M.~Velasco, {\em {Higgs boson
  decays to quarkonia and the $H\bar{c}c$ coupling}\/},
  \href{http://dx.doi.org/10.1103/PhysRevD.88.053003}{Phys. Rev. {\bf D88}
  (2013) no.~5, 053003},
\href{http://arxiv.org/abs/1306.5770}{{\tt arXiv:1306.5770 [hep-ph]}}.

\bibitem{Isidori:2013cla}
G.~Isidori, A.~V. Manohar, and M.~Trott, {\em {Probing the nature of the
  Higgs-like Boson via $h \to V \mathcal{F}$ decays}\/},
  \href{http://dx.doi.org/10.1016/j.physletb.2013.11.054}{Phys. Lett. {\bf
  B728} (2014)  131--135},
\href{http://arxiv.org/abs/1305.0663}{{\tt arXiv:1305.0663 [hep-ph]}}.

\bibitem{Kagan:2014ila}
A.~L. Kagan, G.~Perez, F.~Petriello, Y.~Soreq, S.~Stoynev, and J.~Zupan, {\em
  {Exclusive Window onto Higgs Yukawa Couplings}\/},
  \href{http://dx.doi.org/10.1103/PhysRevLett.114.101802}{Phys. Rev. Lett. {\bf
  114} (2015) no.~10, 101802},
\href{http://arxiv.org/abs/1406.1722}{{\tt arXiv:1406.1722 [hep-ph]}}.

\bibitem{Koenig:2015pha}
M.~K{\"o}nig and M.~Neubert, {\em {Exclusive Radiative Higgs Decays as Probes
  of Light-Quark Yukawa Couplings}\/},
  \href{http://dx.doi.org/10.1007/JHEP08(2015)012}{JHEP {\bf 08} (2015)  012},
\href{http://arxiv.org/abs/1505.03870}{{\tt arXiv:1505.03870 [hep-ph]}}.

\bibitem{Duarte-Campderros:2018ouv}
J.~Duarte-Campderros, G.~Perez, M.~Schlaffer, and A.~Soffer, {\em {Probing the
  strange Higgs coupling at $e^+e^-$ colliders using light-jet flavor
  tagging}\/},
\href{http://arxiv.org/abs/1811.09636}{{\tt arXiv:1811.09636 [hep-ph]}}.

\bibitem{Soreq:2016rae}
Y.~Soreq, H.~X. Zhu, and J.~Zupan, {\em {Light quark Yukawa couplings from
  Higgs kinematics}\/},  \href{http://dx.doi.org/10.1007/JHEP12(2016)045}{JHEP
  {\bf 12} (2016)  045},
\href{http://arxiv.org/abs/1606.09621}{{\tt arXiv:1606.09621 [hep-ph]}}.

\bibitem{Bishara:2016jga}
F.~Bishara, U.~Haisch, P.~F. Monni, and E.~Re, {\em {Constraining Light-Quark
  Yukawa Couplings from Higgs Distributions}\/},
  \href{http://dx.doi.org/10.1103/PhysRevLett.118.121801}{Phys. Rev. Lett. {\bf
  118} (2017) no.~12, 121801},
\href{http://arxiv.org/abs/1606.09253}{{\tt arXiv:1606.09253 [hep-ph]}}.

\bibitem{Aaboud:2017xnb}
{ATLAS Collaboration}, M.~Aaboud et al., {\em {Search for exclusive Higgs and
  $Z$ boson decays to $\phi\gamma$ and $\rho\gamma$ with the ATLAS
  detector}\/},  \href{http://dx.doi.org/10.1007/JHEP07(2018)127}{JHEP {\bf 07}
  (2018)  127},
\href{http://arxiv.org/abs/1712.02758}{{\tt arXiv:1712.02758 [hep-ex]}}.

\bibitem{dEnterria:2017dac}
D.~d'Enterria, {\em {Higgs physics at the Future Circular Collider}\/},
  \href{http://dx.doi.org/10.22323/1.282.0434}{PoS {\bf ICHEP2016} (2017)
  434},
\href{http://arxiv.org/abs/1701.02663}{{\tt arXiv:1701.02663 [hep-ex]}}.

\bibitem{Contino:2016spe}
R.~Contino et al., {\em {Physics at a 100 TeV pp collider: Higgs and EW
  symmetry breaking studies}\/},
  \href{http://dx.doi.org/10.23731/CYRM-2017-003.255}{CERN Yellow Report (2017)
  no.~3, 255--440},
\href{http://arxiv.org/abs/1606.09408}{{\tt arXiv:1606.09408 [hep-ph]}}.

\bibitem{Borgonovi:2642471}
L.~Borgonovi, S.~Braibant, S.~Di~Micco, E.~Fontanesi, P.~Harris, C.~Helsens,
  D.~Jamin, M.~L. Mangano, G.~Ortona, M.~Selvaggi, A.~Sznajder, M.~Testa, and
  M.~Verducci, {\em {Higgs measurements at FCC-hh}\/},   CERN-ACC-2018-0045,
  CERN, Geneva, Oct, 2018.
\newblock \url{https://cds.cern.ch/record/2642471}.

\bibitem{Plehn:2015cta}
M.~L. Mangano, T.~Plehn, P.~Reimitz, T.~Schell, and H.-S. Shao, {\em {Measuring
  the Top Yukawa Coupling at 100 TeV}\/},
  \href{http://dx.doi.org/10.1088/0954-3899/43/3/035001}{J. Phys. {\bf G43}
  (2016) no.~3, 035001},
\href{http://arxiv.org/abs/1507.08169}{{\tt arXiv:1507.08169 [hep-ph]}}.

\bibitem{ALEPH:2005ab}
{The LEP Electroweak Working Group}, {the SLD Electroweak Group}, {the SLD
  Heavy Flavour Group}, and {the ALEPH, DELPHI, L3, OPAL and SLD
  Collaborations}, {\em {Precision electroweak measurements on the Z
  resonance}\/},  \href{http://dx.doi.org/10.1016/j.physrep.2005.12.006}{Phys.
  Rept. {\bf 427} (2006)  257--454},
\href{http://arxiv.org/abs/hep-ex/0509008}{{\tt arXiv:hep-ex/0509008
  [hep-ex]}}.

\bibitem{Azzi:2019yne}
{HL-LHC, HE-LHC Working Group Collaboration}, P.~Azzi et al., {\em {Standard
  Model Physics at the HL-LHC and HE-LHC}\/},
\href{http://arxiv.org/abs/1902.04070}{{\tt arXiv:1902.04070 [hep-ph]}}.

\bibitem{deBlas:2016ojx}
J.~de~Blas, M.~Ciuchini, E.~Franco, S.~Mishima, M.~Pierini, L.~Reina, and
  L.~Silvestrini, {\em {Electroweak precision observables and Higgs-boson
  signal strengths in the Standard Model and beyond: present and future}\/},
  \href{http://dx.doi.org/10.1007/JHEP12(2016)135}{JHEP {\bf 12} (2016)  135},
\href{http://arxiv.org/abs/1608.01509}{{\tt arXiv:1608.01509 [hep-ph]}}.

\bibitem{hepfitsite}
{{\tt HEPfit} Collaboration}, http://hepfit.roma1.infn.it.

\bibitem{Jarlskog:1990kt}
C.~Jarlskog, {\em {Neutrino Counting at the $Z$ Peak and Right-handed
  Neutrinos}\/},
\href{http://dx.doi.org/10.1016/0370-2693(90)91873-A}{Phys. Lett. {\bf B241}
  (1990)  579--583}.

\bibitem{Baak:2014ora}
{Gfitter Group Collaboration}, M.~Baak, J.~C\'{u}th, J.~Haller, R.~Hoecker,
  A.and~Kogler, K.~M\"{o}nig, M.~Schott, and J.~Stelzer, {\em The global
  electroweak fit at NNLO and prospects for the LHC and ILC\/},
  \href{http://dx.doi.org/10.1140/epjc/s10052-014-3046-5}{Eur. Phys. J. {\bf
  C74} (2014)  3046},
\href{http://arxiv.org/abs/1407.3792}{{\tt arXiv:1407.3792 [hep-ph]}}.

\bibitem{dEnterria:2015kmd}
D.~d'Enterria and P.~Z. Skands, eds., {\em {Proceedings, High-Precision
  $\alpha_s$ Measurements from LHC to FCC-ee}}, CERN.
\newblock CERN, Geneva, 2015.
\newblock \href{http://arxiv.org/abs/1512.05194}{{\tt arXiv:1512.05194
  [hep-ph]}}.
\newblock
\url{http://lss.fnal.gov/archive/2015/conf/fermilab-conf-15-610-t.pdf}.
\newblock

\bibitem{Abe:2005nqa}
{SLD Collaboration}, K.~Abe et al., {\em {Measurement of the branching ratio of
  the Z0 into heavy quarks}\/},
  \href{http://dx.doi.org/10.1103/PhysRevD.71.112004}{Phys. Rev. {\bf D71}
  (2005)  112004},
\href{http://arxiv.org/abs/hep-ex/0503005}{{\tt arXiv:hep-ex/0503005
  [hep-ex]}}.

\bibitem{Janot:2015gjr}
P.~Janot, {\em {Direct measurement of $\alpha_{QED}(m_{Z}^{2})$ at the
  FCC-ee}\/},  \href{http://dx.doi.org/10.1007/JHEP02(2016)053,
  10.1007/JHEP11(2017)164}{JHEP {\bf 02} (2016)  053},
  \href{http://arxiv.org/abs/1512.05544}{{\tt arXiv:1512.05544 [hep-ph]}}.
[Erratum: JHEP11,164(2017)].

\bibitem{dEnterria:2016rbf}
D.~d'Enterria and M.~Srebre, {\em {$\alpha_s$ and $\rm V_{cs}$ determination,
  and CKM unitarity test, from W decays at NNLO}\/},
  \href{http://dx.doi.org/10.1016/j.physletb.2016.10.012}{Phys. Lett. {\bf
  B763} (2016)  465--471},
\href{http://arxiv.org/abs/1603.06501}{{\tt arXiv:1603.06501 [hep-ph]}}.

\bibitem{Buchmuller:1985jz}
W.~Buchmuller and D.~Wyler, {\em {Effective Lagrangian Analysis of New
  Interactions and Flavor Conservation}\/},
\href{http://dx.doi.org/10.1016/0550-3213(86)90262-2}{Nucl. Phys. {\bf B268}
  (1986)  621--653}.

\bibitem{Grzadkowski:2010es}
B.~Grzadkowski, M.~Iskrzynski, M.~Misiak, and J.~Rosiek, {\em {Dimension-Six
  Terms in the Standard Model Lagrangian}\/},
  \href{http://dx.doi.org/10.1007/JHEP10(2010)085}{JHEP {\bf 10} (2010)  085},
\href{http://arxiv.org/abs/1008.4884}{{\tt arXiv:1008.4884 [hep-ph]}}.

\bibitem{Deflorian:2016spz}
{LHC Higgs Cross Section Working Group Collaboration}, D.~de~Florian et al.,
  {\em {Handbook of LHC Higgs Cross Sections: 4. Deciphering the Nature of the
  Higgs Sector}\/},
\href{http://arxiv.org/abs/1610.07922}{{\tt arXiv:1610.07922 [hep-ph]}}.

\bibitem{Grojean:2013nya}
C.~Grojean, E.~Salvioni, M.~Schlaffer, and A.~Weiler, {\em {Very boosted Higgs
  in gluon fusion}\/},  \href{http://dx.doi.org/10.1007/JHEP05(2014)022}{JHEP
  {\bf 05} (2014)  022},
\href{http://arxiv.org/abs/1312.3317}{{\tt arXiv:1312.3317 [hep-ph]}}.

\bibitem{Franceschini:2017xkh}
R.~Franceschini, G.~Panico, A.~Pomarol, F.~Riva, and A.~Wulzer, {\em
  {Electroweak Precision Tests in High-Energy Diboson Processes}\/},
  \href{http://dx.doi.org/10.1007/JHEP02(2018)111}{JHEP {\bf 02} (2018)  111},
\href{http://arxiv.org/abs/1712.01310}{{\tt arXiv:1712.01310 [hep-ph]}}.

\bibitem{Azatov:2017kzw}
A.~Azatov, J.~Elias-Miro, Y.~Reyimuaji, and E.~Venturini, {\em {Novel
  measurements of anomalous triple gauge couplings for the LHC}\/},
  \href{http://dx.doi.org/10.1007/JHEP10(2017)027}{JHEP {\bf 10} (2017)  027},
\href{http://arxiv.org/abs/1707.08060}{{\tt arXiv:1707.08060 [hep-ph]}}.

\bibitem{Grojean:2018dqj}
C.~Grojean, M.~Montull, and M.~Riembau, {\em {Diboson at the LHC vs LEP}\/},
  \href{http://dx.doi.org/10.1007/JHEP03(2019)020}{JHEP {\bf 03} (2019)  020},
\href{http://arxiv.org/abs/1810.05149}{{\tt arXiv:1810.05149 [hep-ph]}}.

\bibitem{Banerjee:2018bio}
S.~Banerjee, C.~Englert, R.~S. Gupta, and M.~Spannowsky, {\em {Probing
  Electroweak Precision Physics via boosted Higgs-strahlung at the LHC}\/},
\href{http://arxiv.org/abs/1807.01796}{{\tt arXiv:1807.01796 [hep-ph]}}.

\bibitem{DiLuzio:2018jwd}
L.~Di~Luzio, R.~Gr{\"o}ber, and G.~Panico, {\em {Probing new electroweak states
  via precision measurements at the LHC and future colliders}\/},
  \href{http://dx.doi.org/10.1007/JHEP01(2019)011}{JHEP {\bf 01} (2019)  011},
\href{http://arxiv.org/abs/1810.10993}{{\tt arXiv:1810.10993 [hep-ph]}}.

\bibitem{Lee:2018fxj}
S.~J. Lee, M.~Park, and Z.~Qian, {\em {Probing New Physics by the Tail of the
  Off-shell Higgs in $V_LV_L$ Mode}\/},
\href{http://arxiv.org/abs/1812.02679}{{\tt arXiv:1812.02679 [hep-ph]}}.

\bibitem{Banfi:2018pki}
A.~Banfi, A.~Bond, A.~Martin, and V.~Sanz, {\em {Digging for Top Squarks from
  Higgs data: from signal strengths to differential distributions}\/},
  \href{http://dx.doi.org/10.1007/JHEP11(2018)171}{JHEP {\bf 11} (2018)  171},
\href{http://arxiv.org/abs/1806.05598}{{\tt arXiv:1806.05598 [hep-ph]}}.

\bibitem{Bishara:2016kjn}
F.~Bishara, R.~Contino, and J.~Rojo, {\em {Higgs pair production in
  vector-boson fusion at the LHC and beyond}\/},
  \href{http://dx.doi.org/10.1140/epjc/s10052-017-5037-9}{Eur. Phys. J. {\bf
  C77} (2017) no.~7, 481},
\href{http://arxiv.org/abs/1611.03860}{{\tt arXiv:1611.03860 [hep-ph]}}.

\bibitem{Contino:2006qr}
R.~Contino, L.~Da~Rold, and A.~Pomarol, {\em {Light custodians in natural
  composite Higgs models}\/},
  \href{http://dx.doi.org/10.1103/PhysRevD.75.055014}{Phys. Rev. {\bf D75}
  (2007)  055014},
\href{http://arxiv.org/abs/hep-ph/0612048}{{\tt arXiv:hep-ph/0612048
  [hep-ph]}}.

\bibitem{Farina:2016rws}
M.~Farina, G.~Panico, D.~Pappadopulo, J.~T. Ruderman, R.~Torre, and A.~Wulzer,
  {\em {Energy helps accuracy: electroweak precision tests at hadron
  colliders}\/},  \href{http://dx.doi.org/10.1016/j.physletb.2017.06.043}{Phys.
  Lett. {\bf B772} (2017)  210--215},
\href{http://arxiv.org/abs/1609.08157}{{\tt arXiv:1609.08157 [hep-ph]}}.

\bibitem{Barbieri:2004qk}
R.~Barbieri, A.~Pomarol, R.~Rattazzi, and A.~Strumia, {\em {Electroweak
  symmetry breaking after LEP-1 and LEP-2}\/},
  \href{http://dx.doi.org/10.1016/j.nuclphysb.2004.10.014}{Nucl. Phys. {\bf
  B703} (2004)  127--146},
\href{http://arxiv.org/abs/hep-ph/0405040}{{\tt arXiv:hep-ph/0405040
  [hep-ph]}}.

\bibitem{Alves:2014cda}
D.~S.~M. Alves, J.~Galloway, J.~T. Ruderman, and J.~R. Walsh, {\em {Running
  Electroweak Couplings as a Probe of New Physics}\/},
  \href{http://dx.doi.org/10.1007/JHEP02(2015)007}{JHEP {\bf 02} (2015)  007},
\href{http://arxiv.org/abs/1410.6810}{{\tt arXiv:1410.6810 [hep-ph]}}.

\bibitem{Chigusa:2018vxz}
S.~Chigusa, Y.~Ema, and T.~Moroi, {\em {Probing electroweakly interacting
  massive particles with Drell-Yan process at 100 TeV hadron colliders}\/},
  \href{http://dx.doi.org/10.1016/j.physletb.2018.12.011}{Phys. Lett. {\bf
  B789} (2019)  106--113},
\href{http://arxiv.org/abs/1810.07349}{{\tt arXiv:1810.07349 [hep-ph]}}.

\bibitem{Azatov:2015oxa}
A.~Azatov, R.~Contino, G.~Panico, and M.~Son, {\em {Effective field theory
  analysis of double Higgs boson production via gluon fusion}\/},
  \href{http://dx.doi.org/10.1103/PhysRevD.92.035001}{Phys. Rev. {\bf D92}
  (2015) no.~3, 035001},
\href{http://arxiv.org/abs/1502.00539}{{\tt arXiv:1502.00539 [hep-ph]}}.

\bibitem{vanderBij:1985ww}
J.~J. van~der Bij, {\em {Does Low-energy Physics Depend on the Potential of a
  Heavy Higgs Particle?}\/},
\href{http://dx.doi.org/10.1016/0550-3213(86)90131-8}{Nucl. Phys. {\bf B267}
  (1986)  557--565}.

\bibitem{McCullough:2013rea}
M.~McCullough, {\em {An Indirect Model-Dependent Probe of the Higgs
  Self-Coupling}\/},  \href{http://dx.doi.org/10.1103/PhysRevD.90.015001,
  10.1103/PhysRevD.92.039903}{Phys. Rev. {\bf D90} (2014) no.~1, 015001},
  \href{http://arxiv.org/abs/1312.3322}{{\tt arXiv:1312.3322 [hep-ph]}}.
[Erratum: Phys. Rev.D92,no.3,039903(2015)].

\bibitem{DiVita:2017vrr}
S.~Di~Vita, G.~Durieux, C.~Grojean, J.~Gu, Z.~Liu, G.~Panico, M.~Riembau, and
  T.~Vantalon, {\em {A global view on the Higgs self-coupling at lepton
  colliders}\/},  \href{http://dx.doi.org/10.1007/JHEP02(2018)178}{JHEP {\bf
  02} (2018)  178},
\href{http://arxiv.org/abs/1711.03978}{{\tt arXiv:1711.03978 [hep-ph]}}.

\bibitem{hhxswg}
{\em {\tt LHC Higgs cross section working group, HH sub-group}\/},
  \url{https://twiki.cern.ch/twiki/bin/view/LHCPhysics/LHCHXSWGHH}.

\bibitem{Borowka:2016ypz}
S.~Borowka, N.~Greiner, G.~Heinrich, S.~P. Jones, M.~Kerner, J.~Schlenk, and
  T.~Zirke, {\em {Full top quark mass dependence in Higgs boson pair production
  at NLO}\/},  \href{http://dx.doi.org/10.1007/JHEP10(2016)107}{JHEP {\bf 10}
  (2016)  107},
\href{http://arxiv.org/abs/1608.04798}{{\tt arXiv:1608.04798 [hep-ph]}}.

\bibitem{Grazzini:2018bsd}
M.~Grazzini, G.~Heinrich, S.~Jones, S.~Kallweit, M.~Kerner, J.~M. Lindert, and
  J.~Mazzitelli, {\em {Higgs boson pair production at NNLO with top quark mass
  effects}\/},  \href{http://dx.doi.org/10.1007/JHEP05(2018)059}{JHEP {\bf 05}
  (2018)  059},
\href{http://arxiv.org/abs/1803.02463}{{\tt arXiv:1803.02463 [hep-ph]}}.

\bibitem{Goncalves:2018qas}
D.~Gon\c{c}alves, T.~Han, F.~Kling, T.~Plehn, and M.~Takeuchi, {\em {Higgs
  boson pair production at future hadron colliders: From kinematics to
  dynamics}\/},  \href{http://dx.doi.org/10.1103/PhysRevD.97.113004}{Phys. Rev.
  {\bf D97} (2018) no.~11, 113004},
\href{http://arxiv.org/abs/1802.04319}{{\tt arXiv:1802.04319 [hep-ph]}}.

\bibitem{Banerjee:2018yxy}
S.~Banerjee, C.~Englert, M.~L. Mangano, M.~Selvaggi, and M.~Spannowsky, {\em
  {$hh$+jet production at 100 TeV}\/},
  \href{http://dx.doi.org/10.1140/epjc/s10052-018-5788-y}{Eur. Phys. J. {\bf
  C78} (2018) no.~4, 322},
\href{http://arxiv.org/abs/1802.01607}{{\tt arXiv:1802.01607 [hep-ph]}}.

\bibitem{Sakharov:1967dj}
A.~D. Sakharov, {\em {Violation of CP Invariance, c Asymmetry, and Baryon
  Asymmetry of the Universe}\/},
  \href{http://dx.doi.org/10.1070/PU1991v034n05ABEH002497}{Pisma Zh. Eksp.
  Teor. Fiz. {\bf 5} (1967)  32--35}.
[Usp. Fiz. Nauk161,61(1991)].

\bibitem{Cohen:1993nk}
A.~G. Cohen, D.~B. Kaplan, and A.~E. Nelson, {\em {Progress in electroweak
  baryogenesis}\/},
  \href{http://dx.doi.org/10.1146/annurev.ns.43.120193.000331}{Ann. Rev. Nucl.
  Part. Sci. {\bf 43} (1993)  27--70},
\href{http://arxiv.org/abs/hep-ph/9302210}{{\tt arXiv:hep-ph/9302210
  [hep-ph]}}.

\bibitem{Morrissey:2012db}
D.~E. Morrissey and M.~J. Ramsey-Musolf, {\em {Electroweak baryogenesis}\/},
  \href{http://dx.doi.org/10.1088/1367-2630/14/12/125003}{New J. Phys. {\bf 14}
  (2012)  125003},
\href{http://arxiv.org/abs/1206.2942}{{\tt arXiv:1206.2942 [hep-ph]}}.

\bibitem{Kuzmin:1985mm}
V.~A. Kuzmin, V.~A. Rubakov, and M.~E. Shaposhnikov, {\em {On the Anomalous
  Electroweak Baryon Number Nonconservation in the Early Universe}\/},
\href{http://dx.doi.org/10.1016/0370-2693(85)91028-7}{Phys. Lett. {\bf 155B}
  (1985)  36}.

\bibitem{Gavela:1993ts}
M.~B. Gavela, P.~Hernandez, J.~Orloff, and O.~Pene, {\em {Standard model CP
  violation and baryon asymmetry}\/},
  \href{http://dx.doi.org/10.1142/S0217732394000629}{Mod. Phys. Lett. {\bf A9}
  (1994)  795--810},
\href{http://arxiv.org/abs/hep-ph/9312215}{{\tt arXiv:hep-ph/9312215
  [hep-ph]}}.

\bibitem{Kajantie:1995kf}
K.~Kajantie, M.~Laine, K.~Rummukainen, and M.~E. Shaposhnikov, {\em {The
  Electroweak phase transition: A Nonperturbative analysis}\/},
  \href{http://dx.doi.org/10.1016/0550-3213(96)00052-1}{Nucl. Phys. {\bf B466}
  (1996)  189--258},
\href{http://arxiv.org/abs/hep-lat/9510020}{{\tt arXiv:hep-lat/9510020
  [hep-lat]}}.

\bibitem{Servant:2018xcs}
G.~Servant, {\em {The serendipity of electroweak baryogenesis}\/},
  \href{http://dx.doi.org/10.1098/rsta.2017.0124}{Phil. Trans. Roy. Soc. Lond.
  {\bf A376} (2018) no.~2114, 20170124},
\href{http://arxiv.org/abs/1807.11507}{{\tt arXiv:1807.11507 [hep-ph]}}.

\bibitem{Curtin:2014jma}
D.~Curtin, P.~Meade, and C.-T. Yu, {\em {Testing Electroweak Baryogenesis with
  Future Colliders}\/},  \href{http://dx.doi.org/10.1007/JHEP11(2014)127}{JHEP
  {\bf 11} (2014)  127},
\href{http://arxiv.org/abs/1409.0005}{{\tt arXiv:1409.0005 [hep-ph]}}.

\bibitem{Kotwal:2016tex}
A.~V. Kotwal, M.~J. Ramsey-Musolf, J.~M. No, and P.~Winslow, {\em
  {Singlet-catalyzed electroweak phase transitions in the 100 TeV frontier}\/},
   \href{http://dx.doi.org/10.1103/PhysRevD.94.035022}{Phys. Rev. {\bf D94}
  (2016) no.~3, 035022},
\href{http://arxiv.org/abs/1605.06123}{{\tt arXiv:1605.06123 [hep-ph]}}.

\bibitem{Huang:2016cjm}
P.~Huang, A.~J. Long, and L.-T. Wang, {\em {Probing the Electroweak Phase
  Transition with Higgs Factories and Gravitational Waves}\/},
  \href{http://dx.doi.org/10.1103/PhysRevD.94.075008}{Phys. Rev. {\bf D94}
  (2016) no.~7, 075008},
\href{http://arxiv.org/abs/1608.06619}{{\tt arXiv:1608.06619 [hep-ph]}}.

\bibitem{Hogan:1986qda}
C.~J. Hogan, {\em {Gravitational radiation from cosmological phase
  transitions}\/},
Mon. Not. Roy. Astron. Soc. {\bf 218} (1986)  629--636.

\bibitem{Kamionkowski:1993fg}
M.~Kamionkowski, A.~Kosowsky, and M.~S. Turner, {\em {Gravitational radiation
  from first order phase transitions}\/},
  \href{http://dx.doi.org/10.1103/PhysRevD.49.2837}{Phys. Rev. {\bf D49} (1994)
   2837--2851},
\href{http://arxiv.org/abs/astro-ph/9310044}{{\tt arXiv:astro-ph/9310044
  [astro-ph]}}.

\bibitem{Caprini:2015zlo}
C.~Caprini et al., {\em {Science with the space-based interferometer eLISA. II:
  Gravitational waves from cosmological phase transitions}\/},
  \href{http://dx.doi.org/10.1088/1475-7516/2016/04/001}{JCAP {\bf 1604} (2016)
  no.~04, 001},
\href{http://arxiv.org/abs/1512.06239}{{\tt arXiv:1512.06239 [astro-ph.CO]}}.

\bibitem{Golling:2016gvc}
T.~Golling et al., {\em {Physics at a 100 TeV pp collider: beyond the Standard
  Model phenomena}\/},
  \href{http://dx.doi.org/10.23731/CYRM-2017-003.441}{CERN Yellow Report (2017)
  no.~3, 441--634},
\href{http://arxiv.org/abs/1606.00947}{{\tt arXiv:1606.00947 [hep-ph]}}.

\bibitem{colliderreach}
G.~Salam and A.~Weiler, {\em {\tt Collider Reach}\/},
  \url{http://collider-reach.web.cern.ch/collider-reach/}.

\bibitem{Jamin:2019mqx}
C.~Helsens, D.~Jamin, M.~L. Mangano, T.~G. Rizzo, and M.~Selvaggi, {\em {Heavy
  resonances at energy-frontier hadron colliders}\/},
\href{http://arxiv.org/abs/1902.11217}{{\tt arXiv:1902.11217 [hep-ph]}}.

\bibitem{Rizzo:2014xma}
T.~G. Rizzo, {\em {Exploring new gauge bosons at a 100 TeV collider}\/},
  \href{http://dx.doi.org/10.1103/PhysRevD.89.095022}{Phys. Rev. {\bf D89}
  (2014) no.~9, 095022},
\href{http://arxiv.org/abs/1403.5465}{{\tt arXiv:1403.5465 [hep-ph]}}.

\bibitem{Han:2013mra}
T.~Han, P.~Langacker, Z.~Liu, and L.-T. Wang, {\em {Diagnosis of a New Neutral
  Gauge Boson at the LHC and ILC for Snowmass 2013}\/},
\href{http://arxiv.org/abs/1308.2738}{{\tt arXiv:1308.2738 [hep-ph]}}.

\bibitem{Randall:1999ee}
L.~Randall and R.~Sundrum, {\em {A Large mass hierarchy from a small extra
  dimension}\/},  \href{http://dx.doi.org/10.1103/PhysRevLett.83.3370}{Phys.
  Rev. Lett. {\bf 83} (1999)  3370--3373},
\href{http://arxiv.org/abs/hep-ph/9905221}{{\tt arXiv:hep-ph/9905221
  [hep-ph]}}.

\bibitem{Baur:1987ga}
U.~Baur, I.~Hinchliffe, and D.~Zeppenfeld, {\em {Excited Quark Production at
  Hadron Colliders}\/},
\href{http://dx.doi.org/10.1142/S0217751X87000661}{Int. J. Mod. Phys. {\bf A2}
  (1987)  1285}.

\bibitem{Baur:1989kv}
U.~Baur, M.~Spira, and P.~M. Zerwas, {\em {Excited Quark and Lepton Production
  at Hadron Colliders}\/},
\href{http://dx.doi.org/10.1103/PhysRevD.42.815}{Phys. Rev. {\bf D42} (1990)
  815--824}.

\bibitem{Hill:1994hp}
C.~T. Hill, {\em {Topcolor assisted technicolor}\/},
  \href{http://dx.doi.org/10.1016/0370-2693(94)01660-5}{Phys. Lett. {\bf B345}
  (1995)  483--489},
\href{http://arxiv.org/abs/hep-ph/9411426}{{\tt arXiv:hep-ph/9411426
  [hep-ph]}}.

\bibitem{Gouskos:2642475}
L.~Gouskos, A.~Sung, and J.~Incandela, {\em {Search for stop scalar quarks at
  FCC-hh}\/},   CERN-ACC-2019-0036, CERN, Geneva, Oct, 2018.
\newblock \url{https://cds.cern.ch/record/2642475}.

\bibitem{Aaboud:2018zhk}
{ATLAS Collaboration}, M.~Aaboud et al., {\em {Observation of $H \rightarrow
  b\bar{b}$ decays and $VH$ production with the ATLAS detector}\/},
  \href{http://dx.doi.org/10.1016/j.physletb.2018.09.013}{Phys. Lett. {\bf
  B786} (2018)  59--86},
\href{http://arxiv.org/abs/1808.08238}{{\tt arXiv:1808.08238 [hep-ex]}}.

\bibitem{Sirunyan:2018kst}
{CMS Collaboration}, A.~M. Sirunyan et al., {\em {Observation of Higgs boson
  decay to bottom quarks}\/},
  \href{http://dx.doi.org/10.1103/PhysRevLett.121.121801}{Phys. Rev. Lett. {\bf
  121} (2018) no.~12, 121801},
\href{http://arxiv.org/abs/1808.08242}{{\tt arXiv:1808.08242 [hep-ex]}}.

\bibitem{Abercrombie:2015wmb}
D.~Abercrombie et al., {\em {Dark Matter Benchmark Models for Early LHC Run-2
  Searches: Report of the ATLAS/CMS Dark Matter Forum}\/},
\href{http://arxiv.org/abs/1507.00966}{{\tt arXiv:1507.00966 [hep-ex]}}.

\bibitem{Sirunyan:2017dnz}
{CMS Collaboration}, A.~M. Sirunyan et al., {\em {Search for Low Mass Vector
  Resonances Decaying to Quark-Antiquark Pairs in Proton-Proton Collisions at
  $\sqrt{s}=13$~TeV}\/},
  \href{http://dx.doi.org/10.1103/PhysRevLett.119.111802}{Phys. Rev. Lett. {\bf
  119} (2017) no.~11, 111802},
\href{http://arxiv.org/abs/1705.10532}{{\tt arXiv:1705.10532 [hep-ex]}}.

\bibitem{Aaboud:2018zba}
{ATLAS Collaboration}, M.~Aaboud et al., {\em {Search for light resonances
  decaying to boosted quark pairs and produced in association with a photon or
  a jet in proton-proton collisions at $\sqrt{s}=13$ TeV with the ATLAS
  detector}\/},  \href{http://dx.doi.org/10.1016/j.physletb.2018.09.062}{Phys.
  Lett. {\bf B788} (2019)  316--335},
\href{http://arxiv.org/abs/1801.08769}{{\tt arXiv:1801.08769 [hep-ex]}}.

\bibitem{ATLAS:2018hzj}
{ATLAS Collaboration}, T.~A. collaboration,
{\em {Search for boosted resonances decaying to two b-quarks and produced in
  association with a jet at $\sqrt{s}=13$ TeV with the ATLAS detector}\/}, .

\bibitem{Lin:2019uvt}
T.~Lin, {\em {TASI lectures on dark matter models and direct detection}\/},
\href{http://arxiv.org/abs/1904.07915}{{\tt arXiv:1904.07915 [hep-ph]}}.

\bibitem{Hooper:2018kfv}
D.~Hooper, {\em {TASI Lectures on Indirect Searches For Dark Matter}\/},
\href{http://arxiv.org/abs/1812.02029}{{\tt arXiv:1812.02029 [hep-ph]}}.

\bibitem{Hui:2016ltb}
L.~Hui, J.~P. Ostriker, S.~Tremaine, and E.~Witten, {\em {Ultralight scalars as
  cosmological dark matter}\/},
  \href{http://dx.doi.org/10.1103/PhysRevD.95.043541}{Phys. Rev. {\bf D95}
  (2017) no.~4, 043541},
\href{http://arxiv.org/abs/1610.08297}{{\tt arXiv:1610.08297 [astro-ph.CO]}}.

\bibitem{Ali-Haimoud:2019khd}
A.~Kashlinsky et al., {\em {Electromagnetic probes of primordial black holes as
  dark matter}\/},
\href{http://arxiv.org/abs/1903.04424}{{\tt arXiv:1903.04424 [astro-ph.CO]}}.

\bibitem{Hook:2018dlk}
A.~Hook, {\em {TASI Lectures on the Strong CP Problem and Axions}\/},
\href{http://arxiv.org/abs/1812.02669}{{\tt arXiv:1812.02669 [hep-ph]}}.

\bibitem{tasi6}
{\em {{\em J. Conrad, 2018 TASI lectures on} "Neutrinos"}\/},
  \url{https://sites.google.com/a/colorado.edu/tasi-2018-wiki/}.

\bibitem{Kolb:1990vq}
E.~W. Kolb and M.~S. Turner, {\em {The Early Universe}\/},
Front. Phys. {\bf 69} (1990)  1--547.

\bibitem{Griest:1989wd}
K.~Griest and M.~Kamionkowski, {\em {Unitarity Limits on the Mass and Radius of
  Dark Matter Particles}\/},
\href{http://dx.doi.org/10.1103/PhysRevLett.64.615}{Phys. Rev. Lett. {\bf 64}
  (1990)  615}.

\bibitem{Blum:2014dca}
K.~Blum, Y.~Cui, and M.~Kamionkowski, {\em {An Ultimate Target for Dark Matter
  Searches}\/},  \href{http://dx.doi.org/10.1103/PhysRevD.92.023528}{Phys. Rev.
  {\bf D92} (2015) no.~2, 023528},
\href{http://arxiv.org/abs/1412.3463}{{\tt arXiv:1412.3463 [hep-ph]}}.

\bibitem{Mahbubani:2017gjh}
R.~Mahbubani, P.~Schwaller, and J.~Zurita, {\em {Closing the window for
  compressed Dark Sectors with disappearing charged tracks}\/},
  \href{http://dx.doi.org/10.1007/JHEP06(2017)119,
  10.1007/JHEP10(2017)061}{JHEP {\bf 06} (2017)  119},
  \href{http://arxiv.org/abs/1703.05327}{{\tt arXiv:1703.05327 [hep-ph]}}.
[Erratum: JHEP10,061(2017)].

\bibitem{Saito:2019rtg}
M.~Saito, R.~Sawada, K.~Terashi, and S.~Asai, {\em {Discovery reach for wino
  and higgsino dark matter with a disappearing track signature at a 100 TeV
  $pp$ collider}\/},
\href{http://arxiv.org/abs/1901.02987}{{\tt arXiv:1901.02987 [hep-ph]}}.

\bibitem{Curtin:2013fra}
D.~Curtin et al., {\em {Exotic decays of the 125 GeV Higgs boson}\/},
  \href{http://dx.doi.org/10.1103/PhysRevD.90.075004}{Phys. Rev. {\bf D90}
  (2014) no.~7, 075004},
\href{http://arxiv.org/abs/1312.4992}{{\tt arXiv:1312.4992 [hep-ph]}}.

\bibitem{Liu:2016zki}
Z.~Liu, L.-T. Wang, and H.~Zhang, {\em {Exotic decays of the 125 GeV Higgs
  boson at future $e^+e^-$ lepton colliders}\/},
  \href{http://dx.doi.org/10.1088/1674-1137/41/6/063102}{Chin. Phys. {\bf C41}
  (2017) no.~6, 063102},
\href{http://arxiv.org/abs/1612.09284}{{\tt arXiv:1612.09284 [hep-ph]}}.

\bibitem{Liu:2017zdh}
J.~Liu, L.-T. Wang, X.-P. Wang, and W.~Xue, {\em {Exposing Dark Sector with
  Future Z-Factories}\/},
  \href{http://dx.doi.org/10.1103/PhysRevD.97.095044}{Phys. Rev. {\bf D97}
  (2018) no.~9, 095044},
\href{http://arxiv.org/abs/1712.07237}{{\tt arXiv:1712.07237 [hep-ph]}}.

\bibitem{Curtin:2015fna}
D.~Curtin and C.~B. Verhaaren, {\em {Discovering Uncolored Naturalness in
  Exotic Higgs Decays}\/},
  \href{http://dx.doi.org/10.1007/JHEP12(2015)072}{JHEP {\bf 12} (2015)  072},
\href{http://arxiv.org/abs/1506.06141}{{\tt arXiv:1506.06141 [hep-ph]}}.

\bibitem{Curtin:2017bxr}
D.~Curtin, K.~Deshpande, O.~Fischer, and J.~Zurita, {\em {New Physics
  Opportunities for Long-Lived Particles at Electron-Proton Colliders}\/},
  \href{http://dx.doi.org/10.1007/JHEP07(2018)024}{JHEP {\bf 07} (2018)  024},
\href{http://arxiv.org/abs/1712.07135}{{\tt arXiv:1712.07135 [hep-ph]}}.

\bibitem{Mangano:2016jyj}
M.~L. Mangano et al., {\em {Physics at a 100 TeV pp Collider: Standard Model
  Processes}\/},  \href{http://dx.doi.org/10.23731/CYRM-2017-003.1}{CERN Yellow
  Report (2017) no.~3, 1--254},
\href{http://arxiv.org/abs/1607.01831}{{\tt arXiv:1607.01831 [hep-ph]}}.

\bibitem{Hinchliffe:2015qma}
I.~Hinchliffe, A.~Kotwal, M.~L. Mangano, C.~Quigg, and L.-T. Wang, {\em
  {Luminosity goals for a 100-TeV pp collider}\/},
  \href{http://dx.doi.org/10.1142/S0217751X15440029}{Int. J. Mod. Phys. {\bf
  A30} (2015) no.~23, 1544002},
\href{http://arxiv.org/abs/1504.06108}{{\tt arXiv:1504.06108 [hep-ph]}}.

\bibitem{Schaumann:2015fsa}
M.~Schaumann, {\em {Potential performance for Pb-Pb, p-Pb and p-p collisions in
  a future circular collider}\/},
  \href{http://dx.doi.org/10.1103/PhysRevSTAB.18.091002}{Phys. Rev. ST Accel.
  Beams {\bf 18} (2015) no.~9, 091002},
\href{http://arxiv.org/abs/1503.09107}{{\tt arXiv:1503.09107
  [physics.acc-ph]}}.

\bibitem{Dainese:2016gch}
A.~Dainese et al., {\em {Heavy ions at the Future Circular Collider}\/},
  \href{http://dx.doi.org/10.23731/CYRM-2017-003.635}{CERN Yellow Report (2017)
  no.~3, 635--692},
\href{http://arxiv.org/abs/1605.01389}{{\tt arXiv:1605.01389 [hep-ph]}}.

\bibitem{Jowett2017}
J.~Jowett et al., {\em {The 2016 proton-nucleus run of the LHC}\/},
\newblock Proceedings of IPAC2017, Copenhagen, Denmark (2017), TUPVA014  ,
  \url{https://doi.org/10.18429/JACoW-IPAC2017-TUPVA014}.

\end{thebibliography}\endgroup

\end{document}